\pdfoutput=1
\documentclass[usenatbib]{mn2e}
\usepackage[letterpaper,totalwidth=480pt,totalheight=680pt]{geometry}
\usepackage{amsmath,amssymb,afterpage} \usepackage{ifthen} \usepackage{comment} %\citestyle{aa}
\newboolean{usepdf}
\usepackage{ifpdf} \ifpdf \pdfoutput=1 \pdftrue \fi

\newboolean{includeversions}

\setboolean{includeversions}{false}

\newboolean{blackandwhite} %\setboolean{blackandwhite}{true}

\ifpdf \setboolean{usepdf}{true} \else \setboolean{usepdf}{false} \fi

\ifthenelse {\boolean{usepdf}}{ \usepackage[pdftex,final]{graphicx} \usepackage[pdftex,draft]{hyperref} }{ \usepackage[dvips,final]{graphicx} \usepackage[dvips]{hyperref} } \graphicspath {{figures/}{.}} % parenthetical group commands which sometimes end up capitalized by 
\newcommand{\kps}{\, {\rm km}/{\rm s}}

\newcommand{\cm}{\, {\rm cm}}

\newcommand{\um}{\, {\mu \rm m}}

\newcommand{\Myr}{\,\mathrm{Myr}}
\newcommand{\Gyr}{\,\mathrm{Gyr}}

\newcommand{\kpc}{\, {\rm kpc}}

\newcommand{\au}{\, {\rm AU}}

\newcommand{\Angstrom}{\, {\rm \AA}}
\newcommand{\degrees}{^{\circ}}

\newcommand{\Msun}{\,\mathrm{M}_{\odot}}
\newcommand{\sfrunit}{\,\mathrm{M}_{\odot}/\mathrm{yr}}

\newcommand{\pc}{\,\mathrm{pc}}
\providecommand{\halpha}{{\mathrm{H}\alpha}}
\providecommand{\hbeta}{{\mathrm{H}\beta}}
\renewcommand{\halpha}{{\mathrm{H}\alpha}}
\renewcommand{\hbeta}{{\mathrm{H}\beta}}

\providecommand{\oii}{[O\mbox{\sc ii}]}
\newcommand{\mcrx}{\textit{Sunrise}}

\providecommand{\la}{\lesssim}

\providecommand{\hii}{\mbox{H\,{\sc ii}}}
 \usepackage {astrojournals} % MN requirements
\renewcommand{\um}{\, {\umu \rm m}} \renewcommand{\mcrx}{{\sc sunrise}} \newcommand{\mapiii}{{\sc mappingsiii}}

\title [High-Resolution Panchromatic Spectral Models of Galaxies]{High-Resolution Panchromatic Spectral Models of Galaxies including Photoionisation and Dust}

\author[Jonsson, Groves \& Cox] { Patrik Jonsson,$^1$\thanks{E-mail address: patrik@ucolick.org} Brent~A. Groves,$^2$ %\thanks{E-mail address: brent@strw.leidenuniv.nl}
\& T.~J. Cox$^3$ %\thanks{E-mail address: tcox@cfa.harvard.edu}
\\ $^1$Santa Cruz Institute for Particle Physics, University of California, Santa Cruz, CA 95064, USA\\ $^2$Sterrewacht Leiden, Leiden University, Niels Bohrweg 2, 2333CA, Leiden, The Netherlands\\ $^3$Harvard-Smithsonian Center for Astrophysics, 60 Garden St., Cambridge, MA 02138, USA }

\begin {document}

\date{\ifthenelse {\boolean{includeversions}}{ Draft Version: \$Id: newmodels.txt 1045 2009-06-11 16:00:29Z patrik $ $ }{} } \pagerange{\pageref{firstpage}--\pageref{lastpage}} \pubyear{2009}

\maketitle

\label{firstpage}

\begin{abstract}
%NLX% end exclude from vocabulary builder
An updated version of the dust radiation transfer code \mcrx ,
including models for star-forming regions and a self-consistent
calculation of the spatially dependent dust and PAH emission, is
presented. Given a hydrodynamic simulation of a galaxy, this model can
calculate a realistic 2-dimensional ultraviolet--submillimeter spectral
energy distribution of the galaxy, including emission lines from \hii\
regions, from any viewpoint. To model the emission from star-forming
regions, the \mapiii \ photoionization code is used.  The high
wavelength resolution ($\sim 1000$ wavelengths) is made possible by the
polychromatic Monte-Carlo algorithm employed by \mcrx . From the 2-D
spectral energy distributions, images in any filter bands or integrated
galaxy SEDs can be created. Using a suite of hydrodynamic simulations
of disc galaxies, the output broad-band images and spectral energy
distributions are compared with observed galaxies from the
multiwavelength SINGS and SLUGS galaxy surveys. Overall, the output
spectral energy distributions show a good match with observed galaxies
in colours ranging from GALEX far-UV to SCUBA submillimeter
wavelengths.  The only possible exception is the $160 \um / 850 \um$
colour, which the simulations underestimate by a factor $\sim 5$
compared to the SINGS sample. However, the simulations here agree with
the SLUGS galaxies, which consistently have significantly larger
amounts of cold dust than the SINGS galaxies. The \mcrx \ model can be
used to generate simulated observations of arbitrary hydrodynamic
galaxy simulations. In this way, predictions of galaxy formation
theories can be directly tested against observations of galaxies.
%NLX% exclude from vocabulary builder
\end{abstract}


%NLX% end exclude from vocabulary builder

\section{Introduction}

The full spectral energy distribution of a galaxy contains information
on all the constituents of the galaxy; stars (both old and currently
forming), the hot and cold gas heated by stellar light or collisions,
the thermally emitting dust associated with the gas, and possibly even
emission from an active galactic nucleus. A model that wants to
describe and disentangle the panchromatic spectral energy distribution
(SED) of galaxies must necessarily include all these sources of
emission.

The necessity of such multi-component models has become more pressing
with the recent increase of multi-wavelength, large-area galaxy
surveys, including those with truly panchromatic wavelength coverage
such as the AEGIS survey \citep{davisetal07aegis}. These surveys are
complemented by the growing number of 2D spatially resolved spectra of
galaxies from integral field spectrographs. These data can give crucial
clues into the dynamic state and resolved star-formation history of
galaxies, but interpreting these high-dimensional data sets is not
straightforward.

Existing models for calculating galaxy SEDs have mostly been focused on
two approaches: Either calculating the broad continuum shapes of galaxy
SEDs based on an assumed global dust distribution
\citep[e.g.][]{silvaetal98, ferraraetal99, charlotfall00,
bianchietal00, tuffsetal04, pierinietal04, dacunhaetal08,
bianchi08trading}, or focusing on more detailed modelling of dusty
star-forming regions assuming that the SED of a galaxy is dominated by
its star-forming regions \citep[e.g.][]{wittgordon96, wittgordon00,
gordonetal01, misseltetal01,grovesetal08sparam}.

Neither of these approaches will suffice if the goal is to model the
detailed, spatially- and spectrally-resolved appearance of simulated
galaxies as they evolve in time. This is especially true if the
observational indicators that are used to determine star-formation rate
and star-formation history, such as emission and absorption lines, are
included. A model with such a goal must take into account the
differential obscuration of various stellar populations, the effect of
extra attenuation of light originating inside molecular cloud
complexes, and the global distribution of stars and dust. It must also
connect these effects with the global dynamical state of the galaxy.

Complementary to, and enhancing, the galaxy surveys, hydrodynamic
simulations of galaxies have furthered our understanding of the
dynamical and spatial state of galaxies.  These simulations have
advanced to the point that realistic spiral galaxies are formed
self-consistently in cosmological contexts both using smooth particle
hydrodynamics \citep{governatoetal08} and adaptive-mesh refinement
codes \citep{ceverinoklypin09}. Adaptive-mesh code galaxy simulations
have recently attained resolutions of a few parsec, beginning to
explicitly resolve the multiphase structure of the ISM
\citep{ceverinoklypin09, kimetal09}.  However, even with realistic
simulated galaxies, a principal obstacle to testing theory against
observations is that simulations trace mass, while observations trace
light.  To draw full benefit from these new multi-wavelength surveys,
theoretical models that predict the spectral energy distributions of
simulated galaxies are necessary.

Galaxy-scale radiative transfer calculations, using the geometry of
stars and dust from the simulations, provide a way to link both
observations and simulations and create SEDs that include the effects
of geometry, differential extinction, and galaxy dynamics.  It is only
in recent times that the full SED of simulated galaxies have been
studied, using Monte Carlo Models to calculate the spectrum
\citep{pjthesis-nourl, pjetal05attn, jonsson06sunrise,
chakrabartietal07radishe, lietal07, chakrabartietal08}, yet to date
they have not taken into account the structures on the scales of
star-forming regions nor included emission lines.  The model presented
here remedies these deficiencies by combining galaxy-scale
radiation-transfer calculations done with the \mcrx \ code
\citep{jonsson06sunrise} with the treatment of star-forming regions
using the dust- and photoionization code \mapiii \
\citep{grovesetal08sparam}. As a result, the model can approximate
radiation-transfer effects that occur on scales below the resolution in
the simulated galaxies while still taking into account the
star-formation history and large-scale geometry of the galaxy.

The outline of this paper is as follows. First the model is described,
including a brief summary of the simulations to which the model is
applied for validation and observational comparison. We then validate
the model and follow by a demonstration of the images and spectra
produced by the model, with a comparison to those of observed galaxies.
We conclude with a discussion of the strengths and weaknesses of the
present model, along with planned future extensions and improvements.

\section{Model}
 \label{section-model}

In principle, the construction of a model to calculate the emerging SED
of a galaxy is straightforward: simply solve the physics of scattering
and absorption of starlight by the dust grains (and by gas) in the
galaxy. In practice, such an approach is not feasible for several
reasons: First, current galaxy simulations have a finite spatial
resolution, at most of order $10 \pc$, or in the majority of cases
$\sim 100 \pc$, and a finite mass resolution of, at best, $10^3 \Msun$.
Such resolution is insufficient to resolve the detailed structures of
gas and stars in star-forming regions. Second, even if the galaxy
simulations had sufficient resolution, accurate radiation-transfer
calculations generally require significant amounts of computer time and
would likely be prohibitively expensive except for single cases
(i.e.~single snapshot of one galaxy). Third, on the scales of
star-forming regions radiation affects the dynamics of the gas so
simulations would need to treat the radiation transfer in conjunction
with the gas dynamics, making such simulations even more prohibitively
expensive.

The solution adopted in this work is to include the relevant processes
on scales below the resolution by using a sub-resolution approximation.
This is similar to how star formation and supernova feedback are
treated within the hydrodynamic simulations, and is an effective way of
lowering the computational expense of the simulations (provided of
course that such approximations are reasonable.)

In our model, the transfer of radiation on galaxy scales is handled by
the Monte Carlo radiation transfer code \mcrx \ \citep[described
in][from now on J06]{jonsson06sunrise}. Emission from \hii\ regions and
their surrounding remnant birth clouds, on scales below that resolved
by \mcrx , are calculated using the \mapiii \ photoionization code
\citep{dopitaetal05, grovesetal08sparam}.

The remaining multi-phase structure of the ISM, not including the
star-forming regions, is also approximated by a sub-resolution model
using the multiphase model of \citet{springelhernquist03} to determine
the fraction of dust that resides in high-density clumps (``molecular
clouds''). These clumps generally have a low volume-filling fractions
and contribute little to the overall opacity in typical galaxies and
are excluded in the current version of the code (see Section
\ref{section_interface} for details).

\subsection{\mcrx }
 \label{section_sunrise}

The Monte Carlo method for solving radiation transfer problems has
gained in popularity in recent years due to the rapidly increasing
capabilities of computers. There are many Monte Carlo radiation
transfer codes in use \citep[][to name a few that are applied to
galaxies]{gordonetal01, jonsson06sunrise, chakrabartietal07radishe,
lietal07, bianchi08trading} but the basic idea is the same: to solve
the radiation transfer problem the way nature does, by simply emitting
enough photons that the space of possible photon histories is well
sampled.  The simplest implementation of the Monte-Carlo
radiative-transfer algorithm follows a single photon through the
medium.  This photon is emitted in a random direction and can then
scatter and/or be absorbed.  Eventually the photon leaves the medium in
some direction and can be captured by an external observer. This method
is in general very inefficient, and in practice the calculation differs
from this description in many ways. \mcrx \ uses many of the standard
techniques for increasing the efficiency of the Monte-Carlo method, and
most of these were described in detail in J06.  Here we review the
essential parts of the algorithm and in the following subsections
outline the improvements made to the model since, including the IR
emission by dust. \mcrx \ is a free, publicly available software that
can be applied to any hydrodynamic galaxy
%NLX% exclude from vocabulary builder 
simulations\footnote{The \mcrx\ web site is \url{http://sunrise.familjenjonsson.org}.}.
%NLX% end exclude from vocabulary builder 
The results presented here detail version 3.01.

The basic algorithm of \mcrx \ is as follows: as input, a number of
snapshots at different time-steps of a hydrodynamic galaxy simulation
(e.g.~merging galaxies) are saved and are used to generate the geometry
of the problem, and for each of these the 2D spectrum at given camera
angles is calculated by \mcrx .  First, a stellar population synthesis
model is used to calculate the SEDs of the emitting sources.  Second,
the adaptive grid needed for the radiative transfer calculations is
generated. Third, the Monte-Carlo radiative transfer calculations are
done, generating the optical-UV images. Finally, the thermal heating of
the dust and resulting IR emission is calculated.

The sources of radiation within the galaxy simulations are the
``stellar particles''. These stellar particles represent a coeval
population of stars (see Section~\ref{section_simulations} for
details). Based on the coeval-age, mass, and metallicity of these
particles, an appropriate SED is selected from a library of single
stellar population SEDs computed with the population synthesis model
Starburst99 \citep{leithereretal99, vazquezleitherer05}. If the
particle is younger than $10 \Myr$, it is presumed to have been
attenuated and modified by the surrounding gas and dust of its ``birth
cloud'', and the emission is taken from the \mapiii \ models of
star-forming regions.  The interface between these subresolution models
and the galaxy-scale simulation is described in
Section~\ref{section_interface}.  The photons emitted from the particle
originate from a random location within a radius $r$ from the particle
centre.  This is done to avoid point-source effects and to model that
the particles represent a collection of stars. Within this work the
radius was set to $100 \pc$, the gravitational softening length of the
galaxy simulations.

As described in J06, the ray-tracing of the \mcrx \ code is performed
on a grid, requiring the translation of the density information of the
hydrodynamic simulation.  To do this, the density of dust is estimated
from the galaxy simulations assuming that a constant fraction of the
metals in the gas is in the form of dust grains. This fraction is set
to 0.4 based on \citet{dwek98}, but the results are not particularly
sensitive to moderate changes in this parameter.

The dust models used for the diffuse ISM are based on the work of
\citet{weingartnerdraine01}, including the updates in
\citet[][hereafter DL07]{draineli07}.  This model includes graphite and
silicate grains with a distribution of sizes.  Carbonaceous grains with
sizes ($a < 100 \Angstrom$) are assumed to have characteristics of
Polycyclic Aromatic Hydrocarbons (PAHs).  Grain cross-sections are
taken from \citet{lidraine01} for the graphite and silicate, and
\citet{draineli07} for the PAHs (though see Section
\ref{section_ir-emission} for differences concerning dust and PAH
emission). For the simulations described here
(Section~\ref{section_simulations}) we assume Milky-Way type dust and a
PAH mass fraction of $q_{ \rm { PAH } } = 4.58 \%$ (MW3.1\_60 in
\citealt{draineli07}) to match the approximately solar metallicity of
the models.

An adaptive-mesh refinement grid is used to describe the dust geometry,
making it possible to resolve small-scale structure over galaxy scales.
The criteria used to determine the structure of the grid are similar to
those described in J06, with relative density variations of  $\sigma_{
\rho_m } / \! \left \langle \rho_m \right \rangle < 0.1$ (the reader is
referred to J06 for a detailed description of the grid refinement
algorithm).

An additional requirement that the V-band optical depth across a grid
cell be less than a given tolerance value, $\tau_{ \rm { tol } }$, was
also used. The purpose of this criterion is to ensure that cells are
not optically thick and that the dust temperature does not change
significantly between neighbouring cells (see Section
\ref{section_ir-emission}).  Typically, $\tau_{ \rm { tol } }$ was set
to 1.0, but see Section~\ref{section_validation} for convergence tests.
 Regardless of the optical depth, however, no cells smaller than $150
\pc$ were used because the limited resolution of the hydrodynamic
simulations assures there is no structure on smaller scales.  As the
stellar particles have a radial extent over which their photons are
emitted, it is also unlikely that the dust temperature would vary on
smaller scales than this. We have also verified that decreasing the
minimum cell size did not substantially affect the results.

\begin{figure} \begin {center} \includegraphics*[width=\columnwidth]{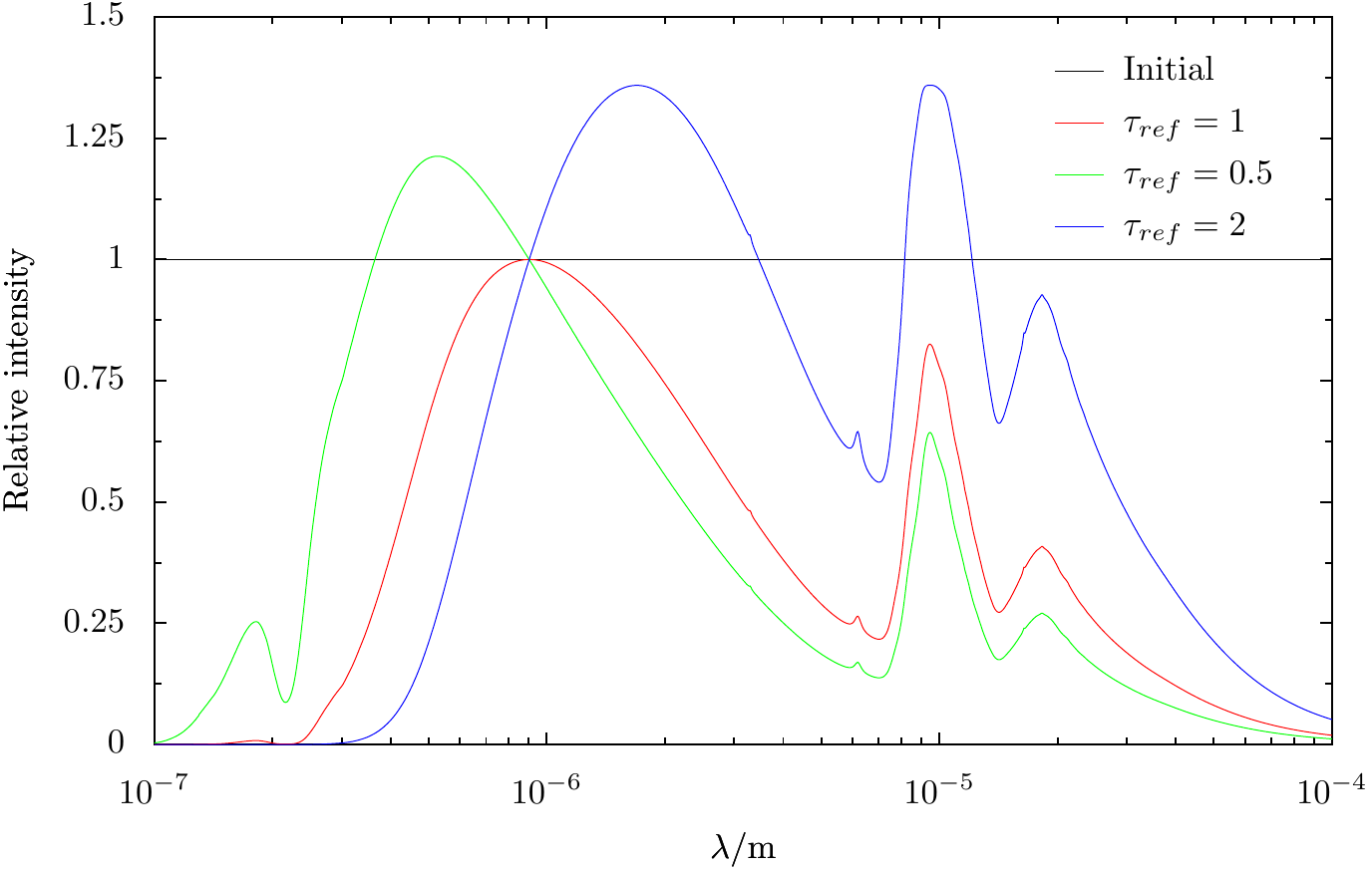} \end {center} \caption{ \label{plot_bias} The effect of the bias factor with respect to the propagation length of the ray. The ray intensity is initially 1 (black line) at all wavelengths, and an interaction optical depth is drawn (at random) at the reference wavelength ($0.9\um$).  On average, the interaction optical depth will be $\tau_{ref}=1$. In this case (red line), the intensity of the ray at wavelengths away from the reference wavelength is decreased as those wavelength are comparatively less likely to interact at the point where $\tau_{ref}=1$.  If an interaction optical depth less than 1, say $\tau_{ref}=0.5$ (green line), is drawn, the shorter wavelengths who interact at shorter depths relative to $\lambda_{\rm{ref}}$ will be boosted while the longer wavelengths are suppressed.  For interaction depths larger than 1 (blue line), the opposite happens. When these are integrated over many rays, the intensity-weighted distribution of interaction optical depths reduce to the expected $e^{-\tau(\lambda)}$ at all wavelengths.  } \end{figure}

The emission and tracing of the rays in the radiation transfer of \mcrx
\ are similar to other Monte Carlo codes and was described in detail in
J06.  One of the unique features and strengths of \mcrx \ is the use of
polychromatic rays.  Unlike in most other Monte Carlo codes, where a
ray only contributes to a single wavelength, the \mcrx \ rays
contribute to all wavelengths simultaneously.  Since the absorption and
scattering probabilities are wavelength dependent, this is accomplished
by an appropriate weighting of the different wavelengths that is
dependent on the ray path.  This bias factor represents the probability
of photons of that wavelength interacting at that depth relative to the
reference wavelength.  The correct bias factors for various situations
were derived in J06.  With polychromatic rays, only one random walk has
to be done for all wavelengths, a significant increase in efficiency
compared to monochromatic calculations, especially for high wavelength
resolution. An example of the bias factor applied to rays after
propagating certain distances is shown in Figure~\ref{plot_bias}.

%NLX% exclude from vocabulary builder
%NLX% end exclude from vocabulary builder

One drawback of using the polychromatic formalism as described in J06
is that the bias factors can potentially be large and thus boost the
contributions from single rays such that the outputs become noisy. In
the current version of \mcrx , this effect is minimised by splitting
rays whose intensity (at any wavelength) has been boosted to some
multiple of the initial value.  The resulting rays, each carrying a
fraction of the intensity of the original ray, then propagate
independently from then on.  An appropriate choice of the reference
wavelength also minimises the range of bias factors encountered.  For
the simulations described here, the threshold intensity for splitting
rays was 10 times the starting intensity, while the reference
wavelength was $0.9 \um$, though it was verified that the results are
unchanged with reference wavelengths ranging from $0.5 \um$ to $1.5
\um$.

Perhaps the most important recent addition to \mcrx \ is the ability to
self-consistently calculate the infrared emission from the dust in the
grid cells. This is a crucial part of the model and is described in
separate in Section~\ref{section_ir-emission}.

\subsection{Emission from Star-Forming Regions}
 \label{section_mappings}

Observations of star-forming galaxies indicate that the extinction
associated with emission lines tends to be greater than that inferred
from the UV light. This suggests that the youngest, most massive stars
associated with the line emission are surrounded by more dust than
slightly older UV-emitting stars \citep{calzetti97, tuffsetal04}. This
differential extinction arises naturally from the fact that star
formation preferentially occurs in regions of high gas density
\citep{kennicutt98}. This density dependence is included in the
hydrodynamic models, and hence this differential extinction is already
partly accounted for in \mcrx . However, clumping on scales below the
resolution of the galaxy simulations is naturally not present. The
molecular complexes out of which stars form, and thus dust directly
associated with the newly formed stars --- the dust within the
left-over ``birth'' clouds of the young stars --- are not resolved.

To deal with the subresolution radiative transfer in these birth
clouds, we use the 1-D photoionization and radiative transfer code
\mapiii \ \citep{grovesthesis}. This code is able to calculate the
complex transfer of stellar light, including the ionizing radiation,
through the surrounding gas and dust of the \hii\ and photodissociation
regions (PDR) associated with young, massive stars, and calculate the
resulting nebular and dust emission, including the stochastic heating
of dust.  For the \mcrx \ models here we specifically use the \mapiii \
starburst region models from \citet{dopitaetal05} \&
\citet{grovesetal08sparam}. These models calculate the radiative
transfer of the radiation from a newly formed massive stellar cluster,
using Starburst99 spectra for the intrinsic stellar emission. The
radiation propagates through a surrounding (spherically symmetric)
\hii\ region and a photodissociation region, whose covering fraction
decreases over time as it is cleared away by the strong winds from the
massive cluster. What distinguishes these models from other starburst
models is that they include the evolution of the stellar wind bubble
blown by the stellar winds over time, constraining the geometry of the
\hii\ region, and reducing the number of physical parameters needed to
describe the emission \citep{dopitaetal05}.

The final result is a spectrum made up of stellar, \hii\ and PDR
emission, from far-UV to mm wavelengths, including about 1800 emission
lines (though most are outside the wavelength range covered by the
present models). The shape of the SED is controlled by five parameters:
the metallicity of the stars and gas (the same set as Starburst99, and
given by the hydrodynamic simulations); the pressure of the surrounding
ISM (given by the hydrodynamic simulations); the compactness of the
stellar clusters ($\cal { C }$); the clearing timescale of the PDR
($\tau_{ \rm { PDR } }$); and the age of the stellar cluster \citep[for
a full description of the parameters and their physical meaning,
see][]{grovesetal08sparam}.

These \mapiii \ models substitute the Starburst99 stellar clusters as
sources in the \mcrx \ calculation for cluster ages less than $10
\Myr$.  By this age, most of the ionizing radiation has been emitted
\citep{dopitaetal06}.  The typical clearing timescale for the ``birth''
clouds surrounding the young stars is also expected to be less than
this, meaning that by $10^7$ years all the gas and dust directly
associated with the stars has been essentially cleared away. What is
left is an exposed stellar cluster (the Starburst99 model) surrounded
by the diffuse gas and dust treated by \mcrx .

The dust properties assumed within the birth cloud models \citep[as
described in][]{dopitaetal05} are similar to those used in \mcrx , thus
are consistent across the birth cloud-diffuse boundary. Grain
cross-sections are taken from \citet{lidraine01} for the graphite and
silicate grains, with a similar power-law distribution. The depletion
of elements onto grains is based on the local interstellar cloud
\citep{kimuraetal03}, and is approximately the same as the
\citet{dwek98} depletion fraction of 0.4. The main difference occurs
with the PAH molecules, where \mapiii \ uses the
\citet{weingartnerdraine01} coronene-sized grains to determine the
opacity, but uses an empirical template for the emission, and includes
the effects of photoelectric heating/losses, as described in
\citet{grovesetal08sparam}.

As discussed in section \ref{section_simulations}, the formation of
stellar particles in the hydrodynamic simulations is a discretized
realisation of the mass of stars formed and does not truly indicate
that a mass of stars equal to the particle mass was formed at the
instant the particle is spawned. Except in massive starbursts, stars
generally form in smaller clusters than the masses of the stellar
particles, which for the large galaxies is $\sim 10^6 \Msun$. The
average cluster size within the \mapiii \ models is a free parameter
connected with the compactness of the star-forming region.  To
integrate the models within \mcrx \ we set this size to a constant
value of $M_{ \rm { cl } } = 10^5 \Msun$, so conceptually a young
stellar particle of mass $M_s$ in the simulations should be thought of
as made up of $M_s / M_{ \rm { cl } }$ separate star clusters occupying
a nearby region of space.

Due to this multiple cluster interpretation of the young stellar
particles (and because realistically individual stellar clusters do not
form instantaneously) we have chosen \citep[as discussed
in][]{grovesetal08sparam} to average over the $10^7$ years, for the
burst, essentially approximating the clusters as continuously forming
stars over this timescale.  Due to this assumption, we alter slightly
the definition of the PDR clearing timescale ($\tau_{ \rm { PDR } }$)
to the PDR covering fraction, the time-averaged fraction of the stellar
cluster solid angle that is ``covered'' by the PDR ($f_{ \rm { PDR }
}$).

With the averaging over age, four free parameters for the \mapiii \
models remain. Two of these (metallicity, ISM pressure) are given by
the hydrodynamic models. The other parameters (PDR covering fraction
and compactness of the stellar clusters) are not directly constrained
by the hydrodynamic models. Therefore, we need to set these to
reasonable values based on empirical evidence. The compactness
parameter ${ \cal C }$ is calculated from our assumed mass of the star
clusters, $M_{ \rm { cl } }$ and the ISM pressure given by the
hydrodynamic model using Equation~(13) in \citet{grovesetal08sparam}.
This means that ${ \cal C }$ is directly dependent upon the ISM
pressure.  For the covering fraction, we assume a fiducial value of
$f_{ \rm { PDR } } = 0.2$, and explore this parameter later in the
paper.

\subsection{Interfacing \mcrx \ with the Subresolution Models}
 \label{section_interface}

An important part of the model is ensuring that the ``boundary
conditions'' between the subresolution \mapiii \ models and the
description of the general ISM are appropriately matched. Because the
\mapiii \ models are associated with young stellar particles but also
contain gas and dust, this mass must be borrowed from the environment
for the $10 \Myr$ lifetime of the \mapiii \ particles. This raises the
concern about a potential ``double-counting'' of gas; when a \mapiii \
particle of mass $m_s$ is created, the emission from the young stars
will for a short time be attenuated by an additional mass of gas
associated with the surrounding \hii\ region and PDR.  If care is not
taken, the photons may then encounter this mass again after entering
the diffuse ISM, in which case there will be an overestimation of the
attenuation.  More severely, if the mass of gas associated with the
\mapiii \ particle is larger than the total amount of gas available in
the vicinity of the newly formed particle, there is in principle not
enough gas surrounding the star particle to make up the gaseous
component of the \mapiii \ particle.

It should be pointed out that this is a problem \emph{in principle}
with any implementation of star formation that wholesale converts
particles of gas into stars. Converting a gas particle into a stellar
particle implies a 100\% efficiency of star formation, which is
unrealistic. In reality, there will always be gas left over from the
formed stars. This issue is further exaggerated by the
\emph{instantaneous feedback} approximation, used in many simulations,
that returns the metal--enriched gas associated with the stars
(i.e.~stellar winds) back to the gas at the moment of the stellar
particle creation. What all this emphasizes is that, on a single
particle basis, the simulations can not be interpreted literally, and
this fact is independent of any radiation-transfer calculations. If the
galaxy simulations lead to situations where the local supply of gas is
completely turned into stars the model will fail in several ways, not
only in the apparent star-formation efficiency but also in calculating
metal enrichment. In such a situation it is not reasonable to expect
the radiative-transfer calculations to be accurate. That said, the
simulations presented here are far from such a regime.

To estimate the potential double counting of mass, we must estimate the
mass associated with the PDR surrounding the \hii\ region (the mass of
ionized gas is negligible). The PDR in the \mapiii \ models used here
is defined to have a hydrogen column density of $10^{ 22 } \cm^{ - 2 }$
(roughly $\tau_{ \rm { V } } = 2$ at Solar metallicity) of material
swept up by the stellar wind. The radius of the PDR is determined by
the solution to the differential equation (13) in \citet{dopitaetal05}.
As we are using the time-averaged formulation of the \mapiii \ models,
the luminosity-weighted time average of $r_{ \rm { PDR } }$ is the
relevant quantity. For the full set of \mapiii \ models this varies
from $\sim 5 \pc$ for low-mass clusters in high-pressure regions up to
$\sim 800 \pc$ for high-mass clusters in low-pressure regions, typical
ranges for massive star-forming regions (Orion Nebula to multiple 30
Doradus).

The mass of the gas in the PDR can be computed as 
\begin{eqnarray}
m_{ \rm { PDR } } & = & 4 \pi r_{ \rm { PDR } }^2 10^{ 22 } \cm^{ - 2 }
f_{ \rm { PDR } } \\ & \approx & 10^9 \Msun \kpc^{ - 2 } r_{ \rm { PDR
} }^2 f_{ \rm { PDR } }.
\end{eqnarray}
 For a typical PDR radius around the $10^5 \Msun$ clusters of $\sim 100
\pc$ and a PDR fraction of $\sim 0.1$, this gives a typical $m_{ \rm {
PDR } }$ of $\sim 10^6 \Msun$. This implies star formation efficiencies
of 10 \%, not unreasonable.

Based on the gas density in the location of the particle, the radius
$r_s$ around the particle that contains the mass $m_{ \rm { PDR } }$ is
calculated ($r_s = 3 m_{ \rm { PDR } } / [ 4 \pi \rho_{ \rm { ISM } }
]$). It is expected that $r_s$ be larger than $r_{ \rm { PDR } }$,
since the former was calculated based on the average density in the
region while the PDR should be made up of higher density gas near the
star-forming region. This is merely a manifestation of the fact that
dense star-forming regions are unresolved in the simulations.

For determining the potential double-attenuation of light, the
interesting comparison is between $r_s$ and $r$, the radius of the
emitting \mapiii \ particle itself. As was mentioned in
Section~\ref{section_sunrise}, the stellar particles have a radial
extent over which their emission originates. This is necessary to avoid
point-source effects and to emphasize that the particles represent a
collection of stars. This is further true for the \hii\ regions and
PDRs which will have a larger extent than the stellar clusters.  The
photons emerging from the \mapiii \ particle will thus enter the ISM a
distance $\sim r$ away from the particle centre.

There are now three possible scenarios: If $r \ll r_s$, then there is
indeed a double-counting issue. The photons will then pass the same
material twice, first in the PDR and then in the general ISM,
potentially leading to an overestimate of the attenuation. (The
question is really more complicated, as the optical depth of a certain
mass of material goes up as it is concentrated to smaller radii. It is
thus not certain that there will be a significant overestimate of the
attenuation, even if the same matter is passed twice. The more
dispersed material will have a much smaller impact on the attenuation.)

Conversely, if $r \gg r_s$, the particle is too big and photons
emerging will ``skip'' some attenuating matter. In this case, the
attenuation of the emission will be underestimated.

The desired situation is thus that $r \approx r_s$, in which case the
radius of emission is well matched to the amount of material
surrounding the particle and the ray encounters the amount of material
that's expected.

In the simulations used here, the particle size $r$ is fixed to $0.1
\kpc$. The $r_s$ for the \mapiii \ particles varies, but is generally
$\sim 0.5 \kpc$. There is thus potential for double-counting, but the
excess optical depths were estimated to be only $\sim 0.2$.
Furthermore, a special run where the radii of the \mapiii \ particles
were set to $r_s$ showed differences in the UV attenuation of $\la 1
\%$ at GALEX wavelengths, so we conclude that this double-counting of
mass does not significantly influence our results.

However, this does point to one of the other issues with the simple
replacement of the stellar particles by \mapiii \ particles: as
mentioned before, \hii\ regions are generally much larger in extent
than the stellar clusters ionizing them. This means that the line
emission will generally be much more diffuse than the stellar particles
can represent, even ignoring the small fraction of diffuse ionizing UV.
Hence in \mcrx \ images, the H$\alpha$ will tend to be more point-like
than in real galaxies.

Another issue is that, due to their 1-D nature, the \mapiii \ models
cannot include the multiphase structure of the ISM, consisting of dense
molecular clouds in a diffuse medium.  This clumping of gas (and dust)
into optically thick clumps (not containing young star clusters) will
decrease the effective optical depth of these regions while also making
the extinction law more grey \citep{wittgordon96}.  In principle, the
adaptive-mesh grid in \mcrx \ could be used to resolve these structures
\citep[as done in][]{bianchi08trading}, but doing so in these
simulations would require a prohibitively large number of grid cells. 
Instead, a simpler, subresolution, approach is used.

A model for this multiphase medium is used as the prescription for
supernova feedback in many GADGET hydro simulations
\citep{springelhernquist03}. This model gives an analytic prescription
for the mass fraction of the dense and diffuse phases of the ISM as a
function of the average gas density.  In the simulations presented here
(in Section~\ref{section_simulations}), it is assumed that only the
diffuse phase contributes to the attenuation of radiation, equivalent
to assuming that the dense phase has a negligibly small volume filling
fraction and that the probability of radiation entering a dense cloud
is negligible.  This is admittedly a crude assumption, and an
improvement which treats these dense clumps using a ``megagrains''
formalism \citep{hobsonpadman93, varosidwek99} is underway.  In the
simulations of quiescently star-forming disk galaxies presented here,
however, the gas densities are regulated by supernova feedback and
never reach significantly higher than the threshold density above which
the multiphase medium develops. Consequently, the mass in the dense
phase is small and the multiphase assumption has a negligible effect on
the results.

\subsection{Calculation of Infrared Emission}
 \label{section_ir-emission}

At UV--optical wavelengths, the emission from stars and nebulae
dominates the SED of galaxies, and dust emission can be ignored. At
wavelengths longer than about $3 \um$ this is no longer the case and
the infrared emission from dust grains and PAH molecules has to be
considered. To model this emission, the local radiation field intensity
needs to be determined and the resulting local thermal emission from
the grains calculated, with both varying as a function of location in
the galaxies.

The infrared emission from the \hii- and photodissociation regions in
the \mapiii \ models includes the effect of thermal fluctuations
undergone by very small grains \citep{grovesthesis}.  The emission from
polycyclic aromatic hydrocarbons (PAHs) that dominates the mid-infrared
emission in galaxies is treated as a special case.  A certain fraction
of the carbon dust is assumed to be in the form of PAH molecules. Their
absorption is calculated based on the cross-sections in
\citet{lidraine01}, and the fraction of this  energy not lost to
photoelectric effects is then emitted as a fixed template made up of
Lorentzian profiles designed to fit Spitzer IRS observations of PAH
emission \citep{dopitaetal05, grovesetal08sparam}, assuming energy
conservation.

Unlike the version described in J06, the current version of \mcrx \ has
the capability to calculate the dust temperatures of the various dust
species and sizes in the grid cells.  With this addition, the infrared
emission from the dust grains will self-consistently reflect the
distribution of radiation intensities heating the grains at various
locations in the galaxy.  Unlike in the \mapiii \ models, the
calculation of the emission SED of the grains is done assuming thermal
equilibrium for all grain sizes, neglecting the effect of thermal
fluctuations.  Emission from PAHs use the same template method as the
\mapiii \ models, but only for a fraction $f_t$ of the grains smaller
than ($a < 100 \Angstrom$). The emission is calculated using the same
template spectrum used by the \mapiii \ code \citep{grovesetal08sparam}
assuming energy conservation.  This approach is considerably less
computationally intensive than a full calculation involving thermal
fluctuations while yielding reasonably realistic results.  $f_t$, the
fraction of small grains emitting through the template was treated as a
free parameter, with a value of $f_t = 0.5$ found to be the most
reasonable and taken as our fiducial value for the simulations
discussed here (see Section~\ref{section_sensitivities} for a
discussion on this parameter).

Neglecting the thermal fluctuations of small grains can potentially
affect the accuracy of the calculated SED.  For this purpose, an
alternative way of calculating the dust emission spectrum using the
precomputed SED templates of DL07 can also be used. These emission
spectra are based on the full temperature distribution of the different
size dust grains in the model, parametrized by the intensity of the
radiation heating the dust. For each grid cell, the template most
closely matching the radiation intensity in that cell is used as the
emission from that cell. While this method of calculating the emission
SED has the advantage that it is a computationally inexpensive way to
include thermally fluctuating grains, it has the drawback that it does
not take into account the wavelength dependence of the radiation field.
 The templates are calculated assuming that the radiation field is a
scaled version of the local interstellar radiation field of
\citet{mathisetal83} and do not take into account the shape of the
radiation spectrum.  The wavelength dependence of the heating radiation
is potentially significant both because of the differing cross-sections
of grains of different sizes, and the fact that, for a given energy
density in the radiation field, high-energy photons will excite larger
thermal fluctuations than low-energy photons. To estimate the magnitude
of this effect, a full calculation of the temperature distribution of
the grains would be necessary, something that is currently not
possible.

While it is straightforward to calculate the dust heating due to
starlight, correctly estimating the heating due to self-absorption of
the emission from the dust grains themselves requires coupling the
emission and absorption in the grid cells and an iterative solution
becomes necessary.  In the simplest method, the dust temperatures are
recalculated at every step and the new estimate of the dust emission
propagated through the volume leading to a new estimate of the dust
temperatures \citep{misseltetal01, bianchi08trading,
chakrabartietal07radishe}.  The convergence of this method is
ultimately limited by the intrinsic Monte Carlo noise in each estimate
of the radiation field. A more efficient method is to use the radiation
field from the previous estimation as a reference solution and only
solve for the change due to the last iteration \citep{juvela05}. This
method, which is used by \mcrx , is guaranteed to converge as the
updates to the radiation field must decrease for each iteration.
Convergence in very optically thick regions can be slow, but this is
not a concern in the simulations studied here.

The iterative scheme employed by \mcrx \ can be thought of as a hybrid
of the methods that recompute the dust temperatures at each iteration
and the temperature correction scheme used by \citet{bjorkmanwood01}.
In the temperature correction method, each photon-packet absorption is
immediately followed by an update of the dust temperature in the cell
where the absorption takes place and the subsequent re-emission of the
photon packet. In this re-emission, the energy of the photon packet is
conserved, but the \emph{wavelength} of the photon packet is drawn from
the \emph{difference} of the emission SED of the dust in the cell
before and after the absorption event.  The re-emitted photon packet
thus corrects the distribution of the photon packets emitted previously
so it is correct for the current dust temperature in the cell.  As many
photons are traced through the volume, the temperature of the dust in
the grid cells will converge to the equilibrium temperature without
explicit iteration.

A big virtue of the temperature correction scheme is that it is
explicitly energy conserving.  However, as outlined in
\citet{chakrabartietal07radishe}, its efficiency is limited by the fact
that only explicit absorption events contribute to the dust temperature
calculation.  A more efficient way of estimating the dust temperature
is to use the estimate of the radiation intensity from the total path
length of photons traversing the cell, also known as ``continuous
absorption'' \citep{lucy99}, but this method is incompatible with the
immediate re-emission method. Instead, a number of rays are traced
through the volume, giving an estimate of the radiation intensity in
all cells.  Based on this estimate, the temperature of the dust in all
grid cells are updated and the \emph{difference} between the current
emission SED of the dust in the grid cell and that of the previous
iteration is used as the source SED.  From energy conservation, it can
be seen that this source SED, summed over all grid cells, can never be
larger than that of the previous iteration, so the scheme must
converge.

Since the transfer of these source SEDs is done using the polychromatic
radiation transfer used in \mcrx , it also avoids a major criticism of
the \citet{bjorkmanwood01} method: that it can fail in certain cases
because the probability distribution from which photon re-emission
wavelengths are drawn can become negative for certain wavelengths
\citep{baesetal05}. With the polychromatic algorithm, the emission
wavelength is not drawn from this distribution.  Instead, the entire
spectrum is propagated at once, and there is nothing requiring the
energy be positive at all wavelengths. Indeed, when the DL07 SED
templates are used for the dust emission, part of the source SED during
the iteration to convergence becomes negative but the method still
converges to an equilibrium.

Formally, the algorithm can be described as follows.  Define $I_{ i ,
\lambda }^\star$ as the radiative intensity at wavelength $\lambda$ in
cell $i$ due to \emph{stellar} emission. This quantity is fixed.
Similarly, $I_{ i , \lambda }^k$ is the intensity from dust emission
after iteration $k$. The emission from dust is described by a function
$B$ that converts radiation intensity into an emission spectrum, such
that 
\begin{equation}
L_\lambda = B_\lambda \left ( I_{ \lambda^\prime } \right ).
\end{equation}
 The radiative coupling between cells can be expressed as a matrix $T_{
ij , \lambda }$ describing the contribution to the radiation intensity
in cell $i$ due to dust emission from cell $j$, such that the intensity
in cell $i$ is 
\begin{equation}
I_{ i , \lambda } = \sum_j L_{ j , \lambda } T_{ ij , \lambda }.
\end{equation}
 Because the cross-sections of the dust grains are independent of
temperature, $T_{ ij , \lambda }$ is constant (for a given wavelength).
(In this picture, the Monte Carlo method is just a way of evaluating
the elements of this matrix.  In principle, one could evaluate the
elements of this matrix once and then invert it to solve for the
equilibrium state.  In practice, this is not possible as the number of
elements is the square of the number of grid cells, $\sim 10^{ 12 }$ in
the simulations used here.)

In the simplest calculation, this equation would be iterated to
convergence, i.e. 
\begin{equation}
\label{equation_simple-iteration} I_{ i , \lambda }^{ k + 1 } = \sum_j
L_{ j , \lambda }^k T_{ ij , \lambda } ,
\end{equation}
 which, as mentioned earlier, has the undesirable property that the
entire intensity estimate is recalculated for each iteration. Because
the Monte Carlo estimate of $T_{ ij , \lambda }$ in each iteration is
subject to independent random error, the iteration will not converge.

By subtracting two successive iterations from each other, we get 
\begin{equation}
\label{equation_iteration} I_{ i , \lambda }^{ k + 1 } = I_{ i ,
\lambda }^k + \sum_j \left ( L_{ j , \lambda }^k - L_{ j , \lambda }^{
k - 1 } \right ) T_{ ij , \lambda }.
\end{equation}
 This shows that the update to the intensity estimate in each cell
comes from transferring an emission distribution consisting of the
\emph{difference} in luminosity between the two previous iterations.
Unlike the iteration in Equation~\ref{equation_simple-iteration}, each
progressive iteration here retains a memory of the previous evaluations
of $T_{ ij }$, effectively averaging the Monte Carlo variance over the
iterations such that the iteration must converge.

In very optically thick situations, a significant fraction of the
emission in a cell will be absorbed inside the cell itself.
Effectively, this will result in a $T_{ ij }$ with very large diagonal
elements leading to slow convergence.  There are methods to accelerate
the convergence \citep{juvela05}, but these are not needed in the
situations presented here.

The iteration is terminated when all cells pass the convergence
criterion requiring that 
\begin{eqnarray}
\label{equation_tolerance1} I_{ i , \lambda }^k / I_{ i , \lambda
}^\star & < & { \rm { tol } }_r = 10^{ - 2 } , { \mbox { and that } }
\\ \label{equation_tolerance2} \frac { \sum_\lambda I_{ i , \lambda }^k
\sigma_\lambda } { \sum_{ j , \lambda } L_{ j , \lambda } \Delta
\lambda } & < & { \rm { tol } }_a = 10^{ - 6 }
\end{eqnarray}
 for all grid cells.  Essentially, the first criterion tests whether
the contribution from dust emission is negligible compared to heating
from starlight and the second criterion, important in regions that
receive little starlight, tests whether the dust emission from
iteration $k$ of cell $i$ is negligible compared to the total dust
emission.  In the simulations used here, convergence is obtained with
only two iterations.

\subsection{Galaxy simulations}
\label{section_simulations}

The galaxy simulations used in this work have been described in detail
in earlier papers \citep{tjthesis-nourl, pjthesis-nourl,
coxetal05methods, pjetal05attn, coxetal07minors, rochaetal07,
lotzetal08mergermorph}, but a brief overview will be given here for
context.

The simulations are of 7 different galaxy models run with the GADGET
SPH code, including star formation and feedback \citep{springeletal01,
springel05g2}. The simulations also include mergers of galaxies, where
two of the models are placed on approaching orbits, but in this paper
only isolated galaxies are considered.  In the simulations, gas is
represented by Lagrangian particles.  As stars form, gas is transformed
into collisionless matter.  This is implemented in a stochastic sense
\citep{springelhernquist03}; each gas particle will spawn a number of
stellar particles, with a probability based on the star-formation rate
of the particle.  These ``new star'' particles have masses $\sim 10^4$
-- $10^6 \Msun$, depending on the mass of the simulated galaxy, and can
be thought of as a cluster of coeval stars. However, it is important to
not take this analogy too far. The star particles describe a
discretized conversion of gas into stars, but the presence of a young
star particle in a region should not be literally interpreted as if
$10^6 \Msun$ of young stars just formed at that location.

At low star-formation rates, where only a few of these particles are
present at any given time, this discretization becomes particularly
severe and it is possible to see large fluctuations in the stellar
light. This will be discussed in detail in
Section~\ref{section_results}.

For the simulations to be stable, a model for supernova feedback is a
necessary ingredient, and many different approaches to modelling it
exists.  The supernova feedback model in the simulations
\citep[extensively analyzed in][]{coxetal05methods} works by
artificially pressurizing star-forming regions, in effect setting their
equation of state.

A simple scheme for metal enrichment, where the metals produced by
supernovae are instantaneously recycled into the gas of the particle,
is used. In effect, each particle is a ``closed box model'' that does
not exchange metals with its neighbours. While simple, this method has
several drawbacks. For one, if the entire gas particle is consumed by
star formation, all metals (and hence all dust) is removed from the gas
phase. Also, late gas recycling by e.g.~AGB stars, which deposit gas
far away from the cloud in which they were born, is not included. Codes
with better treatment of metal production and gas recycling from stars
exist \citep[e.g.][]{scannapiecoetal05, stinsonetal06}, and using such
simulations would improve the accuracy of the radiation-transfer
calculation.

The 7 galaxy models are isolated galaxies, where the galaxies are
evolved in isolation for $1 \Gyr$ and snapshots saved every $50 \Myr$
to study how the galaxies evolve.  The galaxies have been modelled
after observed properties of local spiral galaxies and span roughly two
orders of magnitude in stellar mass. They contain a disk of gas and
stars, a stellar bulge, and a dark matter halo. The properties of the
models are summarised in Table~\ref{table_galaxy_models}. There are two
series of models, the ``Sbc''-type models are modelled after local
late-type spirals, while the ``G''-series cover a wider range in mass
and are modelled on median properties from the SDSS.  The metallicity
and age distribution of stars and gas in the galaxy models was chosen
to agree with observations \citep{rochaetal07}.

%NLX% exclude from vocabulary builder
\begin{table*} \begin{minipage}{\textwidth} \caption{The parameters of the simulated galaxies, adopted from \citet{rochaetal07} and \citet{pjetal05attn}.} \label{table_galaxy_models} \begin{tabular}{lccccccccccccc} \hline

Model & $M_{\mbox{vir}}$$^\mathrm{a}$ & $M_{b}$$^\mathrm{b}$ & $R_{d}$$^\mathrm{c}$ & $Z_{d}/\!R_{d}$$^\mathrm{d}$ & $R_{g}/\!R_{d}$$^\mathrm{e}$ & $f_{g}$$^\mathrm{f}$ & $f_{b}$$^\mathrm{g}$ & $R_{b}$$^\mathrm{h}$ & $V_{\mbox{rot}}$$^\mathrm{i}$ & $Z_{1.3}$$^\mathrm{j}$ & $\mathrm{d}Z/\mathrm{d}r^\mathrm{k}$ & % $N_g$$^\mathrm{l}$ &
Age$^\mathrm{m}$ & $\tau^\mathrm{n}$ \\

& ($\!\Msun$) & ($\!\Msun$) &(kpc) & & & & & (kpc) & ($\!\kps$) & (Z$_\odot$) & (dex/kpc) & % $(\mathrm{x}10^4)$ &
(Gyr) & (Gyr) \\

\hline

Sbc+ & $9.28\cdot10^{11}$ & $1.56\cdot10^{11}$ & 7.0 & 0.125 & 3.0 & 0.52 & 0.10 & 0.60 & $210$ & $1.12$ & 0.023 & % 3 &
13.9 & 110 \\

% Sc & $8.90\cdot10^{11}$ & $1.12\cdot10^{11}$ & 4.7 & 0.2 & 4.0 & 0.69
% & 0.00 & 0.00 & $196$ & $1.00$ & 0.043 & % 3 &
% 13.8 & -13\\

Sbc & $8.12\cdot10^{11}$ & $1.03\cdot10^{11}$ & 5.5 & 0.125 & 3.0 & 0.52 & 0.10 & 0.45 & $195$ & $1.00$ & 0.030 & % 3 &
13.9 & -106\\

G3 & $1.16\cdot10^{12}$ & $6.22\cdot10^{10}$ & 2.8 & 0.125 & 3.0 & 0.20 & 0.14 & 0.37 & $192$ & $1.00$ & 0.058 & % 5 &
14.0 & 10 \\

Sbc- & $3.60\cdot10^{11}$ & $4.98\cdot10^{10}$ & 4.0 & 0.125 & 3.0 & 0.52 & 0.10 & 0.40 & $155$ & $0.70$ & 0.041 & % 3 &
13.7 & 124\\

G2 & $5.10\cdot10^{11}$ & $1.98\cdot10^{10}$ & 1.9 & 0.2 & 3.0 & 0.23 & 0.08 & 0.26 & $139$ & $0.56$ & 0.04 & % 3 &
14.0 & 8.2 \\

G1 & $2.00\cdot10^{11}$ & $7.00\cdot10^{9}$ & 1.5 & 0.2 & 3.0 & 0.29 & 0.04 & 0.20 & $103$ & $0.40$ & 0.05 & % 2 &
11.5 & 3.7 \\

G0 & $5.10\cdot10^{10}$ & $1.60\cdot10^{9}$ & 1.1 & 0.2 & 3.0 & 0.38 & 0.01 & 0.15 & $ 67$ & $0.28$ & 0.06 & % 1 &
8.7 & 1.4 \\ \hline \end{tabular}

\medskip $^\mathrm{a}$Virial mass. $^\mathrm{b}$Baryonic mass. $^\mathrm{c}$Stellar disk scalelength.  $^\mathrm{d}$Ratio of stellar-disk scaleheight and scalelength.  $^\mathrm{e}$Ratio of scalelengths of gas and stellar disks. $^\mathrm{f}$Gas fraction (of baryonic mass). $^\mathrm{g}$Bulge fraction (of baryonic mass). $^\mathrm{h}$Bulge scale radius. $^\mathrm{i}$Circular velocity. $^\mathrm{j}$Metallicity at 1.3 scalelengths from the centre (gas and stars). %$^\mathrm{k}$Number of gas particles.
$^\mathrm{k}$Metallicity gradient. $^\mathrm{l}$Age of oldest stars (formation time of bulge and oldest disk stars).  $^\mathrm{m}$Exponential time constant of the star formation rate for the disk stars. \end {minipage} \end{table*}
%NLX% end exclude from vocabulary builder

\section{Model Validation}
 \label{section_validation}

In this section, we demonstrate that the radiation-transfer code is
correct and that the simulations performed are converged, establishing
trust in the model results. A number of tests of \mcrx \ were performed
in J06 and will in general not be repeated here. The exception is the
benchmark problem of \citet{pascuccietal04}, which was only partially
done in J06.

\subsection{The Pascucci et al. (2004) 2D Radiative Transfer Benchmark}

\begin{figure} \begin {center} \includegraphics*[width=0.95\columnwidth]{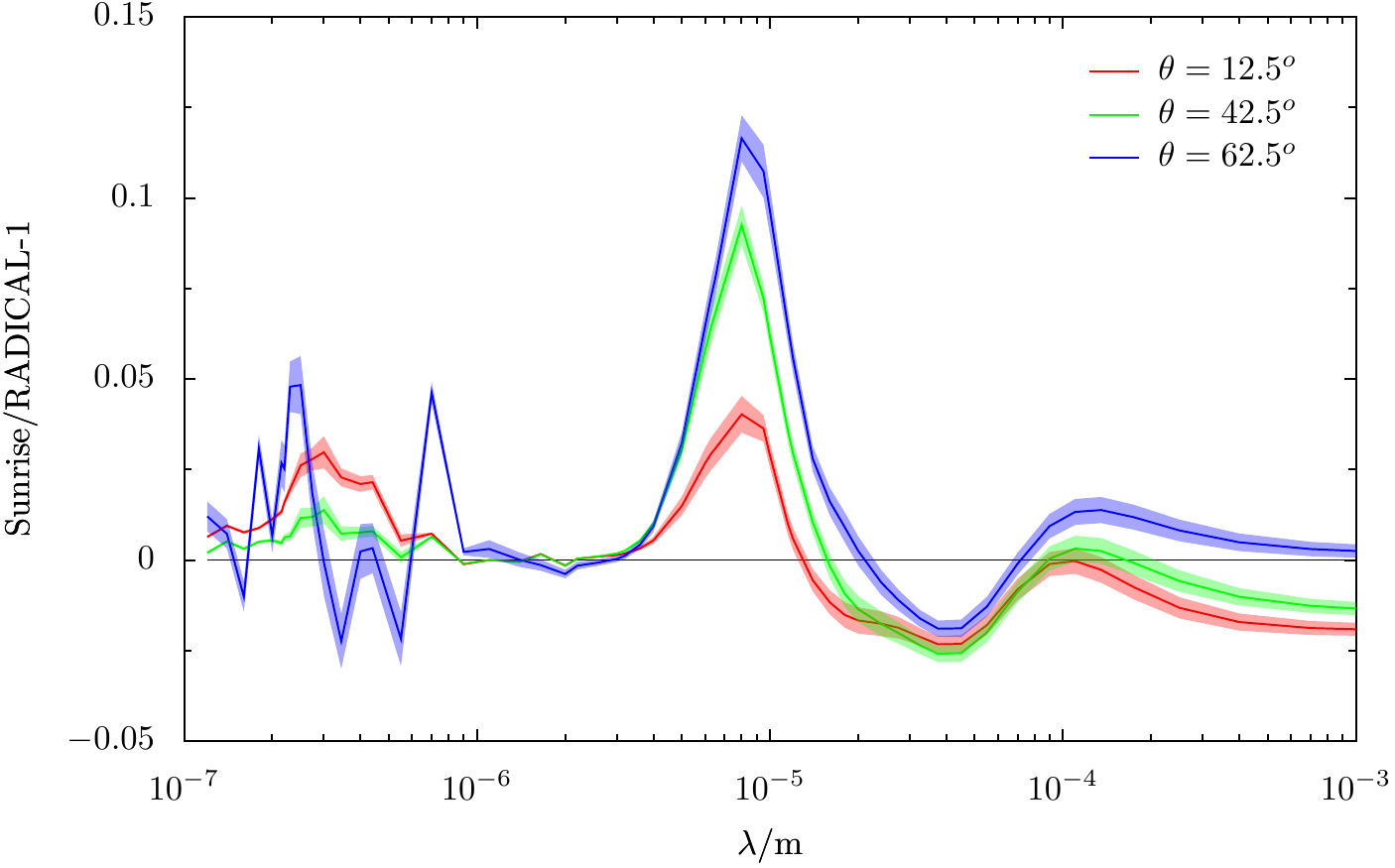} \end {center} \caption{ \label{plot_p04} The ratio of the \mcrx\ SED to the published RADICAL outputs of the 2D axisymmetric radiation transfer benchmark of \citet{pascuccietal04}. The outputs are shown for the three different inclinations in P04, and the shaded region represents the $1\sigma$ variance in the \mcrx\ results as estimated from 5 different runs with independent random number sequences. The agreement is good across all wavelengths, with the largest excursion seen at $8\um$. It is also worth noting that the wiggles in the difference at UV/visual wavelengths are much larger than the uncertainty in the \mcrx\ calculations, so are likely due to stochastic variations in the RADICAL results.  } \end{figure}

To first verify that \mcrx , including the improvements described
above, give correct results, the comparison with the benchmark problem
of \citet{pascuccietal04} that was done in Section~5.5 of J06 is
repeated.  Since infrared emission is now included in \mcrx , the
comparison is done over the entire wavelength range unlike in J06 where
only wavelengths dominated by stellar radiation were included. For
brevity, we also omit the lower-optical depth cases and only present
the results from the most stringent $\tau = 100$ case. For this test
$10^7$ polychromatic rays were used. The difference between the \mcrx \
output SEDs and the RADICAL output of \citet{pascuccietal04} is shown
in Figure~\ref{plot_p04}.

%NLX% exclude from vocabulary builder
%NLX% end exclude from vocabulary builder

In general, the agreement is good, within about 4 percent except at
wavelengths around $8 \um$ where the difference reaches a maximum of 11
percent in the edge-on case.  This agreement is well within the
internal differences between the various codes used in the P04
benchmark and mimics the behaviour of several of these codes. The
sensitivity of the feature around $8 \um$ is likely due to the
importance of resolving the inner edge of the disk where the hottest
dust will be located. It is also worth noting that the short-wavelength
wiggles in the difference must be due to stochastic variations in the
RADICAL results, as they are substantially larger than the uncertainty
in the \mcrx \ results.  The \mcrx \ calculation was done with
polychromatic rays, so there is no inherent stochastic
wavelength-to-wavelength variation in the results unlike for the codes
that perform an independent calculation for each wavelength.  (The
results in J06 differed substantially more from those in P04. The
majority of this difference was due to the incorrect truncation of the
disk at a height of only $100 \au$ instead of $1000 \au$.)

\subsection{Convergence Concerning Number of Rays Traced}

\begin{figure} \begin {center} \includegraphics*[width=0.48\textwidth]{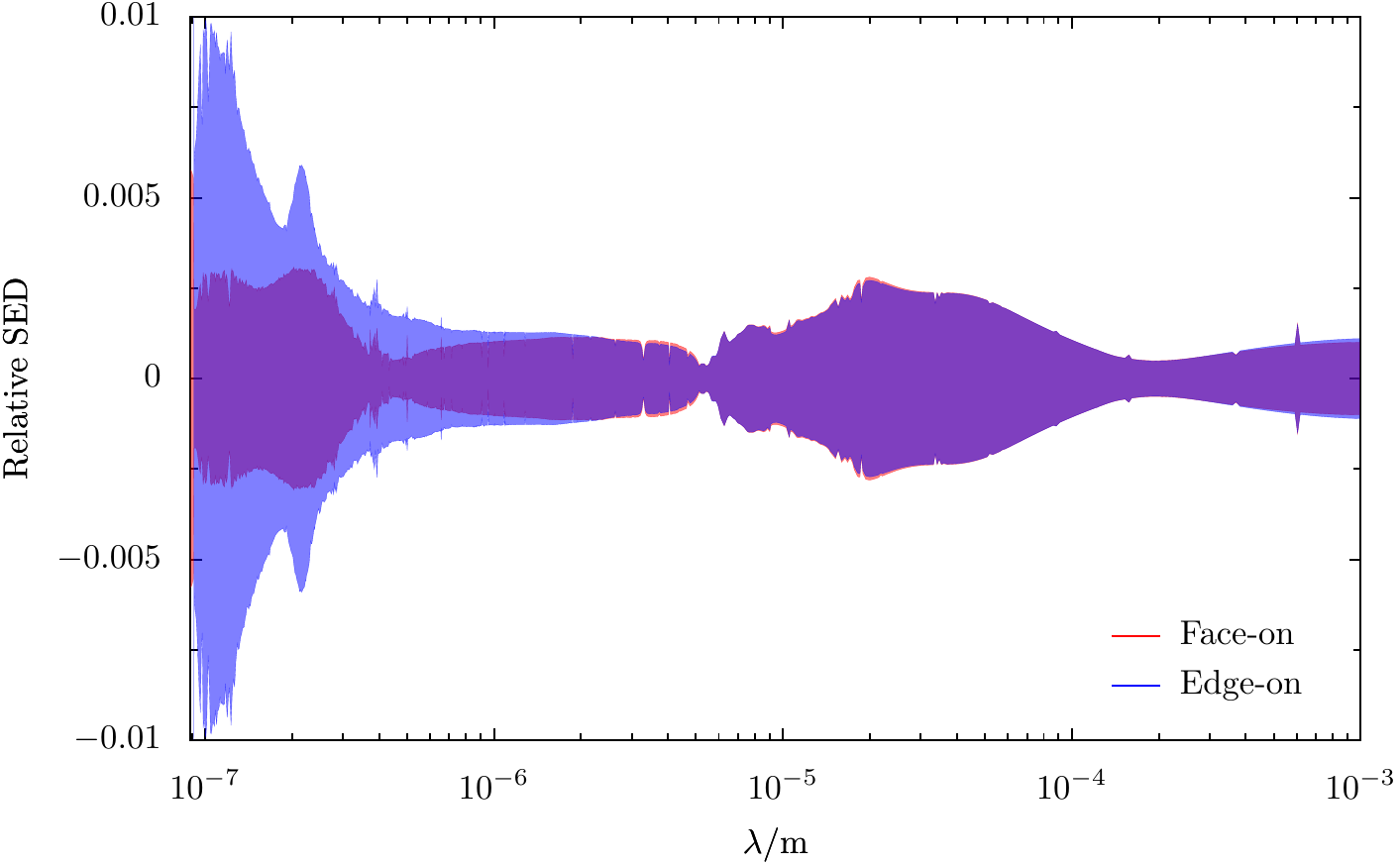} \end {center} \caption{ \label{plot_variance} The stochastic (Monte Carlo) variance in the output spectral energy distributions for the Sbc galaxy, using $10^6$ Monte-Carlo rays. The shaded region indicates the $1\sigma$ variance in the face-on (red) and edge-on (blue) directions, relative to the mean SED. The variance is less than 1\% at all wavelengths, and normally substantially less.  } \end{figure}

Having shown that the radiation-transfer calculation reproduces the
results of a non-trivial test case, we now turn to the galaxy
simulations.  At a most basic level, the convergence of the output
spectral energy distributions with regards to number of rays and grid
resolution is the first test. Figure~\ref{plot_variance} shows the
Monte Carlo $1 \sigma$ variance in the SEDs for the Sbc galaxy viewed
from the two extreme inclinations of face-on and edge-on.  The variance
is less than 1 percent at all wavelengths, and for most wavelengths
substantially less, indicating that the number of rays is sufficient
for a well-constrained integrated SED. The requirements are more
stringent when the spatial dependence of the SED is investigated, a
matter we return to in Section~\ref{section_pixels}.

%NLX% exclude from vocabulary builder
%NLX% end exclude from vocabulary builder

\subsection{Convergence Concerning Grid Resolution}

It is difficult to conclusively show convergence with regard to grid
resolution, because increasing the resolution of the grid much above
what is used for the simulations is impossible due to the exponentially
increasing memory requirements. The edge-on SED is quite sensitive to
accurately resolving the exponentially decreasing density of dust in
the vertical direction. This is also a geometry which the adaptive-mesh
refinement grid, being made up of cubic cells, is unable to capture
efficiently, leading to large numbers of grid cells.

\begin{figure} \begin {center} \includegraphics*[width=0.48\textwidth]{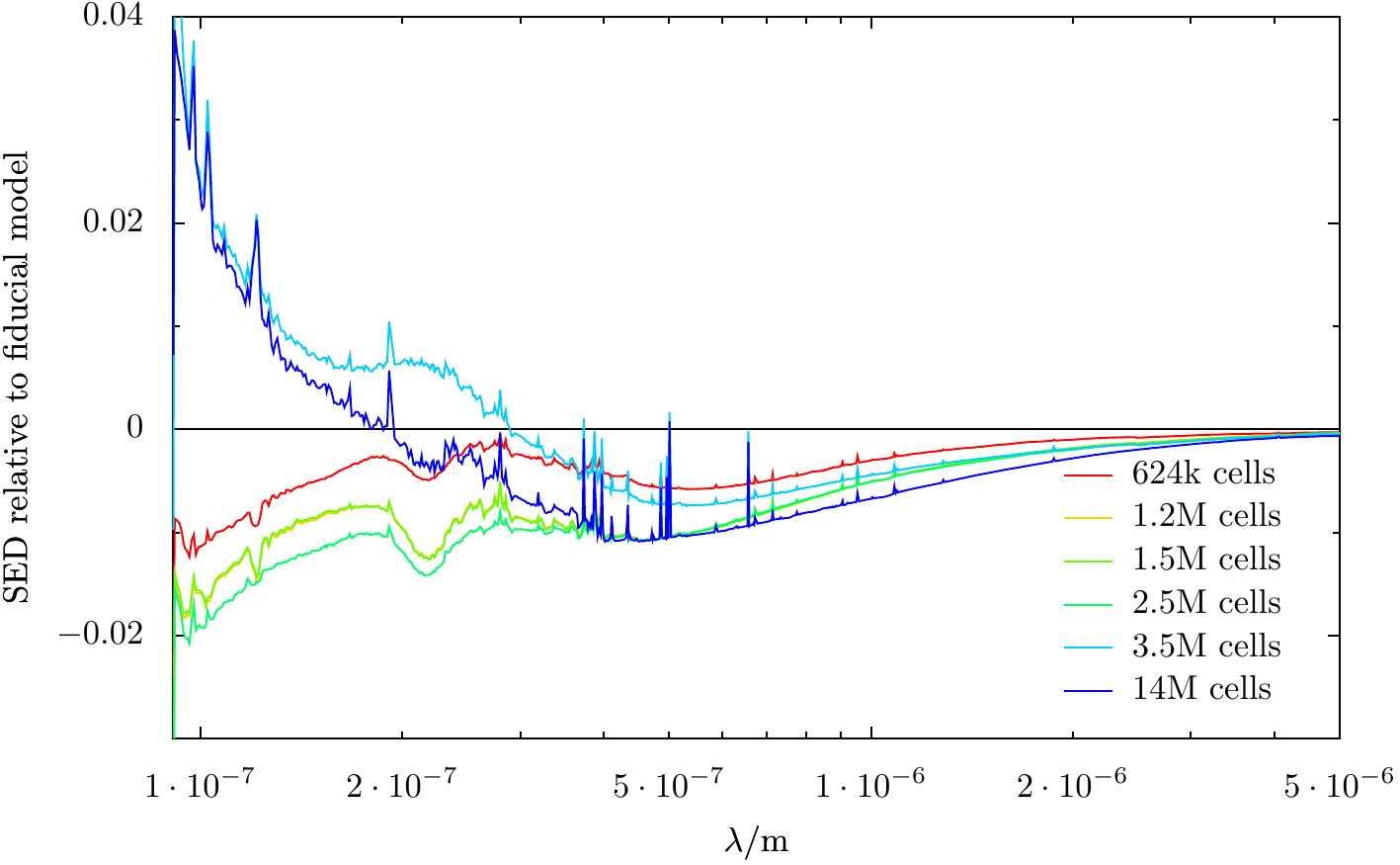} \end {center} \caption{ \label{plot_grid_resolution} The effect on the edge-on SED of the Sbc galaxy from increasing the grid resolution above the fiducial parameters adopted for the simulations presented in the paper. The SEDs are plotted relative to the fiducial model with 590k cells, labelled by the number of cells in the grid. The effect is in most cases within about 1\%, and always less than 4\%. The runs with more than $3\cdot 10^6$ cells differ by up to 5\% at the Lyman limit. For directions more than $15^\circ$ away from edge-on, the variations are less than 1\% for all wavelengths and models. } \end{figure}

To support our current resolution and to demonstrate convergence in the
grid, a number of tests were run on the Sbc galaxy model, where the
resolution was increased through the alteration of the various
parameters governing the structure of the grid.  The fiducial
resolution grid for the Sbc galaxy contains 590k grid cells, and tests
were done with up to 14M cells. The results for the (most sensitive)
edge-on case are shown in Figure~\ref{plot_grid_resolution}. The
maximum deviation is only $\sim 5 \%$ and occurs in the edge-on case
for wavelengths shorter than $0.1 \um$ at the highest resolution cases.
At larger wavelengths and for inclinations more than $15 \degrees$ away
from edge-on, the variations are less than $1 \%$. This variation is
likely due to the exponential decline of dust in the vertical
direction, as well as possible ``lines of sight'' opening up due to the
finer resolution.

As the dust emission SED proved to be converged at resolutions much
lower than the runs shown here, these tests were run without dust
emission to minimize the amount of computer memory needed. Thus, only
wavelengths dominated by stellar emission are shown.

%NLX% exclude from vocabulary builder
%NLX% end exclude from vocabulary builder

The tests shown in Figure~\ref{plot_grid_resolution} demonstrate that
the output SED is converged to within a few percent for our current
resolution given by our standard grid refinement parameters of: $\tau_{
tol } = 1$; a minimum cell size of $150 \kpc$; relative variation in
the dust density $\sigma_{ \rho_{ dust } } / \langle \rho_{ dust }
\rangle < 0.10$; and finally that $n_{ \rm { rays } }$ used to set a
lower limit on the opacity that is likely to affect the results be set
to $10^7$ (see J06 for a detailed description of the grid refinement
parameters).

\subsection{Convergence of the Dust Temperature Iterative Solution}

Another check that must be performed is to verify that the iterative
solution to the dust temperature distribution is converged.  When
increasing the accuracy of the iterative solution by an order of
magnitude by setting the tolerances, ${ \rm { tol } }_r$ and ${ \rm {
tol } }_a$, to one-tenth their values in
Equations~\ref{equation_tolerance1} \& \ref{equation_tolerance2}, the
number of iterations to convergence were not affected. This indicates
that the results are not affected by the accuracy of the iterative
temperature solution, at least for the fairly low optical depths
encountered in these simulations.

\subsection{Verifying Energy Conservation}

As the dust and stellar emission processes being distinct (though
connected) parts of \mcrx , and with the dust emission being based on
the Monte Carlo estimate of the radiation field, it is important to
verify that global energy conservation is in fact maintained to within
a reasonable level in the simulations.  Using too few rays, for
example, will result in a poorly sampled radiation field and thus a
noisy estimate of the dust heating, which could in principle result in
a significant energy non-conservation. Insufficient accuracy of the
iterative solution to the dust temperature distribution would also
result in apparent energy non-conservation.

Checking energy conservation, however, is somewhat non-trivial, because
the emerging flux from the galaxies are only estimated at a finite
number of cameras.  Since the emerging radiation field is not
isotropic, a test of energy conservation will only be meaningful if the
radiation field is sampled at a sufficiently large number of
directions.  For this purpose, a special run was made with 88 cameras
(instead of the 13 typically used for the simulations presented here)
arranged isotropically around the galaxy.  When averaged over these
points, the emerging luminosity from the galaxy did indeed match the
intrinsic stellar luminosity to within $\sim 10^{ - 3 }$.

\subsection{Convergence of Hydrodynamic Simulations}

Once it has been verified that the radiative-transfer calculations are
converged, the question turns to the galaxy simulations themselves: Are
they converged with respect to particle number? It was verified in
\citet{coxetal05methods} that the global star-formation rate in the
simulations is converged with respect to particle number. However,
\citet{lotzetal08mergermorph} noted that the Gini coefficient
calculated from the images of the simulated galaxies indeed was
sensitive to particle number. This effect was due to the fact that
young star particles, which are very luminous and, for moderate
star-formation rates, few in number, carry enough individual light to
affect the Gini coefficient. When the particle number in the simulation
was increased by an order of magnitude, the luminosity of the young
stars were distributed over more particles, and the surface brightness
distribution of the galaxies became more uniform.

%NLX% exclude from vocabulary builder
\begin{figure*} \begin {center} \includegraphics*[width=0.49\textwidth]{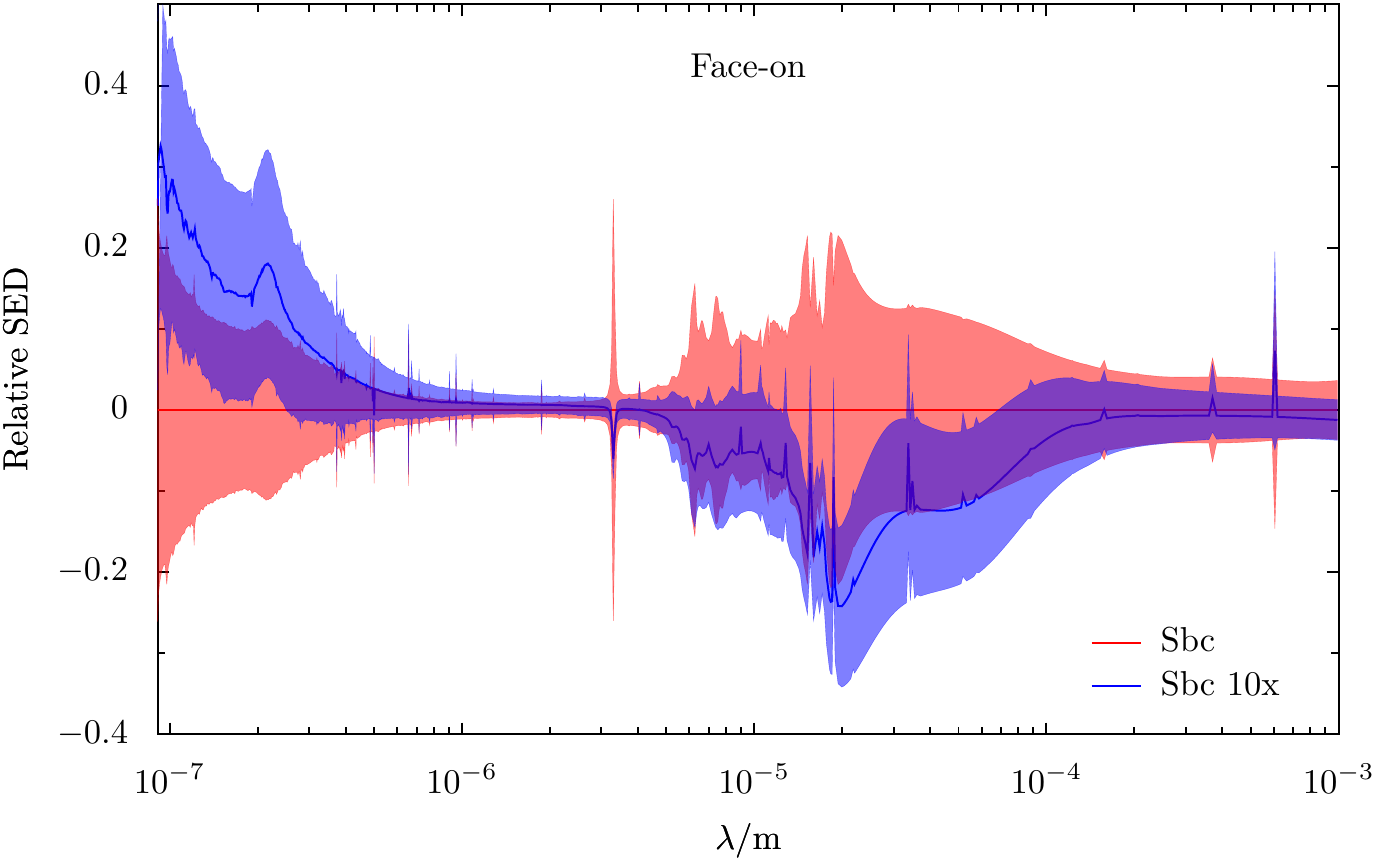} \includegraphics*[width=0.49\textwidth]{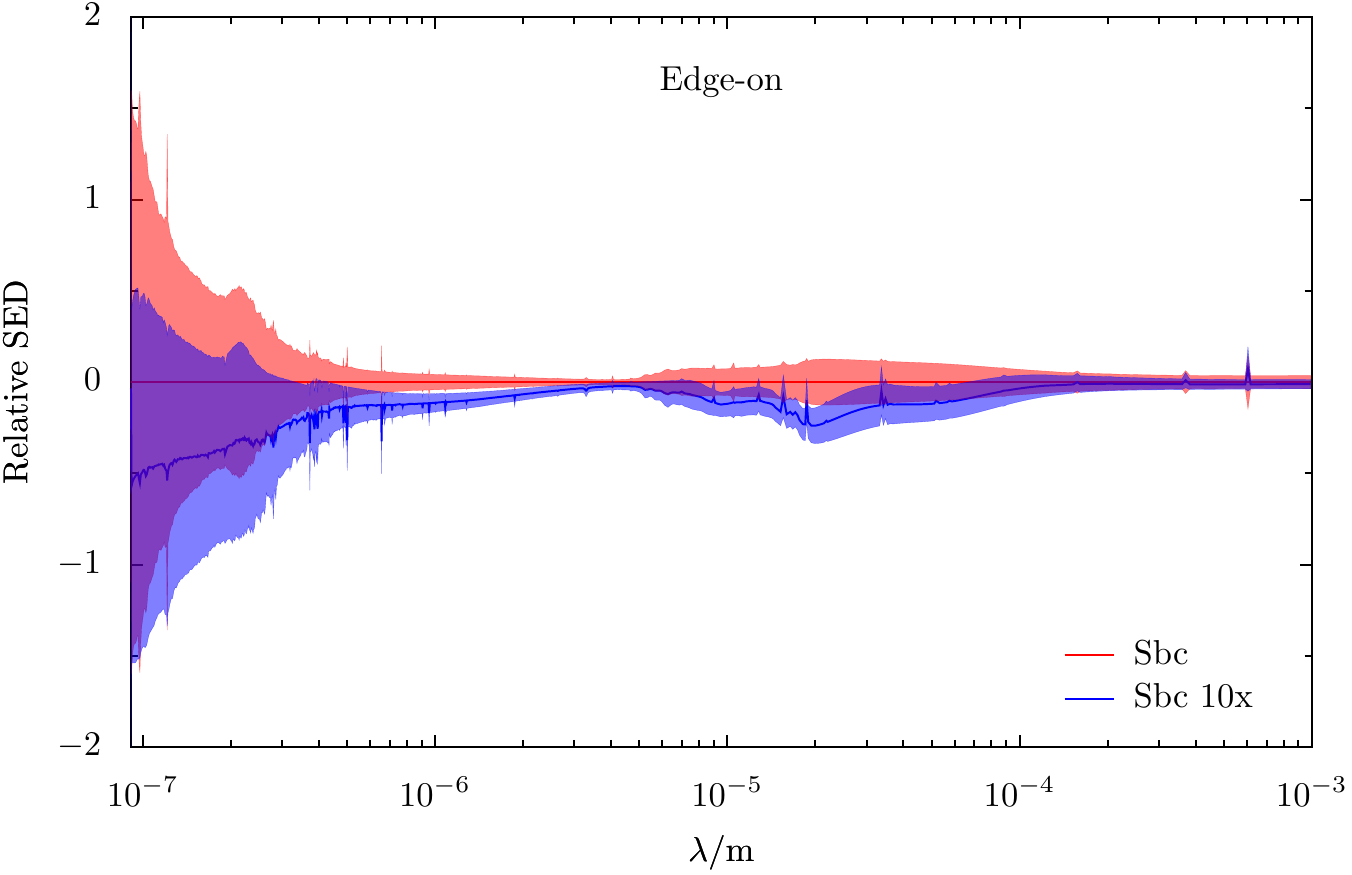} \end {center} \caption{ \label{plot_particle_resolution} The SED of the Sbc galaxy simulated with ten times as many particles (\emph{blue}) relative to the standard-resolution Sbc galaxy (\emph{red}). The left panel shows the face-on SED, the right panel the edge-on SED. In both panels both SEDs have been normalized to the mean (time-averaged) standard-resolution SED over $1\Gyr$, with the shaded region indicating the $1\sigma$ variance of the SEDs over this time. While the SEDs differ systematically by up to 30\% in the face-on and 50\% in the edge-on direction, this difference is generally of the same order as the internal variance of the simulations themselves. As the simulations evolve independently, they will have slightly different dynamical evolution, so such differences are to be expected. The only region where the two simulations appear to differ significantly compared to their internal variance is in the edge-on direction at near-IR wavelengths, where the higher-resolution simulation is about 10\% fainter, 2 -- 3 times their internal variance.} \end{figure*}
%NLX% end exclude from vocabulary builder

With this in mind, it is not obvious whether the SED of the galaxies
can be expected to be converged with respect to particle number. The
SEDs of the Sbc galaxy compared to that of a version with 10 times
larger number of particles is shown in
Figure~\ref{plot_particle_resolution}.  As these are different
simulations, with slightly different dynamical evolutions of the
galaxy, the comparison is done in relation to the variance of the SED
for the galaxies over their $1 \Gyr$ evolution.  The difference between
the standard and 10x higher resolution in the face-on direction is
consistent within about $1.5 \sigma$, though the higher-resolution
galaxy is systematically brighter in the UV and fainter in the
infrared. This indicates that the higher-resolution Sbc simulation has
slightly less dust attenuation compared to the standard-resolution
galaxy. Conversely, in the edge-on direction, the higher-resolution
galaxy is systematically fainter than its standard-resolution
counterpart. This difference is significant in the near-IR where it
amounts to $\sim 10 \%$, 2 -- 3 times larger than their internal
variation. This difference must be due to a difference in the dust
attenuation, as the differences of the intrinsic source SEDs of the two
simulations are much smaller.

Quite likely, the differences in the SEDs originate in the better
ability of the higher-resolution simulation to resolve the vertical
structure of the galaxy disks. A more puffed-up vertical structure in
the lower-resolution simulation would be expected to result in a higher
face-on and lower edge-on attenuation, as seen. Given the sensitivity
of the dust attenuation to resolving the vertical structure of the
galaxy disks, higher-resolution simulations would be desirable. New
hydrodynamic codes such as the unstructured-mesh code AREPO
\citep{springel09arepo} should also be more efficient at resolving
highly flattened structures, and it would be interesting to see the
impact of such simulations on the radiation-transfer calculations.

\section{Results}
\label{section_results}

To determine the validity of the models and to demonstrate its output
we have run \mcrx \ on our 7 simulated galaxies at 21 points in time
during their $1 \Gyr$ evolution (every $50 \Myr$) for 13 different
viewing angles ranging from face-on to edge-on to face-on in the
opposite direction. Each run results in a two-dimensional, far-UV to mm
wavelength SED for the galaxy from that viewing angle, that can be used
to create specific band pass images or integrated galaxy SEDs.

\begin{figure*} \begin {center} \begin {tabular} {lcccc} Galaxy & $0^\circ$ & $63^\circ$ &$85^\circ$ &$90^\circ$ \\ \raisebox{0.07\textwidth}{Sbc+} & \includegraphics*[width=0.20\textwidth]{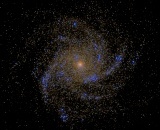} & \includegraphics*[width=0.20\textwidth]{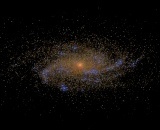} & \includegraphics*[width=0.20\textwidth]{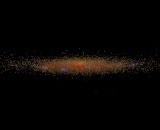} & \includegraphics*[width=0.20\textwidth]{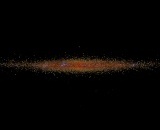} \\ \raisebox{0.07\textwidth}{Sbc} & \includegraphics*[width=0.20\textwidth]{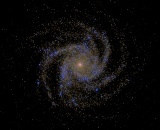} & \includegraphics*[width=0.20\textwidth]{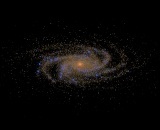} & \includegraphics*[width=0.20\textwidth]{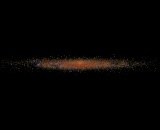} & \includegraphics*[width=0.20\textwidth]{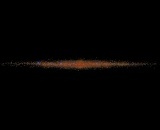} \\ \raisebox{0.07\textwidth}{Sbc$-$} & \includegraphics*[width=0.20\textwidth]{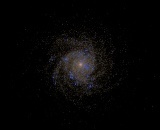} & \includegraphics*[width=0.20\textwidth]{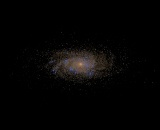} & \includegraphics*[width=0.20\textwidth]{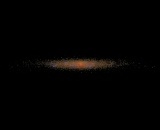} & \includegraphics*[width=0.20\textwidth]{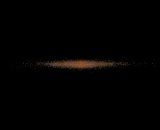} \\ \raisebox{0.07\textwidth}{G3} & \includegraphics*[width=0.20\textwidth]{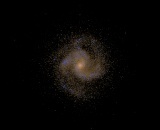} & \includegraphics*[width=0.20\textwidth]{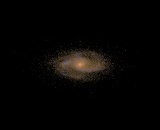} & \includegraphics*[width=0.20\textwidth]{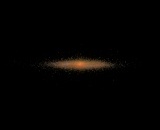} & \includegraphics*[width=0.20\textwidth]{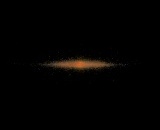} \\ \raisebox{0.07\textwidth}{G2} & \includegraphics*[width=0.20\textwidth]{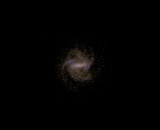} & \includegraphics*[width=0.20\textwidth]{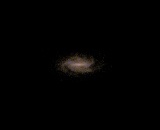} & \includegraphics*[width=0.20\textwidth]{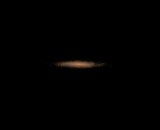} & \includegraphics*[width=0.20\textwidth]{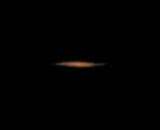} \\ \raisebox{0.07\textwidth}{G1} & \includegraphics*[width=0.20\textwidth]{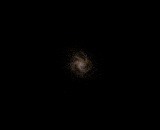} & \includegraphics*[width=0.20\textwidth]{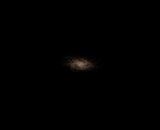} & \includegraphics*[width=0.20\textwidth]{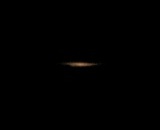} & \includegraphics*[width=0.20\textwidth]{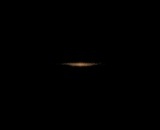} \\ \raisebox{0.07\textwidth}{G0} & \includegraphics*[width=0.20\textwidth]{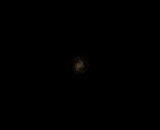} & \includegraphics*[width=0.20\textwidth]{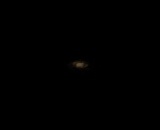} & \includegraphics*[width=0.20\textwidth]{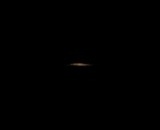} & \includegraphics*[width=0.20\textwidth]{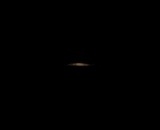} \\ \end{tabular} \end {center} \caption{ \label{plot_galaxy_images} SDSS $urz$ colour composite images of the simulated galaxies at $0.5\Gyr$. Each row shows a different simulated galaxy, while the columns show different inclinations.  The horizontal extent of the images is $60\kpc$. The images were generated with the algorithm of \citet{lupton03}. Bright, blue regions are regions of star formation. In the larger galaxies, a red dust lane is clearly visible in the edge-on view. } \end{figure*}

\begin{figure*} \begin {center} \begin {tabular} {cccc} GALEX FUV & GALEX NUV & B & V \\ \includegraphics*[width=0.22\textwidth]{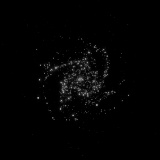} & \includegraphics*[width=0.22\textwidth]{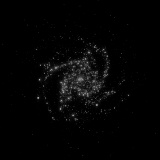} & \includegraphics*[width=0.22\textwidth]{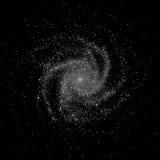} & \includegraphics*[width=0.22\textwidth]{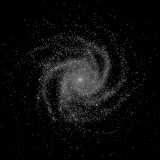} \\ R & I & J & K$_\mathrm{s}$ \\ \includegraphics*[width=0.22\textwidth]{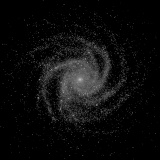} & \includegraphics*[width=0.22\textwidth]{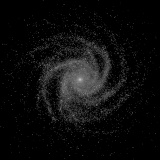} & \includegraphics*[width=0.22\textwidth]{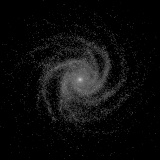} & \includegraphics*[width=0.22\textwidth]{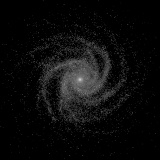} \\ IRAC1 $3.6\micron$ & IRAC2 $4.5\micron$ & IRAC3 $5.8\micron$ & IRAC4 $8.0\micron$ \\ \includegraphics*[width=0.22\textwidth]{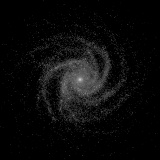} & \includegraphics*[width=0.22\textwidth]{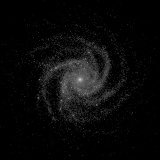} & \includegraphics*[width=0.22\textwidth]{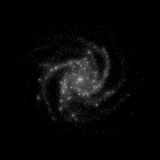} & \includegraphics*[width=0.22\textwidth]{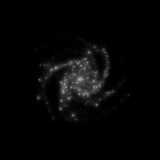} \\ MIPS $24\micron$ & MIPS $70\micron$ & MIPS $160\micron$ & SCUBA $850\micron$ \\ \includegraphics*[width=0.22\textwidth]{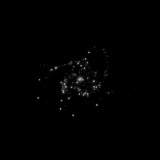} & \includegraphics*[width=0.22\textwidth]{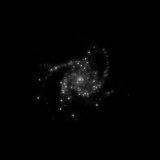} & \includegraphics*[width=0.22\textwidth]{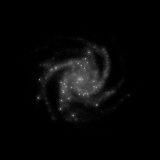} & \includegraphics*[width=0.22\textwidth]{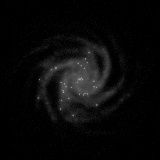} \\ \end{tabular} \end {center} \caption{ \label{plot_multiwavelength} The Sbc galaxy (at $0.5\Gyr$) shown at wavelengths from the GALEX FUV band to the SCUBA $850\micron$ band. The images are $60\kpc$ in scale. Instrumental resolution effects are not considered, and the stretch of each image has been adjusted to the maximum surface brightness in the band.  The difference in morphology between the GALEX/MIPS $24\micron$ bands dominated by the star-forming regions, the optical-NIR bands dominated by the older stellar population, and the FIR MIPS $160\micron$ and SCUBA bands dominated by diffuse dust emission, is striking.  } \end{figure*}

\begin{figure*} \begin {center} \begin {tabular} {cc} Sbc+ & Sbc \\ \includegraphics*[width=0.45\textwidth]{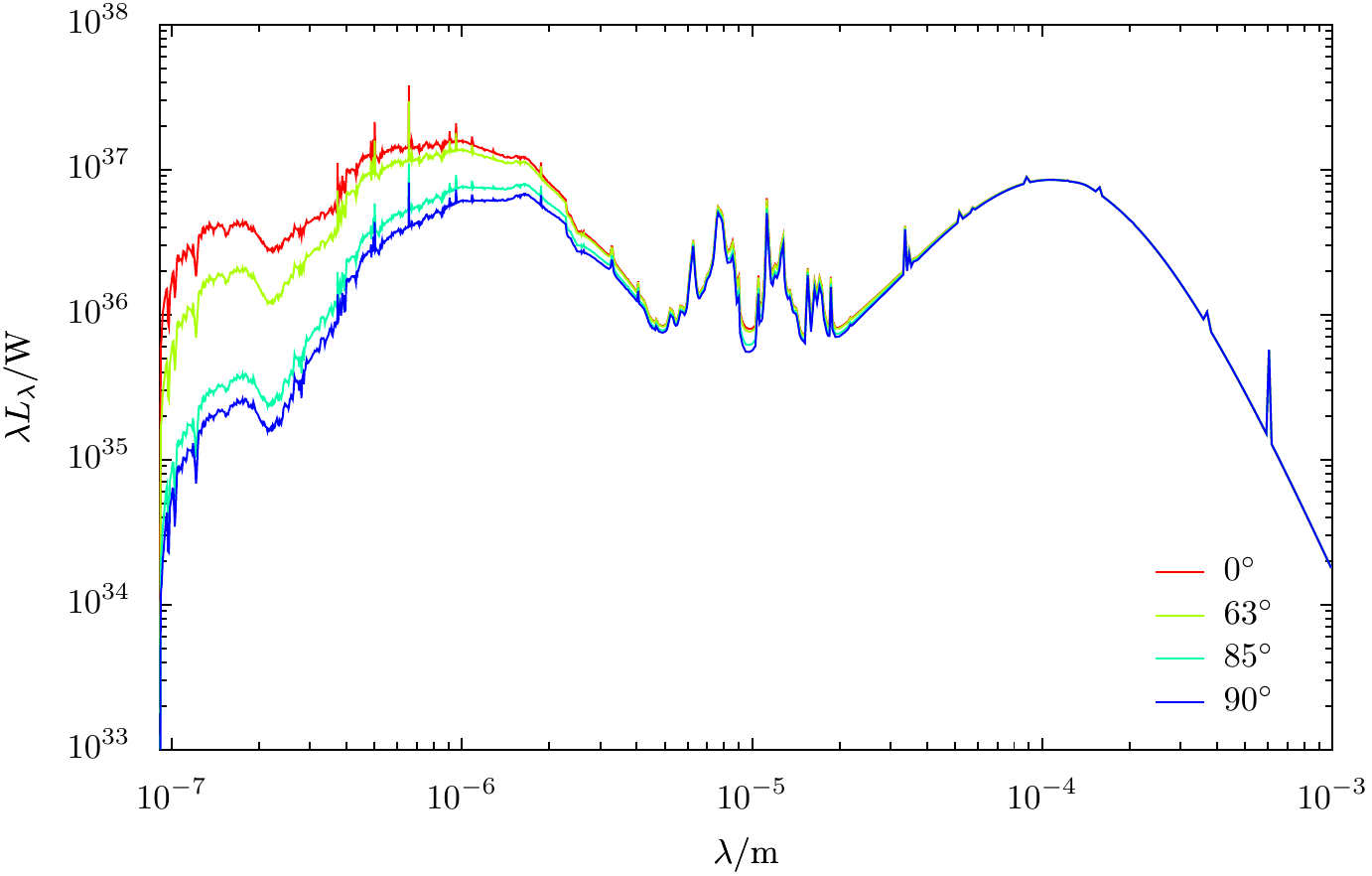} & \includegraphics*[width=0.45\textwidth]{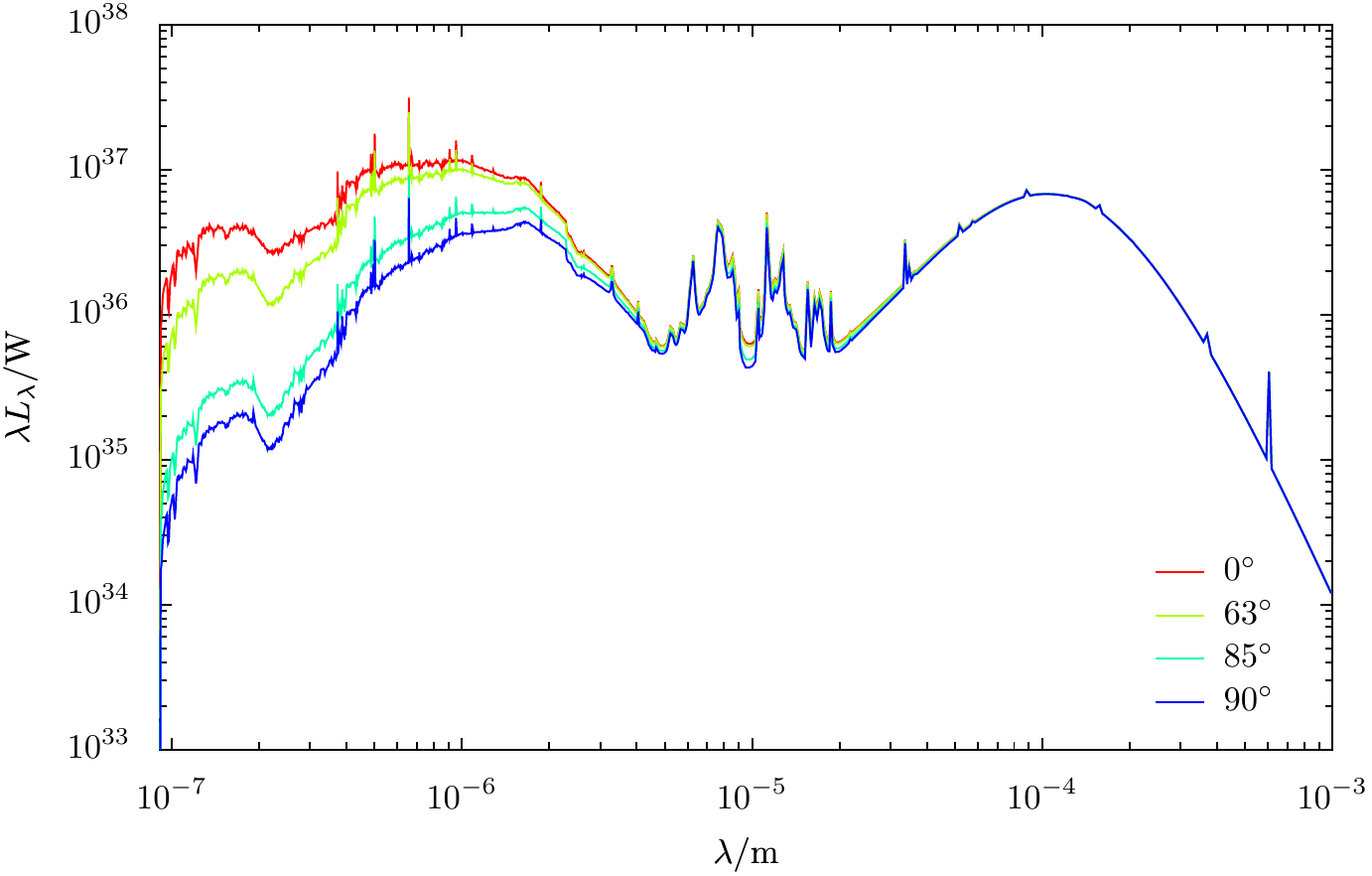} \\ Sbc$-$ & G3 \\ \includegraphics*[width=0.45\textwidth]{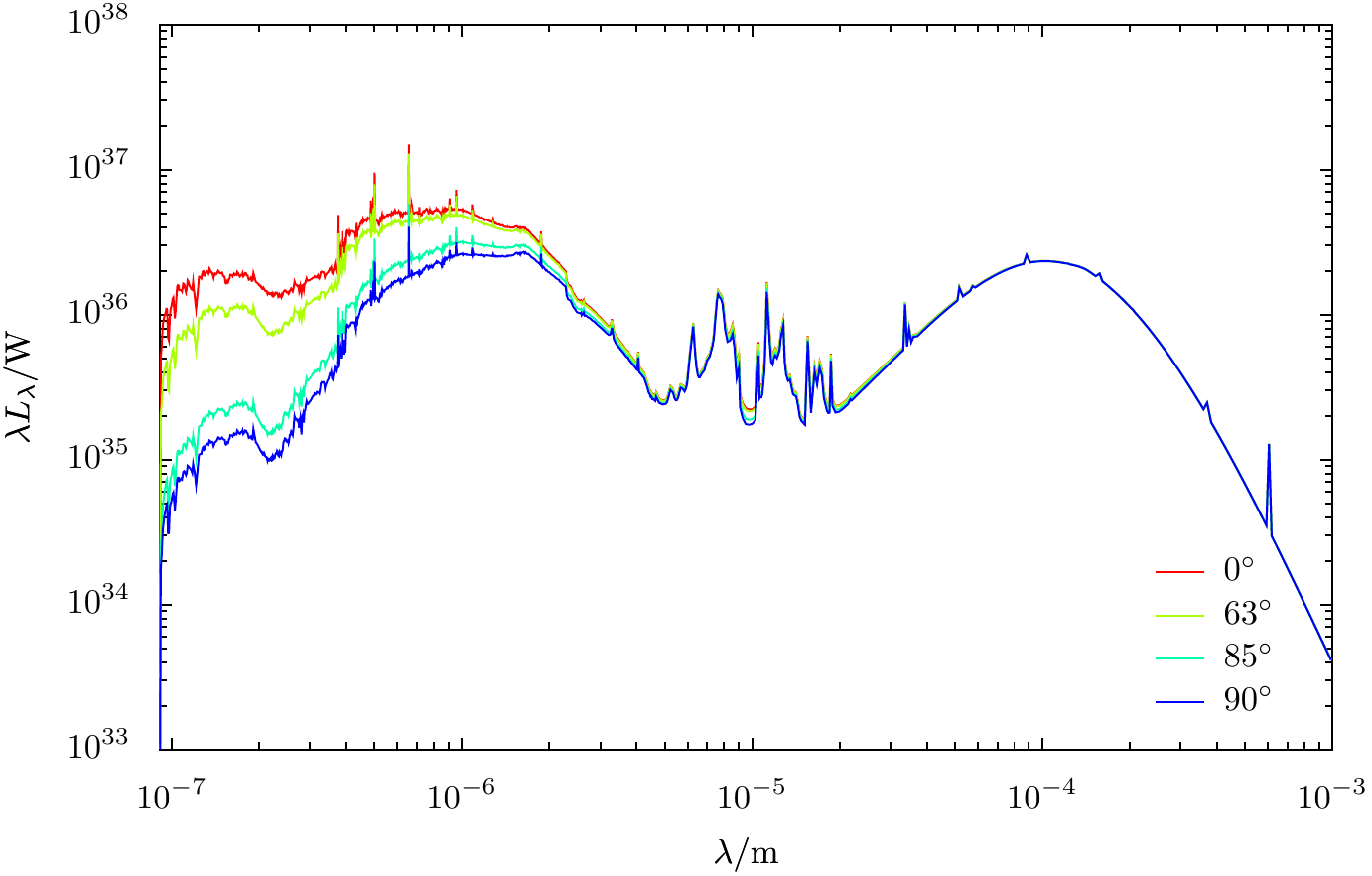} & \includegraphics*[width=0.45\textwidth]{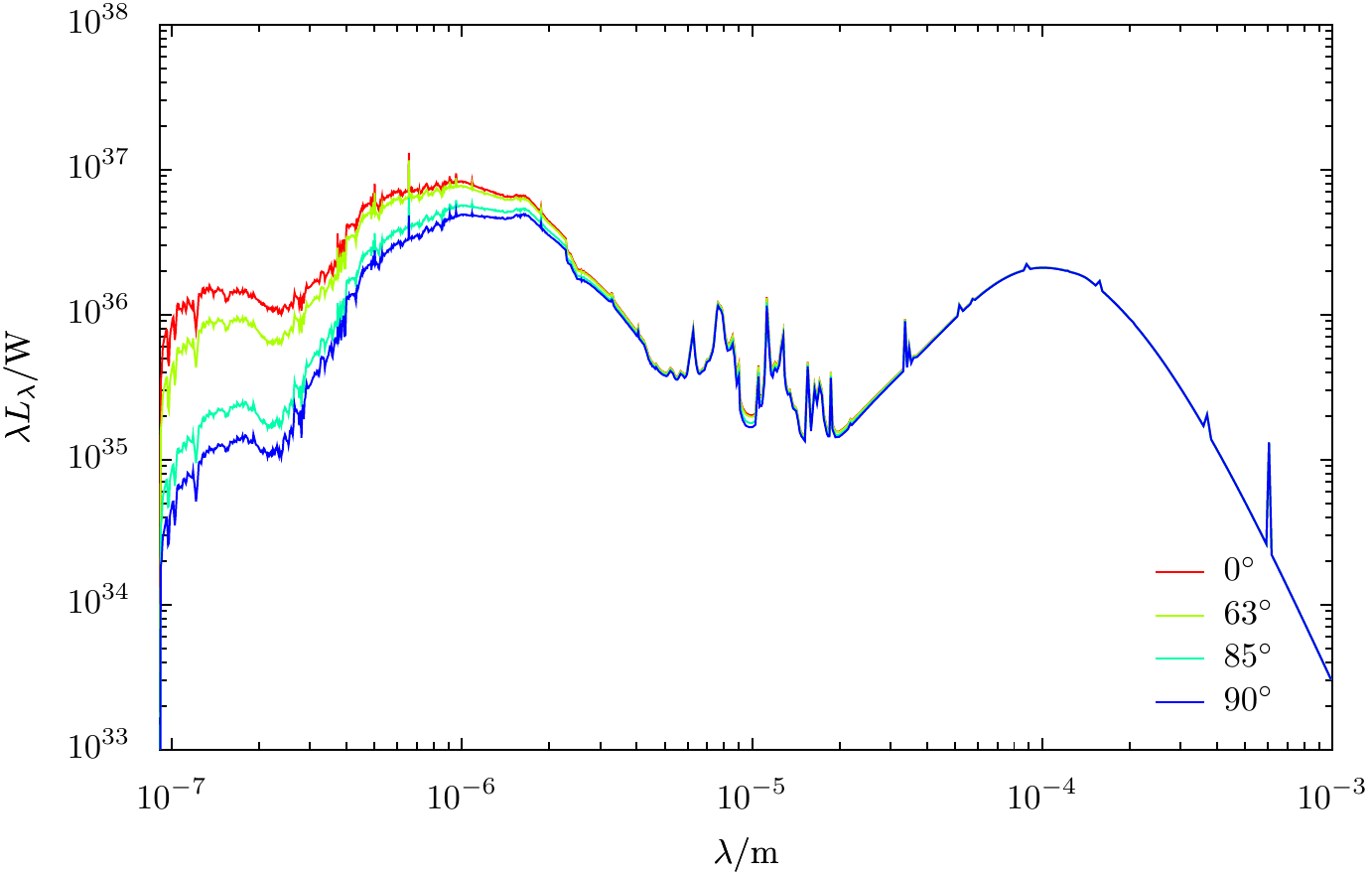} \\ G2 & G1 \\ \includegraphics*[width=0.45\textwidth]{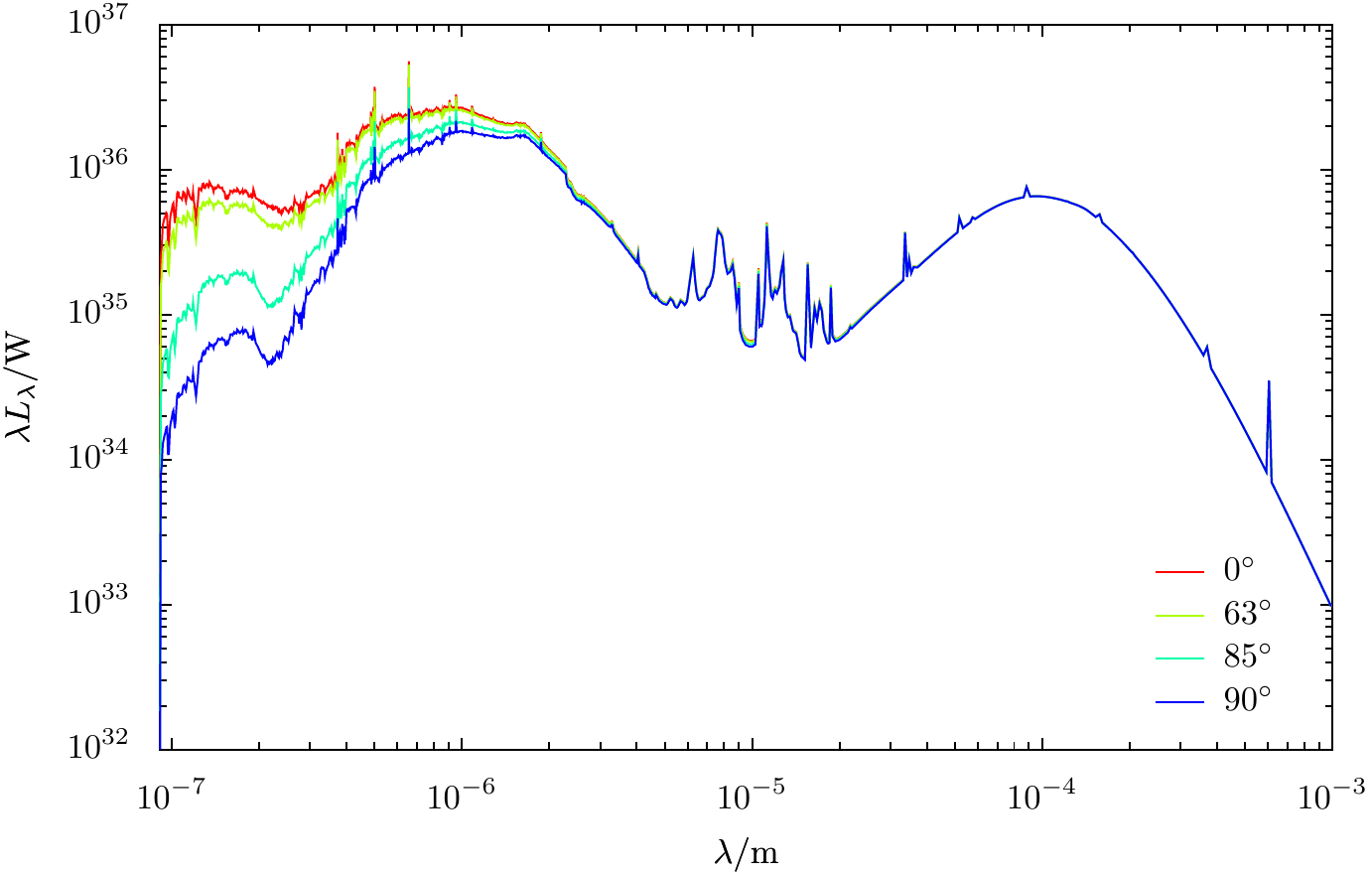} & \includegraphics*[width=0.45\textwidth]{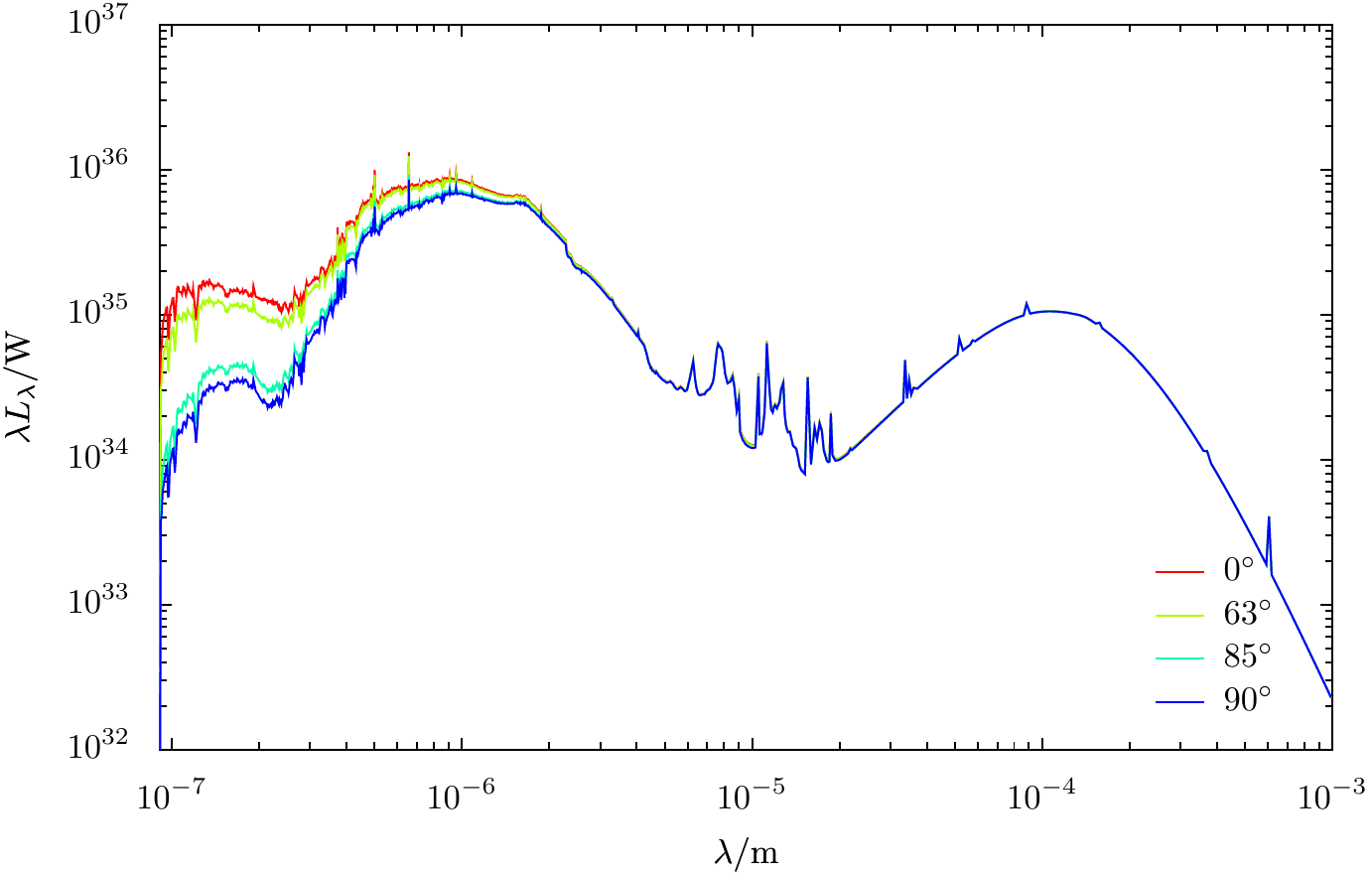} \\ G0 & G0 (time variation, face-on view) \\ \includegraphics*[width=0.45\textwidth]{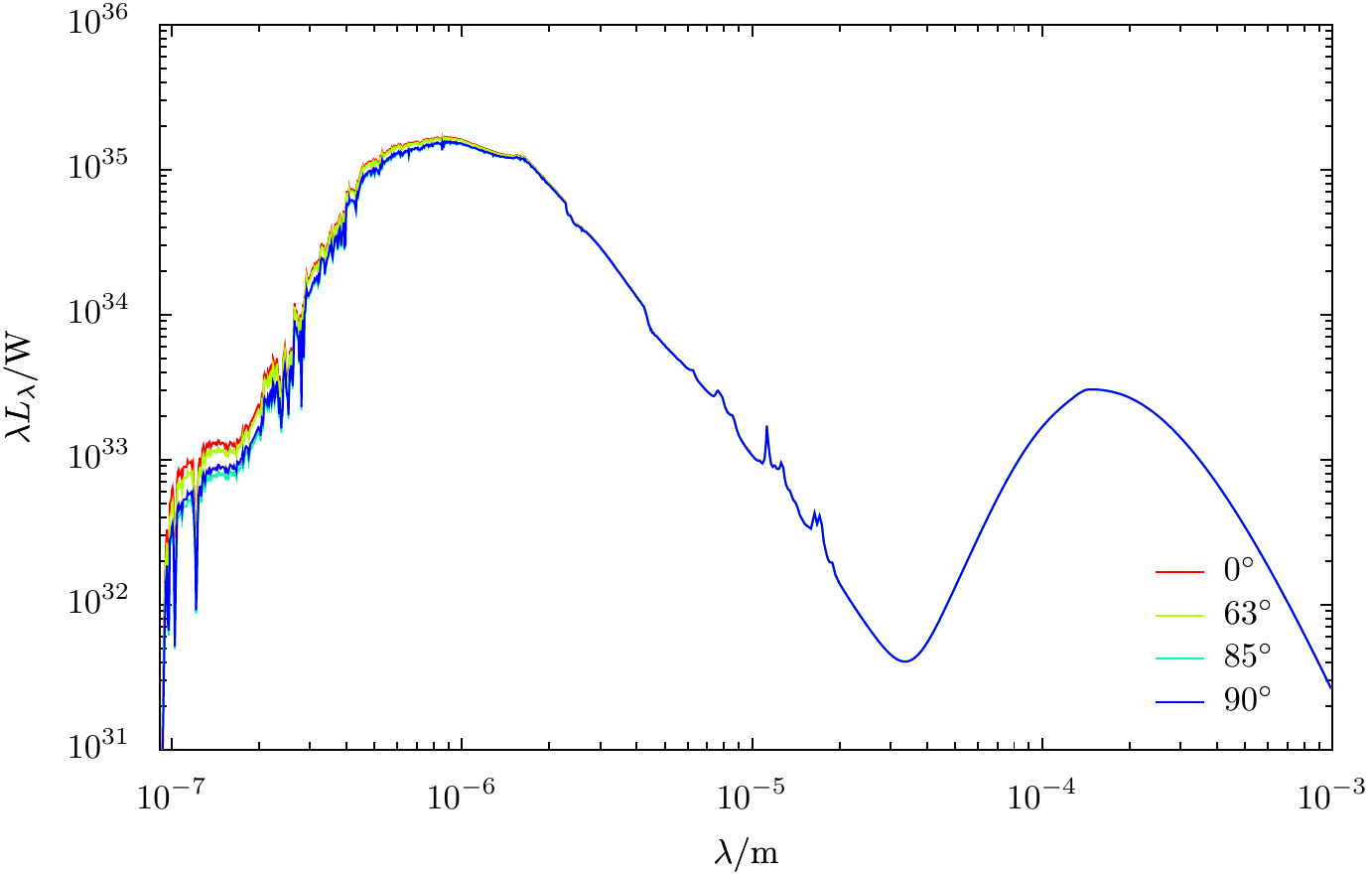} & \includegraphics*[width=0.45\textwidth]{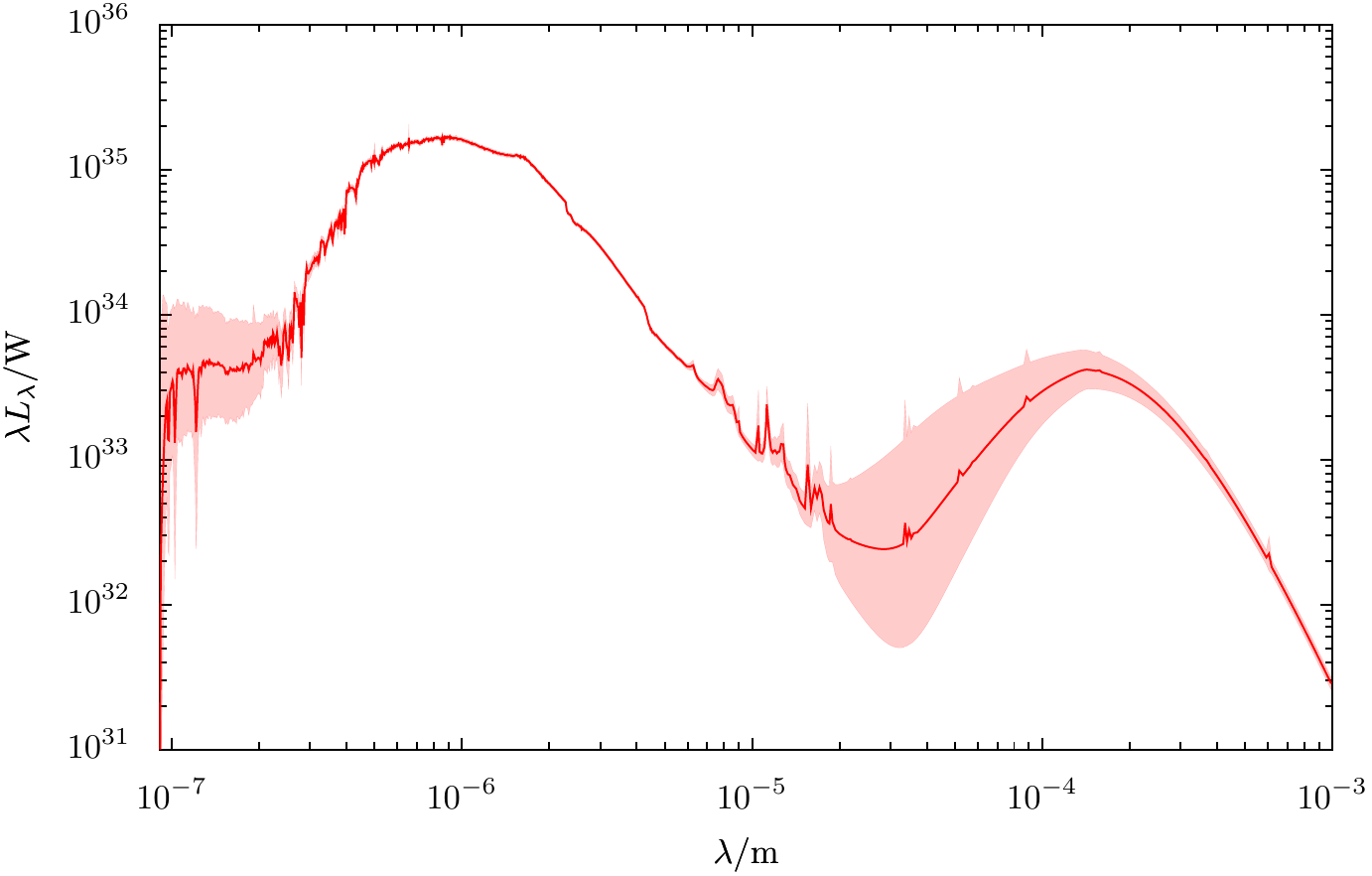} \\ \end{tabular} \end {center} \caption{ \label{plot_seds_bygalaxy} The SEDs of the 7 simulated galaxies for 4 inclinations. The UV attenuation is substantially higher in the edge-on configuration for all but the smallest galaxies. At wavelengths longer than $5\um$, the SED is unaffected by inclination except for the silicate absorption feature at $10\um$. Due to the very low star-formation rate of the G0 galaxy, its SED varies substantially in time, depending on whether a young star particle is present. This variation ($1\sigma$) is shown in the bottom-right panel. } \end{figure*}

To familiarize the reader with the model outputs, we first present a
comprehensive gallery of images and spectra of the 7 simulated
galaxies. Figure~\ref{plot_galaxy_images} shows colour-composite images
of the galaxies at 4 different inclinations, while
Figure~\ref{plot_multiwavelength} shows images of the Sbc galaxy at 16
different passbands from GALEX FUV to SCUBA $850 \um$.
Figure~\ref{plot_seds_bygalaxy} shows the integrated SED of each galaxy
as a function of inclination. 
%% (DONT THINK WE NEED THE FOLLOWING PLOT, SEEING AS WE HAVE (refer to:
%% plot_seds_bygalaxy), BUT DONT FEEL STRONGLY AGAINST IT EITHER)
%% , while Figure~(refer to: plot_seds_bydirection) shows the
%% SEDs of the different galaxy models in the face-on and edge-on
%% directions.

All galaxies in Figure~\ref{plot_galaxy_images} have been evolved for
$0.5 \Gyr$, and the physical extent of the images is $60 \kpc$, which
clearly demonstrates the decrease in scale from the Sbc$+$ to G0
galaxies. The images simulate closely the SDSS ``postage stamps'', as
they use the SDSS $urz$-bands to make the colour composite images using
the algorithm of \citet{lupton03}.  In fact, their appearance is
strikingly similar to real galaxies: Star-forming regions outline the
spiral arms, a yellowish bulge is present in the centre, and, when
viewed edge-on, the large galaxies show a prominent, reddened dust
lane. The smaller G1 and G0 galaxies have noticeably lower surface
brightnesses than the larger ones, and show no obvious signs of dust
attenuation. (Incidentally, this is largely consistent with findings of
\citet{dalcantonetal04} that dust lanes are prevalent in spiral
galaxies only when rotational velocities are greater than $120
kilometres per second$.)

In Figure~\ref{plot_multiwavelength} we concentrate on the Sbc galaxy
(also at $0.5 \Gyr$ with a $60 \kpc$ image size), demonstrating the
strong morphological differences between different wavelength
observations. Note that the resolution in the images is due to the
model itself, with the ``typical'' resolution of instruments that
observe at each wavelength not considered.  In the FUV, the images are
dominated by regions of recent star formation. These trace the spiral
arms with almost no indication of the general shape of the galaxy. This
appearance resembles many of the real galaxies imaged with GALEX,
except that the limited number of young stellar particles in the
simulations give the model galaxies a more ``speckled'' appearance.

In the optical bands, the star-forming regions decline in importance
and the older stellar population dominates the shape of the galaxy.
This continues into the NIR up to wavelengths longer than $\sim 4 \um$,
beyond which PAH emission begins to outline the star-forming regions
again. In the $5.8 \um$ and $8.0 \um$ bands, PAH emission excited by UV
emission from young stars trace the spiral arms and star-forming
regions. In the MIPS $24 \um$ band, emission from hot dust in the
star-forming regions dominates the appearance, while in the far-IR,
cooler, more diffuse dust makes the galaxy look progressively smoother.

Figure~\ref{plot_seds_bygalaxy} supports the previous figures, showing
the full UV to sub-mm SEDs at the same 4 inclinations of
Figure~\ref{plot_galaxy_images}. In all 7 galaxies, the effects of
inclination are clearly visible in the optical-UV, with extinction
increasing with inclination due to the dust in the disks. Conversely,
for all galaxies the IR is little affected by inclination due to its
isotropic emission and low opacity, with the exception of the strong
$\sim 10 \um$ silicate feature in the Sbc galaxies (this feature is
also reflected in the bias factors in Figure~\ref{plot_bias}).  As in
Figure~\ref{plot_galaxy_images} the differences between the galaxies
are visible, with the decrease in both SFR and dust content visible
through the decreasing equivalent width of the emission lines and
decreasing dust effects (extinction and emission) as we go from the
massive Sbc galaxies to the small G-series. In fact, in the smallest G0
galaxy, the SFR is so low that dependence on inclination is dwarfed by
the the time variation, as shown in the bottom-right of
Figure~\ref{plot_seds_bygalaxy}.

\begin{figure*} \begin {center} \includegraphics*[width=0.49\textwidth]{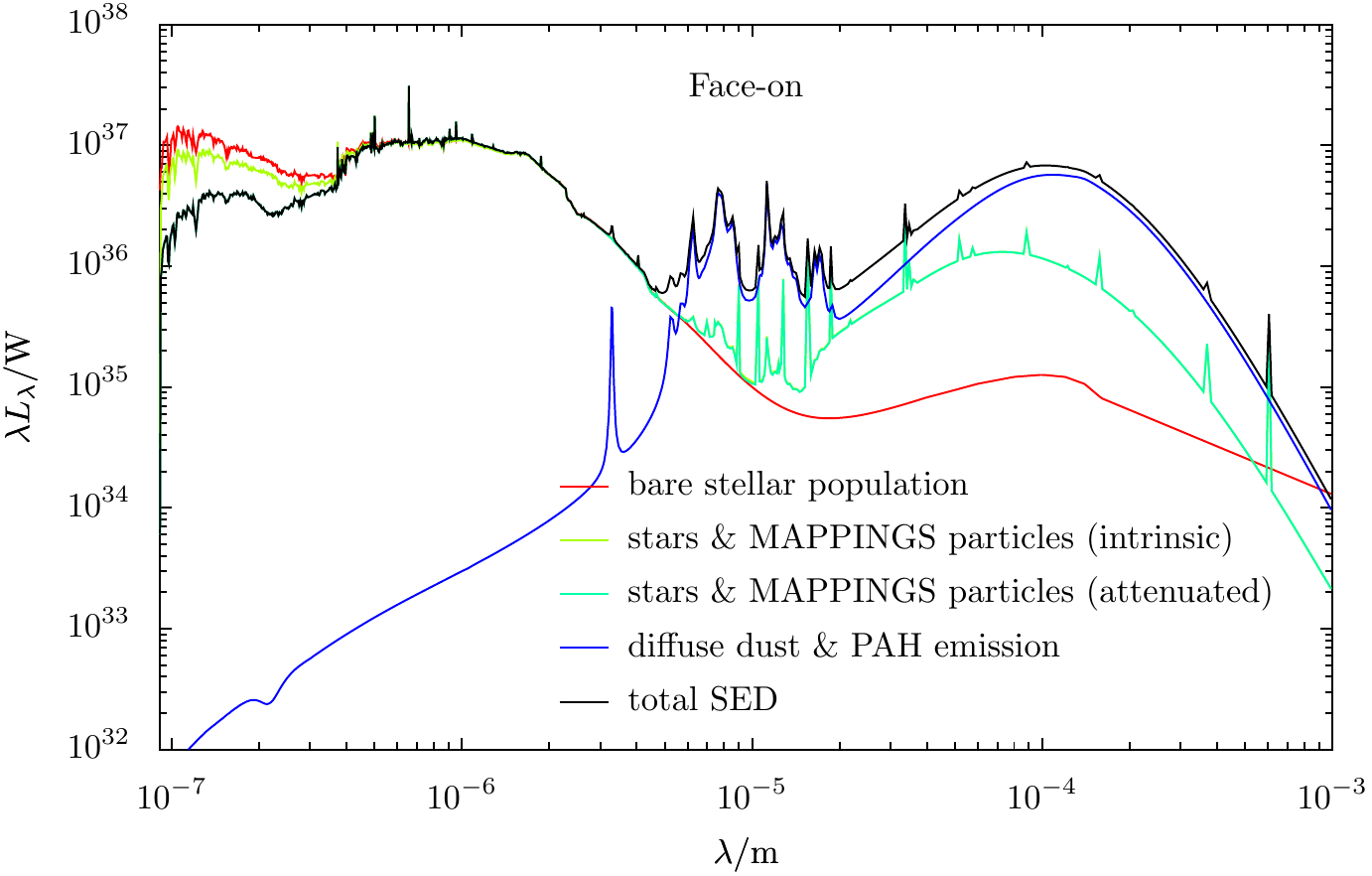} \includegraphics*[width=0.49\textwidth]{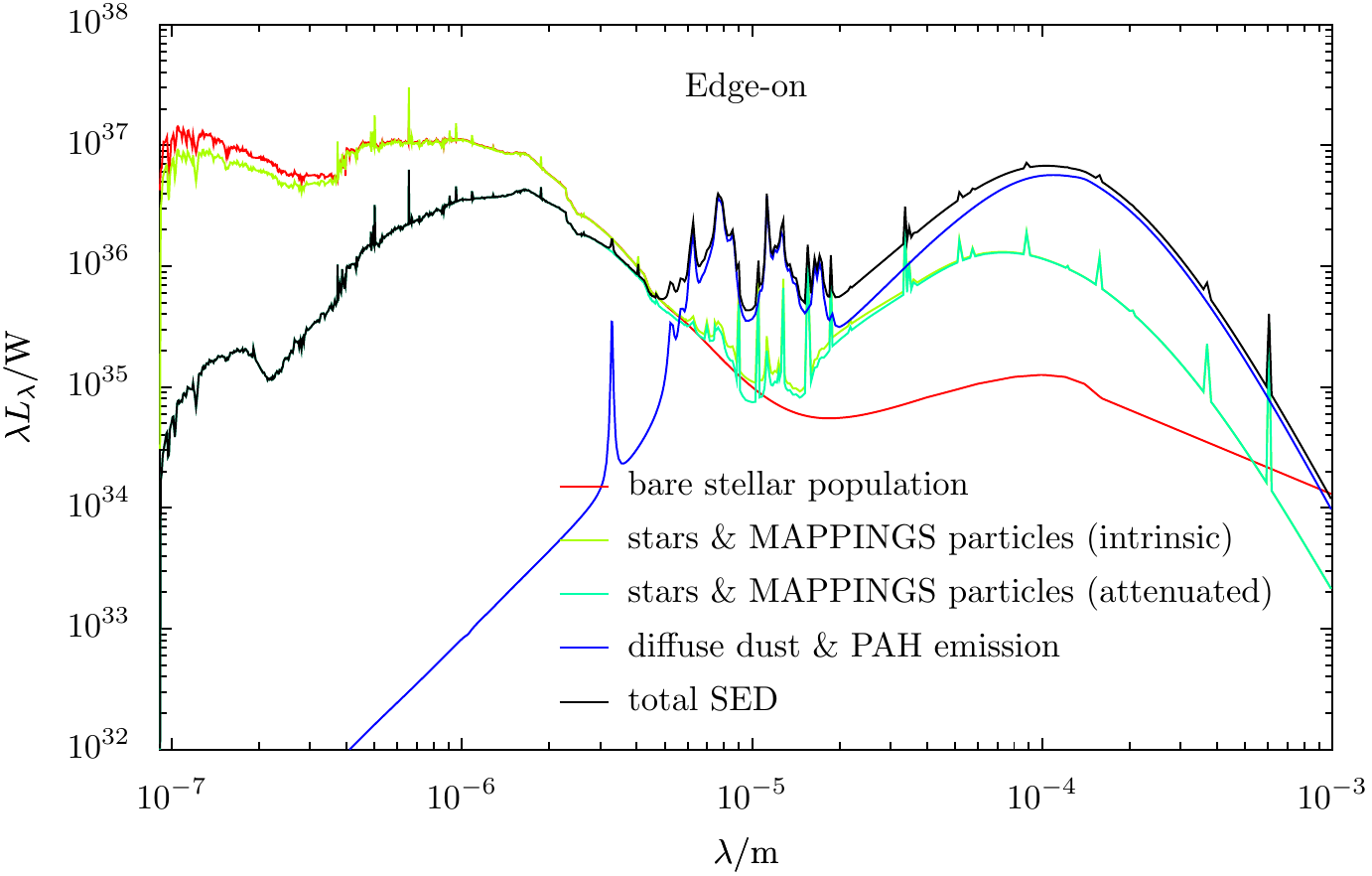} \end {center} \caption{ \label{plot_sed_contributions} The SED of the simulated Sbc galaxy viewed face-on (left panel) and edge-on (right panel), with the different components of the SED separated. The red line shows the intrinsic emission of the bare stellar population; the green line shows the intrinsic emission of the source particles (i.e.~after the young, $<10 \Myr$ old, stellar population has undergone radiative transfer through \mapiii ); the cyan line shows the attenuated source spectrum making it to the observer through the dust in the galaxy. The emission from the diffuse dust (including self-absorption) is shown by the blue line, and the black line is the total emerging spectral energy distribution of the galaxy. It is clear in this figure that the impact of the attenuation of the young stars in the \mapiii\ particles is quite small and that, for this galaxy, the infrared spectrum is dominated by diffuse dust emission at all wavelengths. The contributions from star-forming regions is most important at wavelengths around $20\micron$, the region dominated by hot dust. The strange-looking intrinsic stellar emission at wavelengths $>10\micron$ is from the poorly approximated nebular continuum in Starburst99, but does not affect the results.} \end{figure*}

The contributions from primary emission (the star and \mapiii \
particles) and from heated dust grains in the diffuse ISM that make up
the SEDs in Figure~\ref{plot_seds_bygalaxy} are shown separately for
the Sbc galaxy (edge-on and face-on) in
Figure~\ref{plot_sed_contributions}.  This figure nicely demonstrates
the ``step-by-step'' radiative process, and the wavelength-dependent
relative contributions.  At all wavelengths longer than $\sim 4 \um$,
where PAH emission begins to dominate over stellar emission, the SED is
dominated by emission from the diffuse ISM. The contribution from
star-forming regions, in the form of \mapiii \ particles, is most
important around $20 \um$ where it contributes almost half of the
emission. Since the emission from star-forming regions is concentrated
into discrete regions, these regions have high surface brightness and
are prominent in the images of the MIR emission, as seen in
Figure~\ref{plot_multiwavelength}.

%NLX% exclude from vocabulary builder
%% \setcounter{bottomnumber}{4}
%% \setcounter{topnumber}{4}
%% \setcounter{dbltopnumber}{4}
%% \renewcommand{\topfraction}{0.8}
%% \renewcommand{\dbltopfraction}{0.8}
%% \renewcommand{\bottomfraction}{0.8}
%% \renewcommand{\textfraction}{0.1}
%% \renewcommand{\floatpagefraction}{0.75}
%% \renewcommand{\dblfloatpagefraction}{0.75}
 
%NLX% end exclude from vocabulary builder

%NLX% exclude from vocabulary builder
 %% \begin{figure*}
%% \begin {center}
%% \includegraphics*[width=0.49\textwidth]{faceon-seds}
%% \includegraphics*[width=0.49\textwidth]{edgeon-seds}
%% \end {center}
%% \caption{ \label{plot_seds_bydirection} The SEDs of the simulated
%%   galaxies, but all galaxies plotted for face-on (left) and edge-on
%%   (right) inclinations. (DROP FIGURE?)}
%% \end{figure*}
%NLX% end exclude from vocabulary builder

%NLX% exclude from vocabulary builder
%NLX% end exclude from vocabulary builder

\subsection{Comparing the Simulations with Observed Galaxies}
 \label{section_sings-comparison}

While the simulated galaxies were not created to be exact replicas of
existing galaxies, it is still worthwhile to test whether the models of
the simulated galaxies really produce realistic SEDs. To do this, the
simulated galaxies were compared to the galaxies in the SINGS sample of
nearby galaxies \citep[][hereafter D07]{daleetal06sings}. The D07
catalogue contains photometry in 17 bands from the GALEX FUV to Spitzer
MIPS $160 \um$ and, in some cases, $850 \um$ from SCUBA for 75 nearby
galaxies, making it possible to test the simulations against a
homogeneous dataset over a very broad wavelength range. To extend the
number of galaxies with $850 \um$ SCUBA data, the SLUGS sample of
\citet{willmeretal09} is also included when comparing the far-infrared
and submillimeter data.

As mention in the previous section, the simulations have been observed
at 21 points in time, each viewed from 13 different angles. For the 7
simulated galaxies this amounts to a total of 1,911 SEDs, making it
infeasible to plot individual points. The results are instead presented
as density plots where the shade represents the density of simulated
galaxy observations in that region. The probability of observing the
simulated galaxies has been weighted appropriately based on time and
solid angle, but no attempt has been made to adjust for the abundance
of galaxies of different masses. The simulated data points can thus be
expected to over-emphasize large galaxies over small. However, the
SINGS sample is also not a statistically representative sample of
galaxies, but rather was selected to cover a wide range of
morphologies, luminosities and infrared-to-optical ratios
\citep{daleetal06sings}. The comparison should thus be done to
ascertain whether the simulations lie in a region of parameter space
occupied by real galaxies, rather than matching the detailed
distributions of the simulated and real galaxies.

In all the following plots we use the given data values from D07, but
do not indicate the uncertainties given in the paper, which range from
$< 10 \%$ for the optical-NIR data up to $\sim 50 \%$ for the $850 \um$
data (see D07 for exact values). In addition, aperture correction
factors were applied to the longer wavelengths, as described in D07,
with factors of 2 for some $850 \um$ observations, increasing possible
uncertainties.

\begin{figure*} \begin {center} \begin{tabular*}{\textwidth}{r@{\extracolsep{\fill}}r} \includegraphics*[scale=0.62]{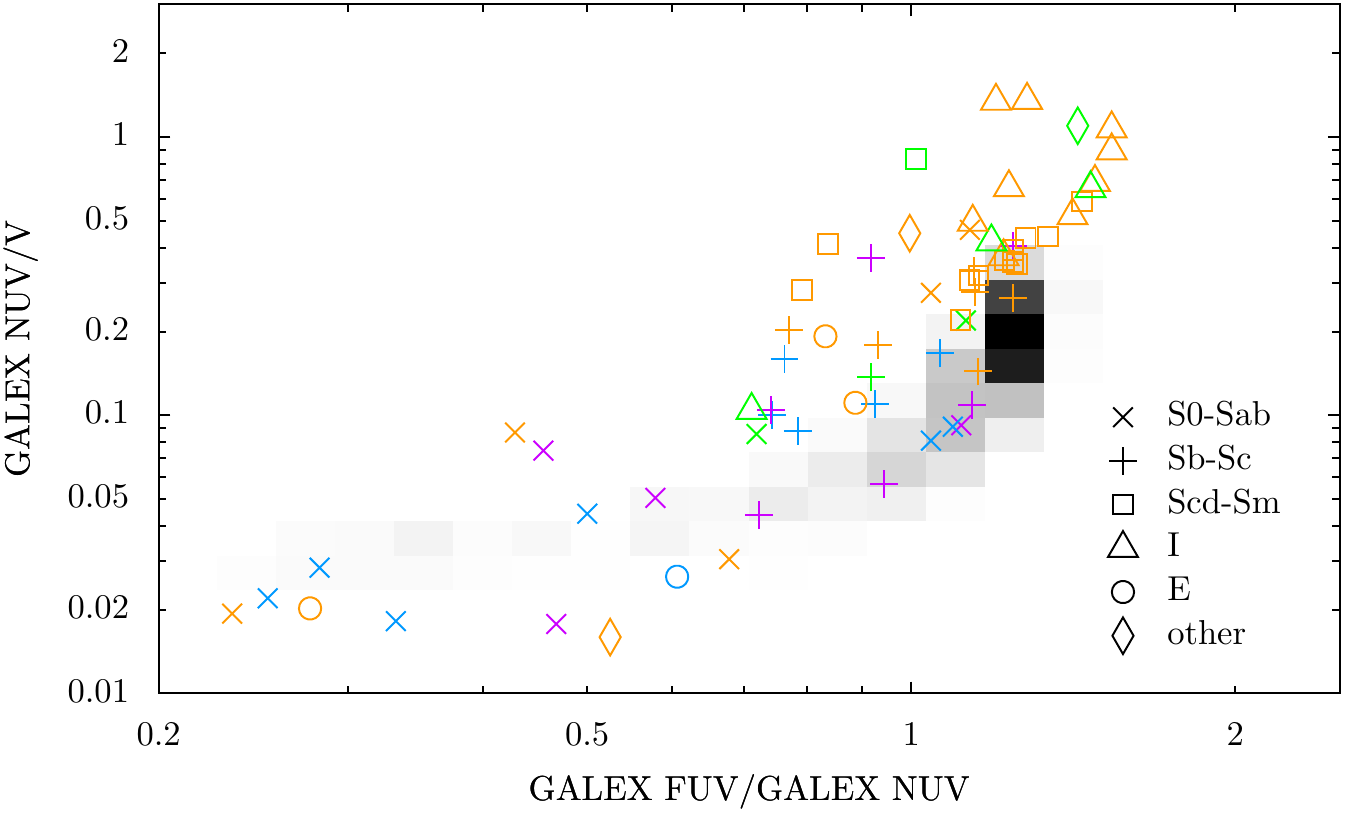} & \includegraphics*[scale=0.62]{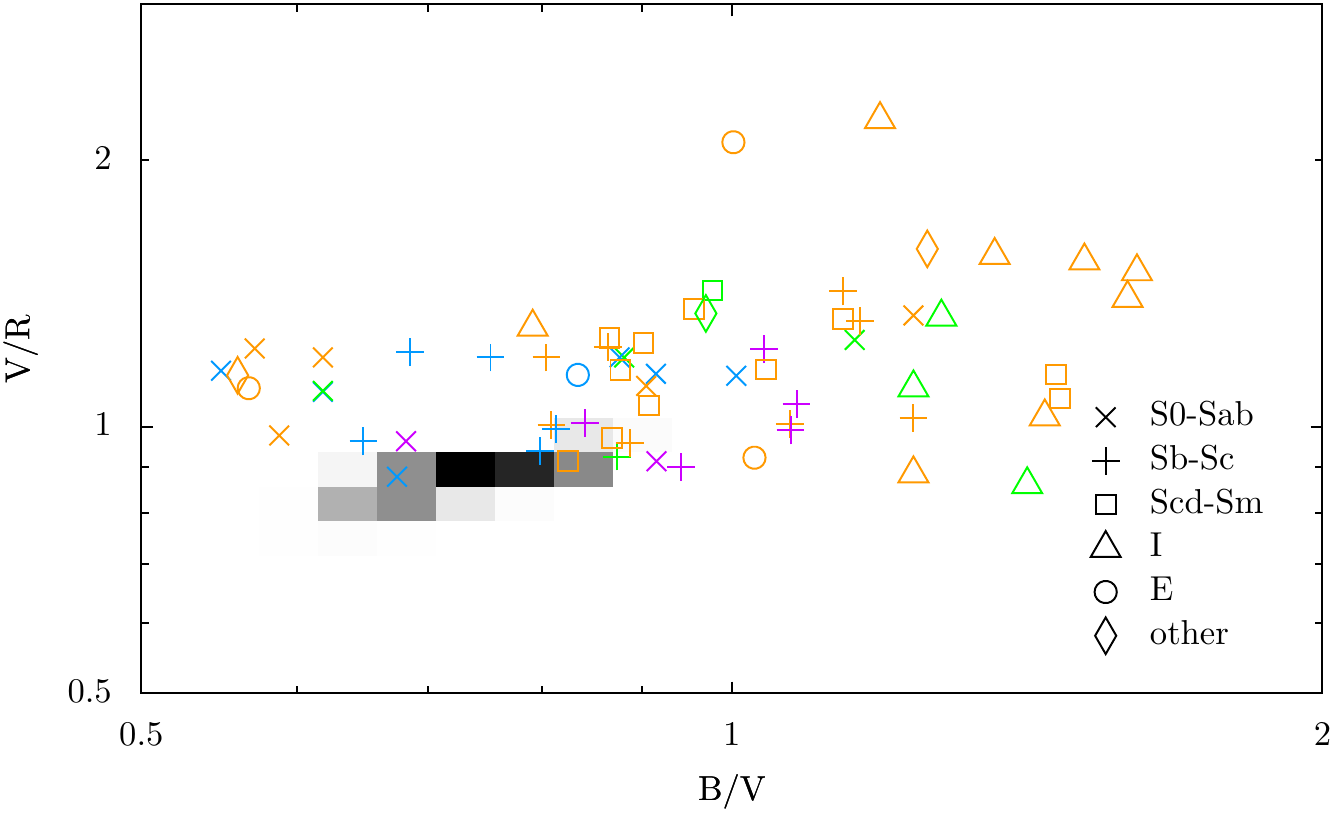} \\ \includegraphics*[scale=0.62]{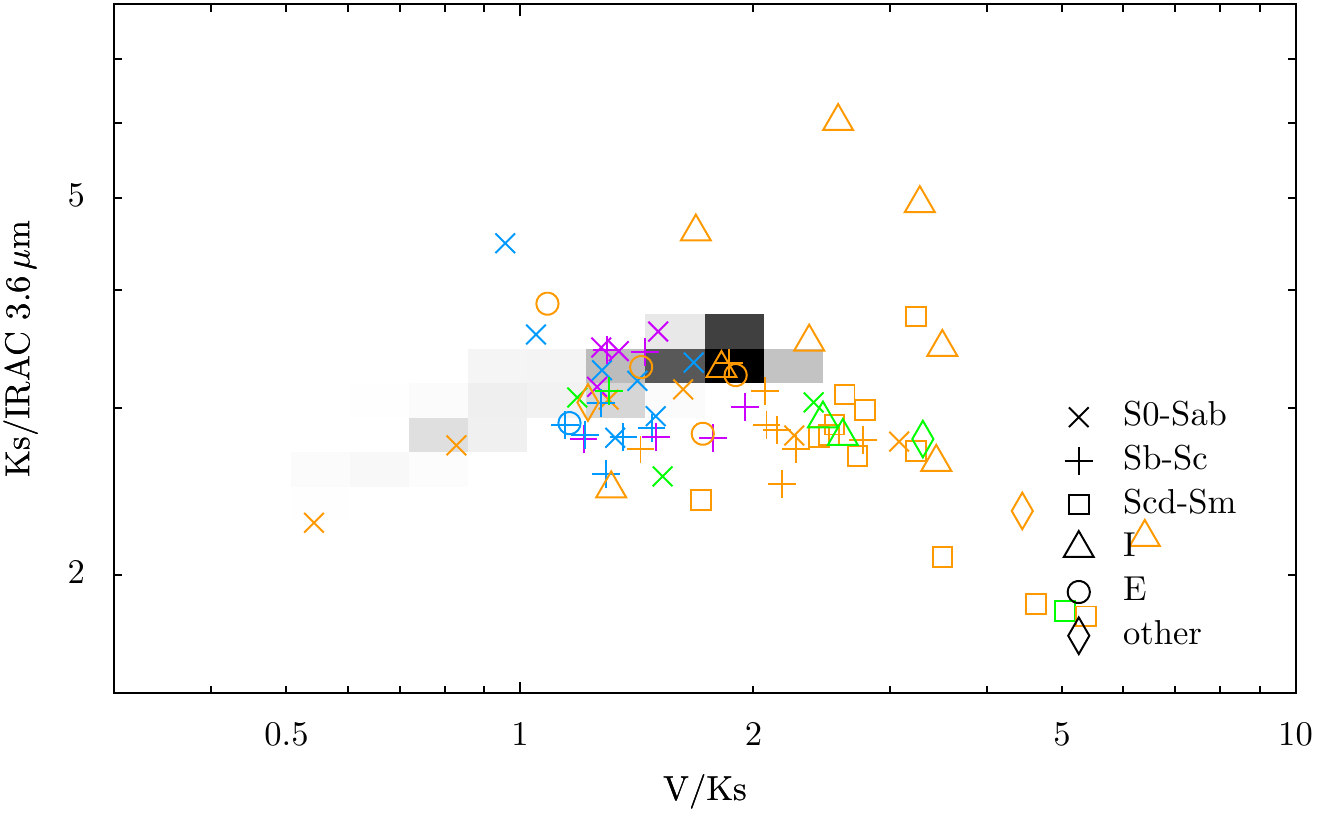} & \includegraphics*[scale=0.62]{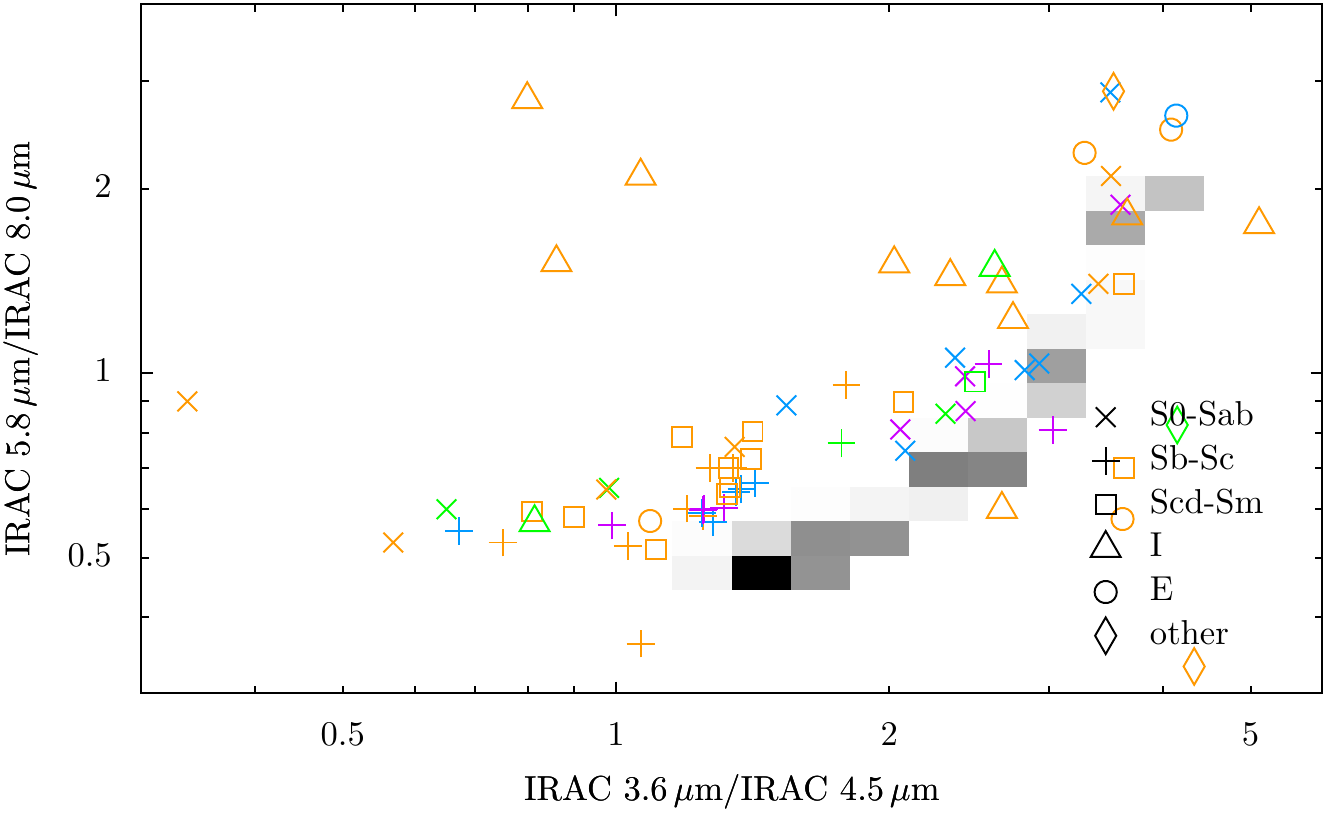} \\ \includegraphics*[scale=0.62]{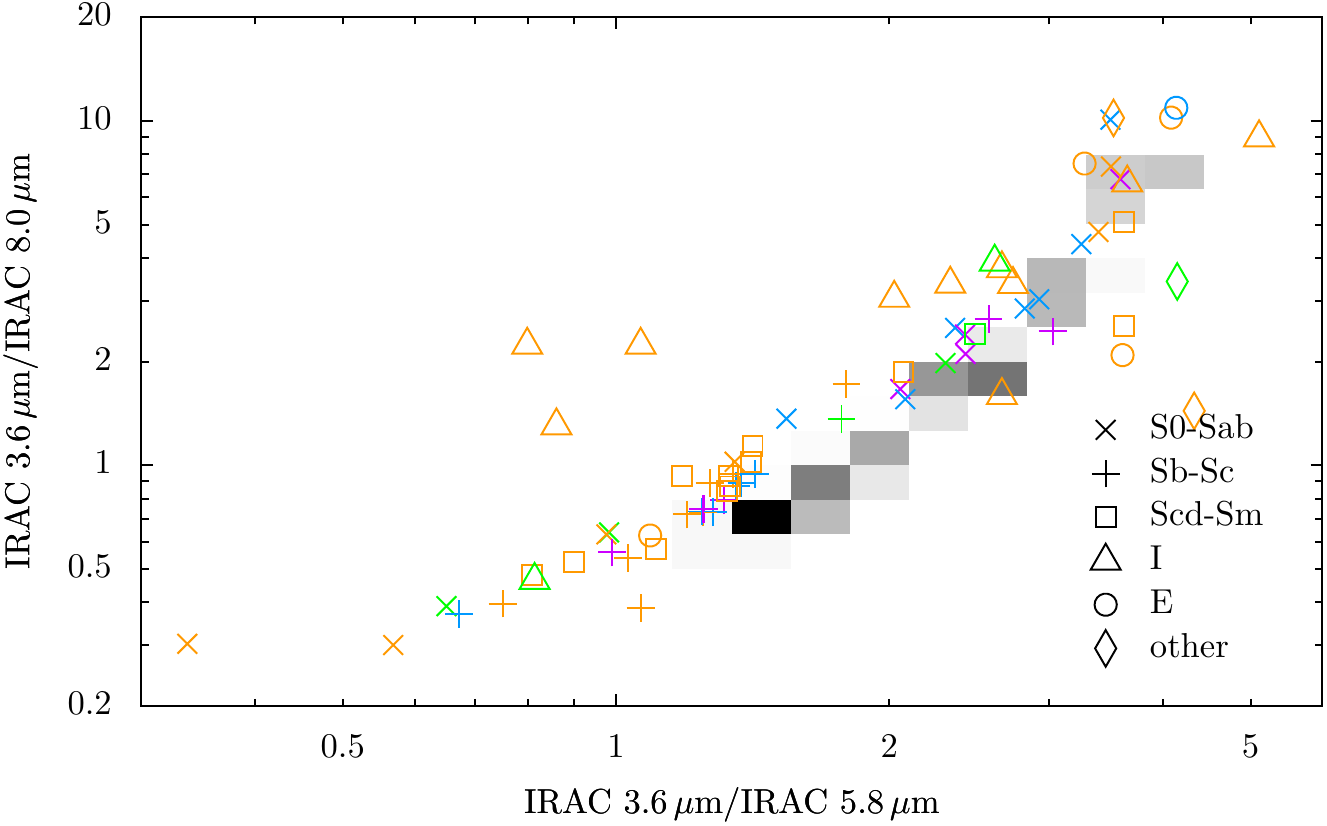} & \includegraphics*[scale=0.62]{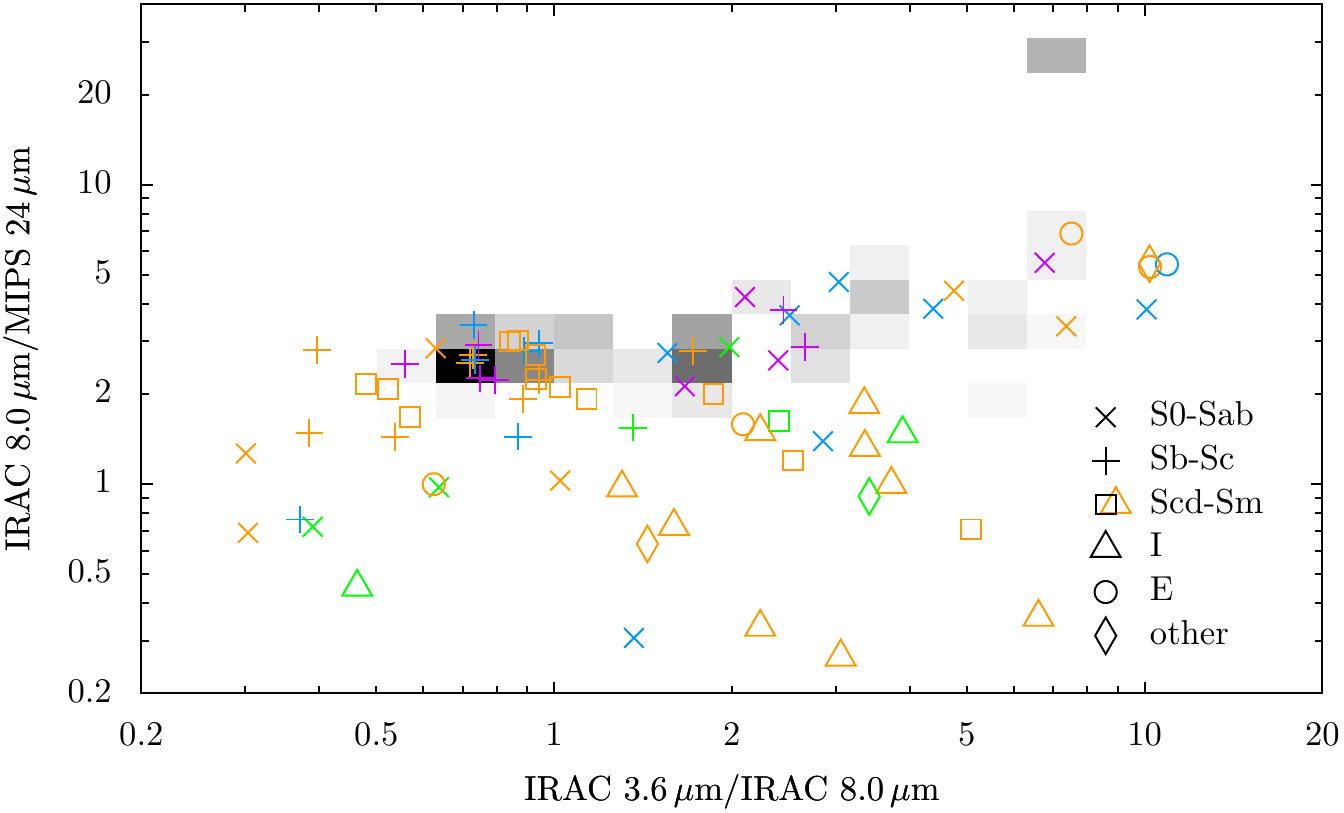} \\ \includegraphics*[scale=0.62]{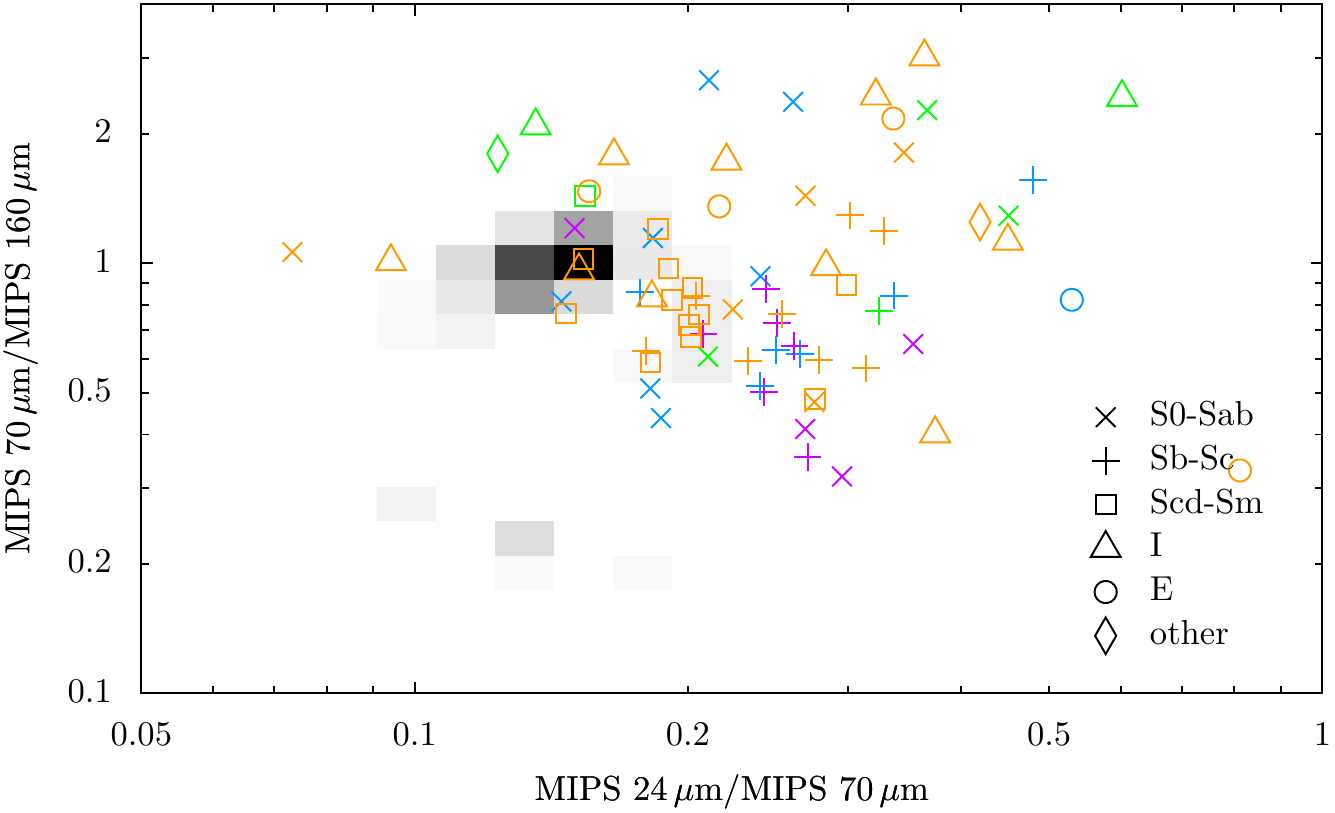} & \includegraphics*[scale=0.62]{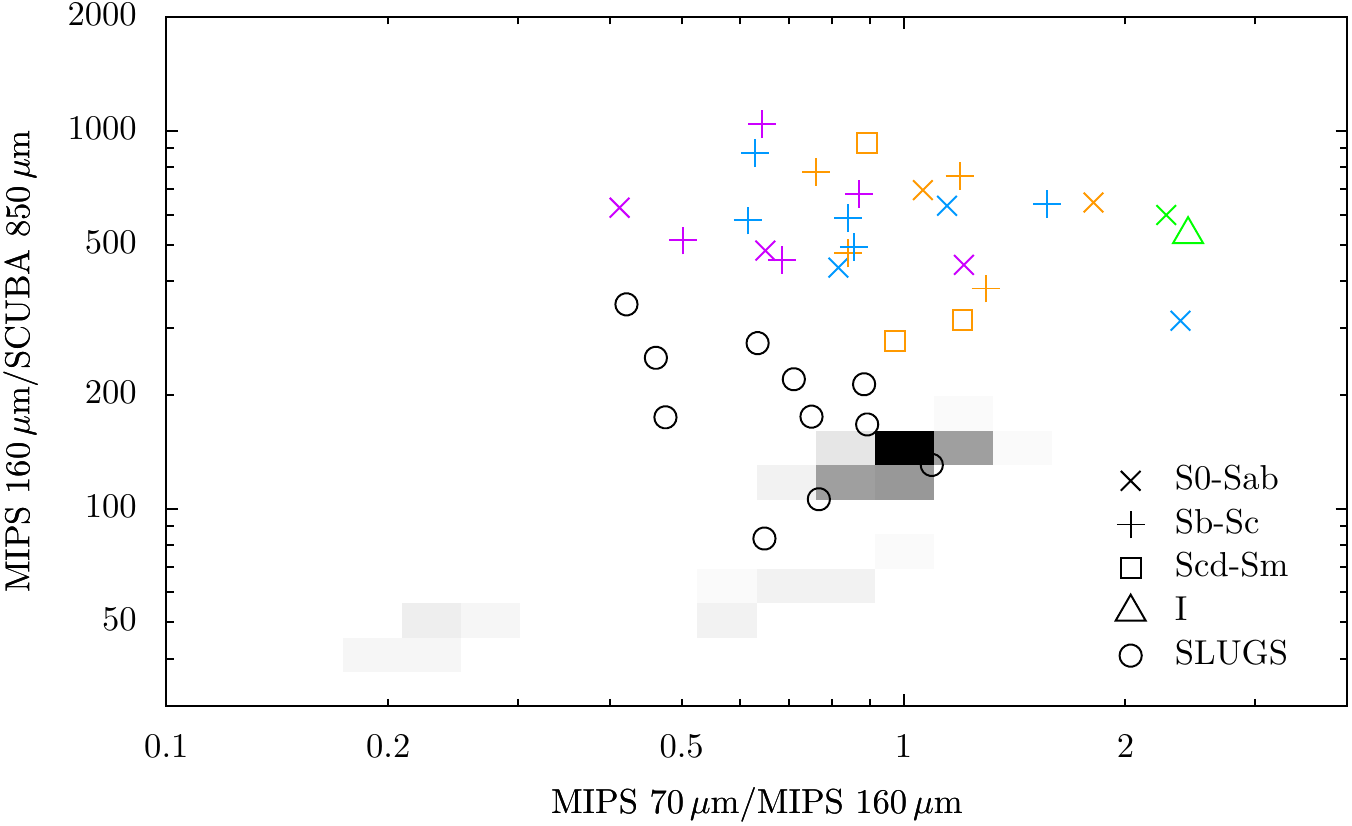} \end{tabular*} \end {center} \caption{ \label{plot_fluxratios} Colour-colour plots of the simulated galaxies (shaded region) in comparison to the SINGS galaxies from \citet{daleetal06sings}. In the $850\micron$ panel, the SLUGS sample from \citet{willmeretal09} is included in addition to the SINGS galaxies. The shade, from white to black, is proportional to the number of simulated galaxy points in the region.  The symbol for the SINGS points indicates the type of the galaxy (as labelled), while the colour indicates nuclear type (orange: starburst, green: LINER, cyan: Seyfert, purple: unclassified). With the exception of the $160/850 \micron$ colours in the lower right panel, the simulations generally occupy the same parameter space as the SINGS spiral galaxies, and in the $160/850\micron$ plot, the simulations occupy the region of the SLUGS sample. The extension to very cool far-IR colours is made up of the G0 galaxy and is due to its very low SFR, as mentioned in the text.} \end{figure*}

\begin{figure*} \begin {center} \includegraphics*[width=0.48\textwidth]{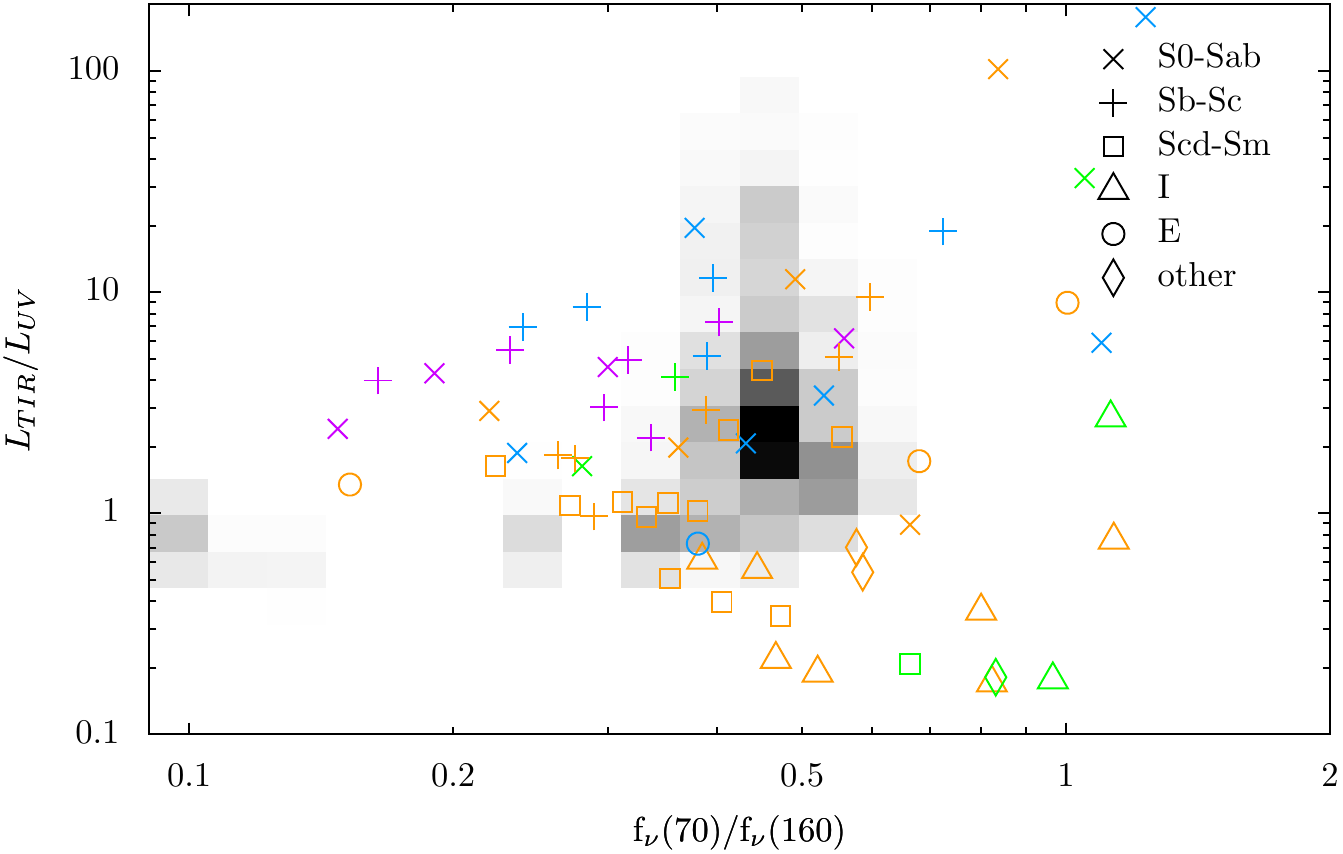} \includegraphics*[width=0.48\textwidth]{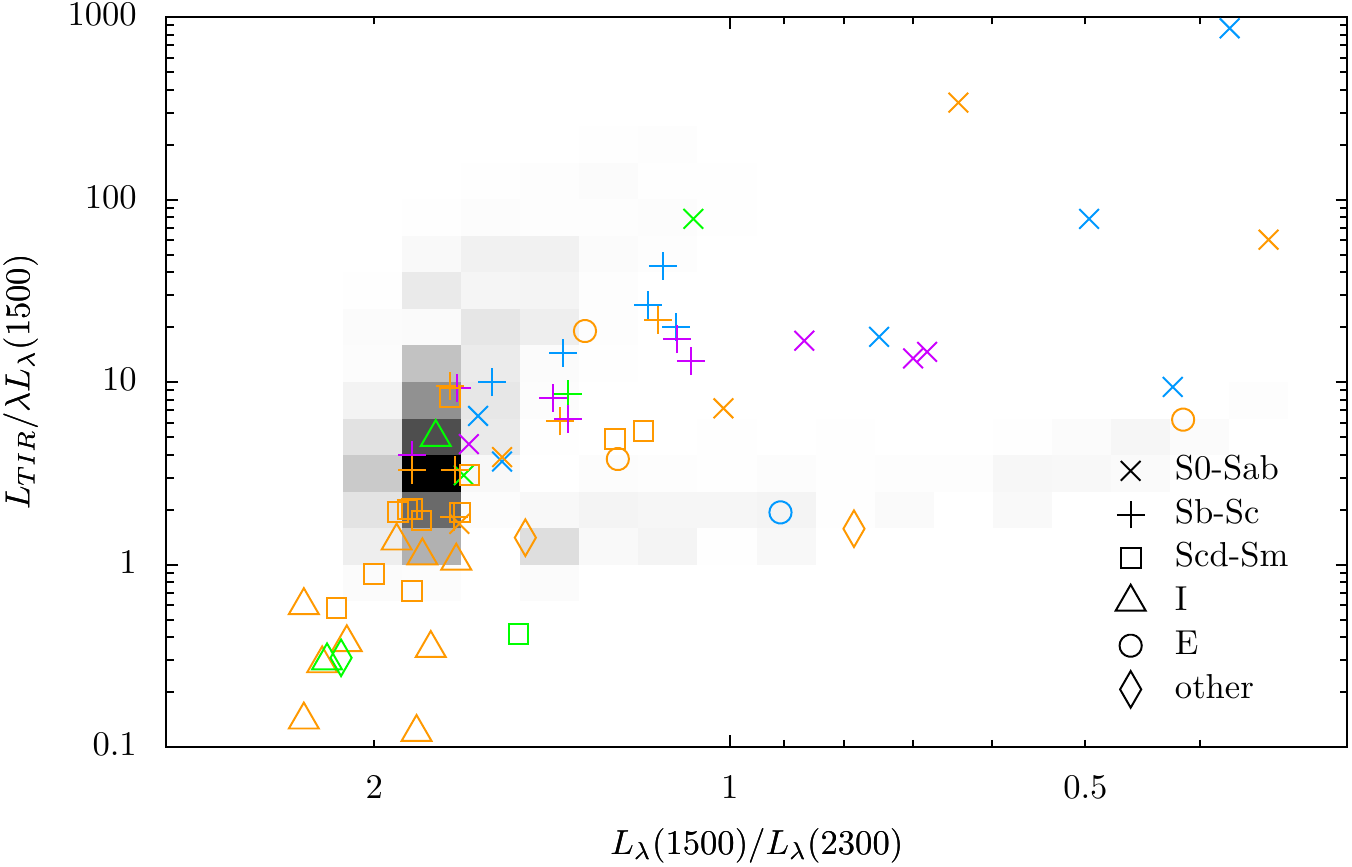} \end {center} \caption{ \label{plot_irxbeta} Comparison between the simulated galaxies and the results of \citet{daleetal06sings}.  The colours and symbols are the same as in Figure~\ref{plot_fluxratios}. The left panel is directly comparable to Figure~6 in D07, showing the ``infrared excess'', total infrared over total ultraviolet luminosity, against the $70/160\micron$ colour. The right panel, a classical ``IRX-$\beta$'' diagram of \citet{mhc99} directly comparable to Figure~12 in D07, shows UV spectral slope against infrared excess. In the IRX-$\beta$ diagram, the simulated galaxies are shifted towards a bluer UV slope compared to the SINGS sample. This is very sensitive to the dust model used, due to their differing ``$2200\Angstrom$ bump'' strengths. The effect of changing the dust model is shown in Figure~\ref{plot_fluxratios-dustmodel}. The simulations with very red UV slopes and low values of IRX, well below the SINGS galaxies, are from the G0 galaxy with its very low star-formation rate.} \end{figure*}

Figure~\ref{plot_fluxratios} shows a series of colour-colour plots
across the full wavelength range, comparing the colours of the
simulated and real galaxies. Figure~\ref{plot_irxbeta} duplicates
Figures~6 and 12 in D07, showing the dependence of the ``infrared
excess'' on the FIR and UV spectral slopes.

%NLX% exclude from vocabulary builder
 
%NLX% end exclude from vocabulary builder

In general, the simulated galaxies do lie in regions of colour-colour
space occupied by SINGS galaxies of type Sb-Sc, which should be
considered the most relevant comparison sample. However, some clear
differences do exist. The most obvious is that the simulated galaxies
seem to be a less diverse sample than the real galaxies, occupying a
smaller range of colour-colour space. This is not unexpected given the
small range of galaxy types considered (see
Section~\ref{section_simulations}). A second, more fundamental,
difference is the visible displacement of the models from the SINGS
galaxies in some colours.

The most glaring difference between the models and the SINGS galaxies
is in the $160 / 850 \um$ flux ratio, where the simulations have
systematically lower flux ratios by about a factor of 4. However, there
are clear inconsistencies between different observational samples in
this colour, as is clear by the location of the SLUGS galaxies of
\citet{willmeretal09}. The SLUGS galaxies, while similar to the SINGS
galaxies in their $70 / 100 \um$ ratios, have systematically lower 
$160 / 850 \um$ flux ratios that are more in agreement with the outputs
from our simulations. This potential discrepancy and its origin will be
discussed further below.

Clear differences with the SINGS sample also exist in the PAH region,
where the simulations have too low $3.6 / 8.0 \um$ and $5.8 / 8.0 \um$
flux ratios. At MIPS wavelengths, the simulations are also displaced
slightly low in $24 / 70 \um$ and high in $70 / 160 \um$ compared to
the Sb-Sc SINGS galaxies.  More subtly, while the simulations do lie in
a region of space occupied by the SINGS galaxies, they appear to be
slightly too blue in the GALEX FUV/NUV colour, slightly too red in $V -
R$, and slightly too blue in $K - 3.6 \um$. This is also seen in the
left panel of Figure~\ref{plot_irxbeta} where the simulations have a
too flat UV slope (too blue UV colours).

In all diagrams in Figures~\ref{plot_fluxratios}
and~\ref{plot_irxbeta}, a faint trail of simulations can be seen
displaced from the main locus, especially in the UV and FIR plots. This
locus arises from the G0 simulation, which has a very low
star-formation rate. As mentioned in Section~\ref{section_simulations},
this low SFR results in strong variations due to the stochastic
implementation of star formation in the simulations. The star-formation
rate of the G0 galaxy is about $10^{ - 3 } \sfrunit$, the mass of the
star particles created in the G0 simulation is $3 \times 10^4 \Msun$. 
This means one star particle is spawned about every $30 \Myr$. Since
the MAPPINGS particles have a lifetime of $10 \Myr$, this means that
there will be on average 1/3 such particles at any given time. When
such a particle is present, it will dominate the UV---and consequently
also the IR---luminosity of the galaxy, but when no such particle is
present, the IR colours will be very cool, thus giving rise to severe
fluctuations in the SED. The simulations are unable to sample the
population of young stars in the G0 galaxy adequately at the current
resolution, so their SEDs are much more uncertain than that of the
other galaxy models.

\begin{figure*} \begin {center} \begin {tabular} {cc} Sbc+ & Sbc \\ \includegraphics*[width=0.45\textwidth]{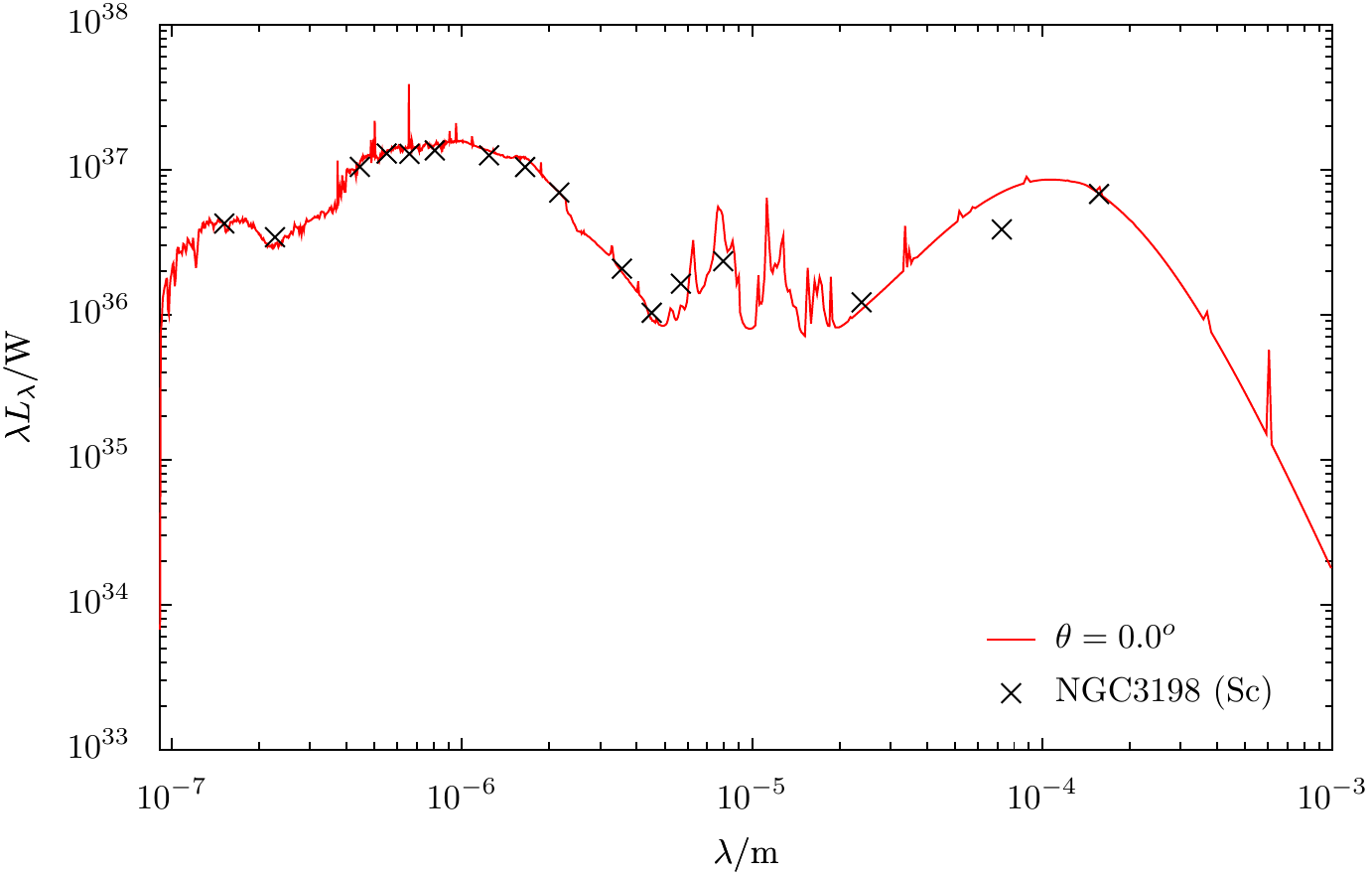} & \includegraphics*[width=0.45\textwidth]{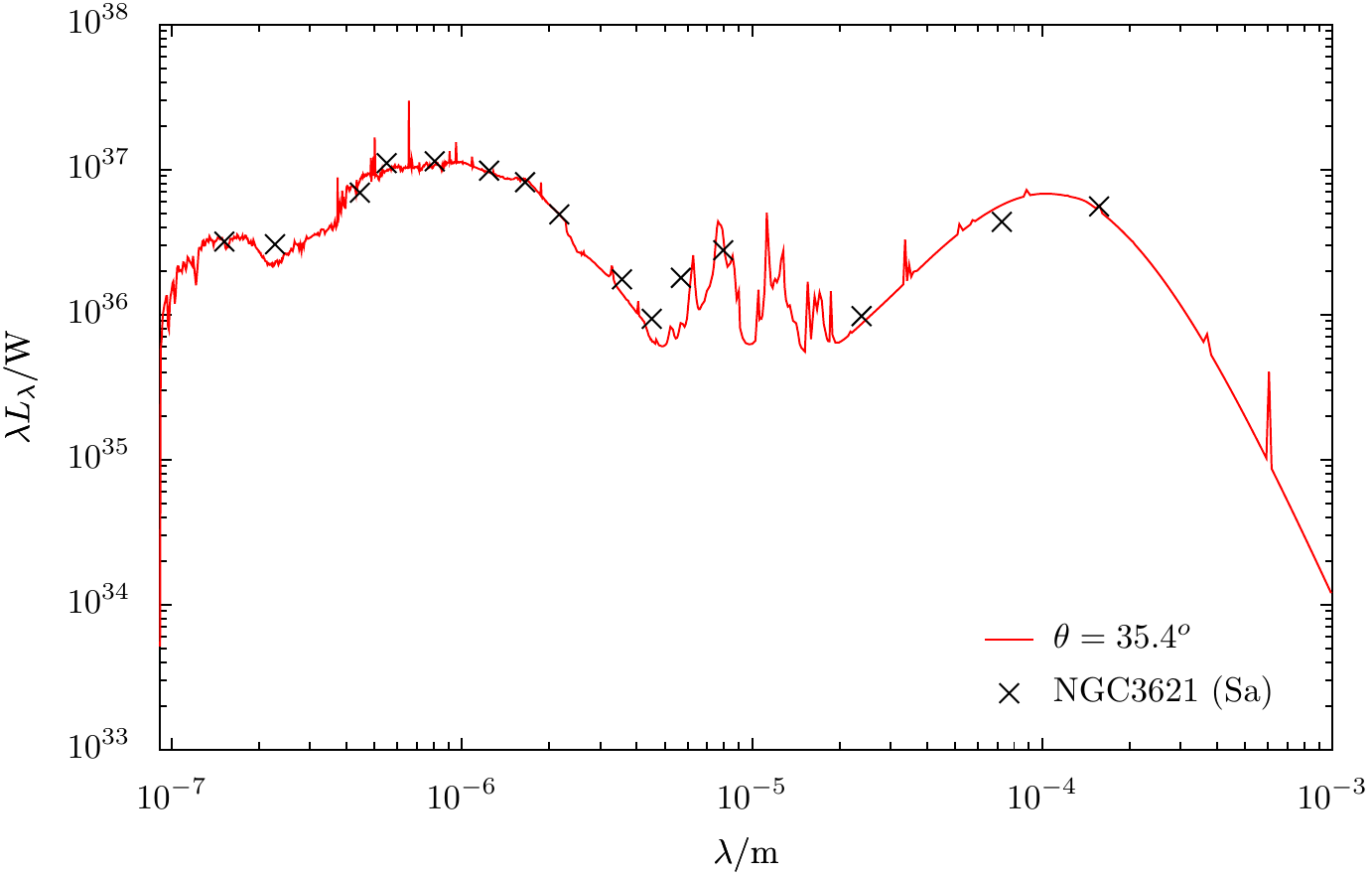} \\ Sbc$-$ & G3 \\ \includegraphics*[width=0.45\textwidth]{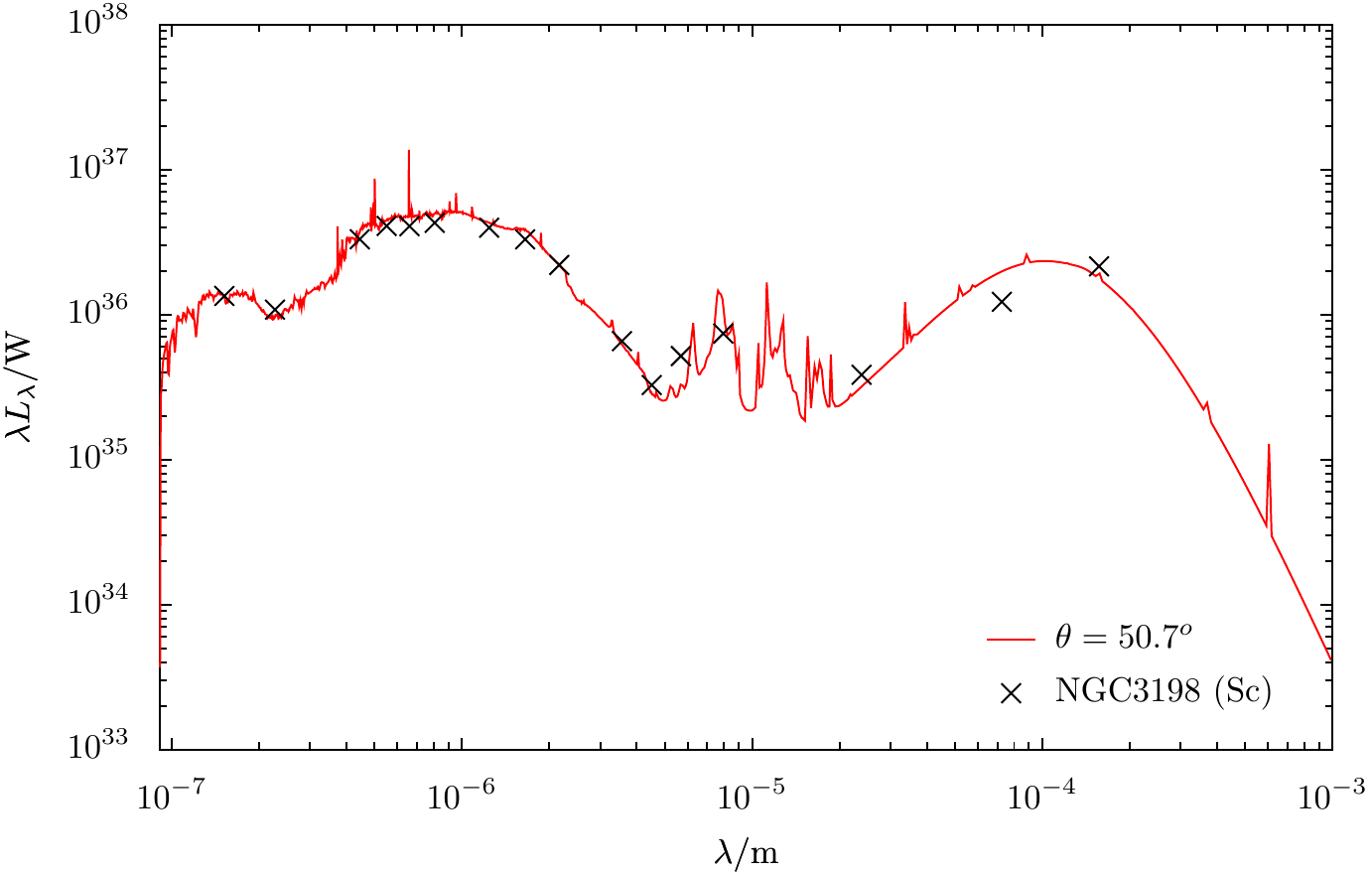} & \includegraphics*[width=0.45\textwidth]{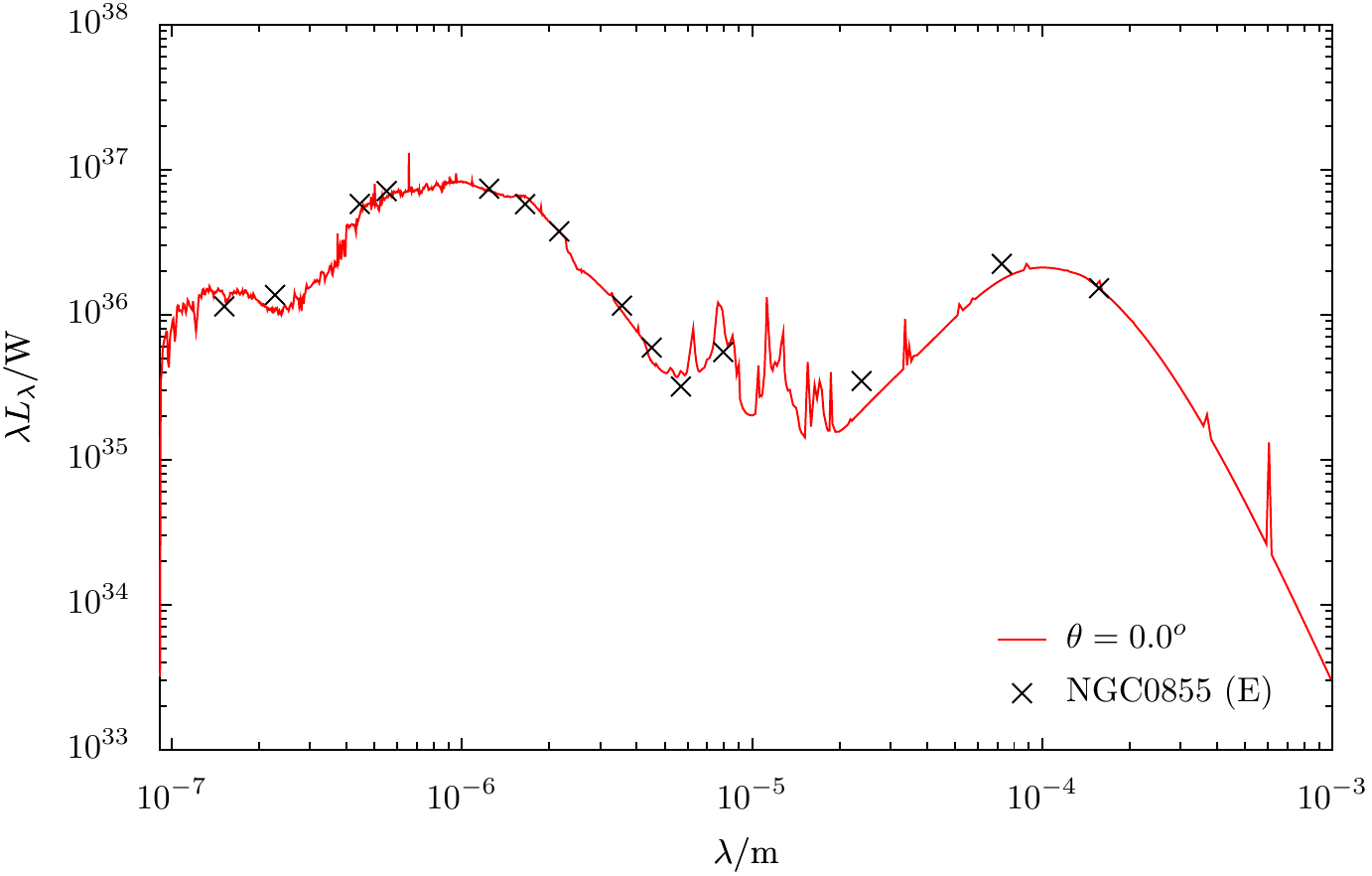} \\ G2 & G1 \\ \includegraphics*[width=0.45\textwidth]{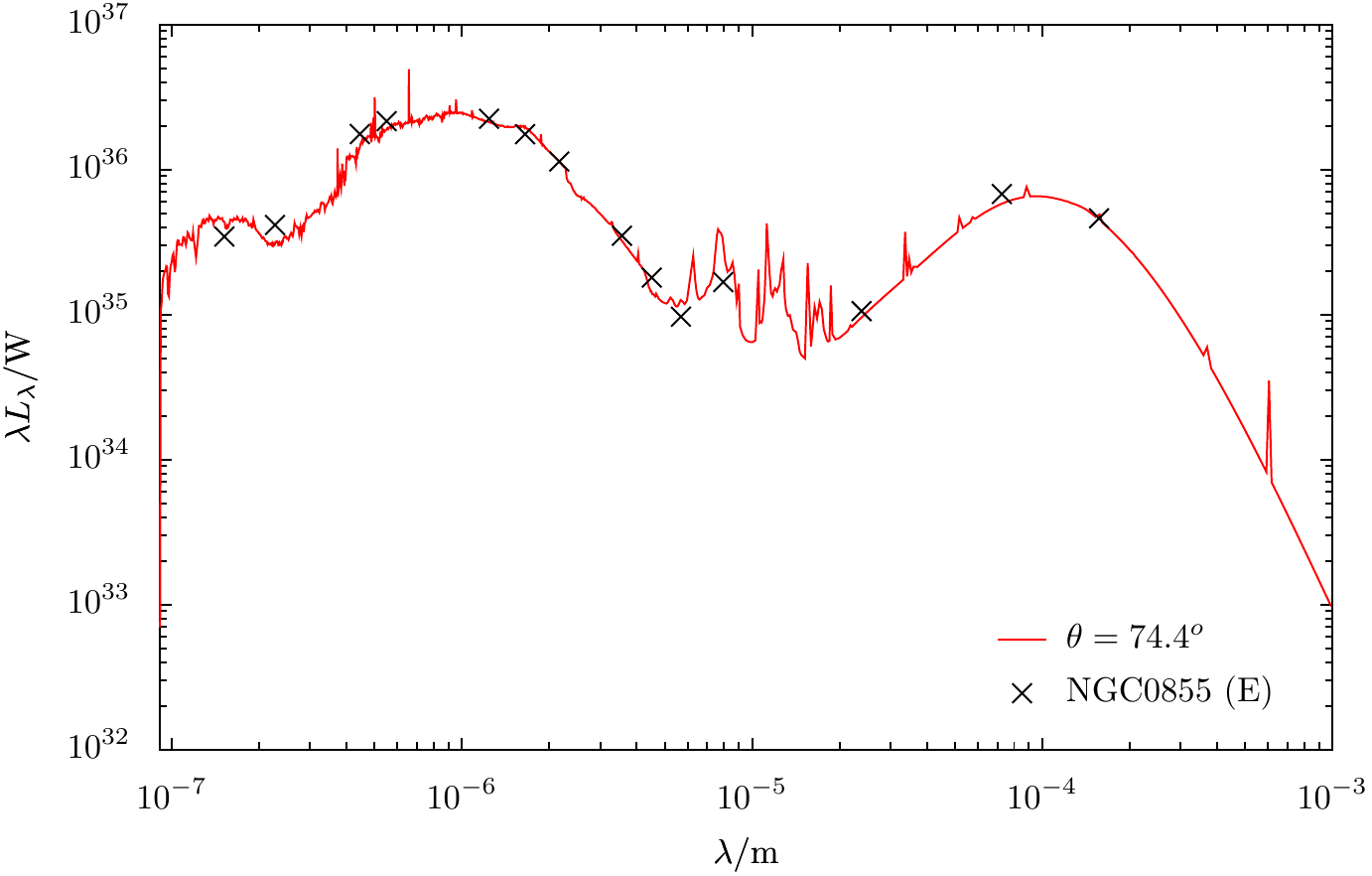} & \includegraphics*[width=0.45\textwidth]{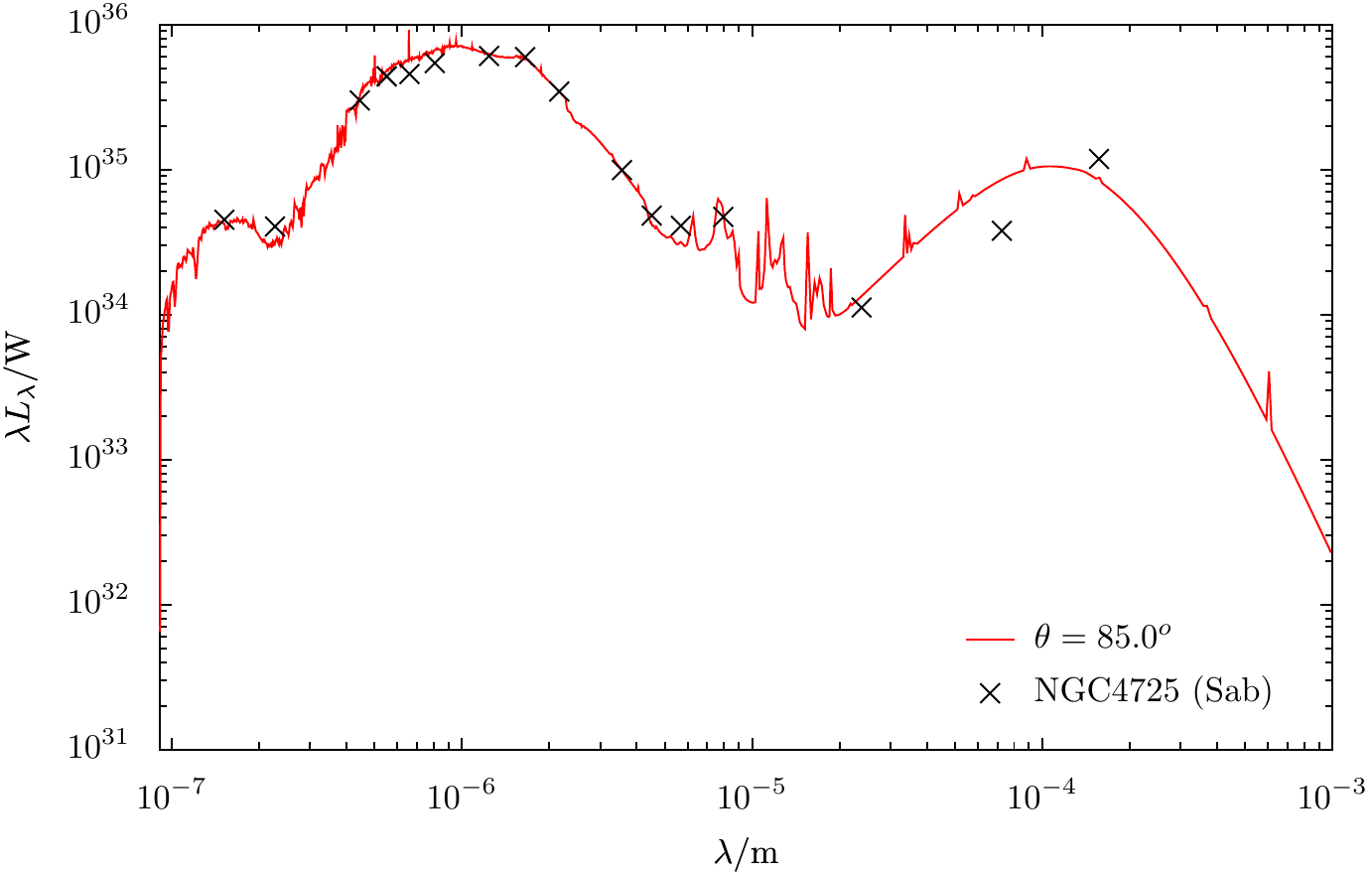} \\ G0 \\ \includegraphics*[width=0.45\textwidth]{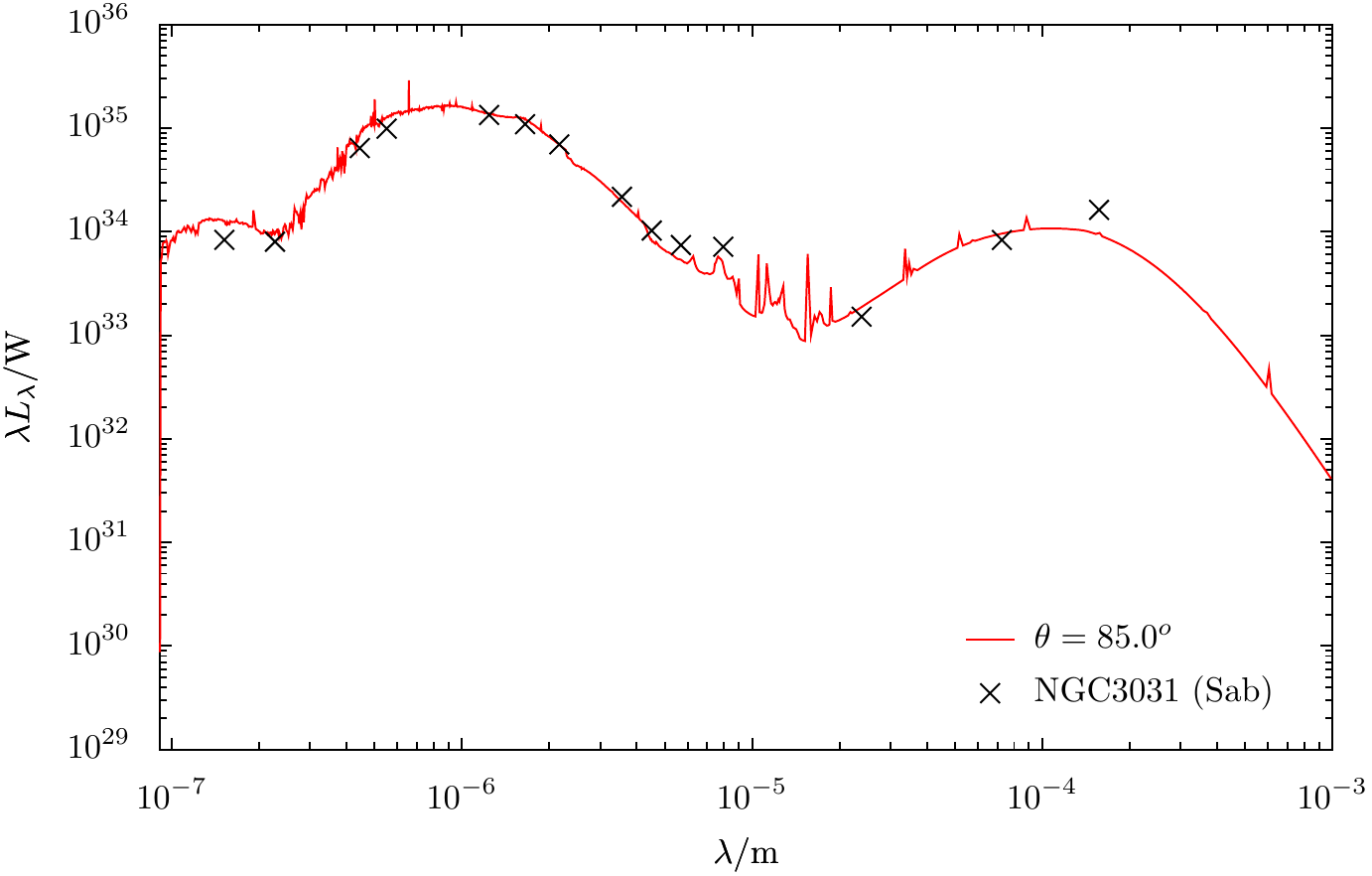} \end{tabular} \end {center} \caption{ \label{plot_sings-bestfit} The simulated galaxies (solid lines) with the best-fitting SINGS galaxy (symbols)  overplotted. After evolving them for $0.5\Gyr$, the SEDs for each galaxy simulation, for various inclinations, were compared to the SEDs of the SINGS galaxies and the inclination that best matched a SINGS galaxy was chosen. All SEDs are normalized to the observations in the K-band. The quoted uncertainties for the SINGS galaxy observations are of the order of, or smaller than, the size of the symbols. } \end{figure*}

A qualitatively different kind of comparison is to compare the SEDs on
a galaxy-by-galaxy basis. Such a comparison is valuable as it can
provide clues to the nature of the systematic differences shown in the
colour-colour plots. For each of the simulations, the SEDs at a time of
$0.5 \Gyr$, for all inclinations, were compared to the SEDs of the
SINGS galaxies after normalizing the fluxes in the K-band. (The
exception is the G0 galaxy. Due to the stochastic effects in its SED,
we extended the matching to all times during the $1 \Gyr$ evolution.)
The best matches are shown in Figure~\ref{plot_sings-bestfit}. Due to
the small number of SINGS galaxies with $850 \um$ data, and the
systematic discrepancy between the simulations and the SINGS sample in
the SCUBA $850 \um$ band, we have not required a $850 \um$ point to be
present.

%NLX% exclude from vocabulary builder
 % \begin{figure*}
%   \begin {center}
%     \begin {tabular} {cc}
%       Sbc+ & Sbc \\
%       \includegraphics*[width=0.45\textwidth]{850-sings-bestfit-Sbcp1s-u4} &
%       \includegraphics*[width=0.45\textwidth]{850-sings-bestfit-Sbc11i4-u4} \\
%       Sbc$-$ & G3 \\
%     \includegraphics*[width=0.45\textwidth]{850-sings-bestfit-Sbcm1s-u4} &
%     \includegraphics*[width=0.45\textwidth]{850-sings-bestfit-G3il-u1} \\
%     G2 & G1 \\
%     \includegraphics*[width=0.45\textwidth]{850-sings-bestfit-G2in-u1} &
%     \includegraphics*[width=0.45\textwidth]{850-sings-bestfit-G1i-u1} \\
%     G0 \\
%     \includegraphics*[width=0.45\textwidth]{850-sings-bestfit-G0i-u1}
%     \end{tabular}
%   \end {center}
%   \caption{ \label{plot_850-sings-bestfit} The simulated galaxies with
%     the best-fitting SINGS galaxy overplotted. Analogous to
%     Figure~\ref{plot_sings-bestfit}, but limited to the
%     SINGS galaxies that have SCUBA $850\micron$ data. }
% \end{figure*}
%NLX% end exclude from vocabulary builder

In general, the agreement is good, surprisingly so as these simulations
were not intended to be exact copies of existing galaxies. The UV-NIR
SED can generally be matched very well, with the discrepancies matching
those found in Figures~\ref{plot_fluxratios} and~\ref{plot_irxbeta}.
For example, the high FUV/NUV ratio can be seen to be due to the model
producing an excessively strong $2200 \Angstrom$ feature in several of
the galaxies. The most obvious discrepancy in the comparisons is that
most SINGS spirals seem to have FIR SEDs which peak at longer
wavelengths (is cooler) than the simulations, leading to the observed
$24 / 70 \um$ and $70 / 160 \um$ offsets.  Somewhat surprising, the
best match to the G3 \& G2 simulations are the elliptical galaxies
NGC855 and NGC4125, which have shorter peak wavelengths more in line
with the simulations. However, these ellipticals are unusual in the
sense that they have detectable neutral gas and blue stellar
populations and are clearly not classical red and dead ellipticals.

% When requiring that the "850 micron" data be present, the fits are
% noticeably worse. The simulations systematically (emphasize:
% under)predict the MIPS "160 micron" point and (emphasize: over)predict
% the "850micron" SCUBA flux. In comparison to the simulations, the
% SINGS galaxies have a significantly steeper long-wavelength tail. This
% may be related to the distribution of radiation intensities heating
% the dust (see Section~(refer to: section_sensitivities)).

\subsection{Emission Lines as Star-Formation Indicators}
 \label{section_lines}

%NLX% exclude from vocabulary builder
\begin{figure*} \begin {center} \includegraphics*[width=0.48\textwidth]{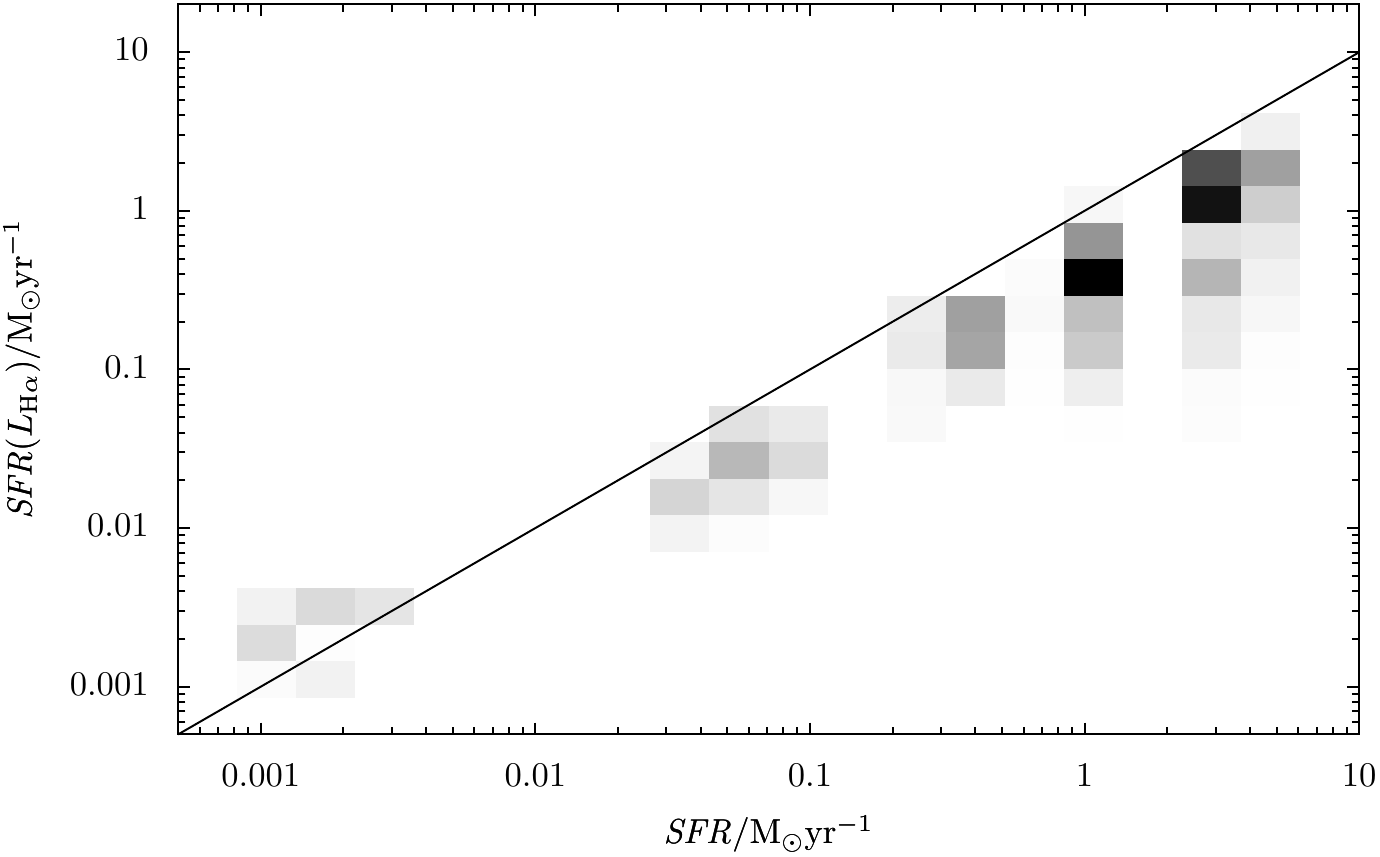} \includegraphics*[width=0.48\textwidth]{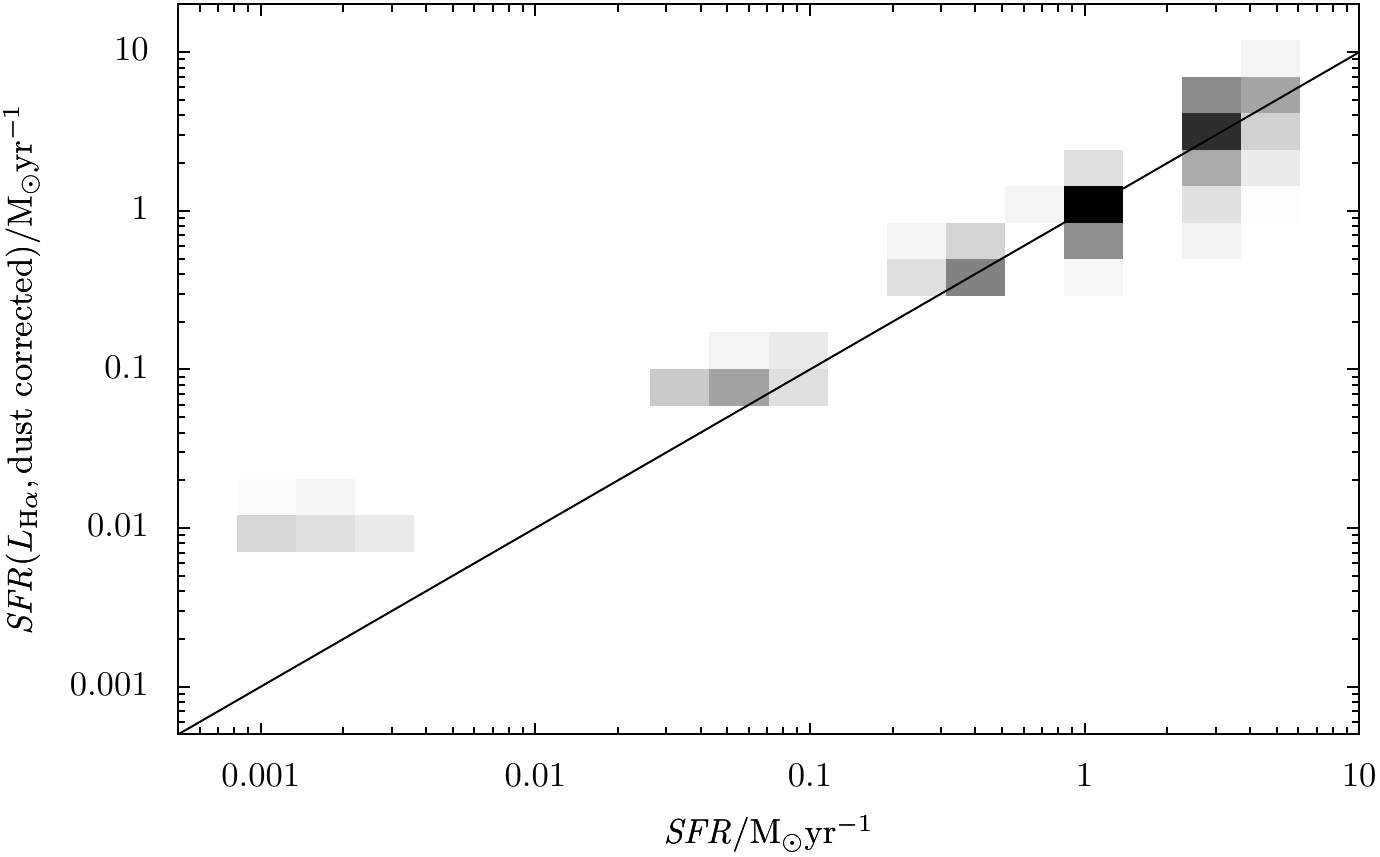} \end {center} \caption{ \label{plot_halpha} $\halpha$-derived star-formation rate compared to the intrinsic SFR in the simulations. The left panel shows the uncorrected $\halpha$ luminosities in the simulations, converted to a star-formation rate using the normalization of \citet{kennicutt98}. In the more massive galaxies, the SFR is underestimated by factors of several. The right panel shows the star-formation based on $\halpha$ luminosities corrected for dust attenuation using the $\halpha/\hbeta$ line ratios using the formula of \citet{calzettietal94}. This works well in correcting the average $\halpha$-predicted SFR, but there is still a scatter of about an order of magnitude for the massive galaxies. The G0 galaxy, with an SFR of $\sim 10^{-3} \sfrunit$, shows an overcorrection of the SFR by an order of magnitude. This is likely due to the stochastic effect of star formation in the simulations, as the smallest non-zero $\halpha$ luminosity attainable is that of one \mapiii\ particle. A substantial fraction of the G0 snapshots have no $\halpha$ emission and thus can not be shown in the graph.} \end{figure*} \begin{figure*} \begin {center} \includegraphics*[width=0.48\textwidth]{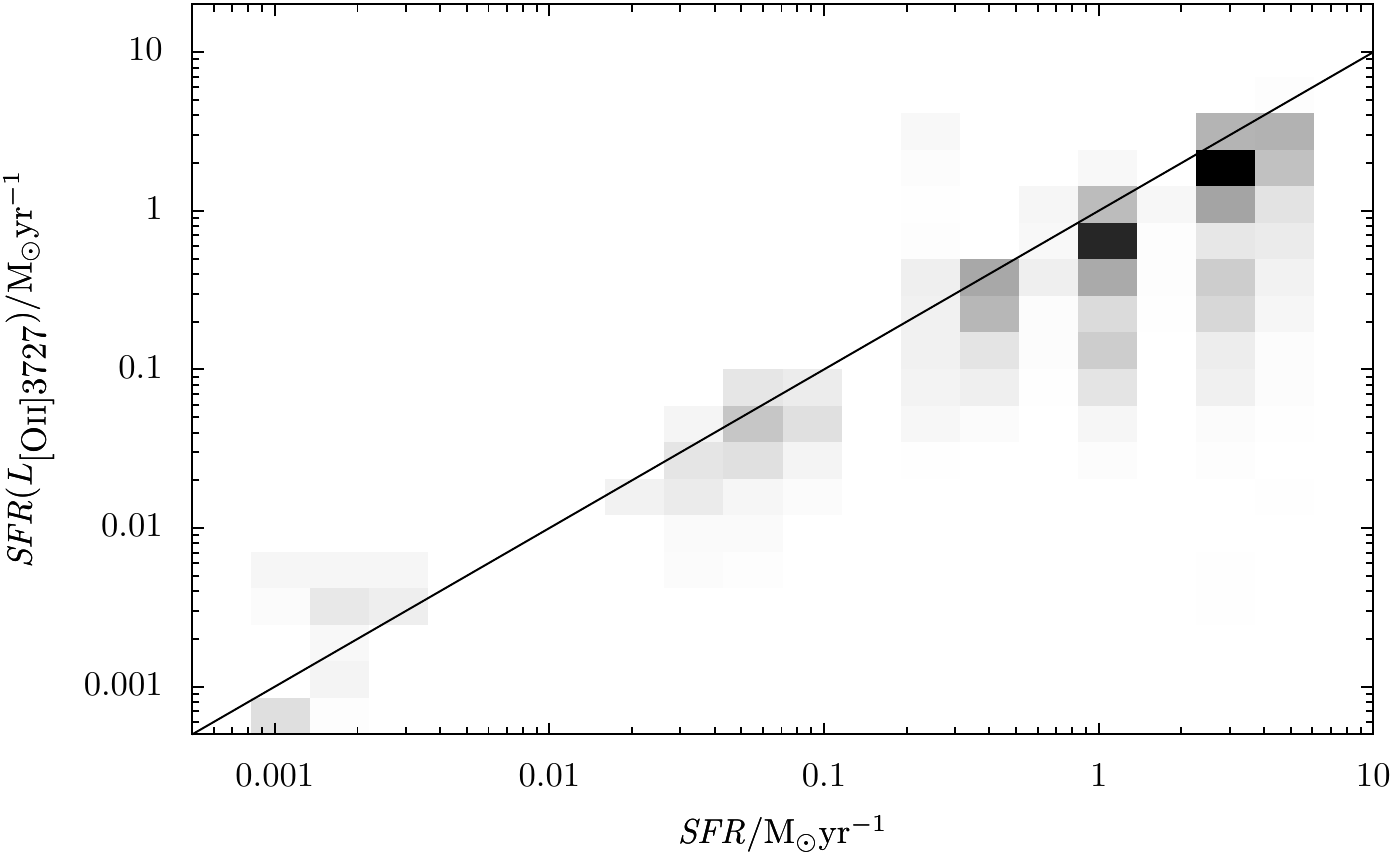} \includegraphics*[width=0.48\textwidth]{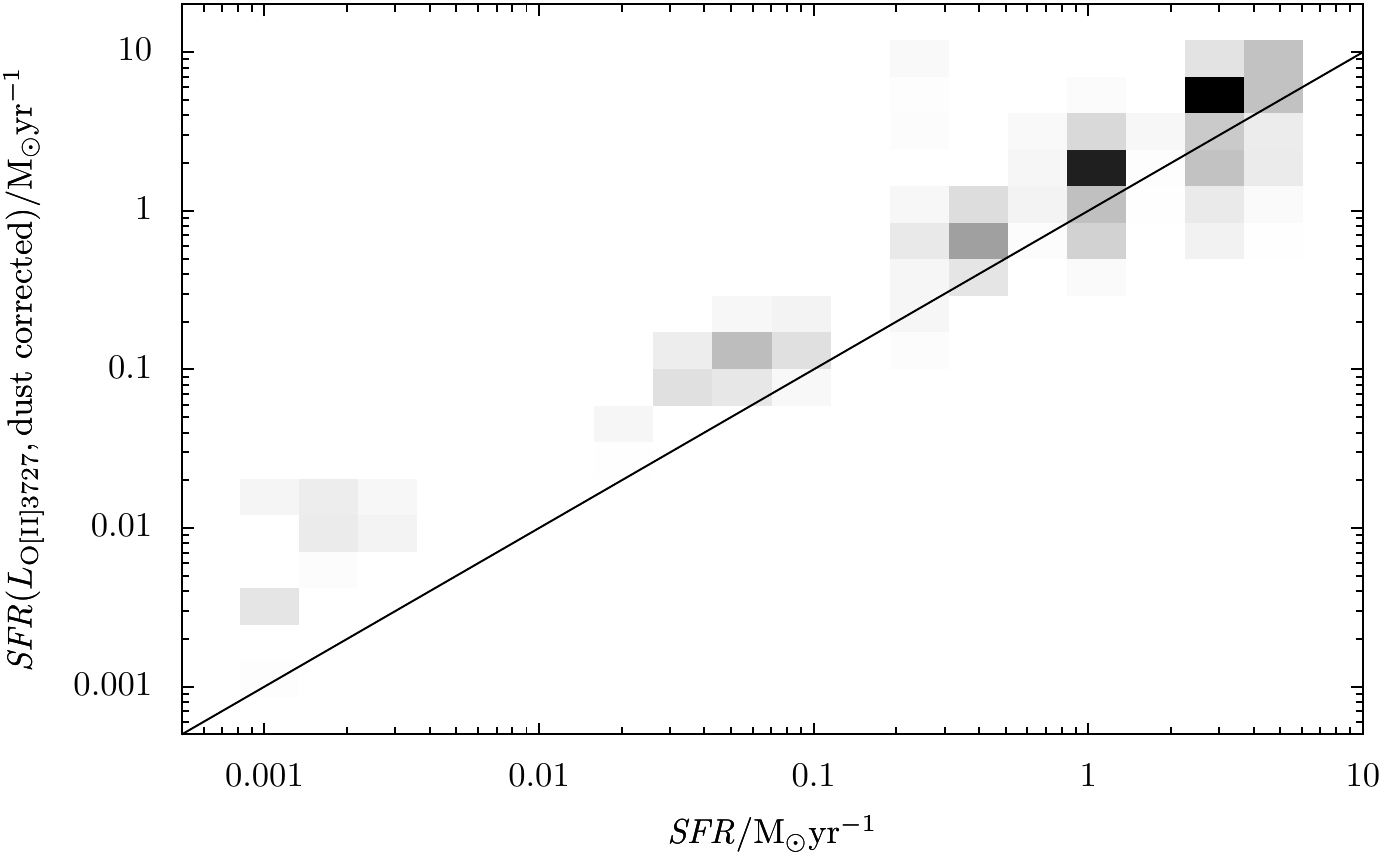} \end {center} \caption{ \label{plot_oii} Star-formation rates derived from the $\oii \lambda 3727$ line compared to the intrinsic SFR in the simulations. The left panel shows the uncorrected $\oii$ luminosities in the simulations, converted to a star-formation rate using the normalization of \citet{kennicutt98}. While SFRs along the upper envelope agree well with the true SFR, there is a substantial scatter towards underestimated SFRs. Compared to the $\halpha$-derived SFR in Figure~\ref{plot_halpha}, the scatter is larger.  The right panel shows the star-formation rates after correcting the $\oii$ luminosities for dust attenuation \citep[at $\halpha$, see][]{kennicutt98} using the $\halpha/\hbeta$ line ratios using the formula of \citet{calzettietal94}. After this correction, the scatter is reduced but there is a tendency to systematically overestimate the SFR.  } \end{figure*}
%NLX% end exclude from vocabulary builder

In the discussion so far, we have focused on broadband features of the
SED. One of the major advantages made feasible with the high wavelength
resolution of our model and the use of the \mapiii \ models of \hii\
regions, however, is the inclusion of emission lines. In the context of
galaxies, the foremost interest in emission lines is from their use as
star-formation rate indicators.  For the $\halpha$ line, this was
already done in \citet{pjthesis-nourl} but, as discussed in
Section~\ref{section_mappings}, the \mapiii \ models include all
important emission lines from \hii\ regions.  Thus we will now compare
the emission line strengths of the simulated galaxies with
star-formation rate calibrations from the literature.

Figure~\ref{plot_halpha} shows the $\halpha$-derived SFR in the
simulations using the calibration from \citet{kennicutt98}. Without
dust correction, the SFR for the massive galaxies is underestimated by
factors of several. When the $\halpha$ luminosity is corrected for dust
attenuation using the $\halpha$/$\hbeta$ line ratio
\citep{calzettietal94}, the SFR is, on average, predicted fairly well,
but the scatter in the derived SFR is still about an order of
magnitude.

Another important star-formation indicator, used for higher-redshift
observations, is the $\oii \lambda 3727 \Angstrom$ line. Because the
luminosity of this line depends on metal abundance and the ionization
state of oxygen atoms, its relationship with star formation rate is
more complicated, but these effects are included in the \mapiii \
models. Figure~\ref{plot_oii} shows the star-formation rate derived
from the luminosity of this line, again using the calibration of
\citet{kennicutt98}. As this line is at much shorter wavelengths, dust
attenuation is more severe and for the larger galaxies, the
star-formation rate can be underestimated by almost two orders of
magnitude. After applying the same dust correction as for the $\halpha$
line (because of the way the SFR from the $\oii$ line was calibrated,
\citealt{kennicutt98} claims the dust correction should be for the
attenuation at $\halpha$) the scatter is reduced to about one order of
magnitude, but there is now a tendency to systematically overestimate
the SFR.

With these proof-of-concept examples of the capabilities of our model,
we defer a systematic study of the sensitivity of emission lines to
dust attenuation, and the prospects of correcting for these dust
effects, to a future paper.

\subsection{Spatially Resolved Quantities}
 \label{section_pixels}

In our presentation of the results so far, we have shown that our model
galaxies have realistic integrated spectra in comparison to real
galaxies. However one of the strengths of \mcrx \ is that it creates
\emph{2-dimensional spectra}. So we now go beyond this and put the
model through some more stringent tests by looking at spatially
resolved spectral quantities. Even if the integrated spectra are
realistic, the agreement may break down when looking at individual
regions in the galaxies.

The spatial variations of the dust emission in the SINGS galaxies was
studied by \citet{bendoetal08}, who investigated emission from PAHs,
measured at $8 \um$, hot dust at $24 \um$, and cold dust at $160 \um$.
They found that the PAH $8 \um / 24 \um$ surface brightness ratio
exhibited significant scatter, but was generally higher in the diffuse
interstellar medium and lower in bright star-forming regions with high
$24 \um$ surface brightness. The PAH $8 \um$ emission was
well-correlated with $160 \um$ emission, with generally larger $8 \um /
160 \um$ surface brightness ratio in regions that are brighter in $160
\um$. They interpreted these results as indicating the PAHs were mostly
associated with the cold, diffuse dust that dominates the $160 \um$
flux.

\begin{figure*} \begin {center} \includegraphics*[width=0.48\textwidth]{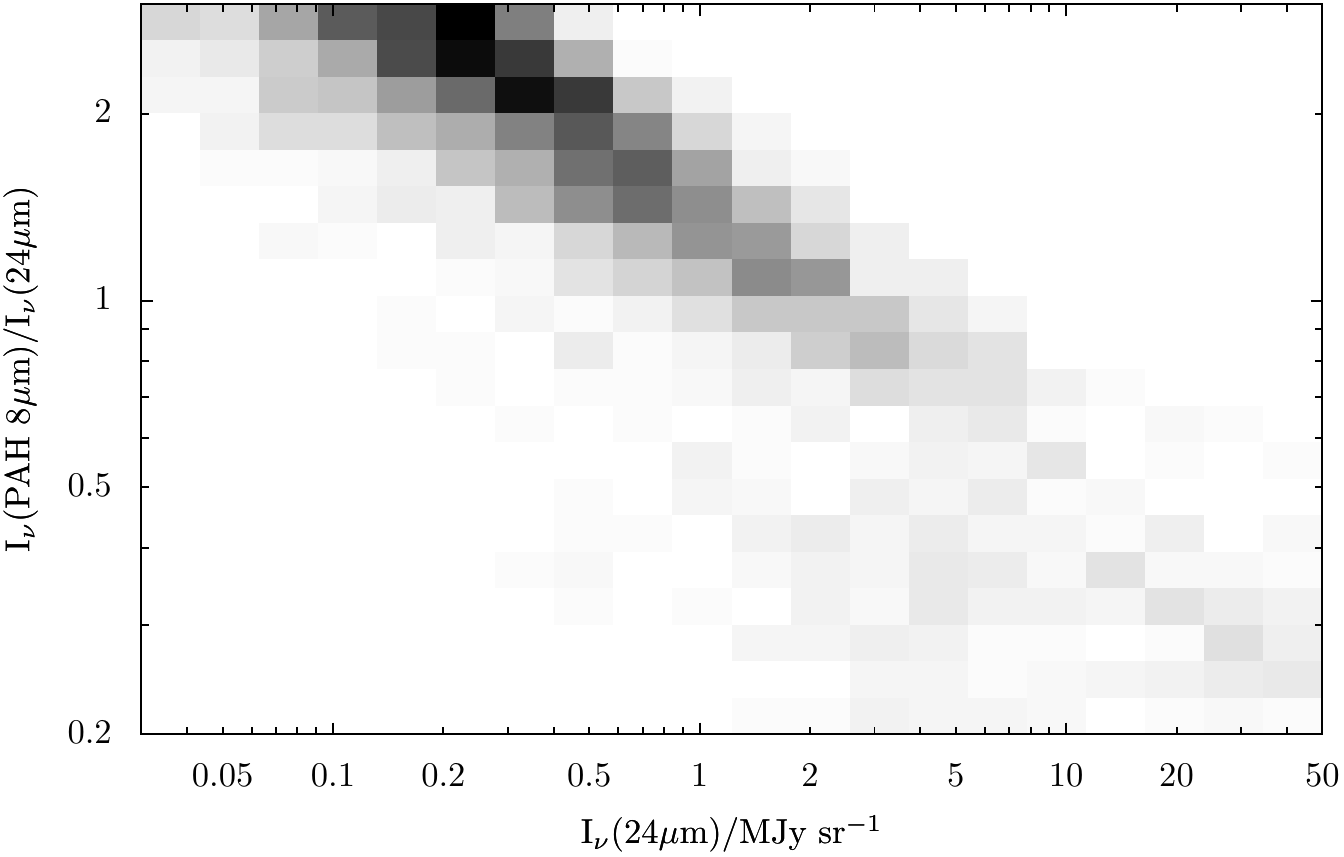} \includegraphics*[width=0.48\textwidth]{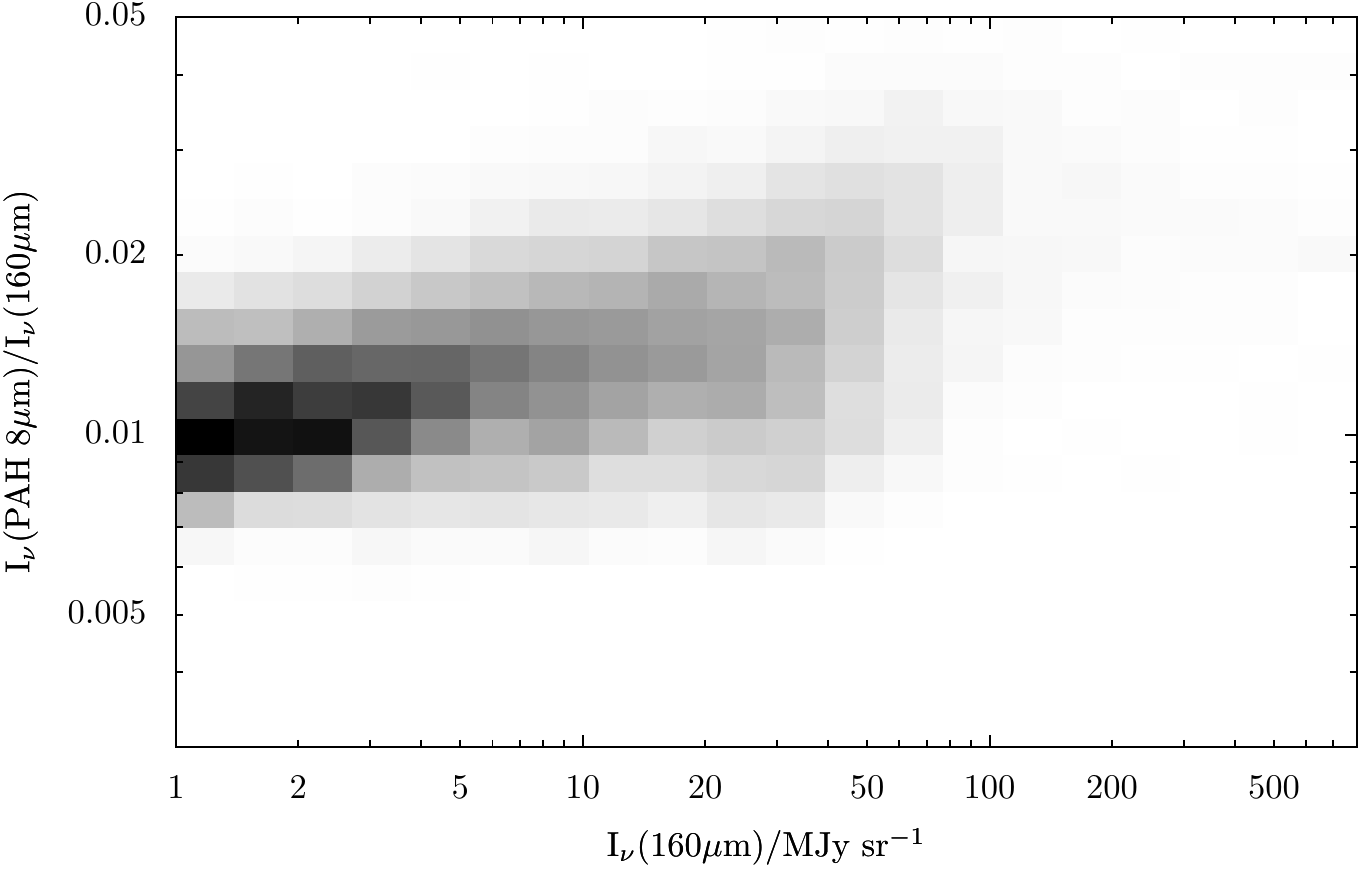} \end {center} \caption{ \label{plot_bendo} Spatially resolved dust-emission colour-magnitude diagrams for the simulated galaxies, shown as a density plot. For each galaxy model, the face-on view at a time of $0.5\Gyr$ is used. The plots are directly comparable with Figures 2 \& 5 in \citet{bendoetal08}. The left plot shows the dependence of the $8/24 \micron$ surface brightness ratio on $24\micron$ surface brightness, the right plot the $8/160\micron$ ratio as a function of $160\micron$ emission. In comparison with the results of \citet{bendoetal08}, the $8/160\micron$ ratio agrees well with the observations. However, the model galaxies have too large $8/24\micron$ ratio, especially in regions with low $24\micron$ surface brightness. This discrepancy is probably indicative of the model lacking stochastically heated dust grains in the diffuse ISM. } \end{figure*}

We have repeated this analysis with our model galaxies. The analysis
was restricted to the face-on galaxies to match the galaxies in the
\citet{bendoetal08} sample, and the pixel size set to $0.5 \kpc$ (the
pixel size in \citet{bendoetal08} varied between 0.7 and $3.6 \kpc$,
but we saw no significant dependence on pixel size in our simulations).
The results are shown in Figure~\ref{plot_bendo} and are directly
comparable to Figures 2 \& 5 in \citet{bendoetal08}. As with the
previous figures, we show the distribution of $8 \um / 24 \um$ and $8
\um / 160 \um$ ratios as greyscale density plots due to the large
number of pixels across the galaxies.

The $8 \um / 160 \um$ ratio agrees well with the observed galaxies, but
the $8 \um / 24 \um$ ratio is too steep, increasing to too large values
in regions with lower $24 \um$ brightness. This is most likely a
manifestation of the \mcrx \ dust model not including stochastically
heated dust grains. These grains contribute strongly to the flux at $24
\um$, which is underestimated. Simulations using the DL07 templates,
which include stochastically heated grains, for emission agree better
with the observational $8 \um / 24 \um$ ratios at low $24 \um$ surface
brightnesses.

%NLX% exclude from vocabulary builder
%NLX% end exclude from vocabulary builder

\subsection{Model Sensitivities}
\label{section_sensitivities}

In the previous sections we have shown how well our models produce
realistic spectra compared to observed galaxies, with some noticeable
offsets. Yet as mentioned in Section \ref{section-model}, there are
several free parameters that can affect the outputs of the models, that
we have chosen to fix to reasonable (and sometimes physically-based)
values.  In this section, we explore how sensitive the model outputs
are to these various free parameters. If the model is insensitive to
the exact value of a parameter, it generally gives an indication that
the model is robust to perturbations in the parameters. On the other
hand, it also means the parameter is largely unconstrained and leaves
the possibility that the parameter may influence the outputs in some
way that has not been tested. Conversely, if the model reacts
sensitively to changing a free parameter, this makes it possible to
tune the parameter but instead leaves open the possibility that the
parameter may not be appropriate in another situation (for example in
merger-driven starburst galaxies as opposed to the isolated,
quiescently evolving galaxies shown here).

In the following figures (\ref{plot_fluxratios-dustmodel} --
\ref{plot_fluxratios-pdr-fraction}) we use
Figures~\ref{plot_fluxratios} and \ref{plot_irxbeta} as bases for
comparison and explore the effects on the Sbc galaxy simulation when
changing a free parameter, keeping the other free parameters fixed to
their fiducial values. In all plots we show the effect of the free
parameter for the 13 different inclinations modelled (face-on to
edge-on to opposite face-on).

The first parameter we explore here is the dust size distribution.
\citet{draineli07} has three classes of distributions: Milky-Way-like,
LMC-like and SMC-like size distributions, with varying amounts of small
carbonaceous and silicaceous grains. Following \citet{draineetal07} and
given the roughly solar metallicity of most of the simulations, we have
used Milky-Way-like dust within our simulations.  Changing the dust
model from this dust distribution to the LMC- or SMC-type size
distributions \citep[shown and discussed in][DL07]{weingartnerdraine01}
has a large effect on the SED, both in the UV and the IR, as shown in
Figure~\ref{plot_fluxratios-dustmodel}. In the UV, both the LMC- and
SMC-type dust have a smaller $2200 \Angstrom$ bump than MW-type dust.
Since the $2200 \Angstrom$ absorption feature sits directly in the
GALEX NUV band, it strongly reduces the reddening effect in the FUV/NUV
bands of dust. Increasing the amount of MW dust, with its strong $2200
\Angstrom$ feature, for example by viewing the galaxy more edge-on,
increases the amount of overall UV absorption but produces very little
reddening in the FUV/NUV bands. SMC-type dust, in contrast, whose
overall opacity is generally lower but has a steeply increasing slope
towards shorter wavelengths, produces strong reddening and less overall
attenuation Both of these effects are seen in the UV colour-colour
diagram and in the ``IR excess-UV slope'' diagram (IRX-$\beta$).  As
the $2200 \Angstrom$ feature is generated by some of the same grains
that also give rise to the PAH emission features in the MIR, LMC and
SMC dust also have weaker PAH emission than MW dust, as can be seen in
the IRAC colour diagrams. From Figure~\ref{plot_fluxratios-dustmodel}
we can consider SMC-type dust unlikely for the SINGS sample based both
on the FUV/NUV colours and on the PAH emission in the IRAC bands
\citep[consistent with][]{draineetal07}. LMC dust on the other hand
seems like just as viable candidate as the MW-type dust, with similar
IRAC colours and slightly better agreement in the FUV/NUV colour,
though the cause for the offset to bluer $24 \um / 70 \um$ colour is
unknown.

Figure~\ref{plot_fluxratios-PAHfraction} shows the effect of changing
$b_C$, the amount of PAH grains in the log-normal components of the
size distributions of DL07. This $b_C$ parameter goes from $6 \times
10^{ - 5 }$ to 0, corresponding to a mass fraction of PAH grains, $q_{
\rm { PAH } }$, of 4.58\% to 0.47\%. Changing the PAH fraction has a
small effect on the SED except in the mid-IR ``PAH'' region. As
expected, models with lower amounts of PAH grains have weaker PAH
emission features, leading to higher $3.6 / 5.8 \um$ and $3.6 / 8.0
\um$ ratios. However, this change is parallel to the locus of observed
galaxies in these colours, so it does not mitigate the offset between
the models and observations in the IRAC bands. More surprisingly,
changing the PAH fraction does not significantly change the strength of
the $2200 \Angstrom$ feature in the NUV band.

The effect of changing the \mapiii \ cluster mass from the fiducial
value of $10^5 \Msun$ to $10^7 \Msun$ is shown in
Figure~\ref{plot_fluxratios-mappings-mcl}. This only affects the SED of
the star-forming regions, and has a small effect on the overall SED
except in the $24 / 70 \um$ colour, where the higher cluster mass leads
to more ``compact'' clusters and hence to a bluer (hotter) colour. This
can be expected, as the $24 \um$ band is where the fraction contributed
by the star-forming regions is the greatest. 
% A higher cluster mass also leads to a
% slightly redder FUV/NUV colour, most likely due to the larger HII\
% regions with bigger cluster masses, leading to a higher (HII\ region)
% dust column. (IM GUESSING HERE, BUT IT COULD EXPLAIN IT)

The other free parameter in the \mapiii \ particles is the PDR
fraction. As with the cluster mass, changing this parameter from its
fiducial value of 0.20 to smaller or larger values has only a small
effect on the overall SED, as shown in
Figure~\ref{plot_fluxratios-pdr-fraction}, most due to the weak overall
contribution of the \mapiii \ particles. A smaller PDR fraction does
result in slightly bluer FUV/NUV and $24 / 70 \um$ colours, as the
\hii\ regions and their hot, young stars are more exposed, while the
change to the PAH region is minimal as it is dominated by diffuse PAH
emission. If the PDR fraction is changed to unity, no light from the
star-forming regions can escape unattenuated, and the colours become
significantly redder, both in the FUV/NUV and $24 / 70 \um$ bands.

The final `tunable' parameter of the model is the PAH emission template
fraction $f_t$, whose variation is shown in
Figure~\ref{plot_fluxratios-pah-template-fraction}. Since this
parameter expressly changes the way the emission in the PAH features is
calculated, its effect is mostly in the IRAC bands. A larger $f_t$
results in stronger PAH features, moving the simulations to redder $3.6
/ 5.8 \um$, $3.6 / 8.0 \um$ and $5.8 / 8.0 \um$ colours. As with the
amount of PAH grains $b_C$, this change is parallel to the locus of the
SINGS galaxies and can not be used to improve the agreement with the
observations. However, $f_t$ also affects the $8.0 / 24 \um$ colour in
such a way that values other than the fiducial $f_t = 0.50$ moves the
simulations away from the region occupied by the Sb-Sc SINGS galaxies,
helping to constrain the value of this parameter.

Another important check is to verify that the use of a thermal
equilibrium approximation for the dust grains does not have a large
influence on the results. For this purpose, the alternative way of
calculating the dust emission spectrum, using the precomputed SED
templates of DL07 as explained in Section~\ref{section_ir-emission},
was also included in \mcrx . We compare the two IR SEDs in
Figure~\ref{plot_dl07-template-sed}. In general the match is good, with
both the PAH features and the IR bump very similar. Yet they disagree
in several details, the most significant being the overall shift of the
DL07 IR bump to shorter (hotter) wavelengths, leading to a visible
decrease in the sub-mm flux. There are also significant offsets around
$20 \um$ and below $5 \um$. The latter offset has little effect on the
overall SED due to the increasing stellar contribution, except in the
$3.3 \um$ PAH feature.

As the IR SEDs are so similar, the change does not have a large impact
on the flux ratios shown in Figure~\ref{plot_fluxratios-dl07-template}.
 The $20 \um$ region does not fall in any of the included filter bands
and is not seen (though this offset would be visible in \emph{Spitzer}
IRS spectra). A large offset ($\sim 25 \%$) in the $3.6 \um$ IRAC band
is clearly visible. This arises from the stronger $3.3 \um$ PAH
emission feature in the DL07 dust relative to the
\citet{grovesetal08sparam} template, which, apart from this feature,
appears to agree very well.  Which of the templates (if either) have
the correct strength of this feature is unclear.

The largest offset in the IR appears in the $70 \um / 160 \um$ colour
(the bottom row of Figure~\ref{plot_fluxratios-dl07-template}) due to
the slight shift in the IR peak. This shift leads to the visibly lower
$850 \um$ flux, slightly lower $160 \um$ flux and slightly increased
$70 \um$ flux in Figure~\ref{plot_dl07-template-sed} and the offset
seen in the colour-colour diagram.  Interestingly, all this means that
using the DL07 templates increases the $160 / 850 \um$ flux ratio by
only about $25 \%$ compared to the standard thermal equilibrium
emission, leaving a still significant offset with the SINGS galaxies,
while making the $70 / 160 \um$ colour too blue to match the SLUGS
sample.

%Including the stochastically heated grains when computing the dust
%emission can be expected to lower the "850 micron" flux and improve
%the discrepancy with the SINGS galaxies, because cold dust will still
%fluctuate to higher temperatures and emit part of their luminosity at
%shorter wavelengths compared to a grain in thermal
%equilibrium.
%(I ACTUALLY DISAGREE WITH THIS - STOCHASTIC DUST SPENDS MORE TIME
%_COOLER_ THEN  EQUILIBRIUM TEMP DUST: HENCE I DONT THINK WE SHOULD
%INCLUDE THIS)

The parameters shown in the Figures mentioned above are not an
exhaustive list, of course. We have also verified that parameters
controlling the \citet{springelhernquist03} multiphase model have
negligible effect on the galaxy SED. As the gas densities in these
quiescently star-forming galaxies are mostly below or around the
threshold density for star formation \citep{coxetal05methods}, there is
little mass at densities where the multiphase medium develops and the
details of this treatment have little effect. The situation is
dramatically different in gas-rich merging galaxies, where the
multiphase model has a strong effect on the emerging SED
\citep{youngeretal08}.

The SED is also surprisingly insensitive to moderate changes in the
overall fraction of metals that are assumed to be in dust grains, which
are largely degenerate with viewing the galaxies at different
inclinations. For higher dust/metal ratios, the maximum infrared excess
is slightly higher, while the UV reddening is almost unchanged.

In addition to the parameters discussed above, a number of other
parameters were examined for completeness. As these show no significant
effects, the figures have been omitted. These additional tests include
the effect of setting the radii of the \mapiii \ particles to $r_s$ (as
discussed in Section~\ref{section_interface}), and the effect of
dropping the \mapiii \ particles altogether (i.e. using only the
Starburst99 models.

%NLX% exclude from vocabulary builder
\begin{figure*} \begin {center} \begin{tabular*}{\textwidth}{r@{\extracolsep{\fill}}r} \includegraphics*[scale=0.62]{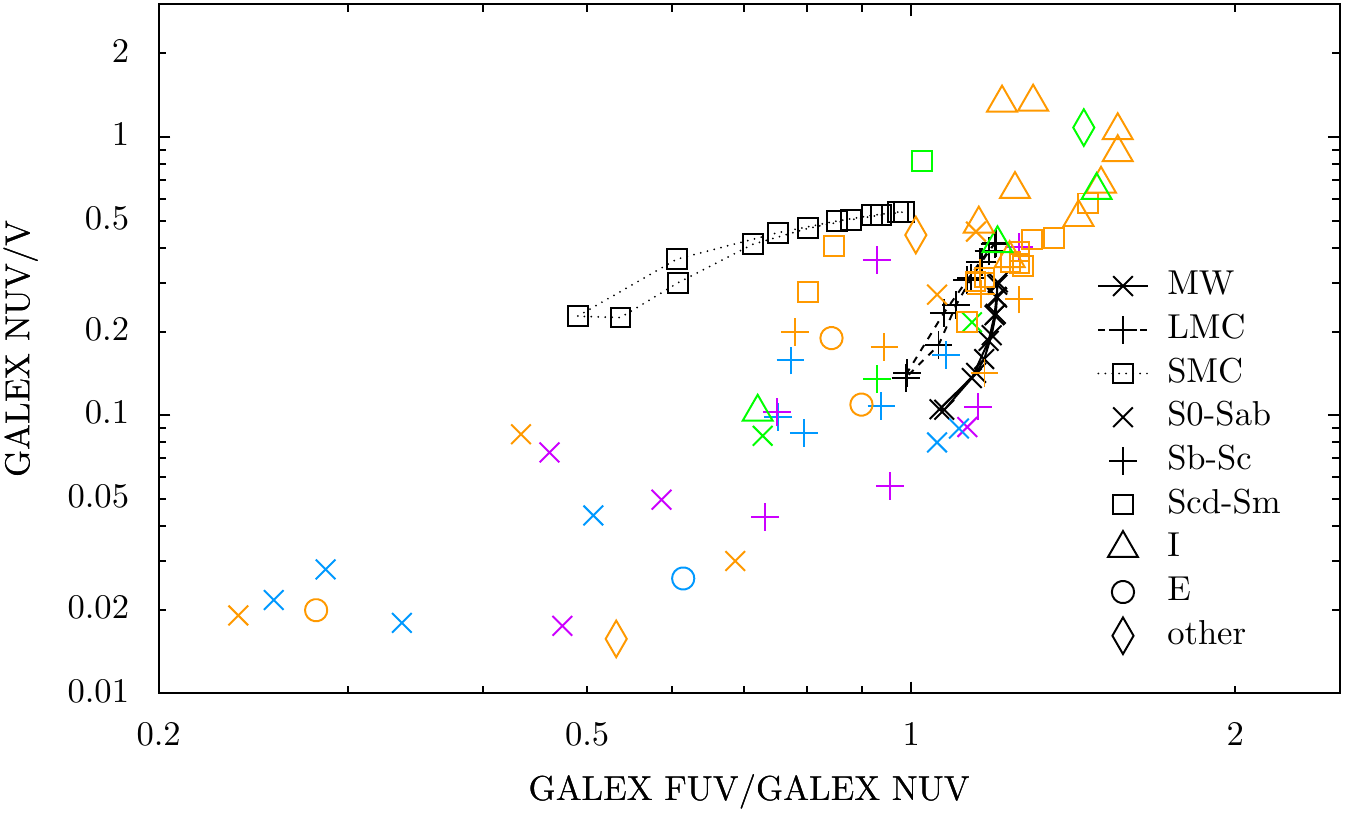} & \includegraphics*[scale=0.62]{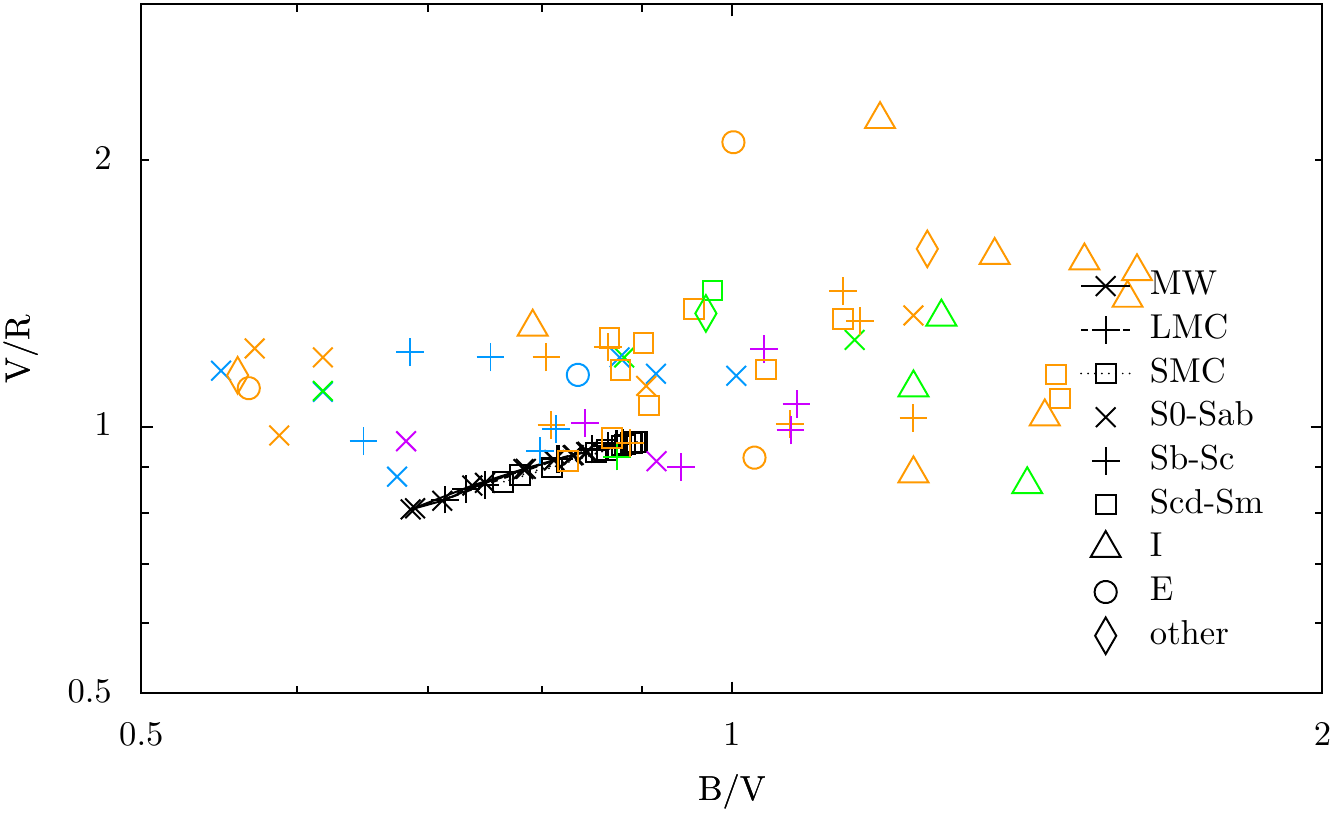} \\ \includegraphics*[scale=0.62]{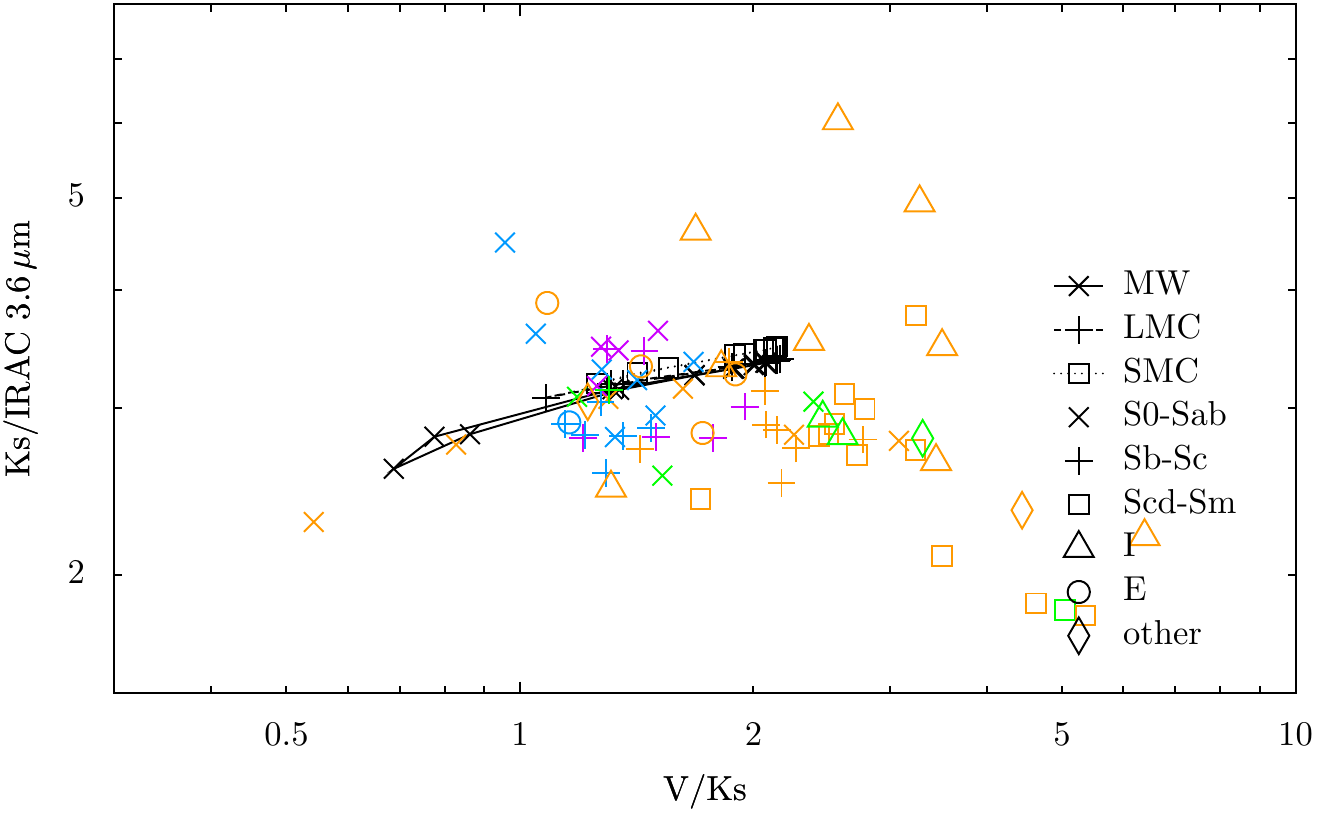} & \includegraphics*[scale=0.62]{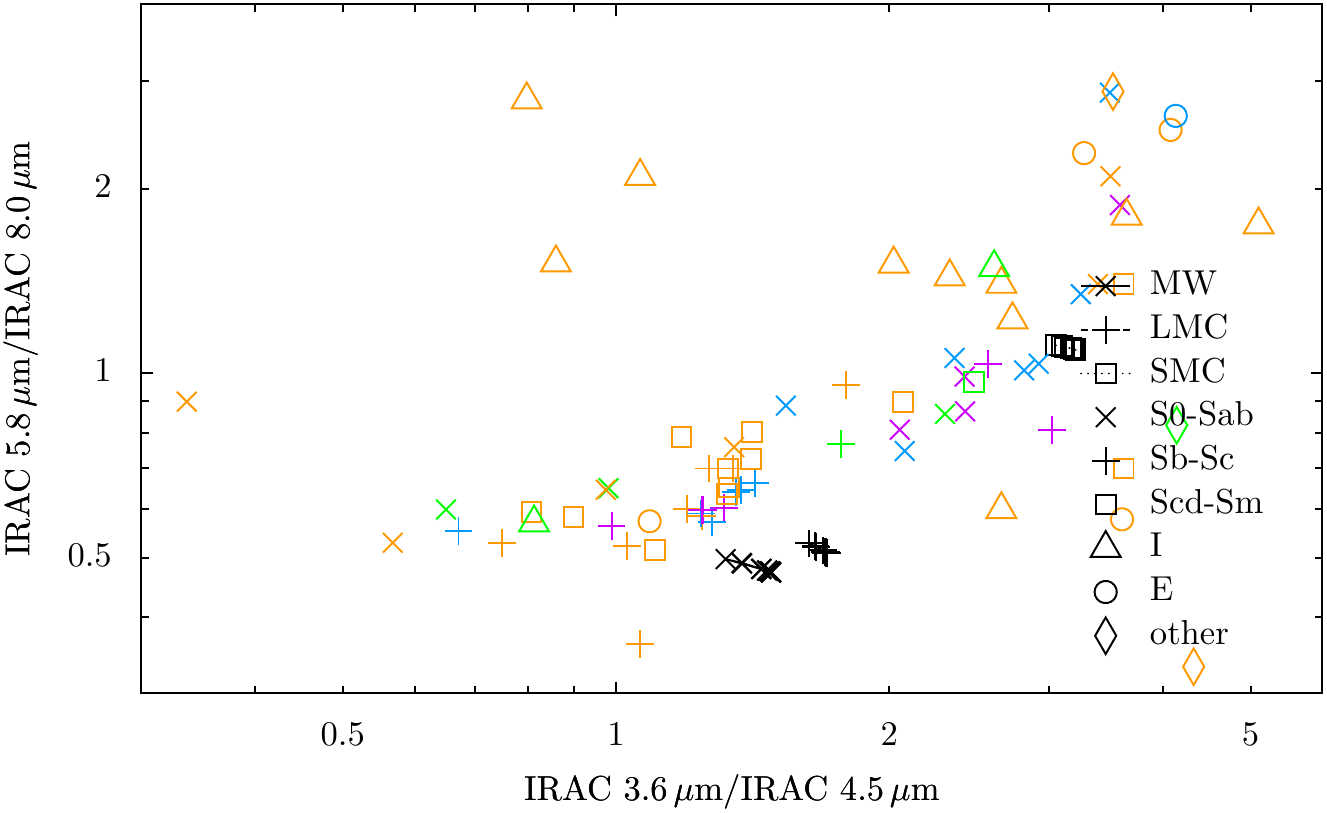} \\ \includegraphics*[scale=0.62]{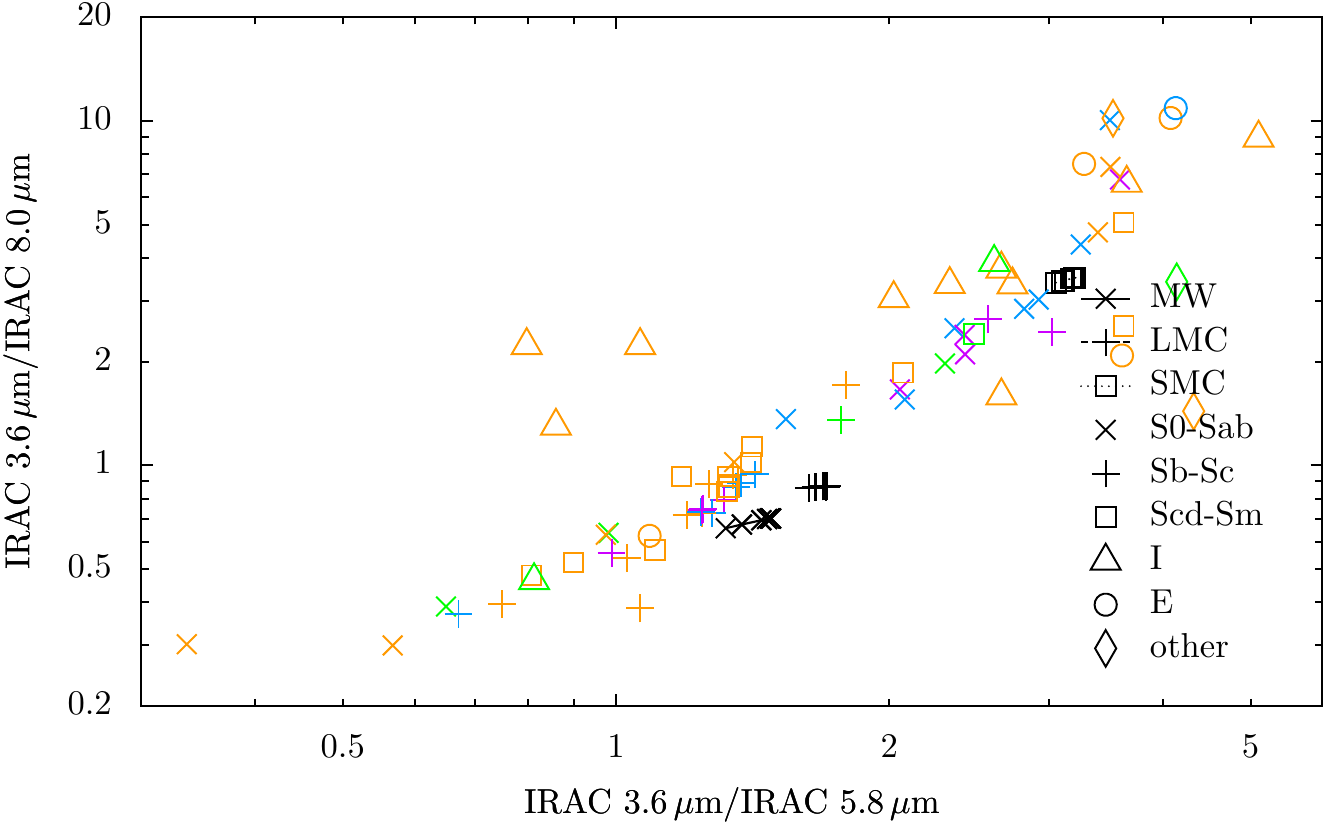} & \includegraphics*[scale=0.62]{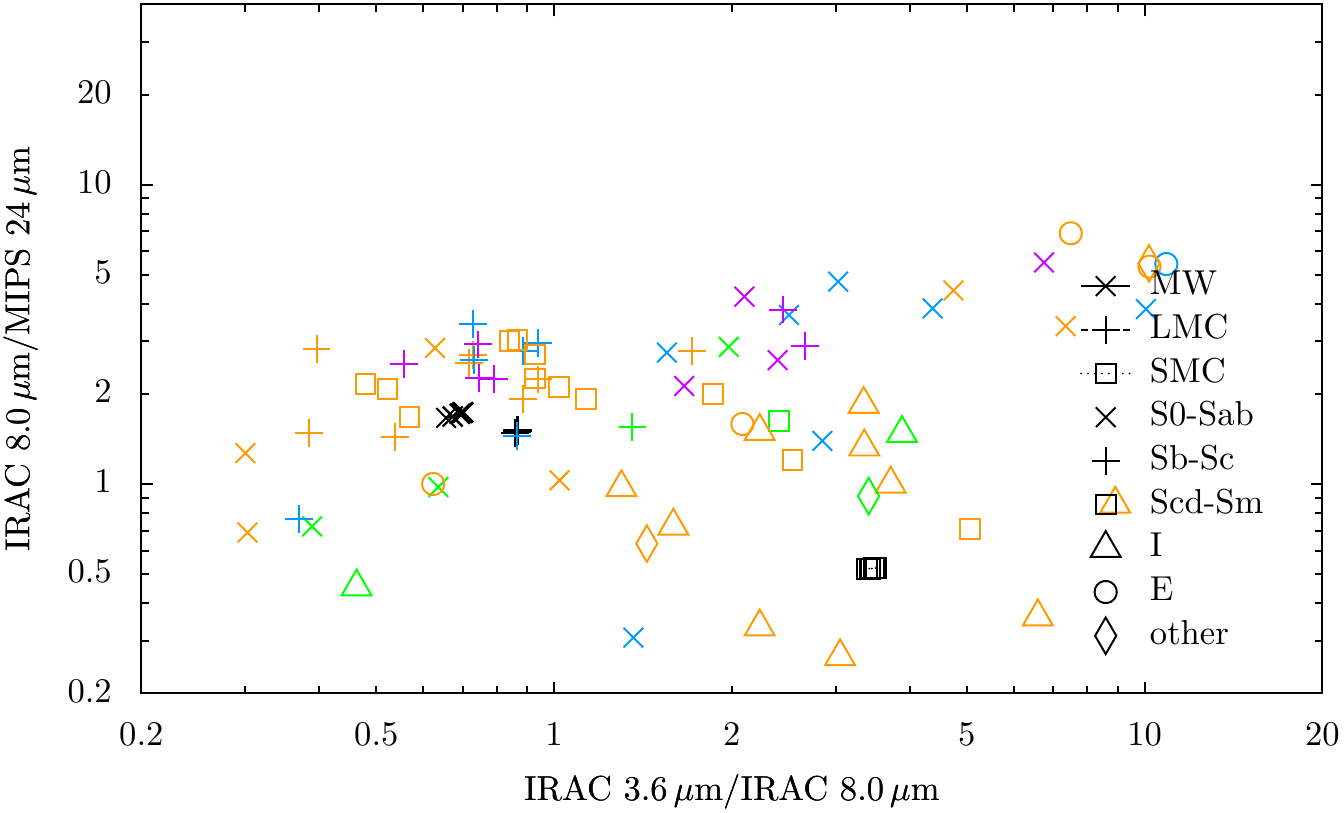} \\ \includegraphics*[scale=0.62]{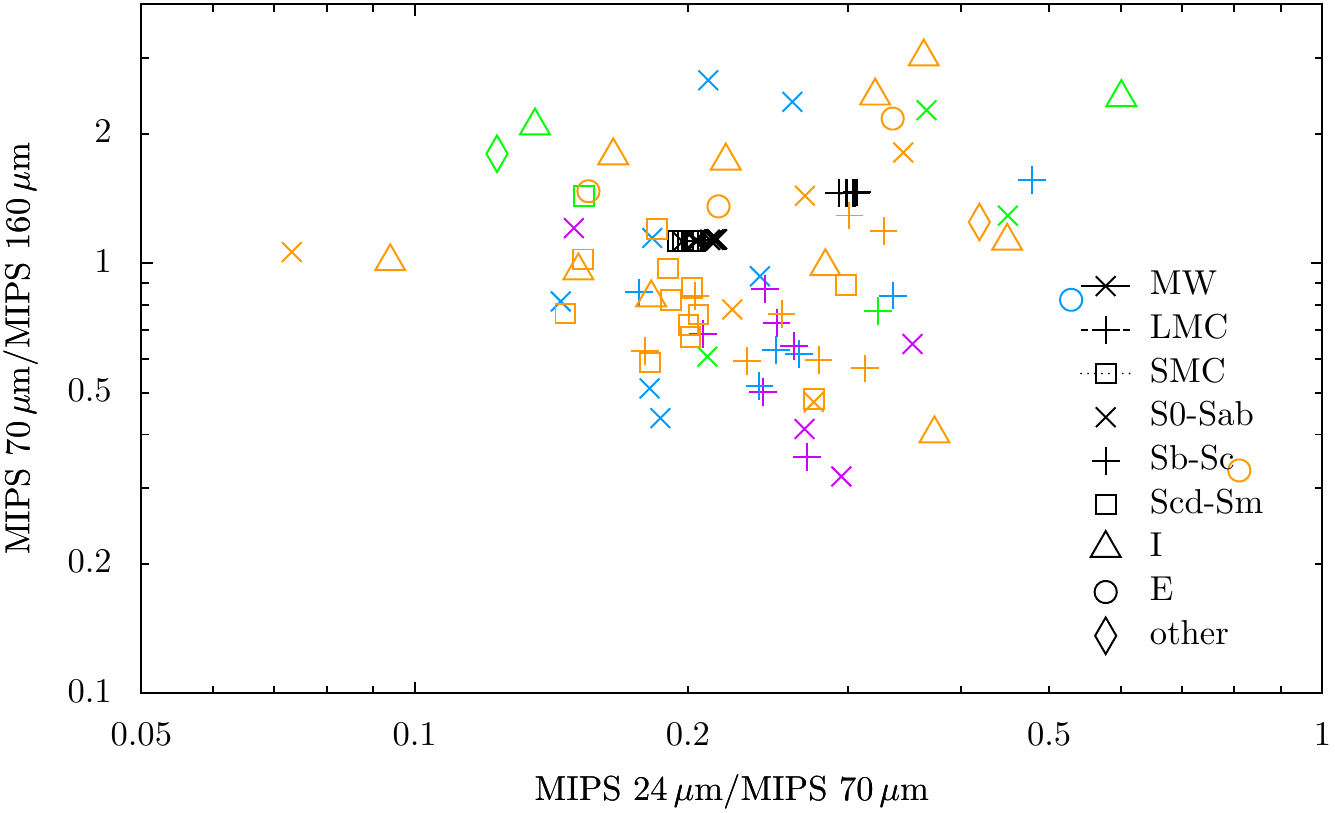} & \includegraphics*[scale=0.62]{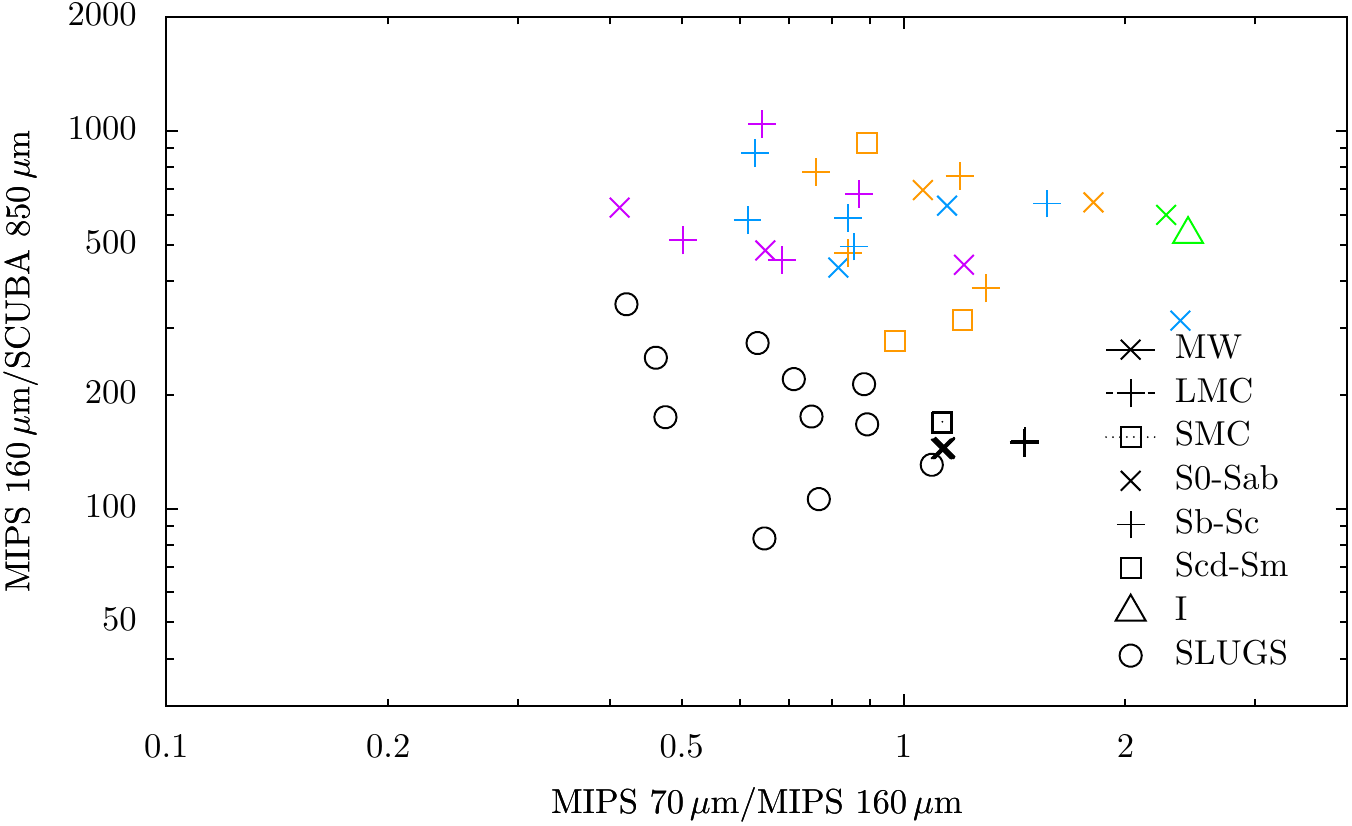} \end{tabular*} \end {center} \caption{ \label{plot_fluxratios-dustmodel} Sensitivity of the Sbc galaxy SED to the dust model used. The simulations are shown as black points for each parameter set, connected by a line showing the dependence on inclination. The SINGS points are as in Figure~\ref{plot_fluxratios}. One of the major differences between different dust models is the presence or absence of the ``$2200\Angstrom$ bump'' in the opacity curve. As this bump sits right in the GALEX NUV band, it strongly affects the NUV flux, which shows up also in the IRX-$\beta$ plot in the lower right. The absence of small carbonaceous grains in the SMC model also strongly affects the PAH fluxes in the IRAC bands.} \end{figure*} \begin{figure*} \begin {center} \begin{tabular*}{\textwidth}{r@{\extracolsep{\fill}}r} \includegraphics*[scale=0.62]{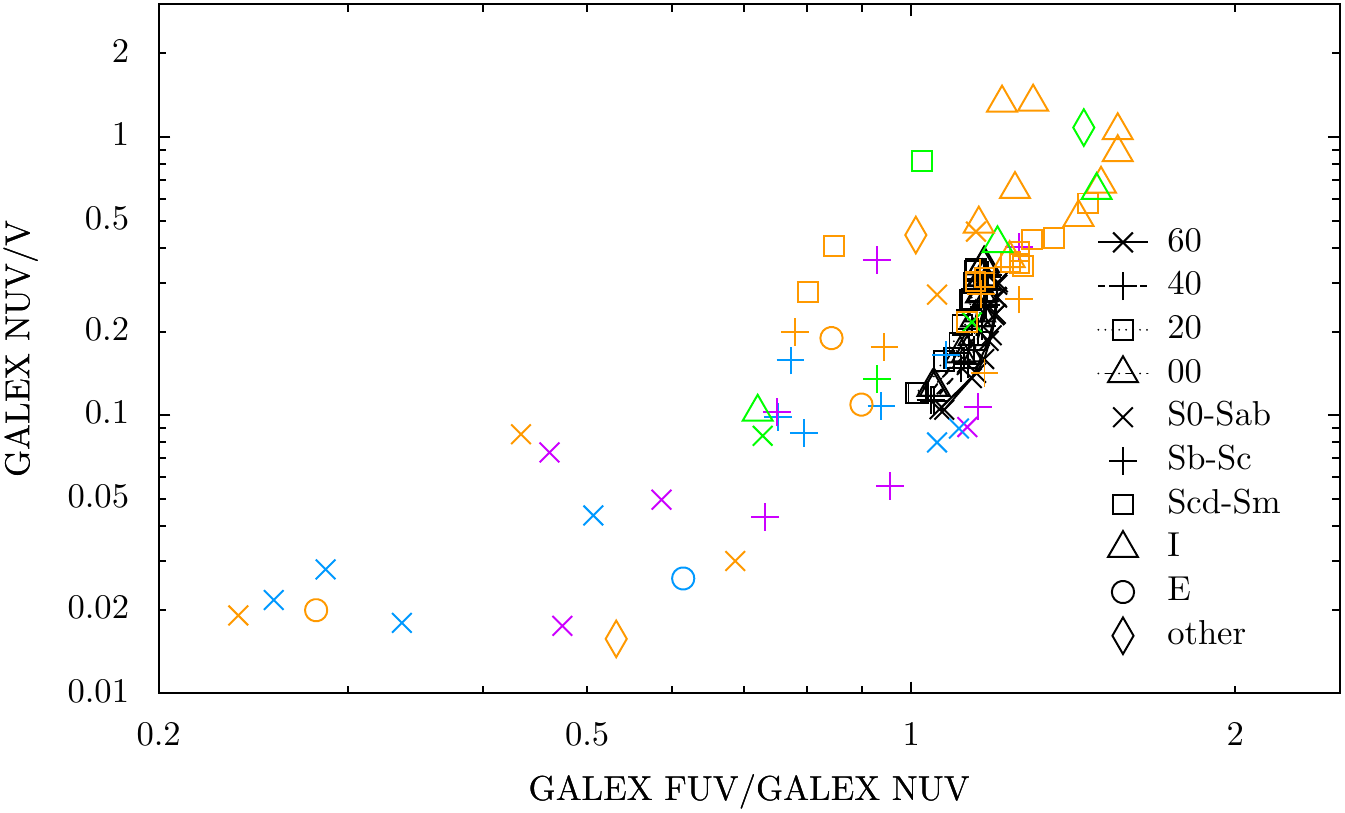} & \includegraphics*[scale=0.62]{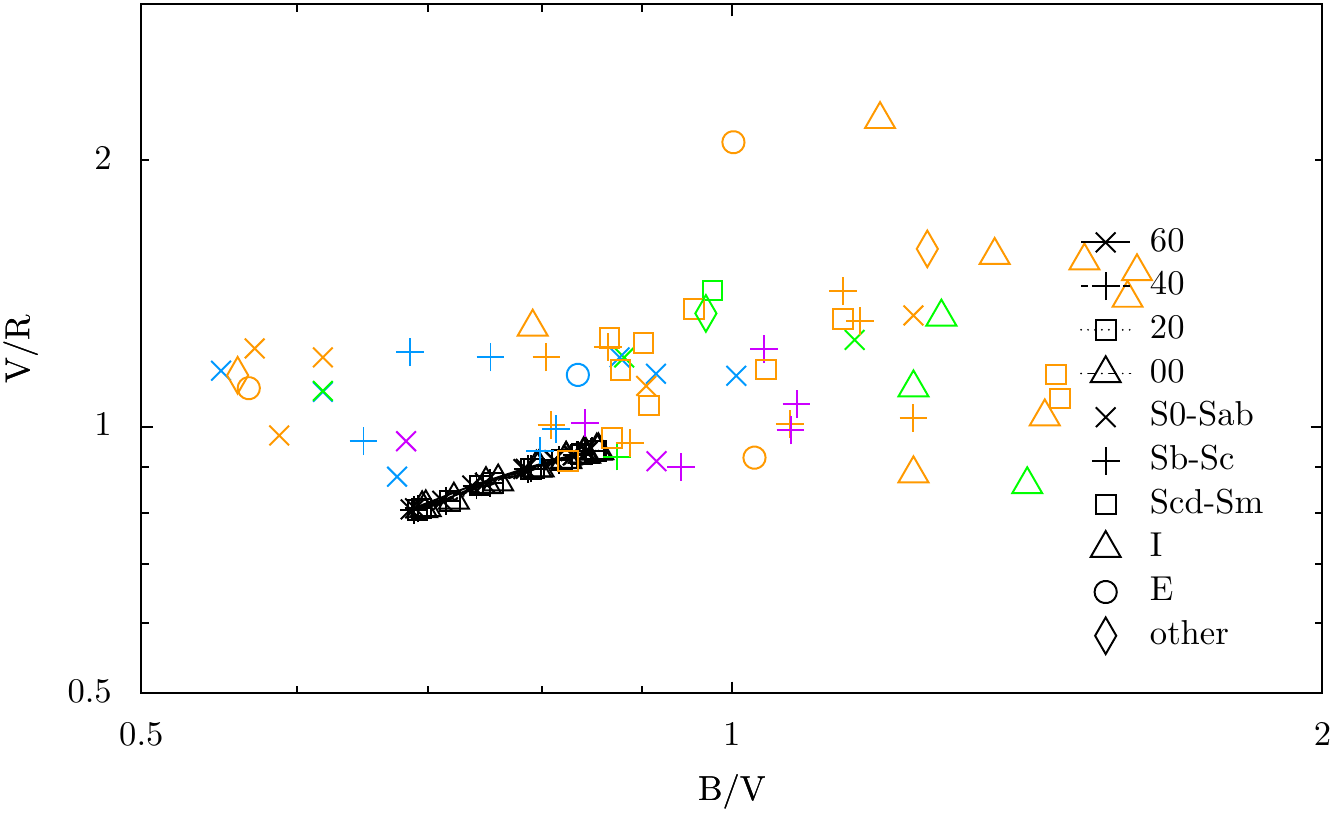} \\ \includegraphics*[scale=0.62]{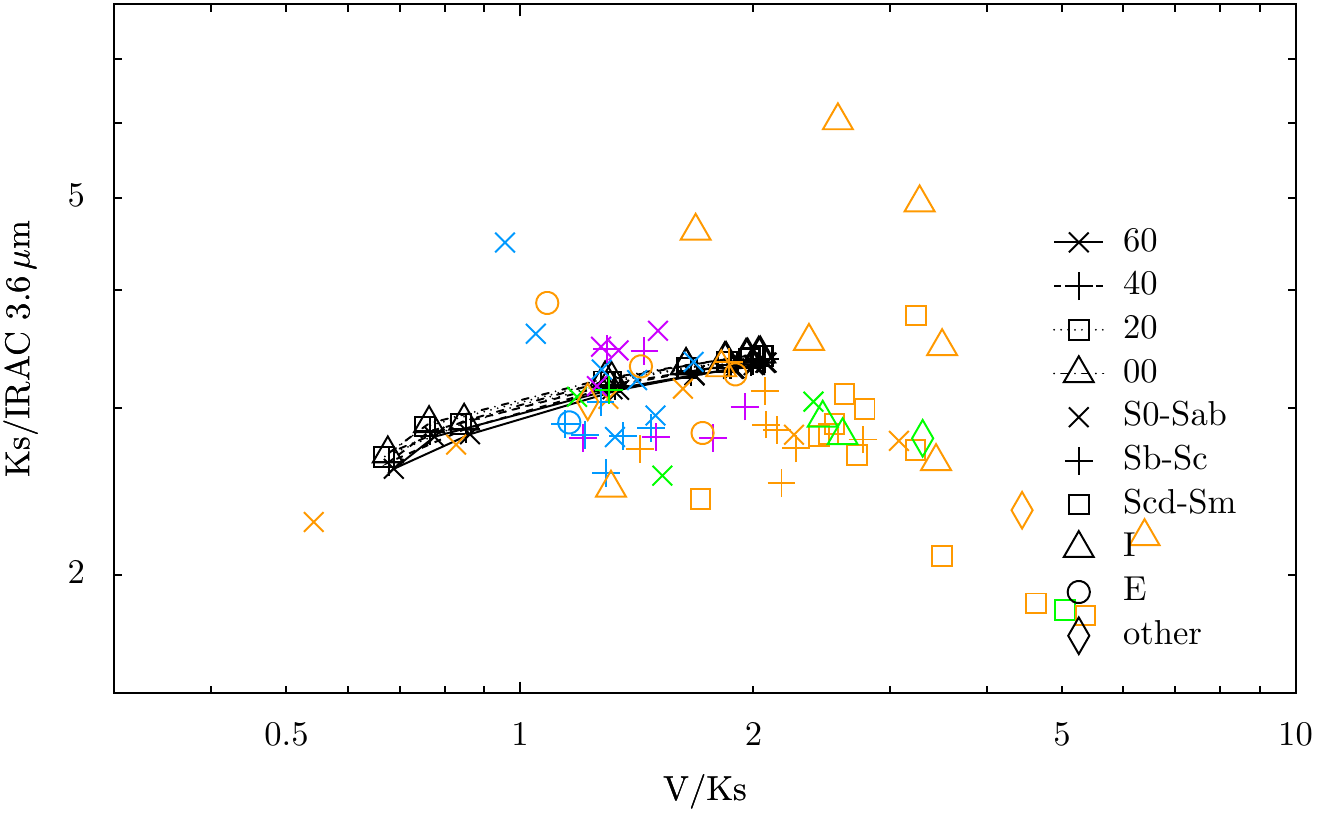} & \includegraphics*[scale=0.62]{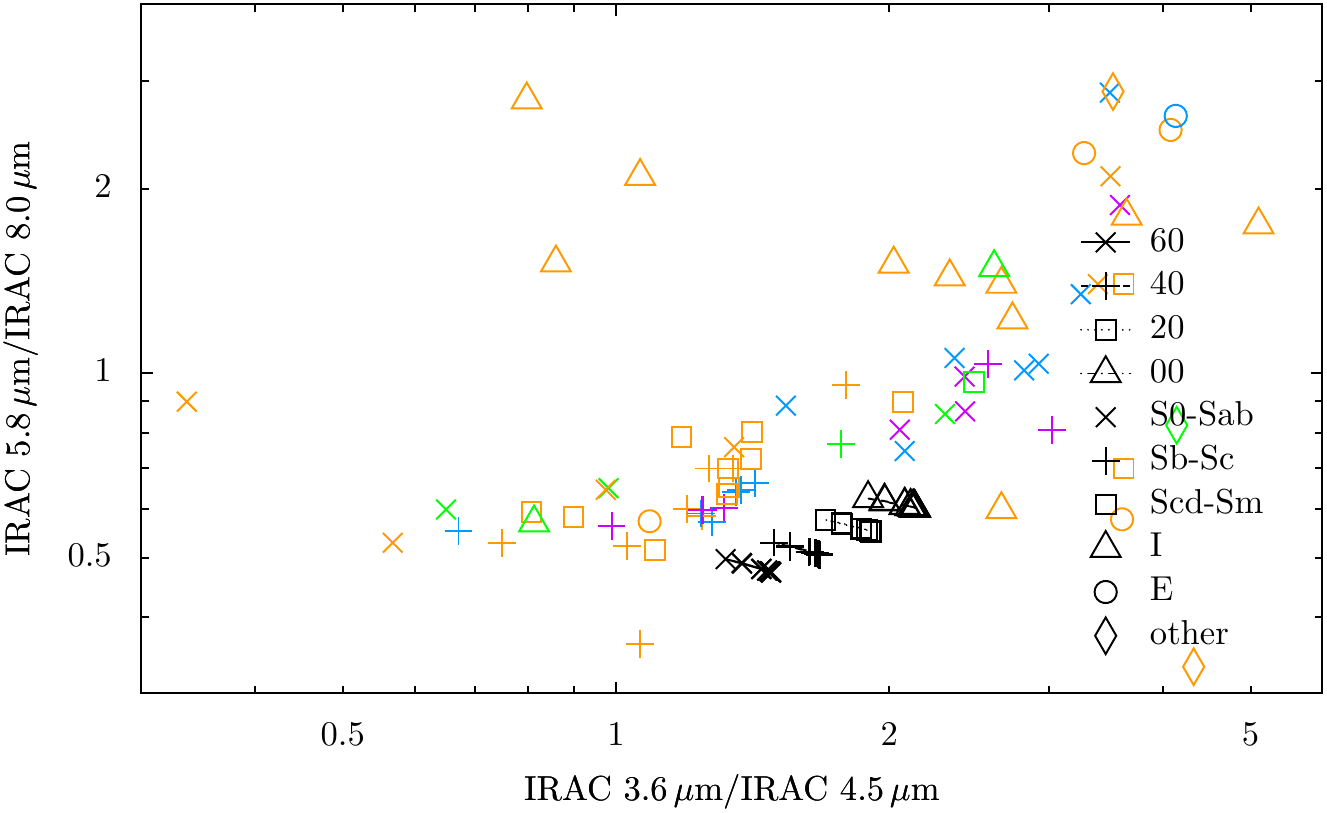} \\ \includegraphics*[scale=0.62]{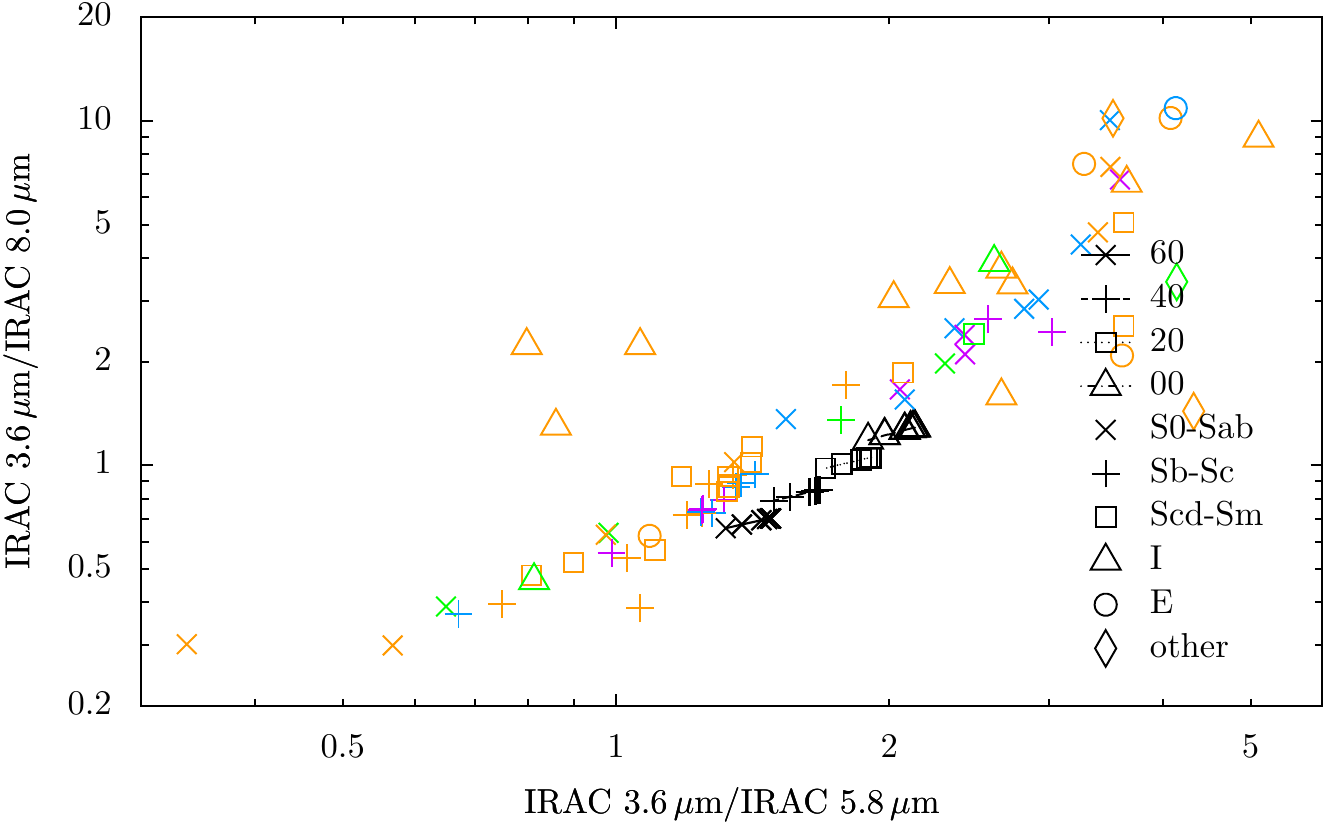} & \includegraphics*[scale=0.62]{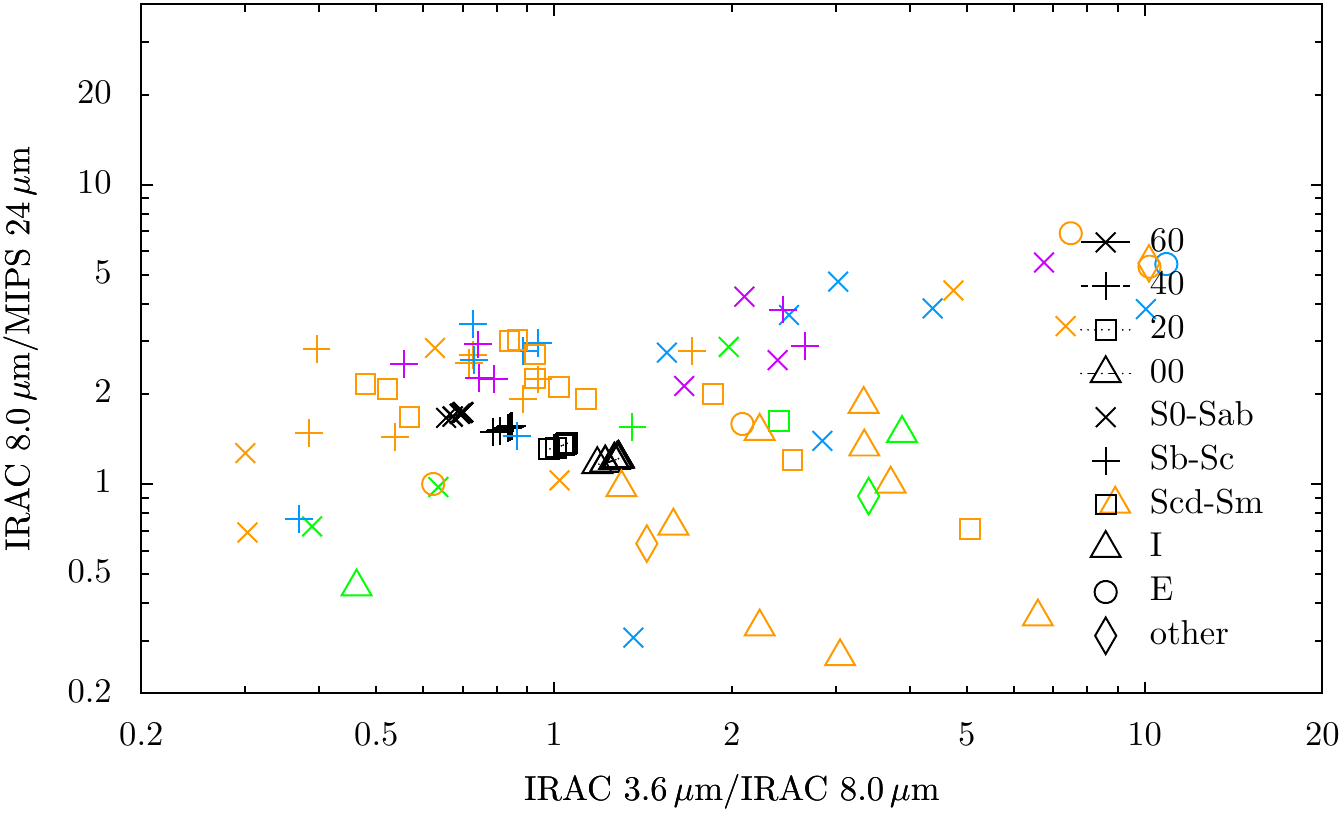} \\ \includegraphics*[scale=0.62]{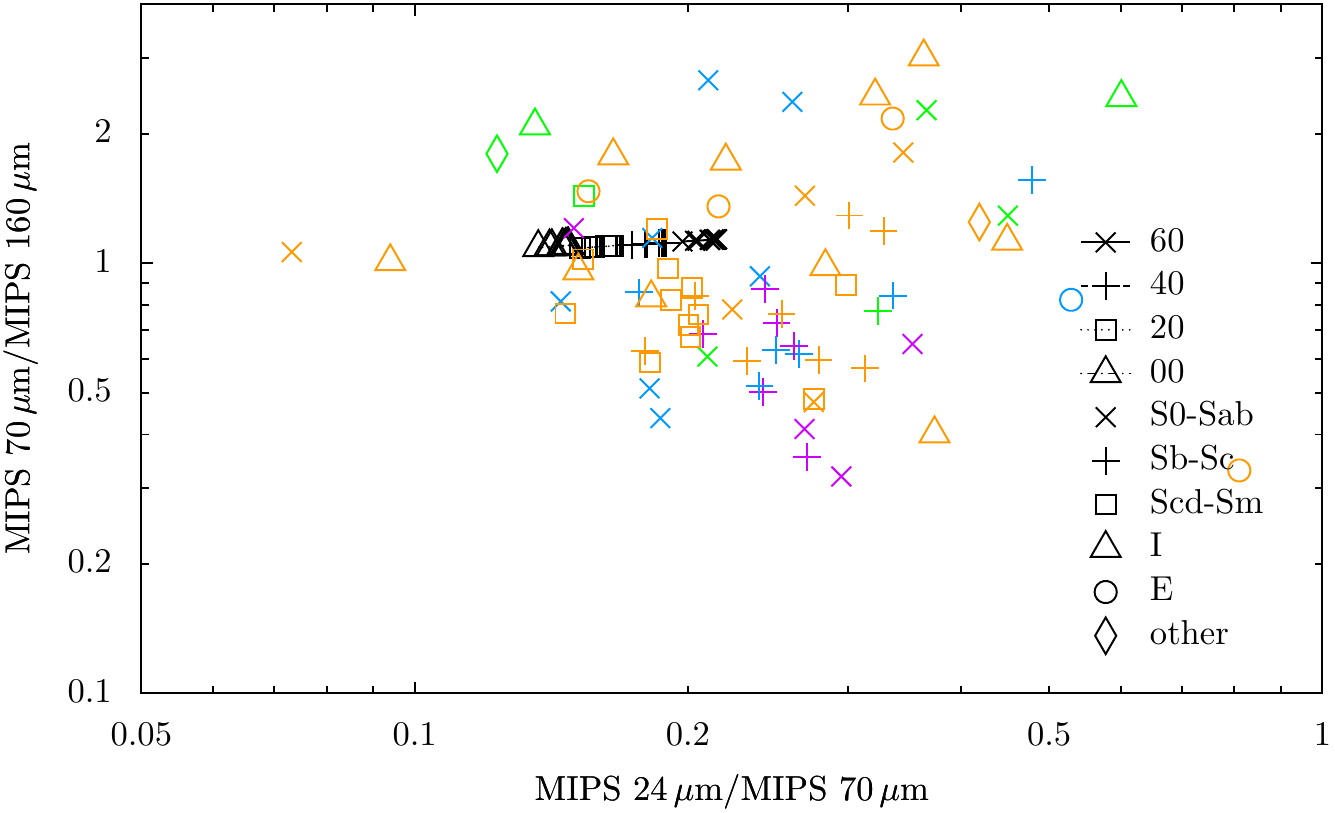} & \includegraphics*[scale=0.62]{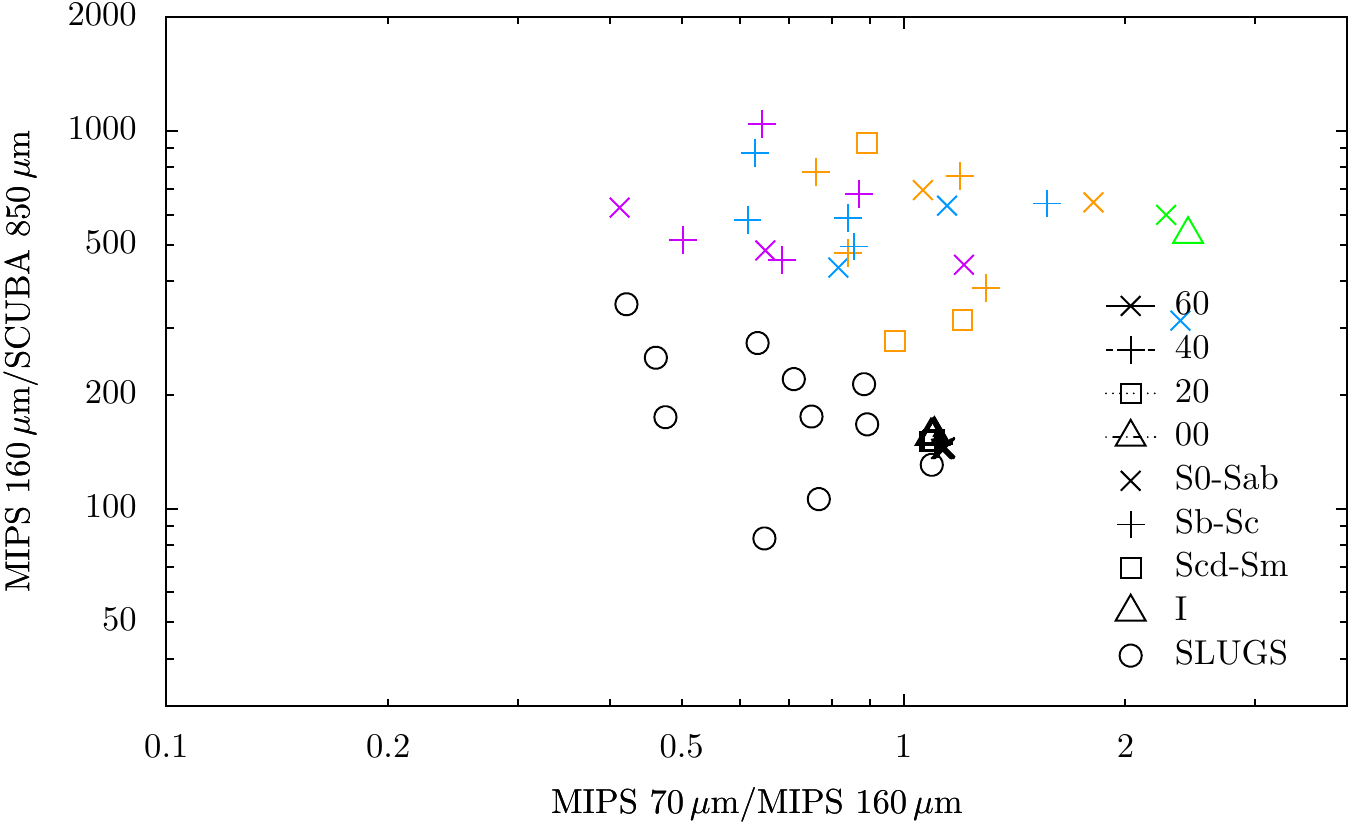} \end{tabular*} \end {center} \caption{ \label{plot_fluxratios-PAHfraction} Sensitivity of the Sbc galaxy SED to $b_C$, the PAH abundance (expressed as C per H nucleus) in the log-normal populations with sizes of $3.5\Angstrom$ and $40\Angstrom$ in the \citet{weingartnerdraine01} and \citet{draineli07} dust models.  Symbols are as in Figure~\ref{plot_fluxratios-dustmodel}. Not surprisingly, this parameter mostly affects the fluxes in the IRAC bands, but the change is parallel to the locus described by the SINGS galaxies.} \end{figure*} \begin{figure*} \begin {center} \begin{tabular*}{\textwidth}{r@{\extracolsep{\fill}}r} \includegraphics*[scale=0.62]{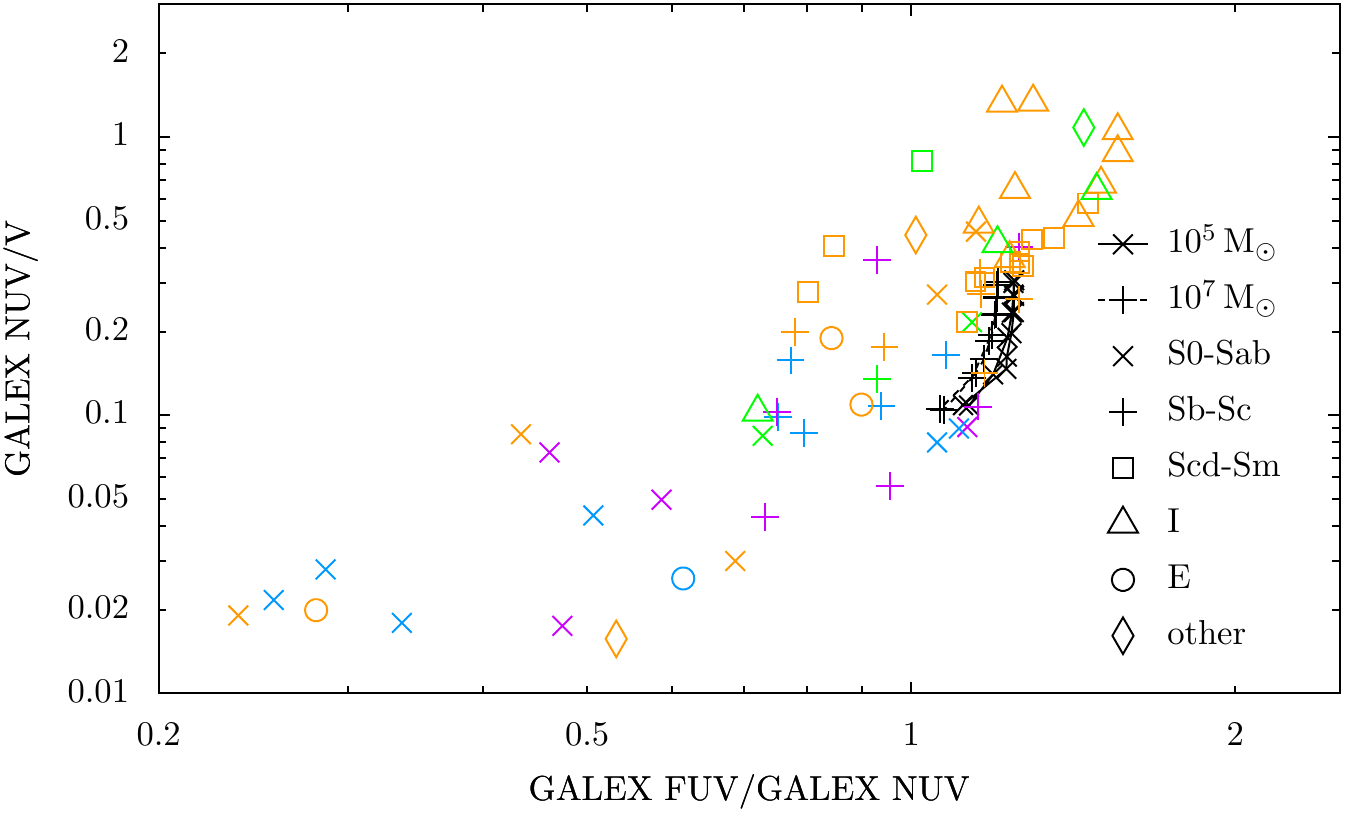} & \includegraphics*[scale=0.62]{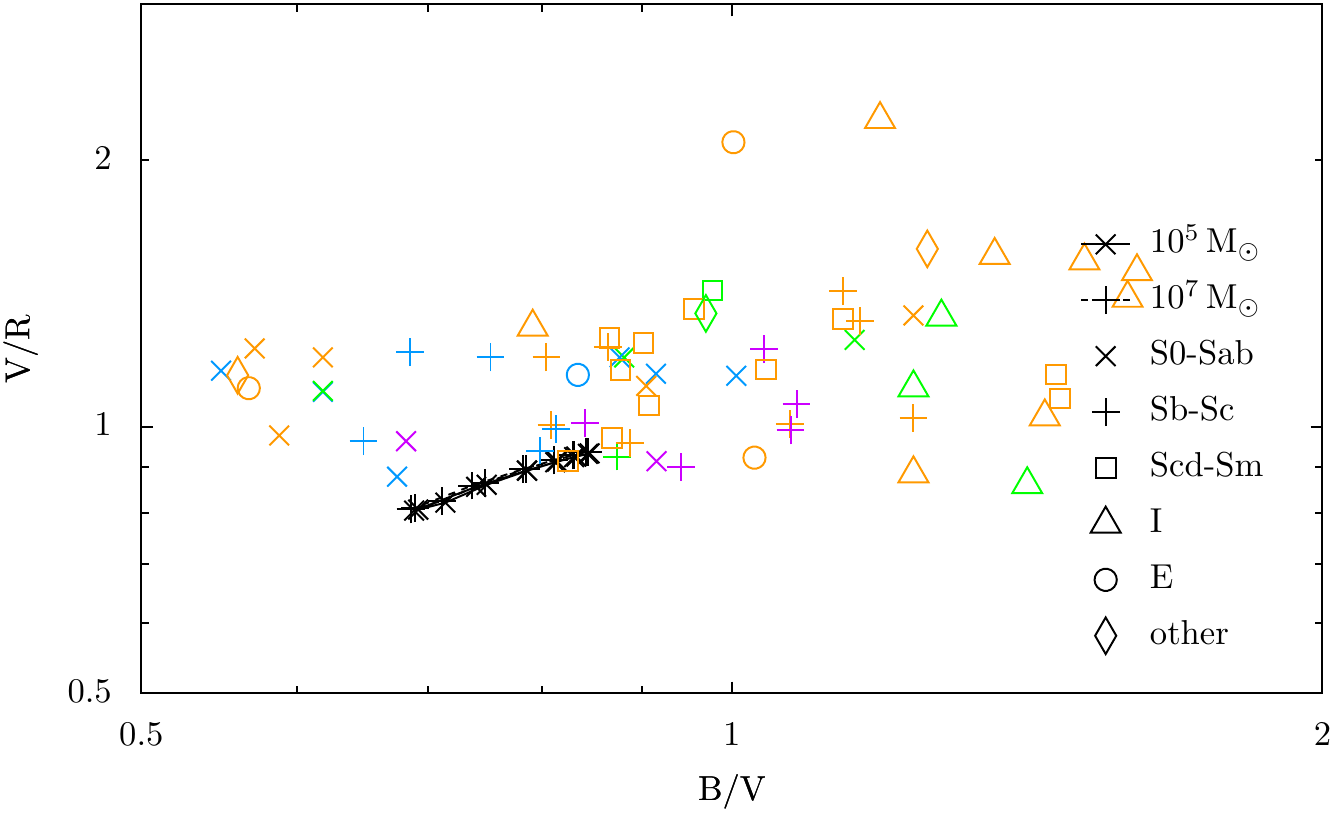} \\ \includegraphics*[scale=0.62]{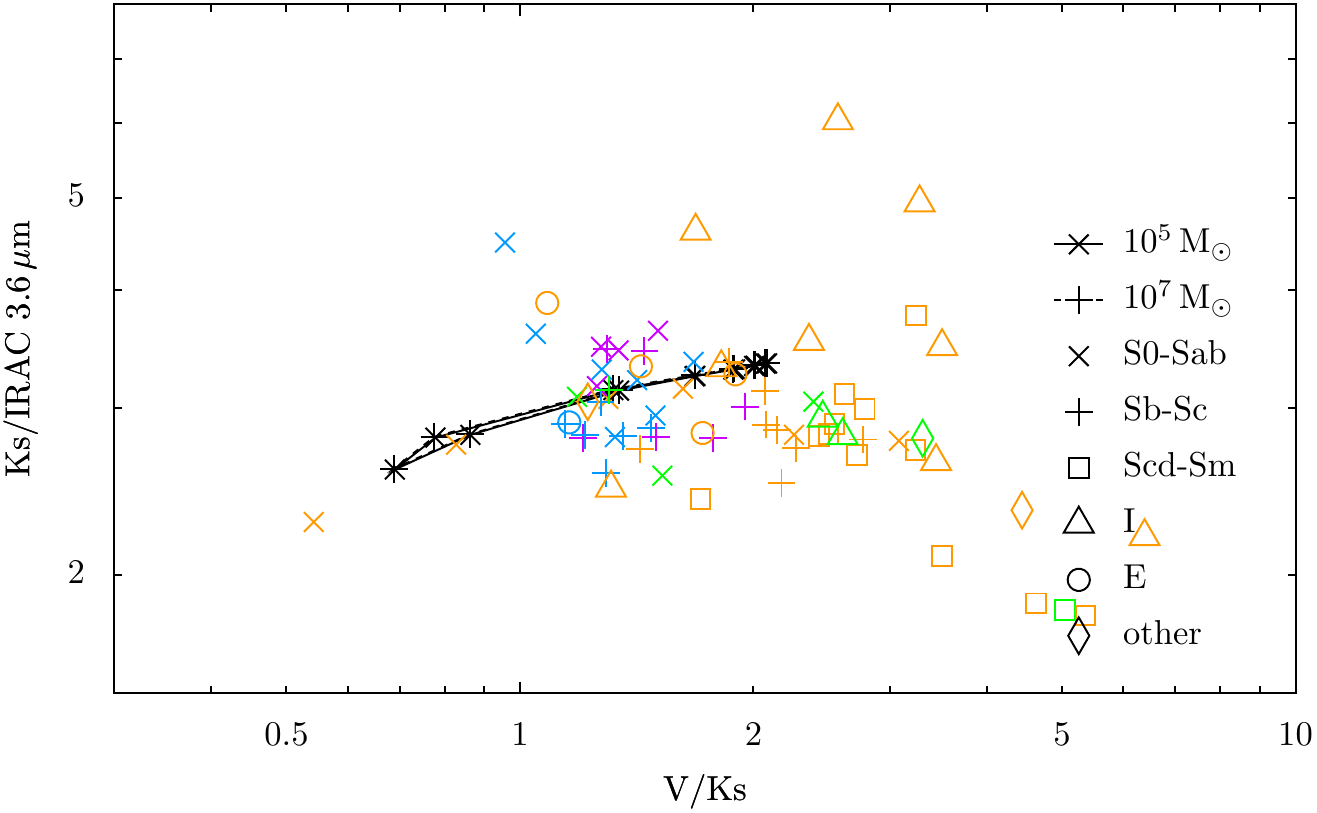} & \includegraphics*[scale=0.62]{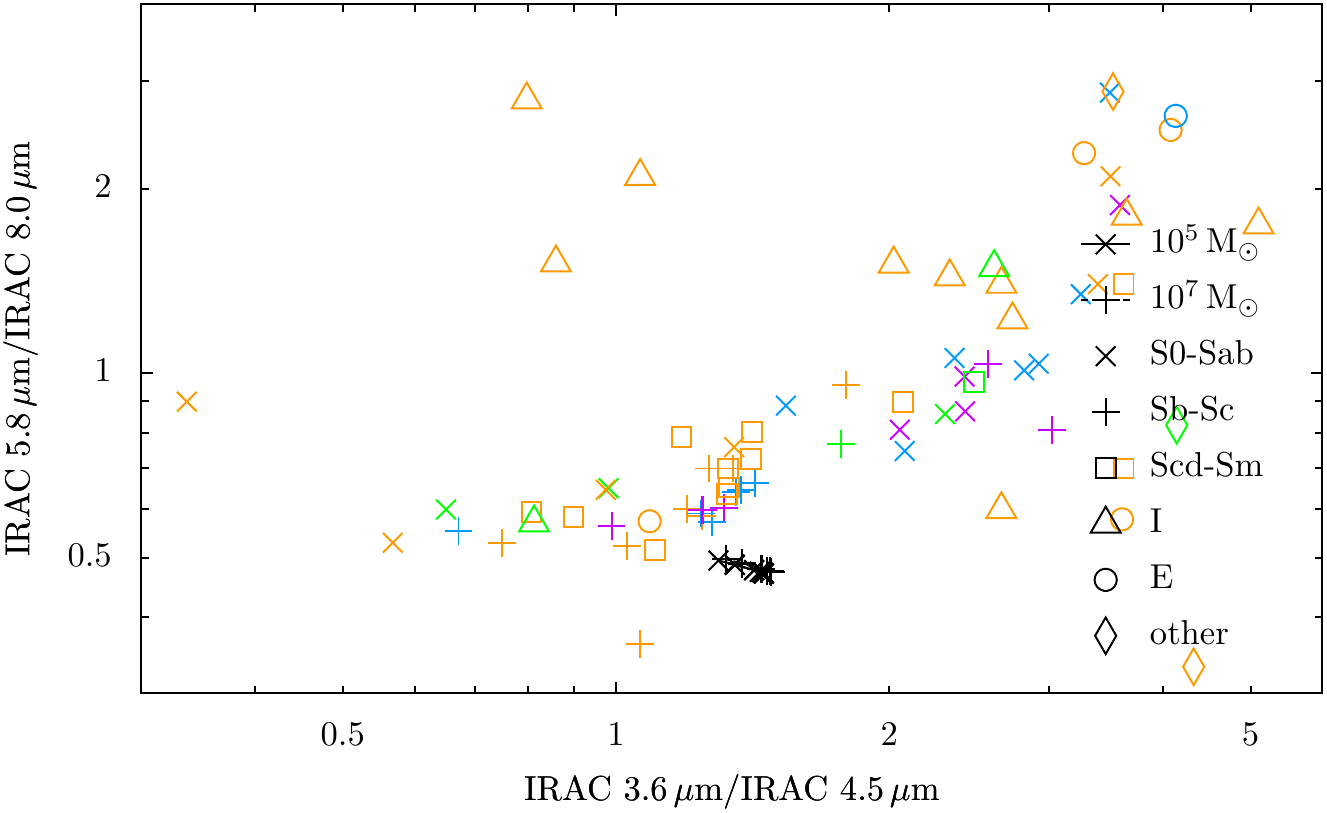} \\ \includegraphics*[scale=0.62]{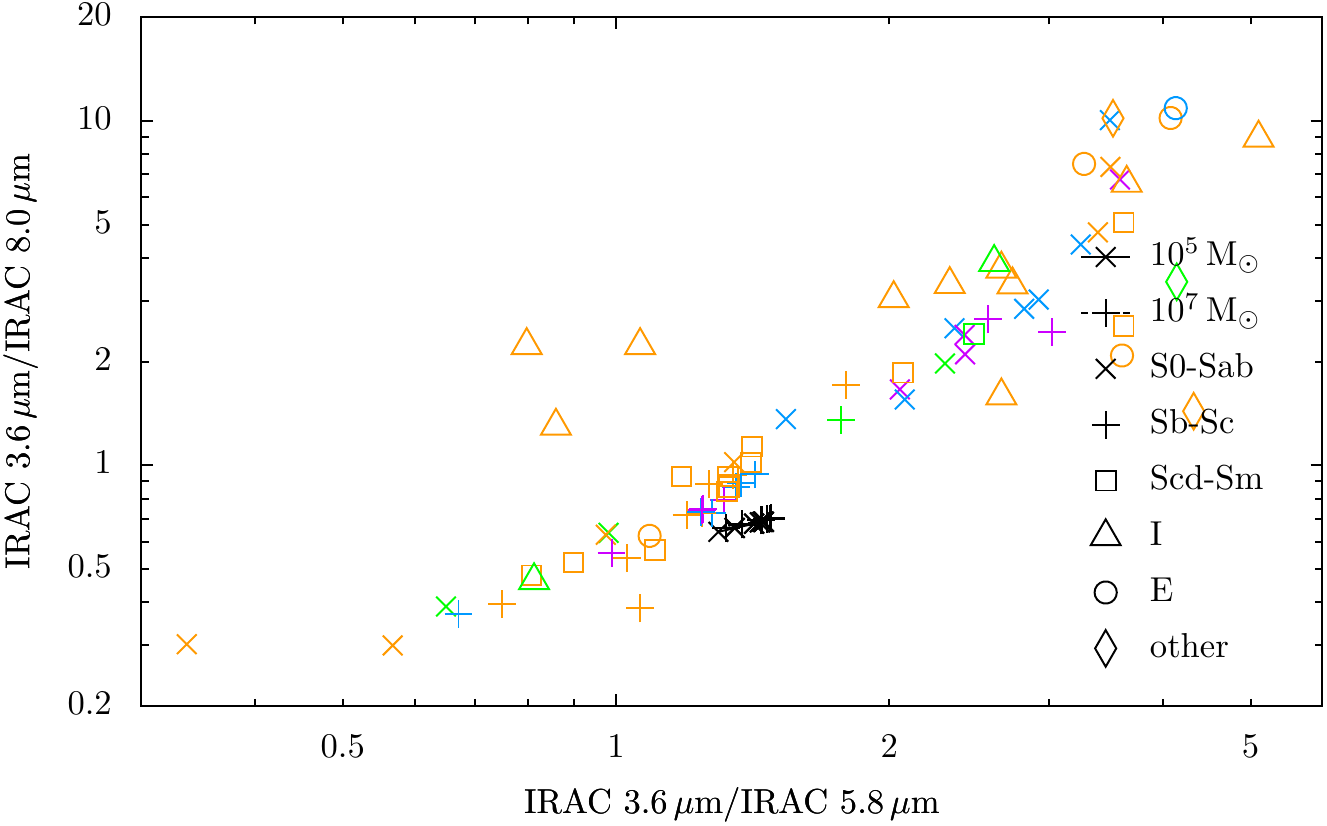} & \includegraphics*[scale=0.62]{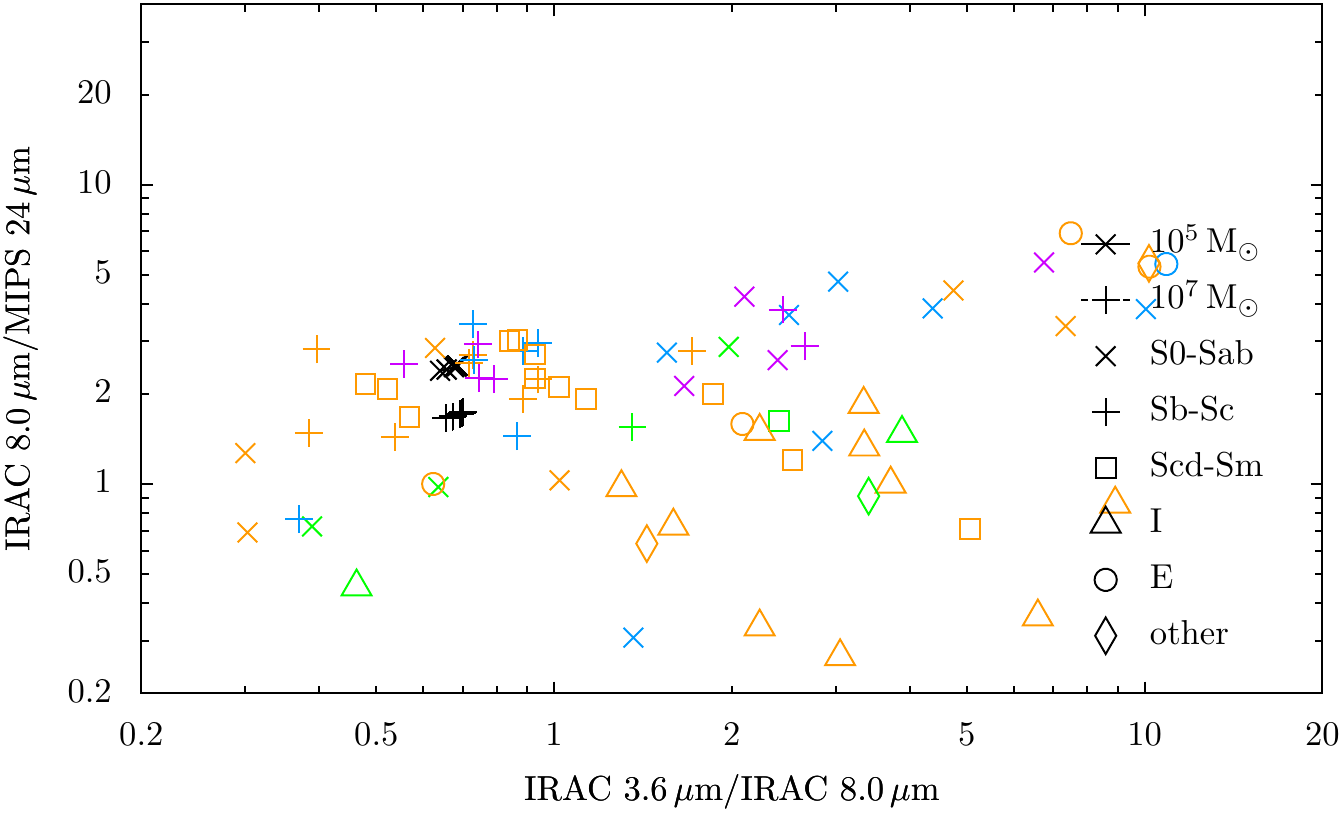} \\ \includegraphics*[scale=0.62]{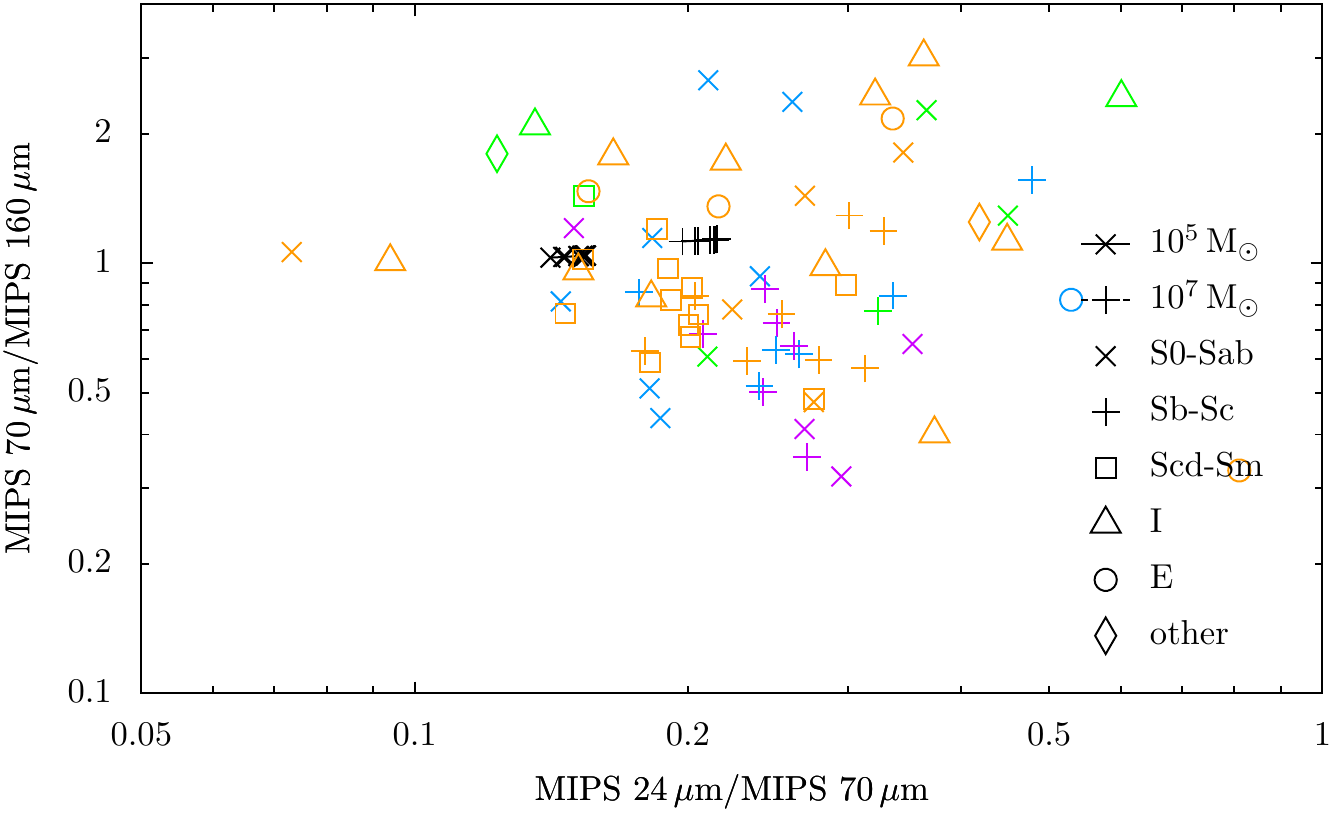} & \includegraphics*[scale=0.62]{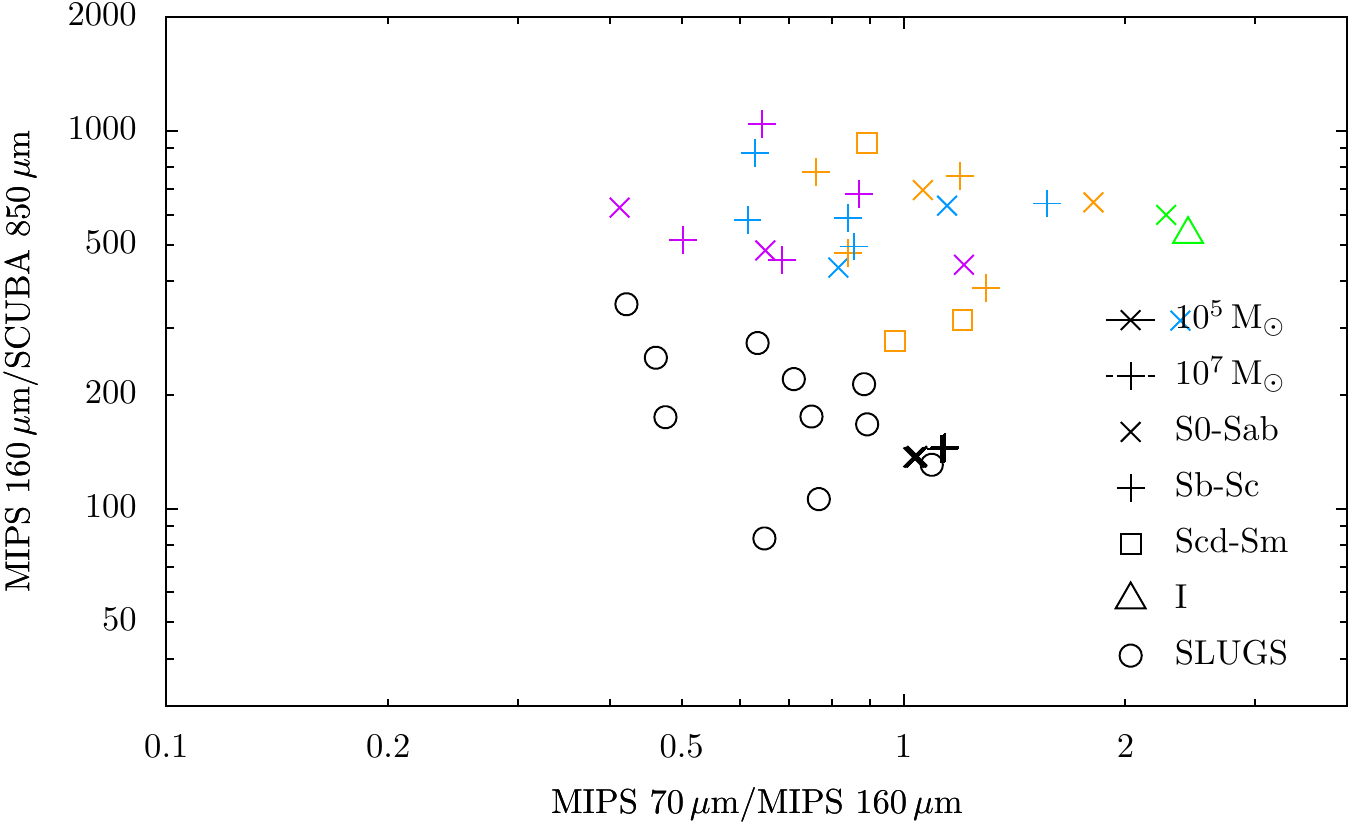} \end{tabular*} \end {center} \caption{ \label{plot_fluxratios-mappings-mcl} Sensitivity of the Sbc galaxy SED to the MAPPINGS cluster mass parameter. Symbols are as in Figure~\ref{plot_fluxratios-dustmodel}. A larger cluster mass mostly results in a slightly hotter $24/70\micron$ flux ratio and slightly redder UV slope.) } \end{figure*} \begin{figure*} \begin {center} \begin{tabular*}{\textwidth}{r@{\extracolsep{\fill}}r} \includegraphics*[scale=0.62]{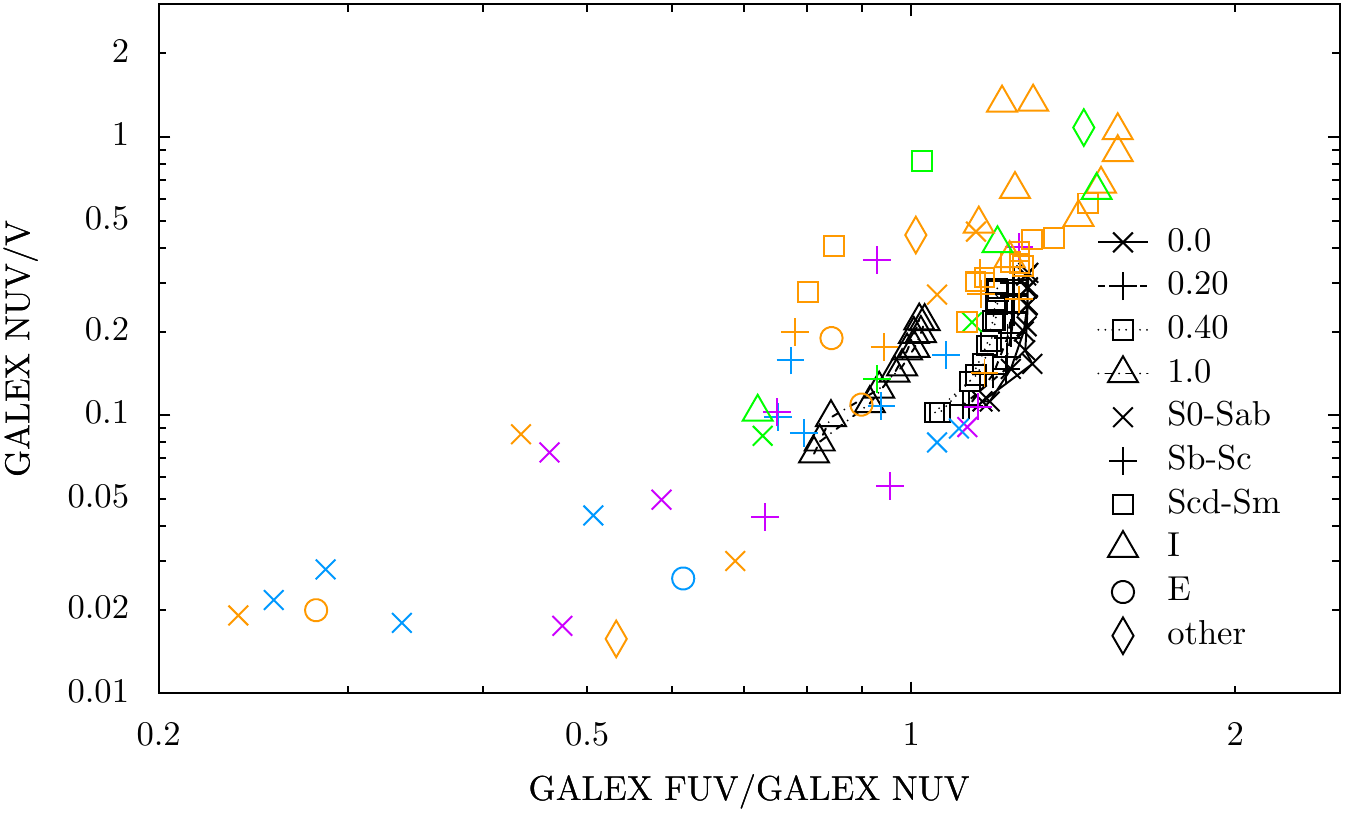} & \includegraphics*[scale=0.62]{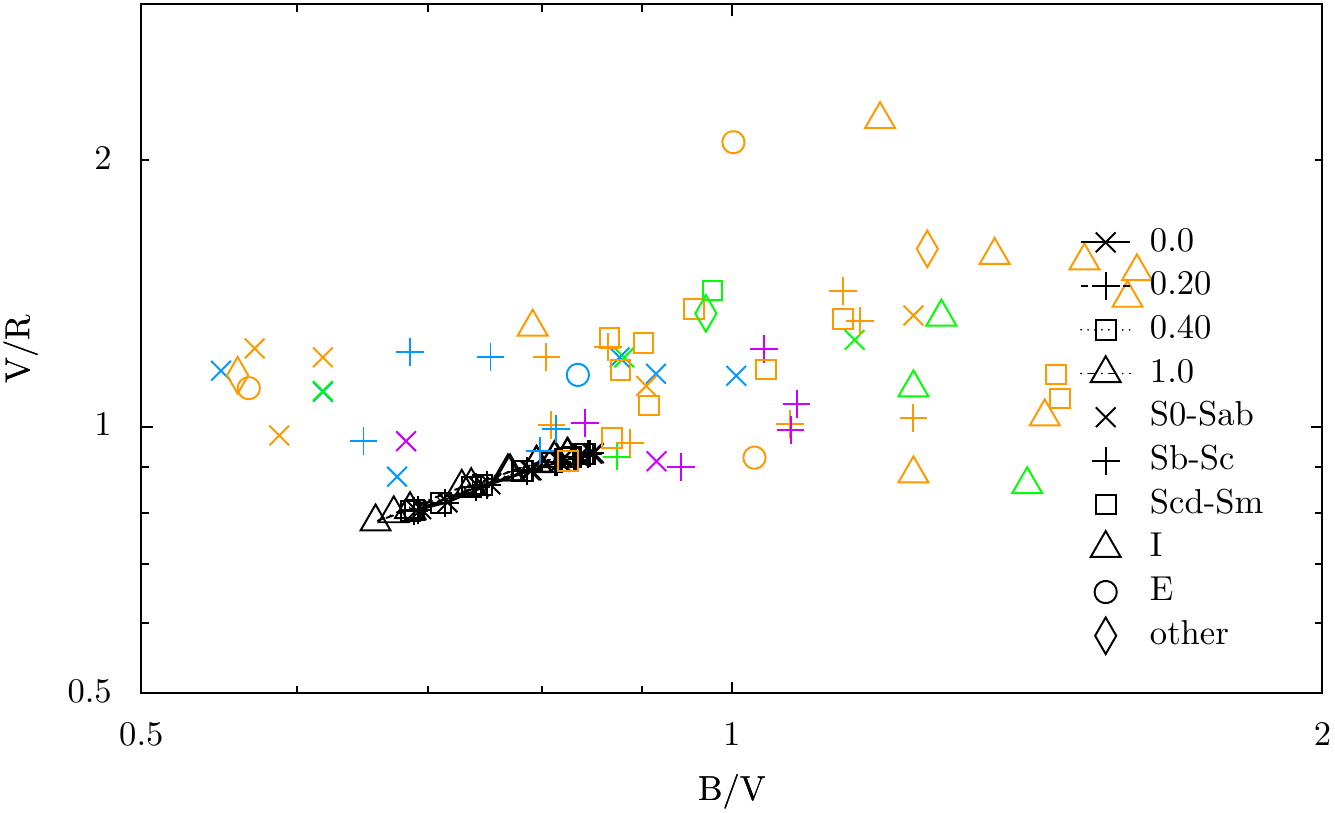} \\ \includegraphics*[scale=0.62]{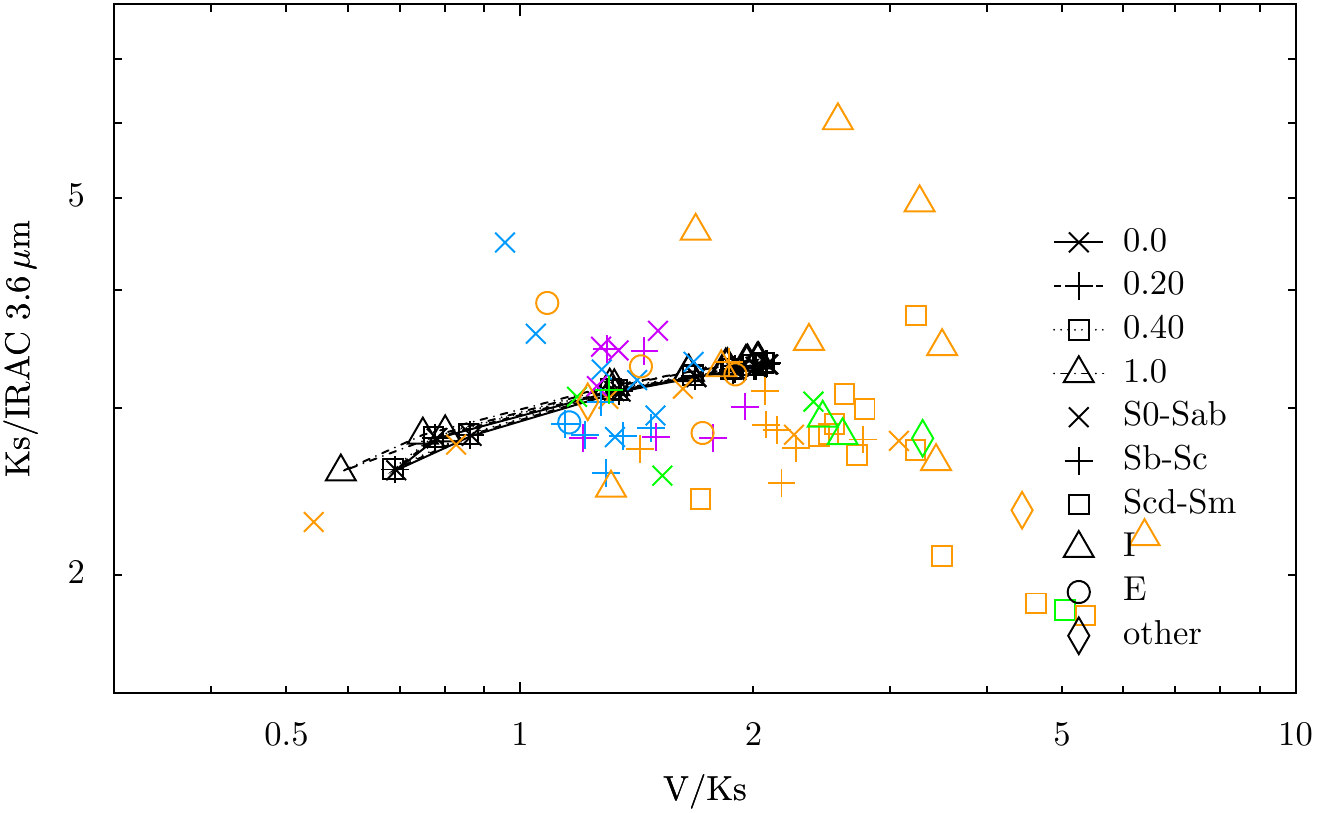} & \includegraphics*[scale=0.62]{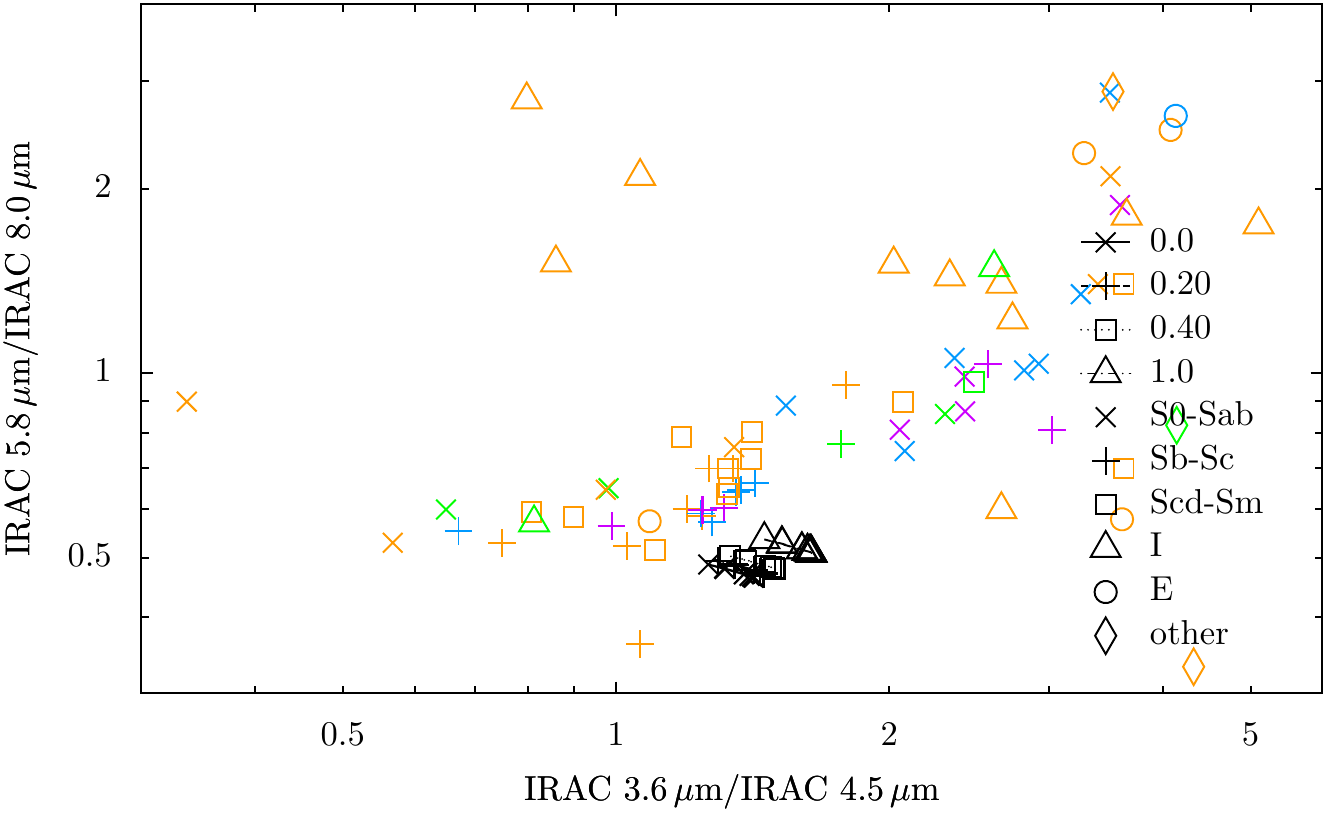} \\ \includegraphics*[scale=0.62]{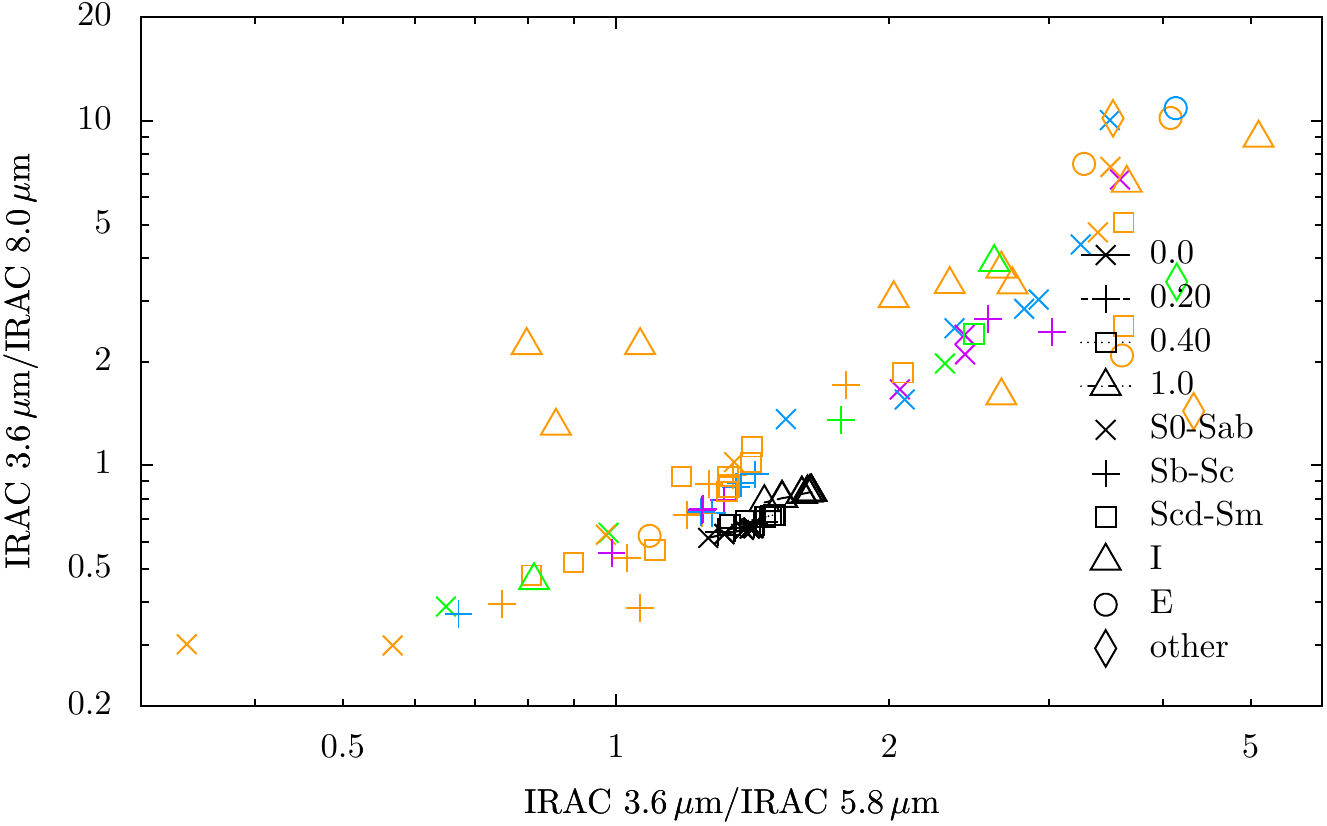} & \includegraphics*[scale=0.62]{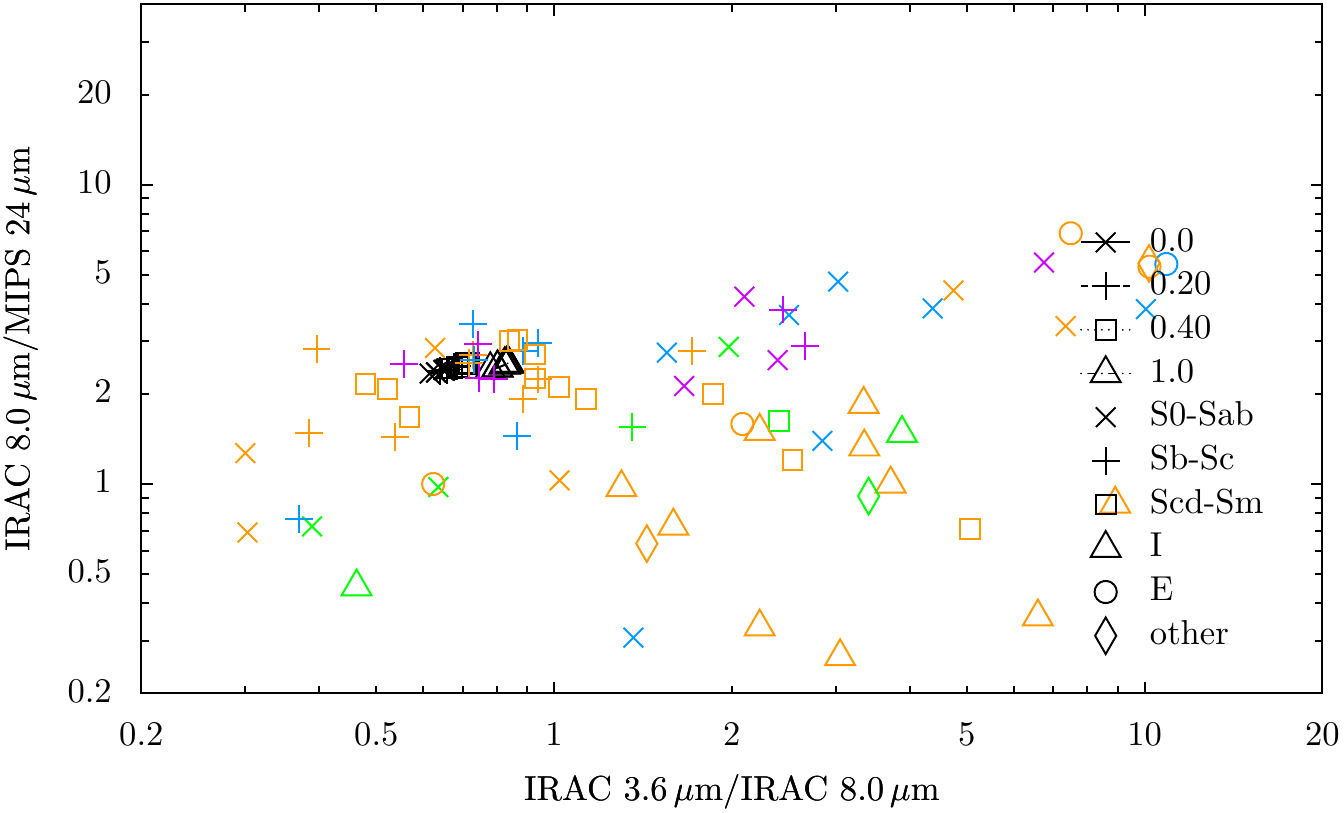} \\ \includegraphics*[scale=0.62]{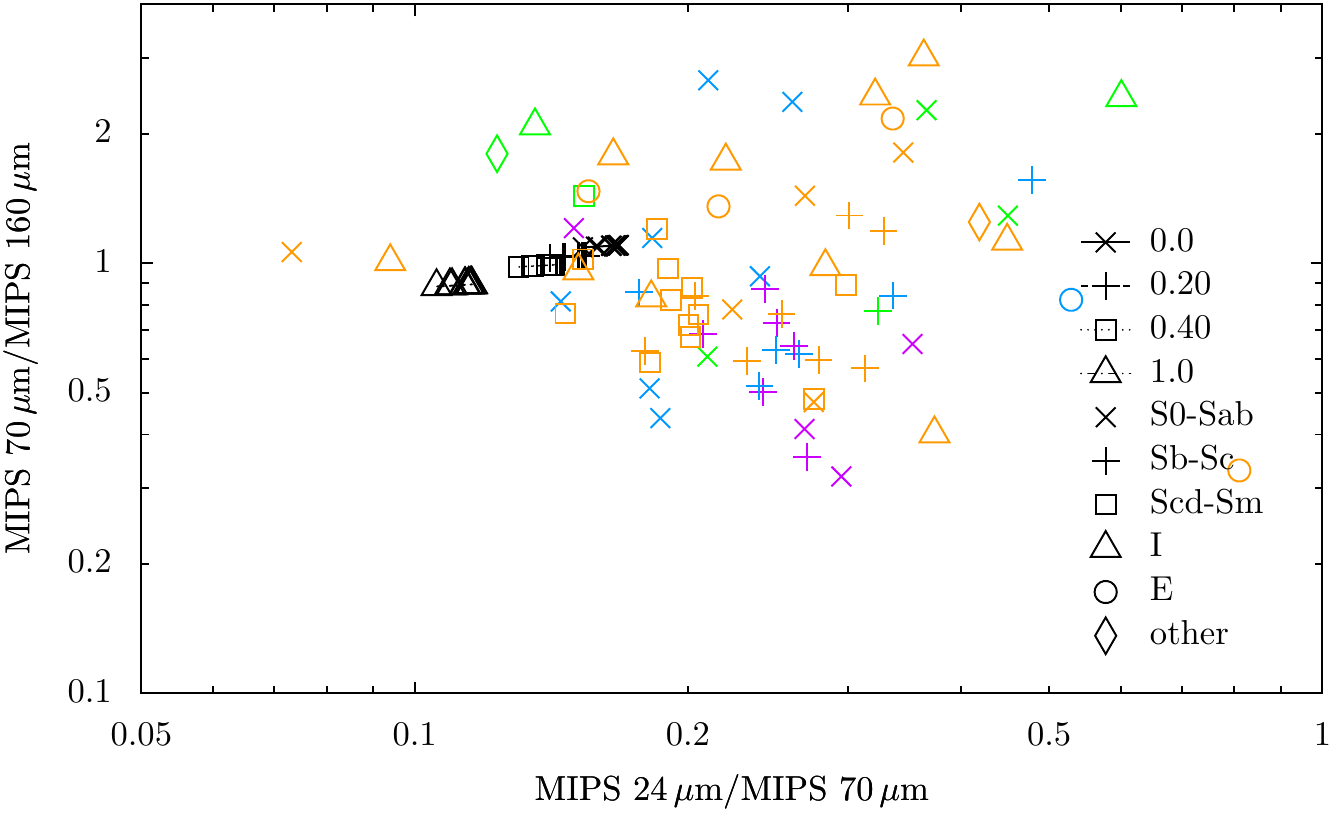} & \includegraphics*[scale=0.62]{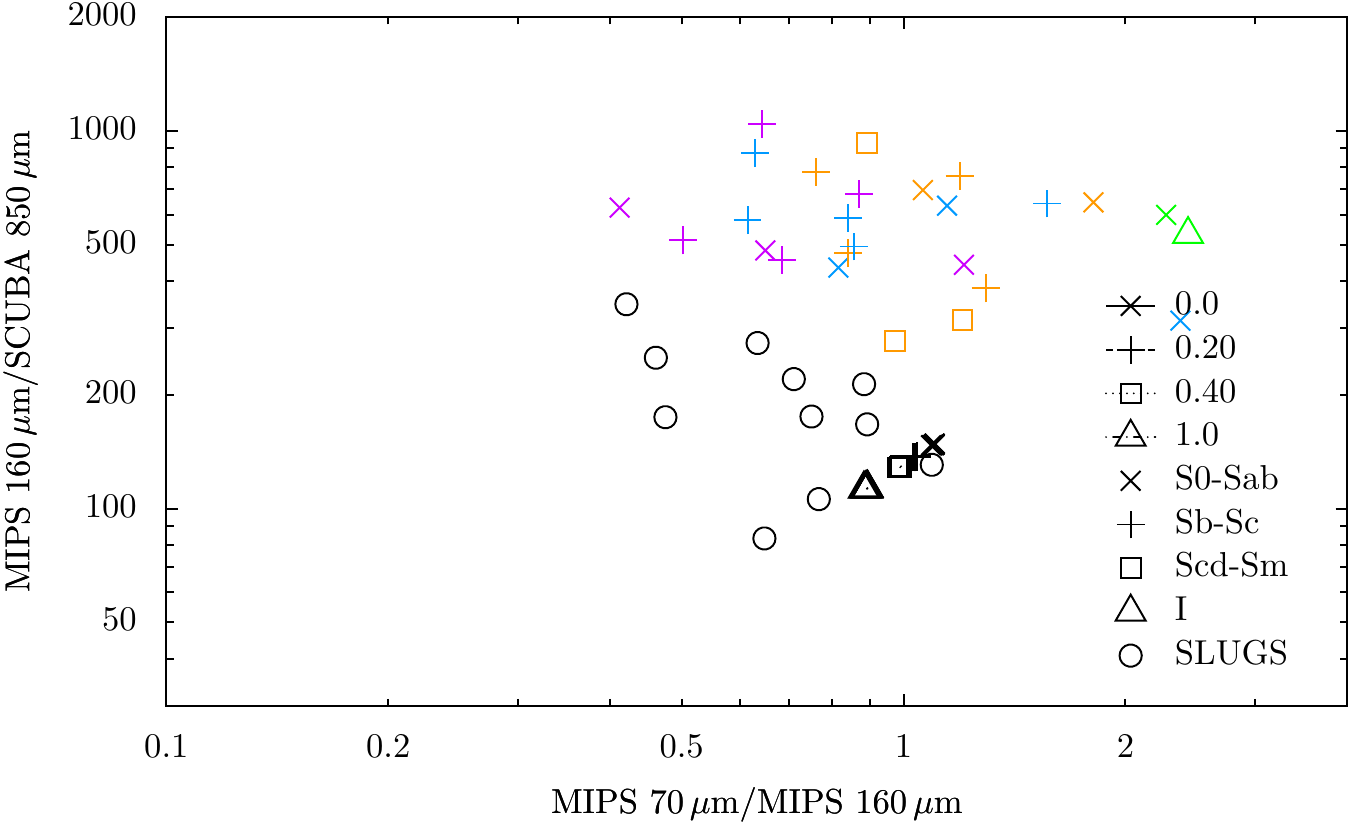} \end{tabular*} \end {center} \caption{ \label{plot_fluxratios-pdr-fraction} Sensitivity of the Sbc galaxy SED to the MAPPINGS PDR fraction parameter. Symbols are as in Figure~\ref{plot_fluxratios-dustmodel}. The PDR fraction generally has a small influence on the SED, which is expected since the MAPPINGS particles are not dominating the SED.} \end{figure*} \begin{figure*} \begin {center} \begin{tabular*}{\textwidth}{r@{\extracolsep{\fill}}r} \includegraphics*[scale=0.62]{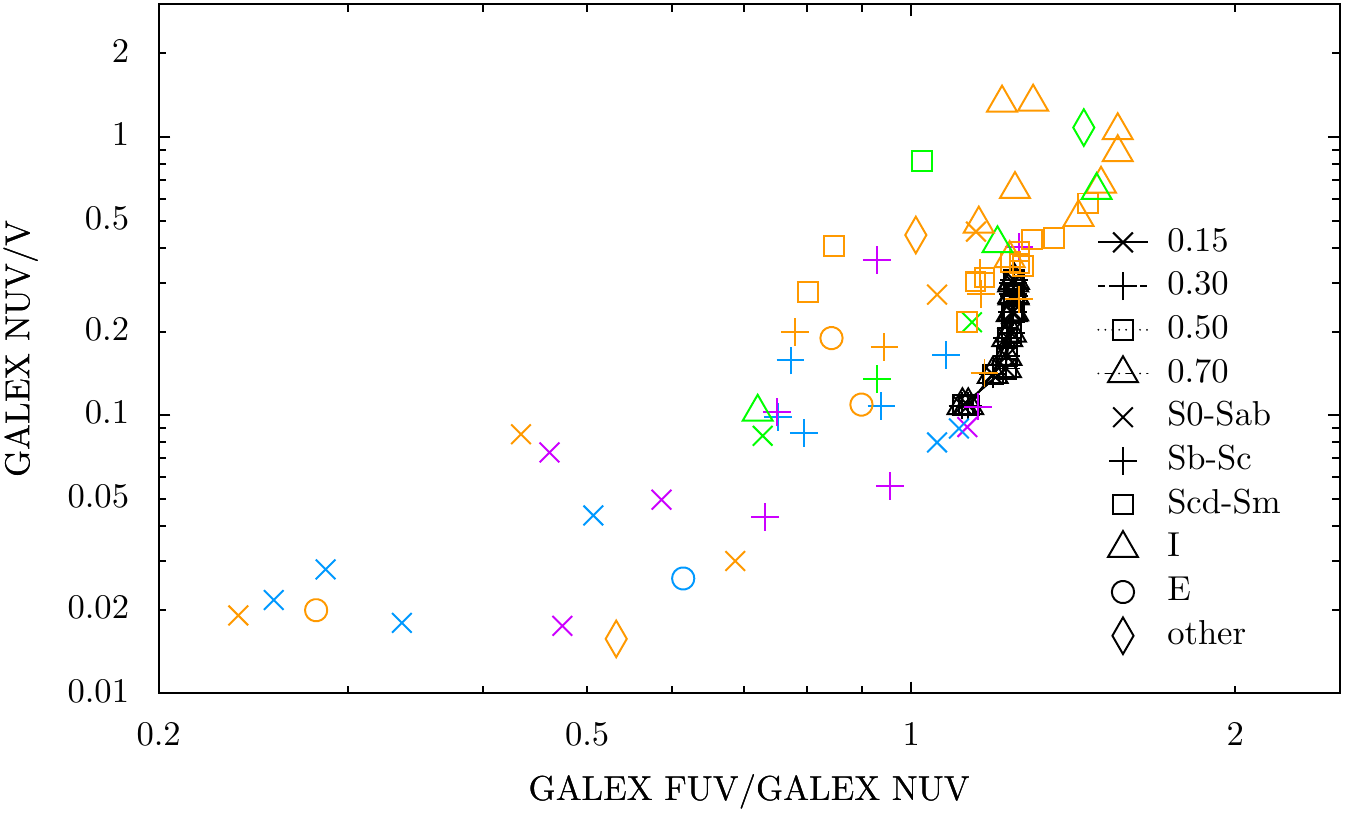} & \includegraphics*[scale=0.62]{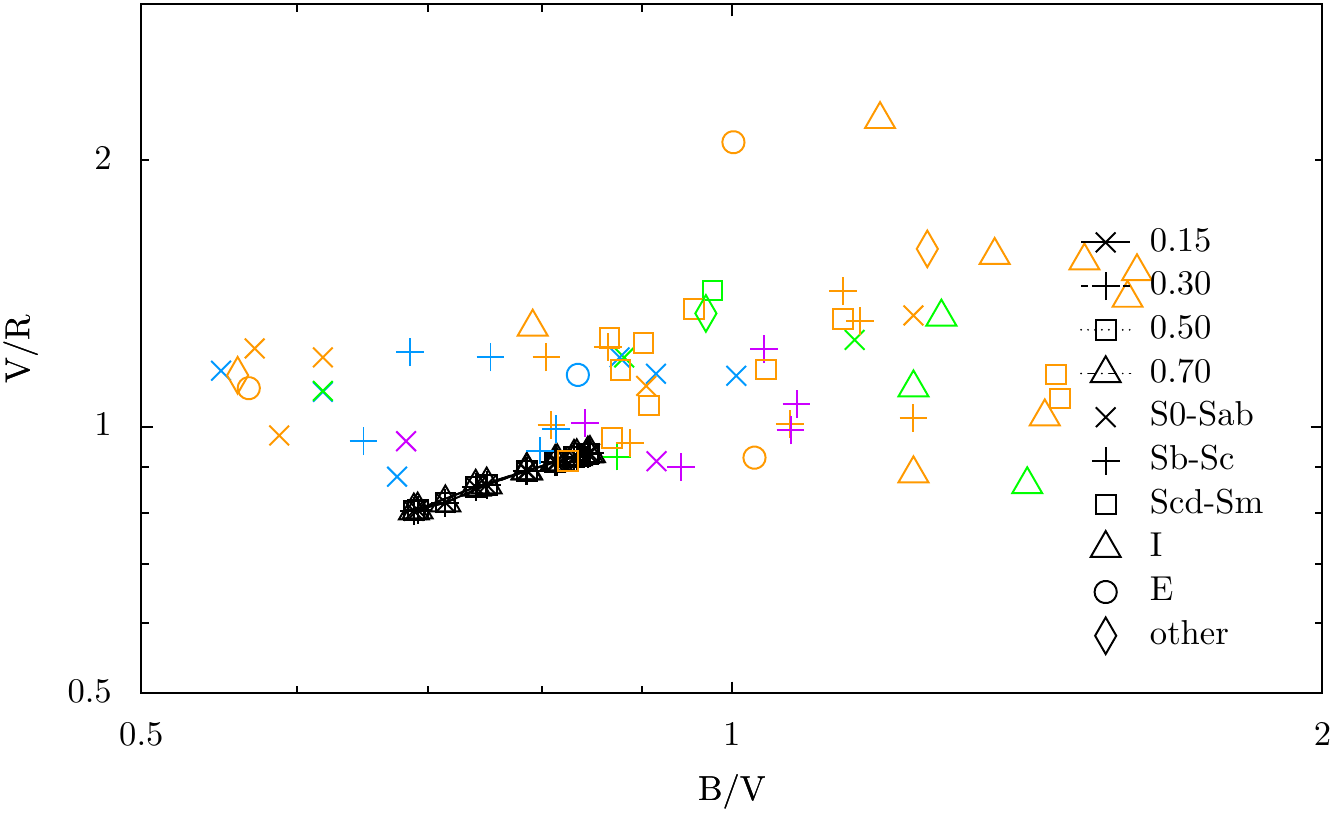} \\ \includegraphics*[scale=0.62]{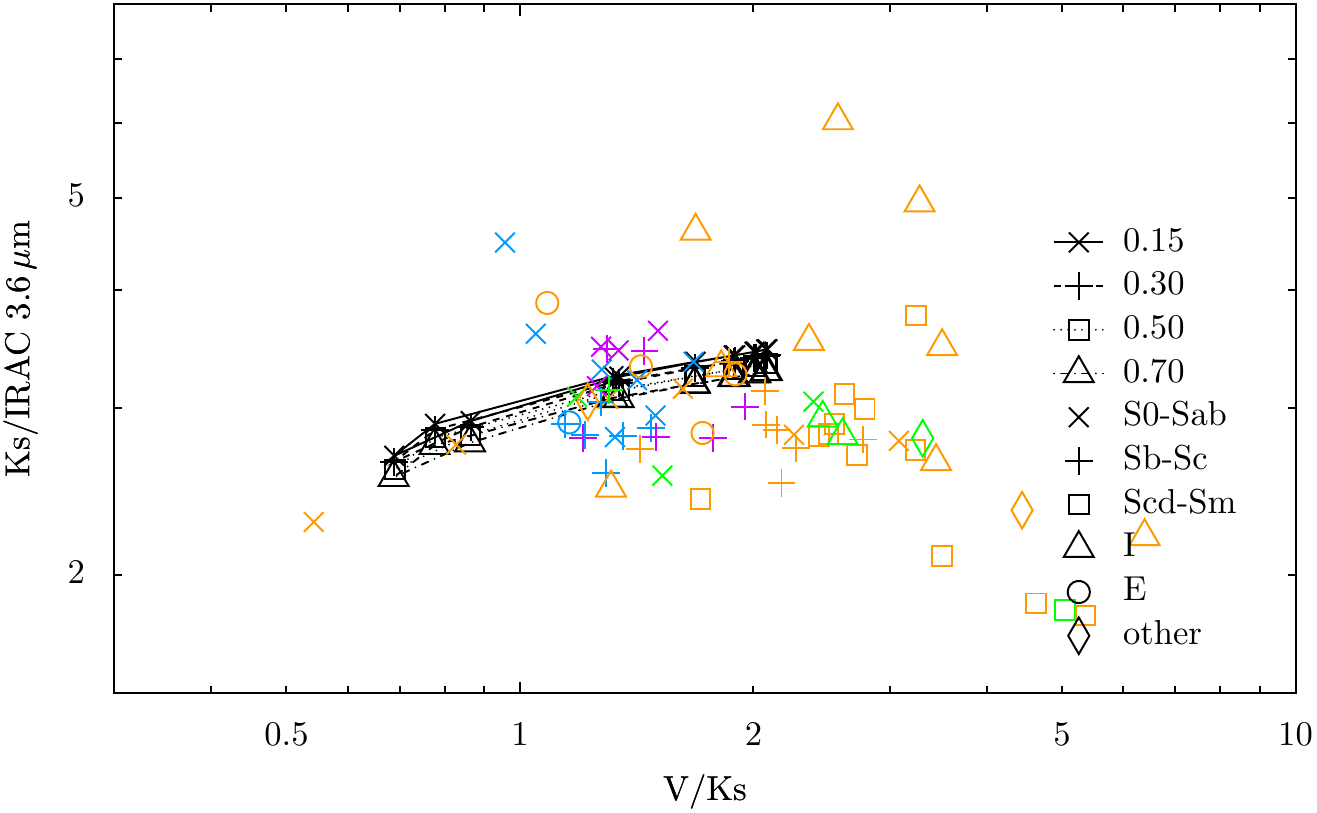} & \includegraphics*[scale=0.62]{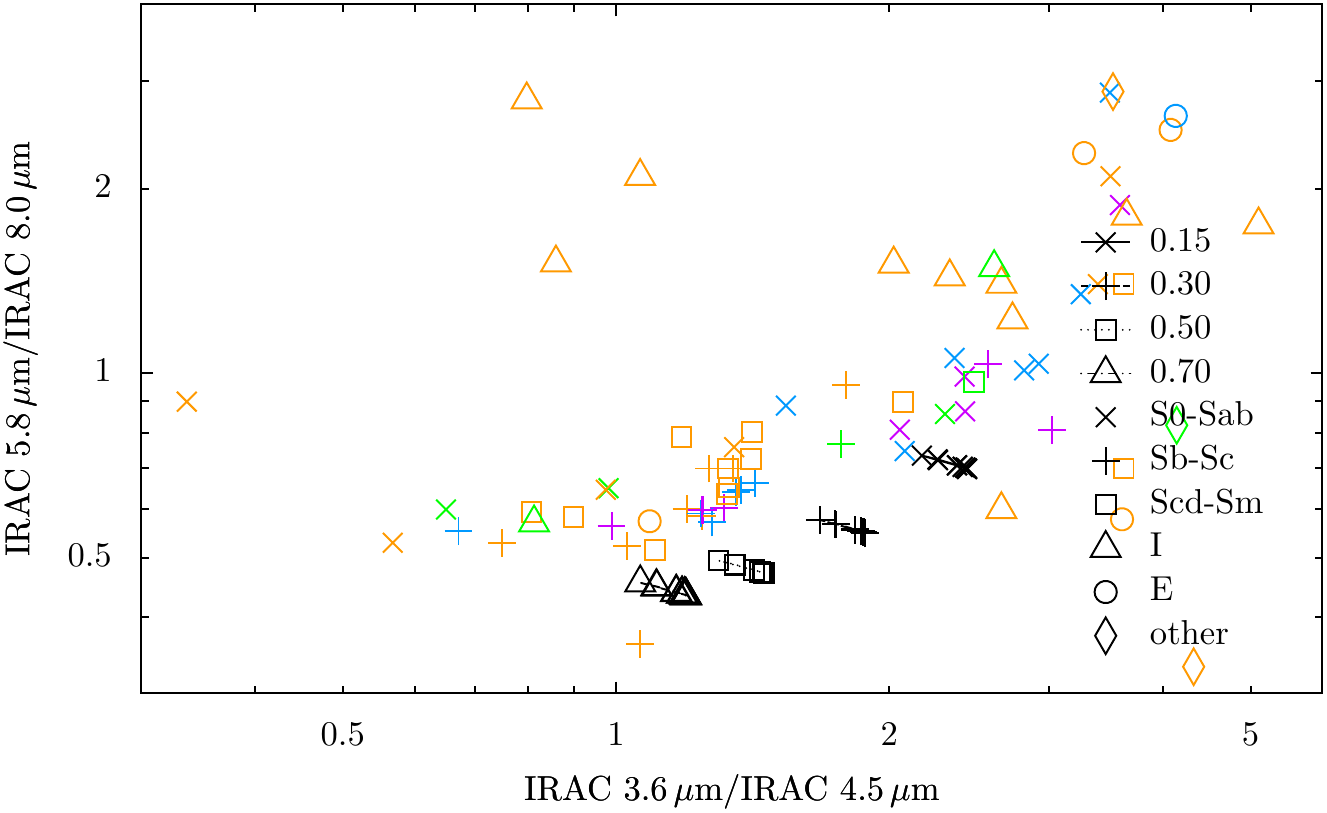} \\ \includegraphics*[scale=0.62]{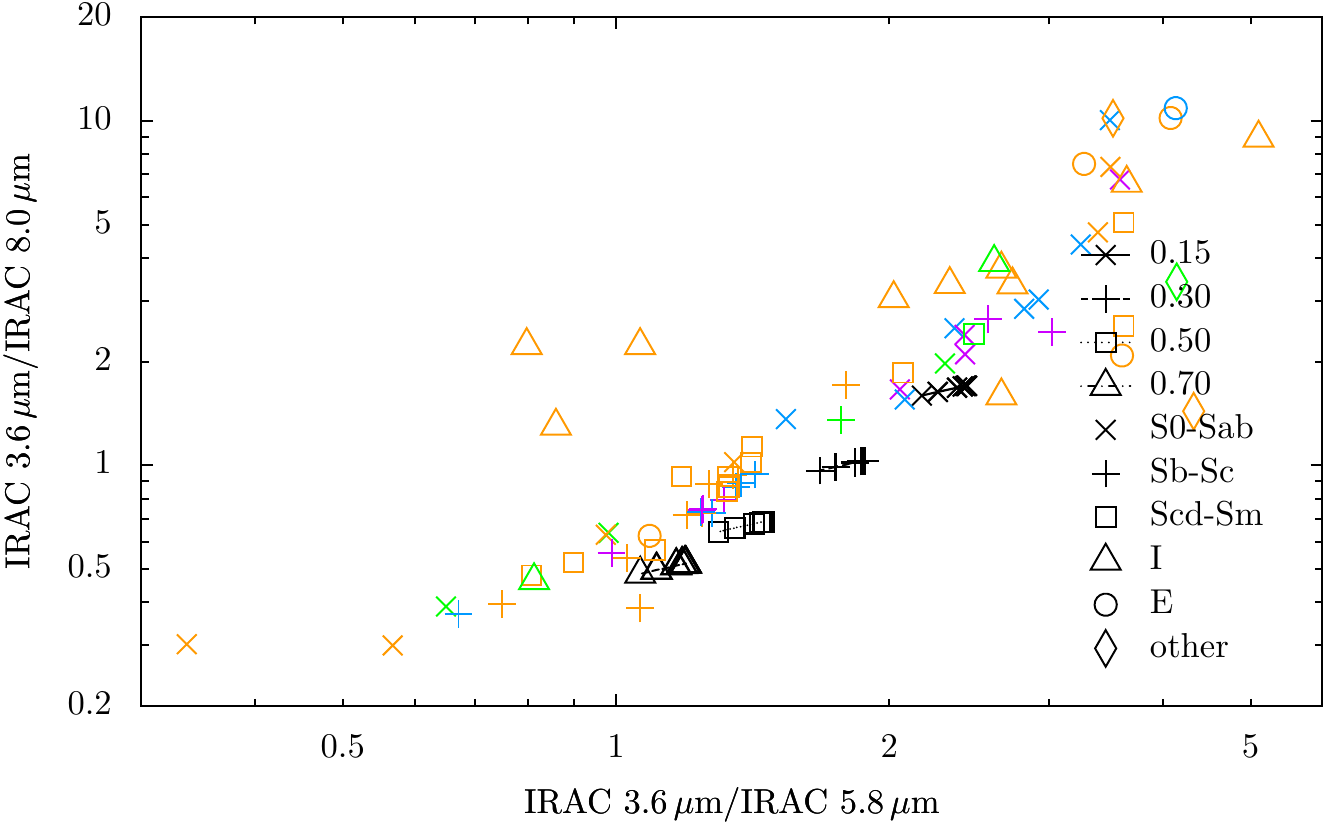} & \includegraphics*[scale=0.62]{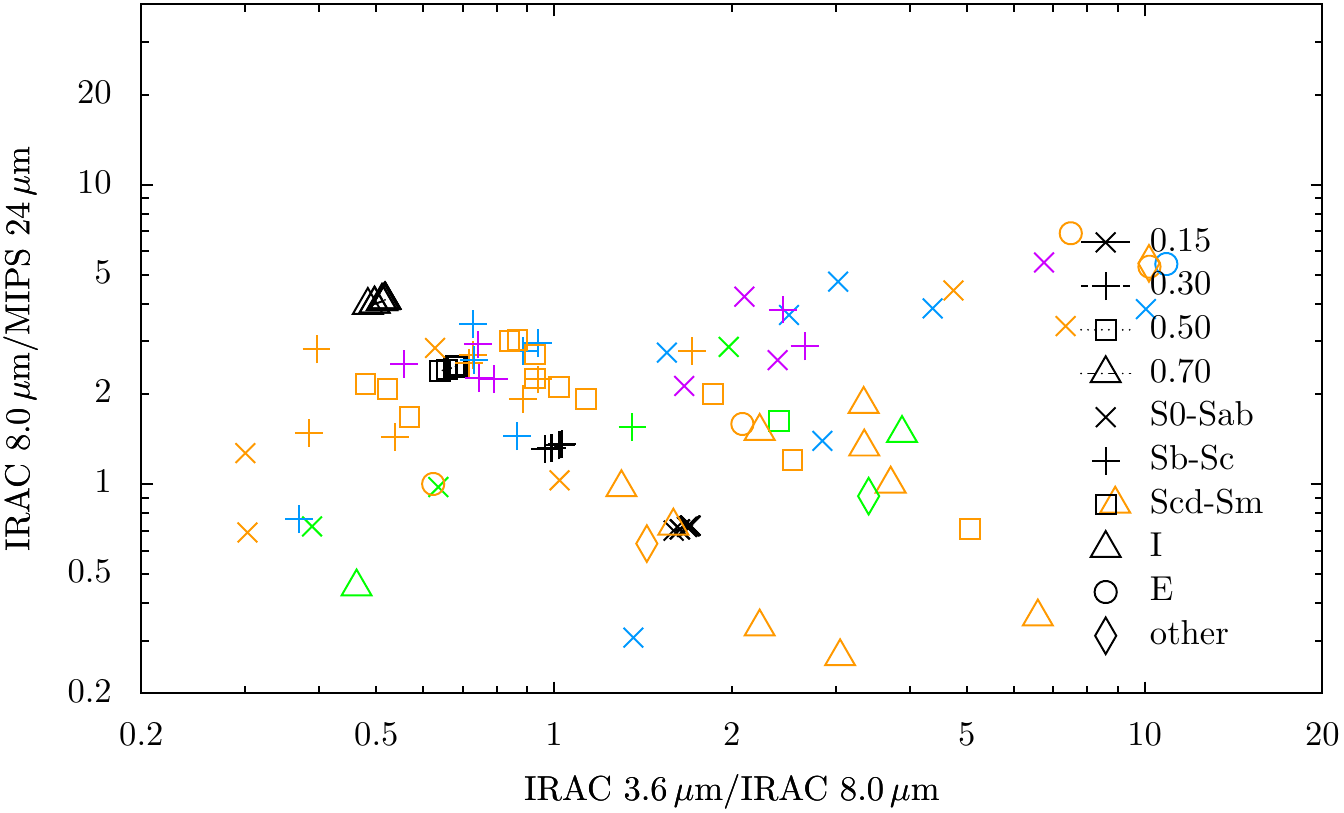} \\ \includegraphics*[scale=0.62]{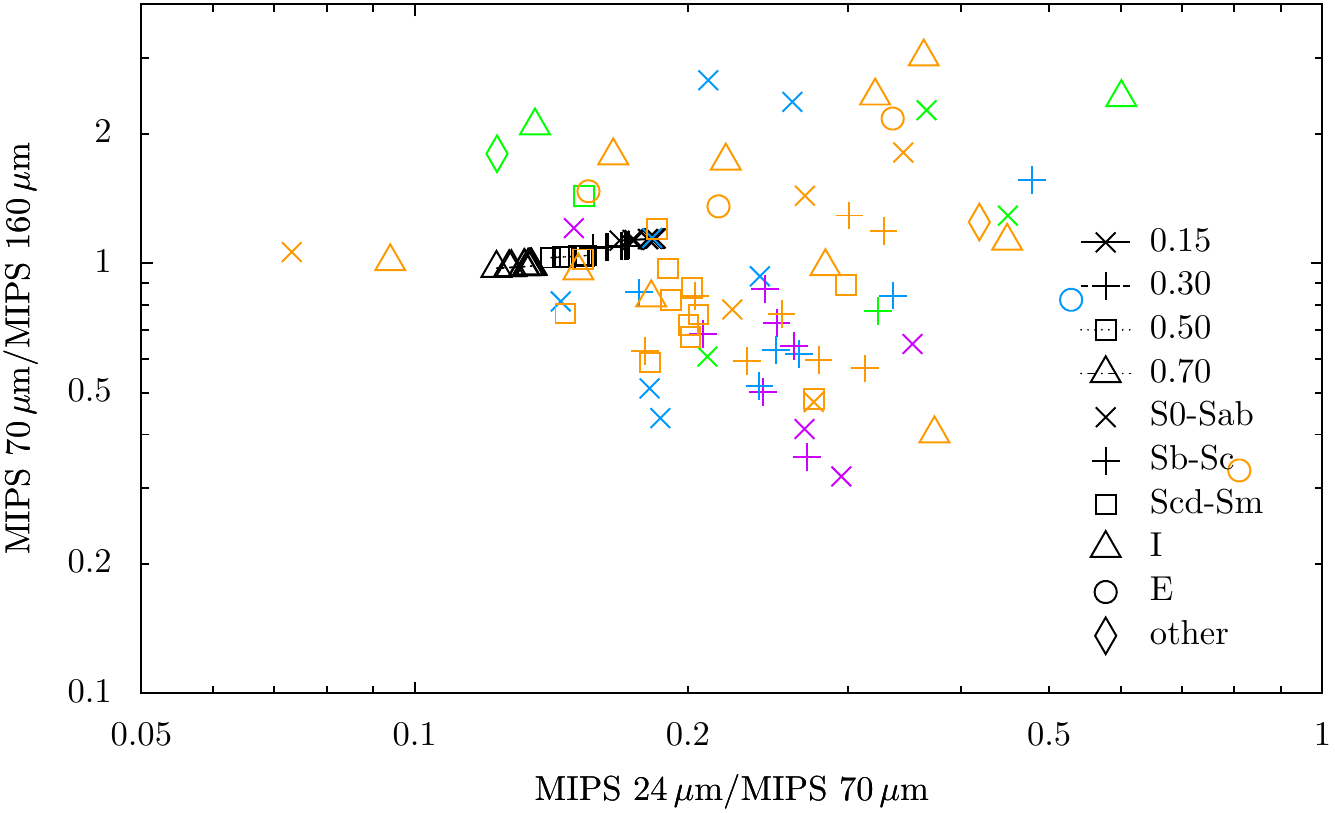} & \includegraphics*[scale=0.62]{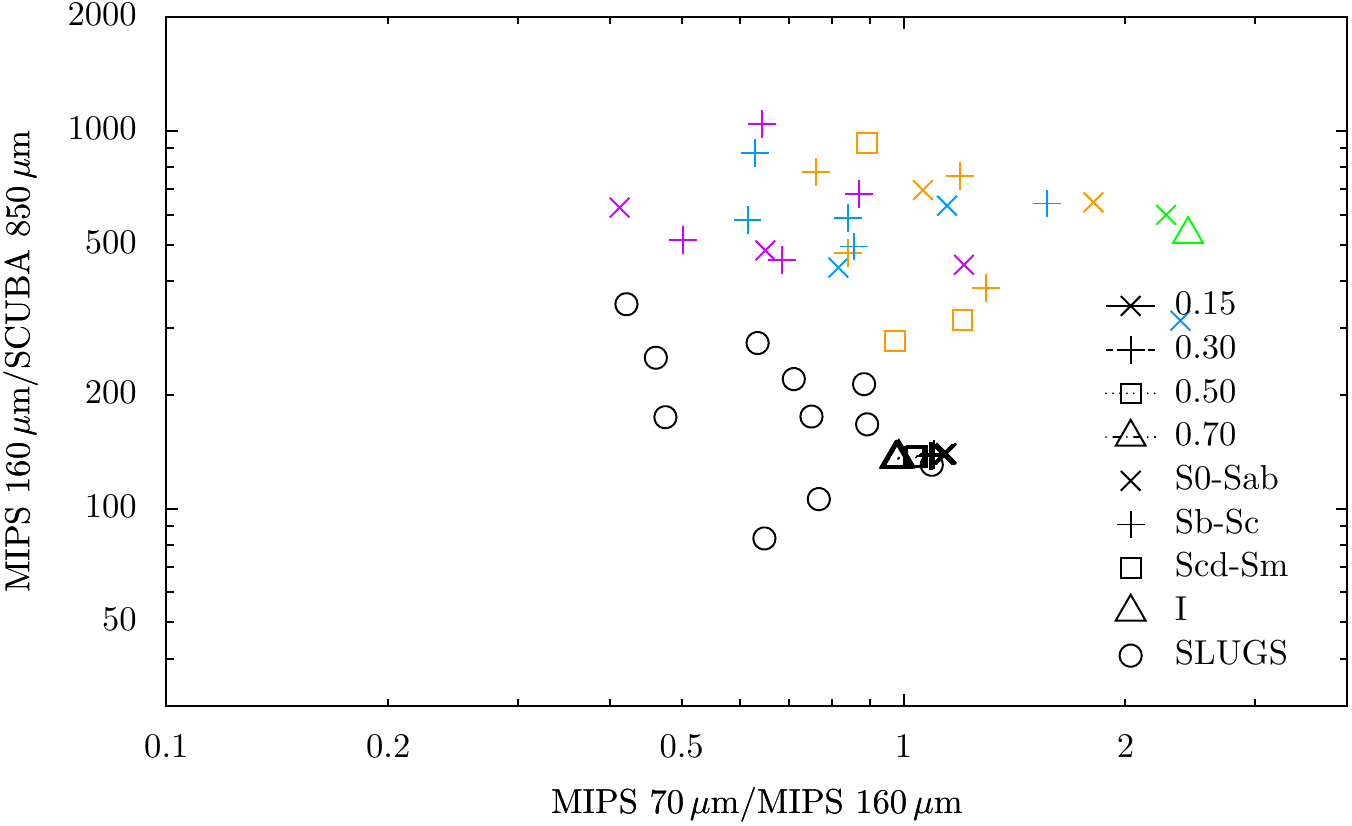} \end{tabular*} \end {center} \caption{ \label{plot_fluxratios-pah-template-fraction} Sensitivity of the Sbc galaxy SED to $f_t$, the fraction of PAH emission which is emitted into the \citet{grovesetal08sparam} template as opposed to based on a modified blackbody in thermal equilibrium. Symbols are as in Figure~\ref{plot_fluxratios-dustmodel}. This parameter, like the amount of PAH grains in Figure~\ref{plot_fluxratios-PAHfraction}, strongly affects the strength of the PAH features in the IRAC bands but in such a way that it moves them parallel to the locus described by the SINGS galaxies. However, the $8.0/24\micron$ flux ratio constrains the parameter to values around 0.50.} \end{figure*}

\begin{figure} \begin {center} \includegraphics*[width=0.98\columnwidth]{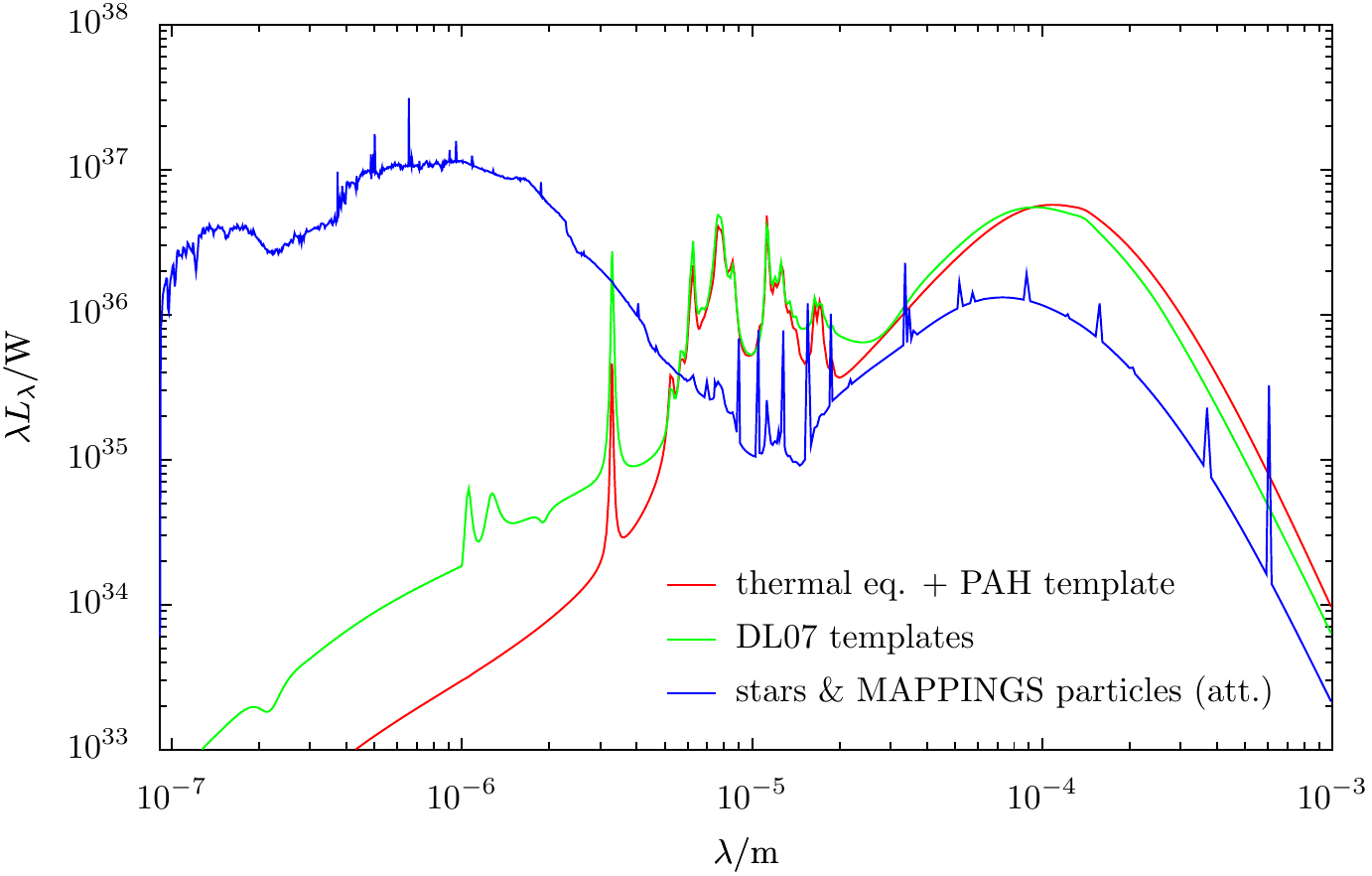} \end {center} \caption{ \label{plot_dl07-template-sed} The SED calculated by using the \citet{draineli07} precomputed emission templates, which include stochastically heated grains (green), on an individual grid cell basis compared to the standard thermal equilibrium approximation + PAH template normally used (red). The thermal equilibrium approximation is surprisingly good, except at wavelengths around $25\micron$, where the stochastically heated grains emit more strongly. The peak of the SED is also at slightly shorter wavelengths with the \citet{draineli07} templates. The output stellar SED is shown in blue for reference.} \end{figure}

\begin{figure*} \begin {center} \begin{tabular*}{\textwidth}{r@{\extracolsep{\fill}}r} \includegraphics*[scale=0.62]{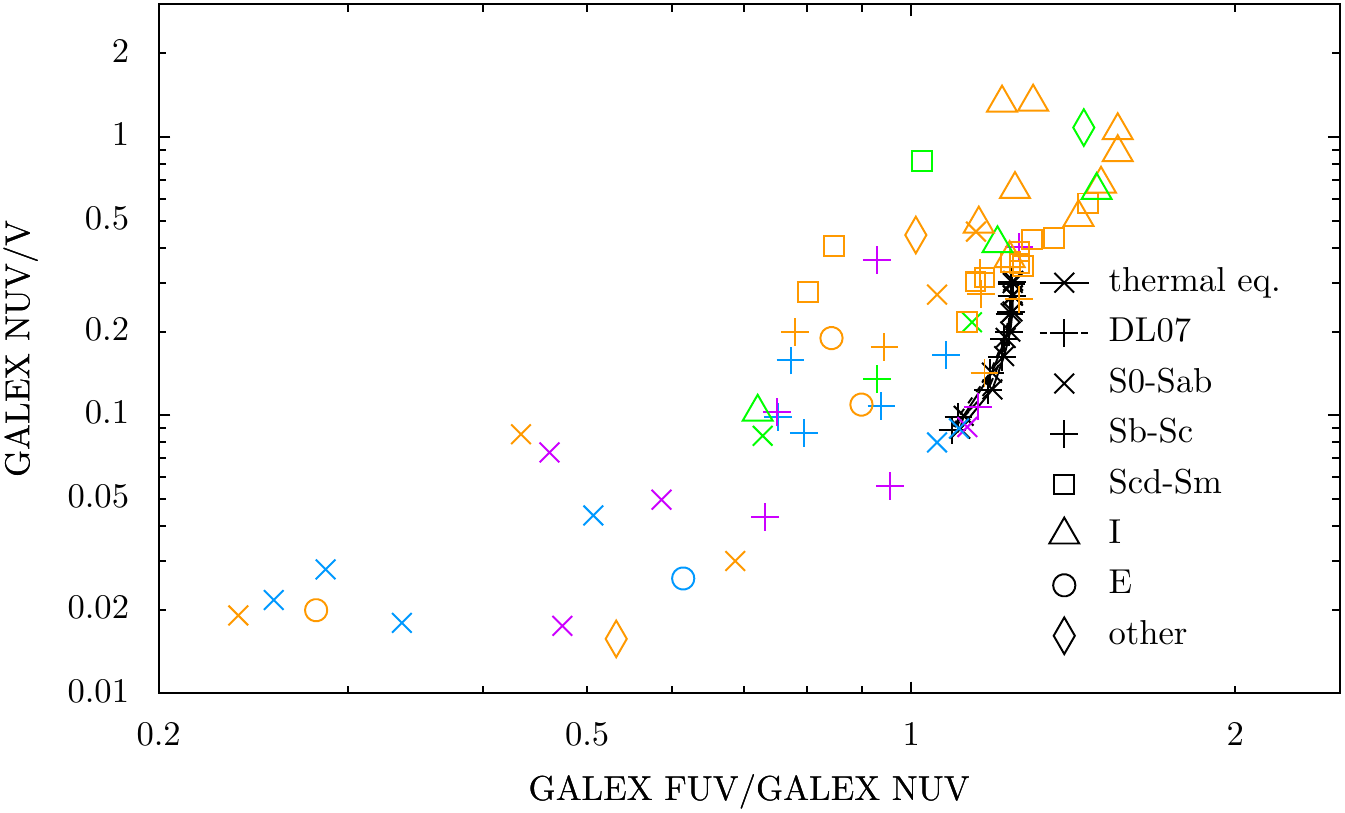} & \includegraphics*[scale=0.62]{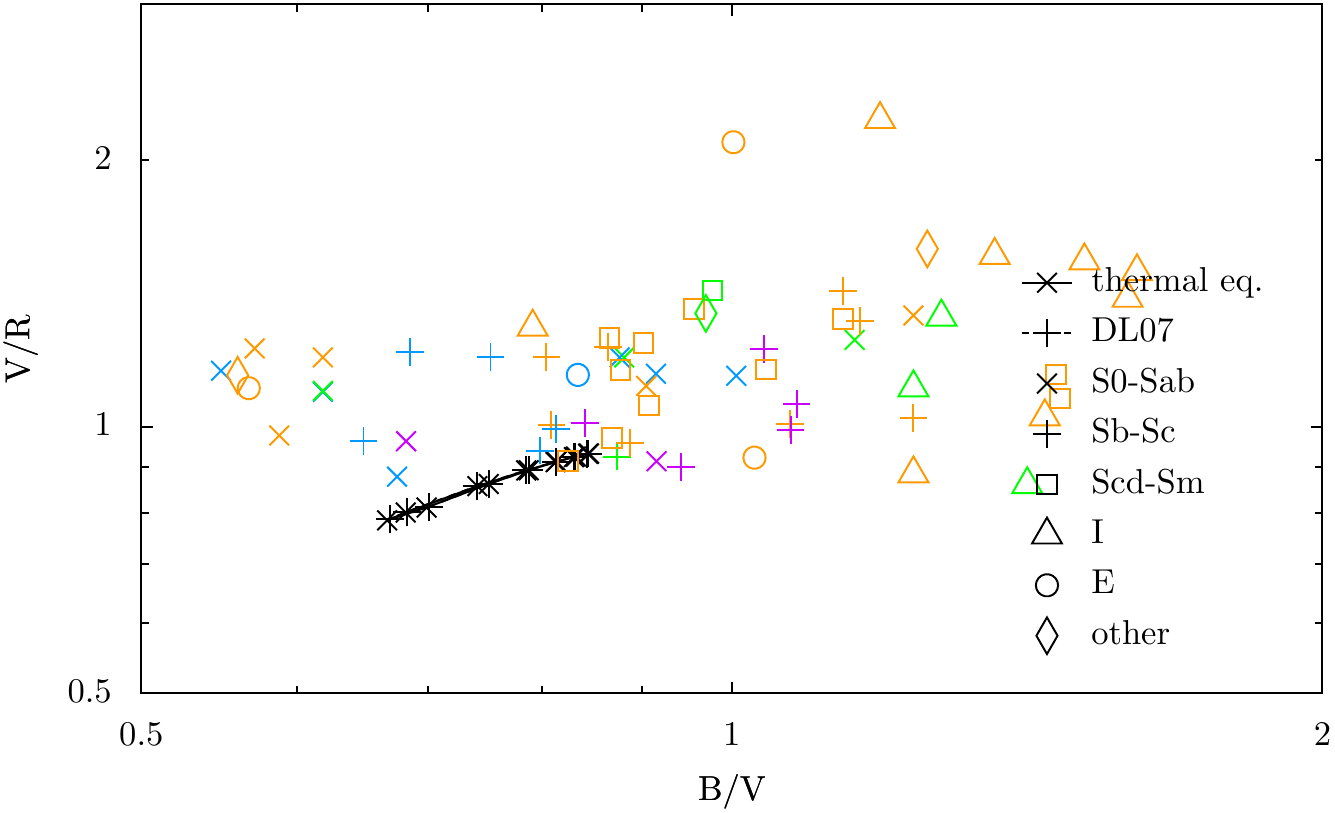} \\ \includegraphics*[scale=0.62]{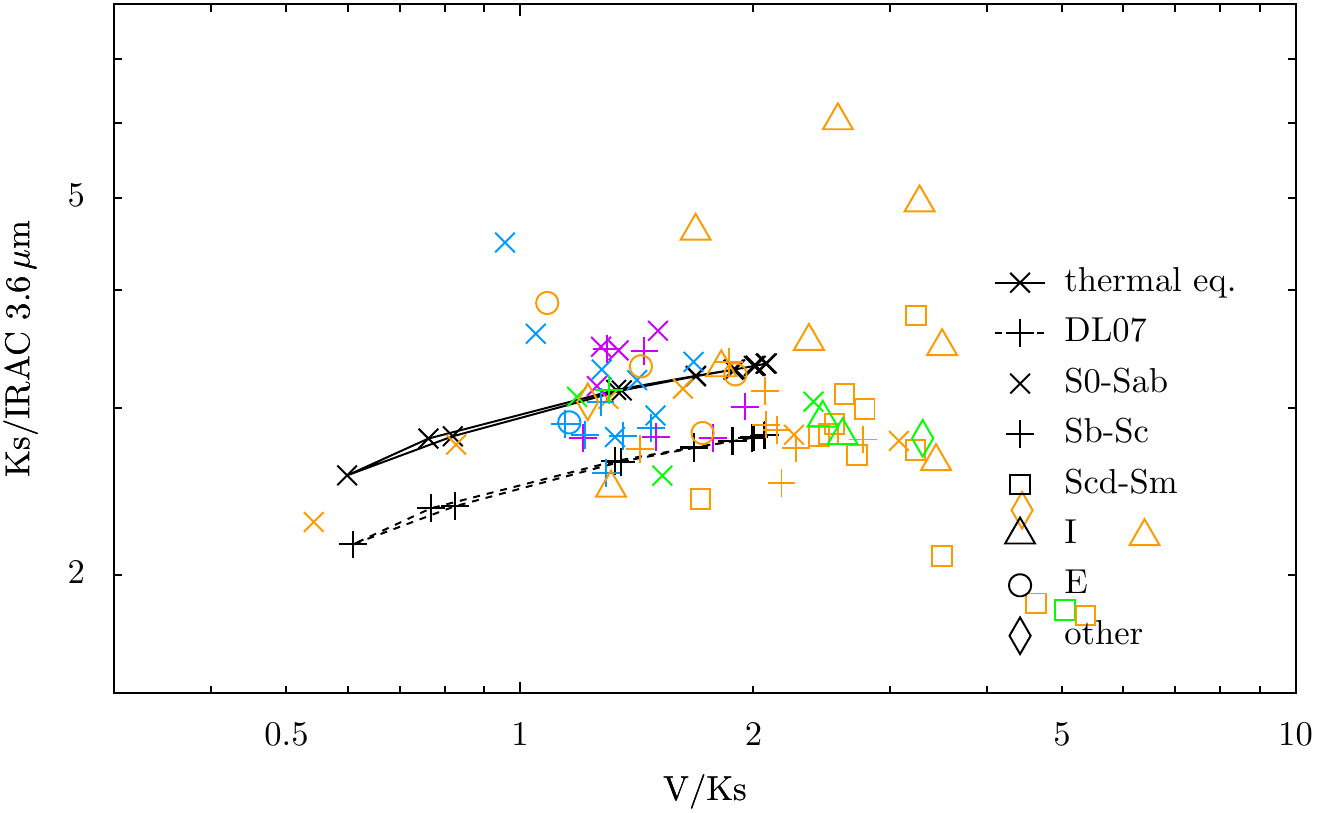} & \includegraphics*[scale=0.62]{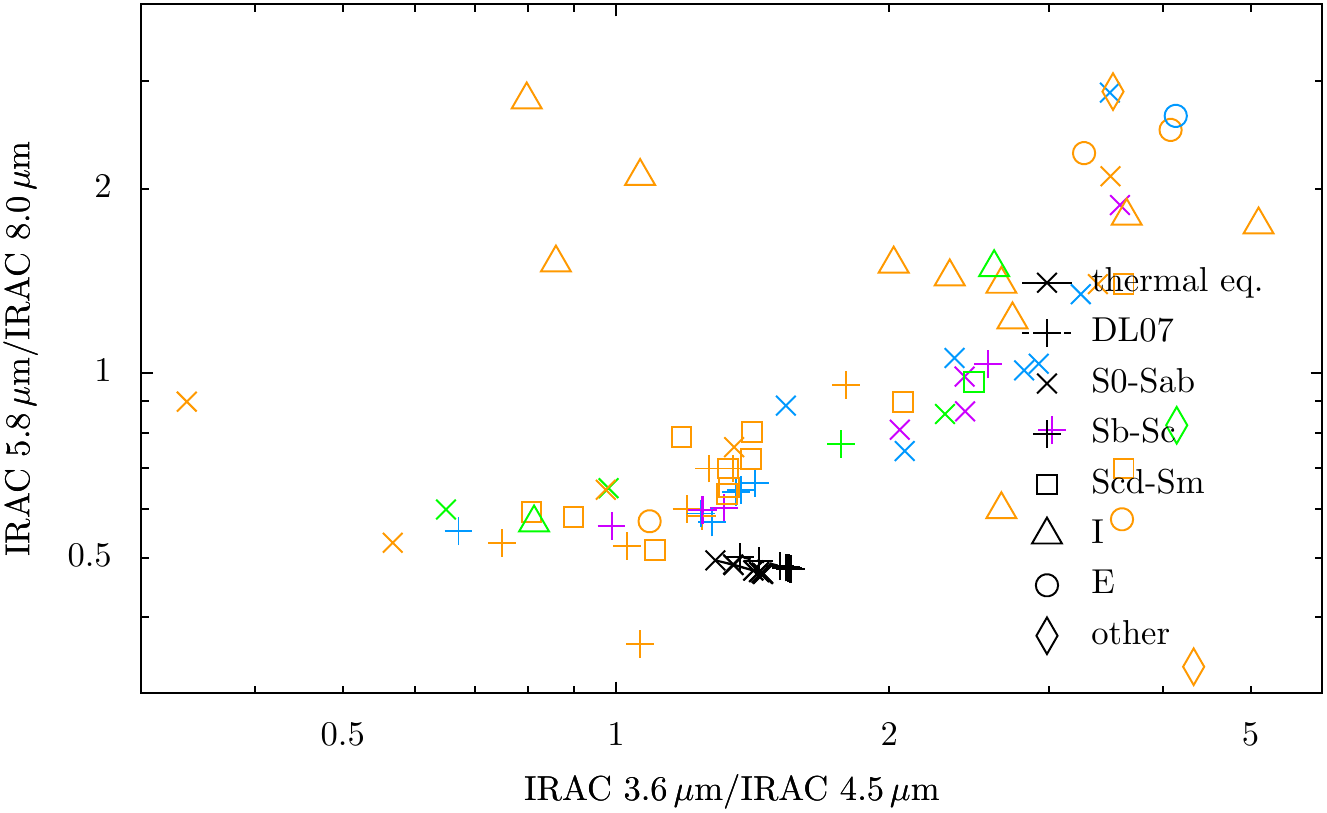} \\ \includegraphics*[scale=0.62]{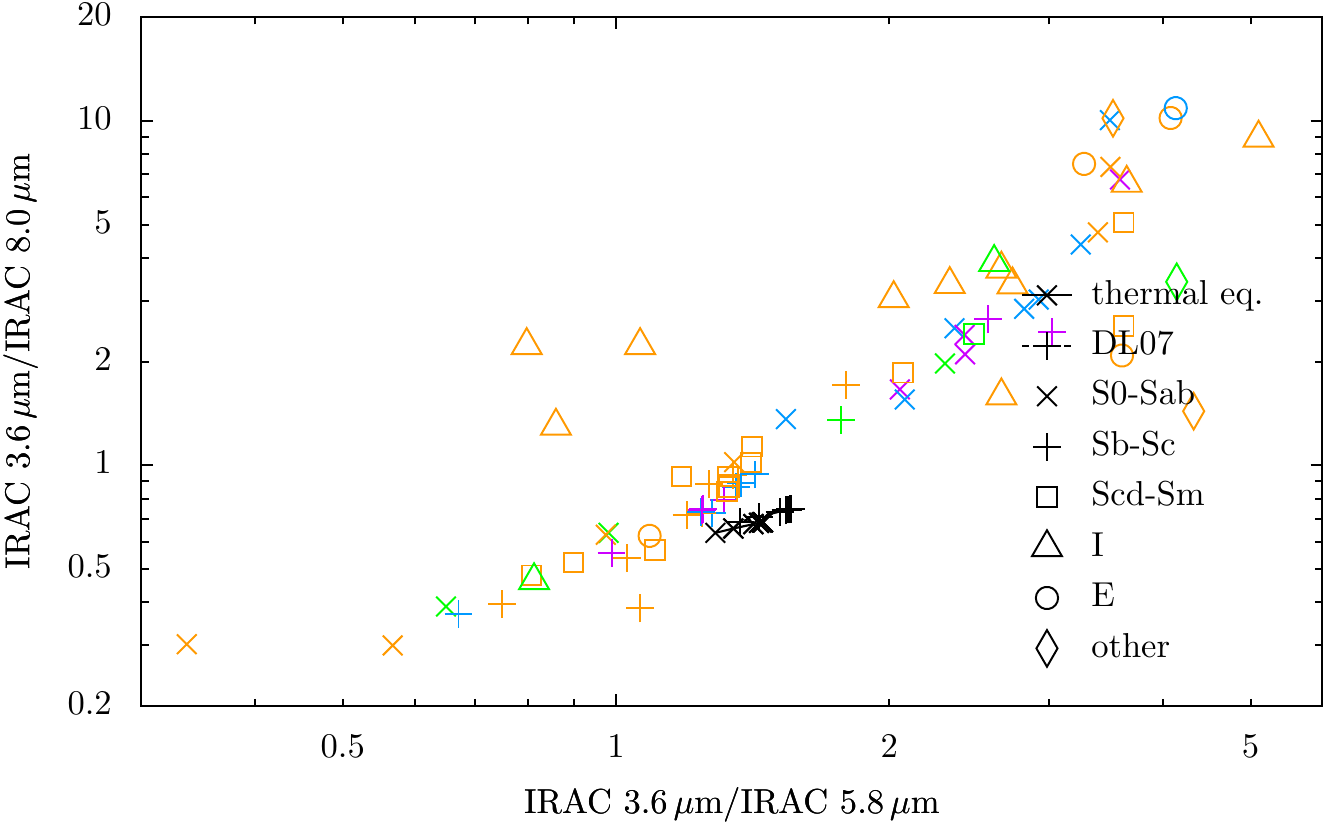} & \includegraphics*[scale=0.62]{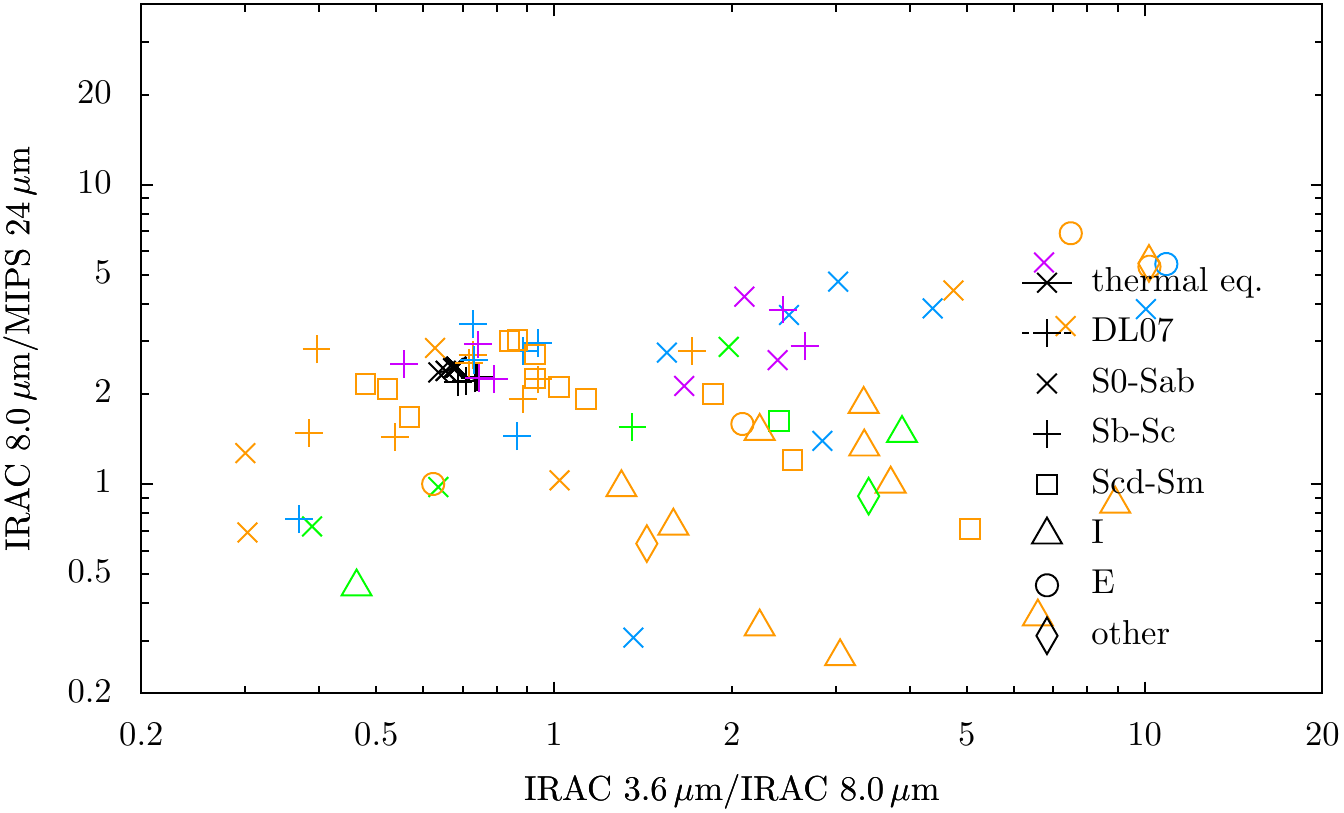} \\ \includegraphics*[scale=0.62]{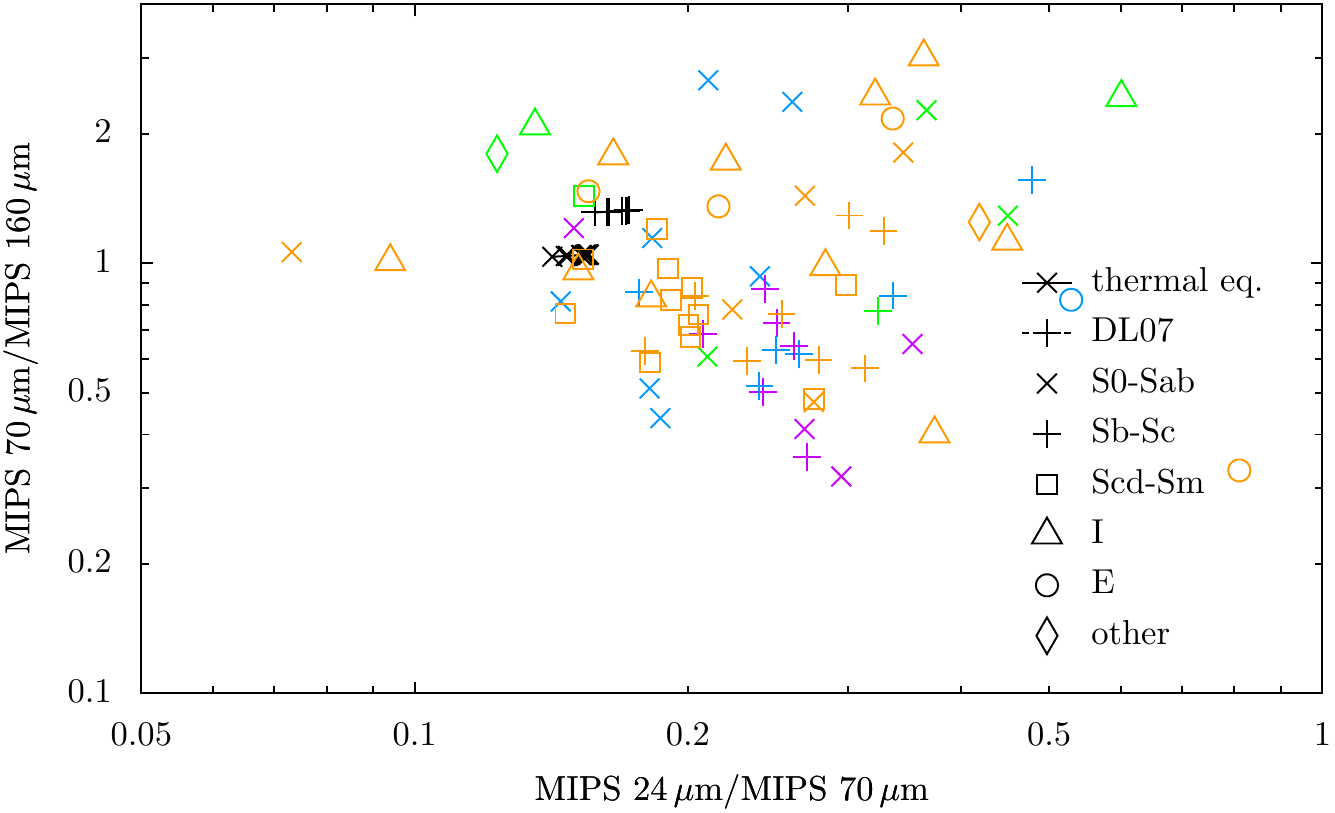} & \includegraphics*[scale=0.62]{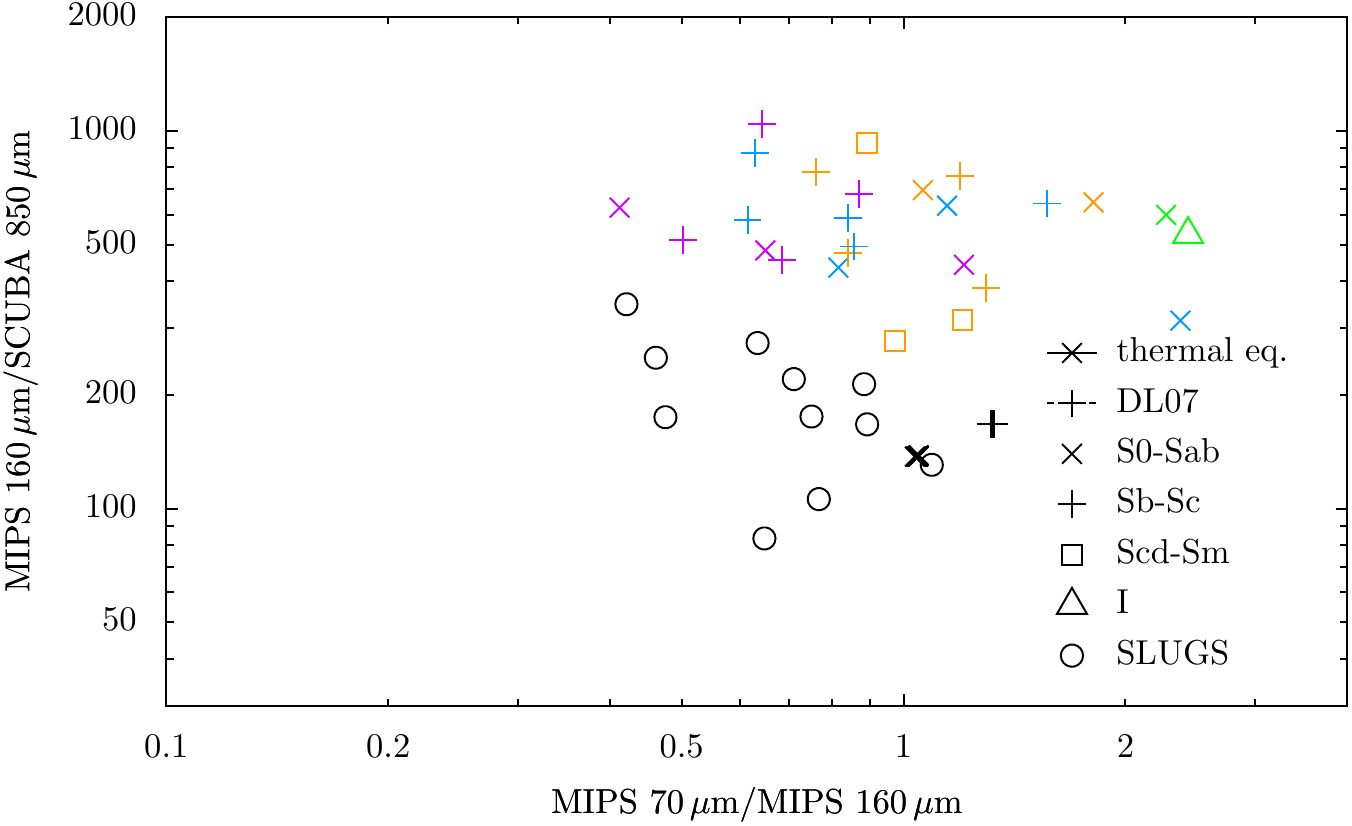} \end{tabular*} \end {center} \caption{ \label{plot_fluxratios-dl07-template} Effect of using the \citet{draineli07} precomputed emission templates, which include stochastically heated grains, instead of the thermal equilibrium approximation and PAH template normally used. Symbols are as in Figure~\ref{plot_fluxratios-dustmodel}. The differences are small, the most significant difference is a significantly redder $\mathrm{K_s} - 3.6 \micron$ arising from a stronger PAH feature at $3.3 \micron$ in the DL07 opacities, and a slightly redder $70/160\micron$ colour.} \end{figure*}
%NLX% end exclude from vocabulary builder

\section{Discussion}
\label{section_discussion}

The previous sections have shown that, given a realistic hydrodynamic
galaxy simulation, \mcrx \ can produce broad-band images that look
remarkably similar to those of real galaxies, such as found in the SDSS
database, and integrated galaxy SEDs that both appear physical and have
colours similar to observed galaxy samples.

Yet there are noticeable differences between the modelled images and
spectra and the data from the observed galaxies that need to be
considered. A lot of the issues with the images arise due to the
resolution of the simulation themselves. The high mass of the stellar
particles means that the radiation field is very concentrated,
especially when considering the youngest populations, leading to the
very ``spotty'' UV and $24 \um$ images in
Figure~\ref{plot_multiwavelength}.

The offsets in the colour-colour diagrams of
Figure~\ref{plot_fluxratios} are more perplexing, requiring a
wavelength-by-wavelength approach. The offset in the UV colours,
visible both in the top left of Figure~\ref{plot_fluxratios} and in the
IRX-$\beta$ diagram in Figure~\ref{plot_irxbeta} can have several
possible causes, the most likely being variation in the type of dust,
as shown in Figure~\ref{plot_fluxratios-dustmodel}. It is clear that
the UV is very sensitive to the type of dust, which results from the
strong dependence of both the UV extinction slope and the $2200
\Angstrom$ feature (which falls directly in the GALEX NUV band) on the
dust size distribution, as discussed in the previous section. Large
values of the MAPPINGS PDR fraction can also, as shown in
Figure~\ref{plot_fluxratios-pdr-fraction}, significantly affect the UV
colours, but such large values are probably unrealistic in isolated
galaxies.

The UV/optical colours are also sensitive to recent star formation
history (SFH), and hence the offset could also be due to differences
between the simulated galaxy SFH and the real galaxy SFHs. Though the
simulations are assumed to have nearly constant or slowly declining
star-formation rates, it is possible the SINGS galaxies have a
sufficiently different SFH that this could explain the offsets in the
optical bands (i.e.~BVR diagram, top right Figure~\ref{plot_fluxratios}
or V-Ks colour).  These offsets result not from the attenuation
calculations of \mcrx \ (which move the galaxies in parallel to the
SINGS sample), but rather directly from the Starburst99 particles. Thus
the offset must be either model SFH or stellar population model
related.

The mid-IR IRAC bands are again a different story. The IRAC1 $3.6 \um$
band is mostly dominated by stellar light, but when compared to the Ks
band, we can see the sensitivity to the other contribution to the band,
the $3.3 \um$ PAH feature (Figure~\ref{plot_dl07-template-sed}), with
this colour suggesting we are in the right ``ballpark'' at least for
this (non-\emph{Spitzer}-IRS observed) PAH feature. The offset in the
IRAC-IRAC colour diagrams is interesting. The IRAC3/IRAC1 ($5.8 \um /
3.6 \um$) and IRAC4/IRAC1 ($8.0 \um / 3.6 \um$) colours are sensitive
to the fraction of PAHs (or at least their emission), as seen in
Figures~\ref{plot_fluxratios-dustmodel},
\ref{plot_fluxratios-PAHfraction} \&
\ref{plot_fluxratios-pah-template-fraction}. These figures suggest that
the dispersion in the SINGS galaxies in these colours may be related to
a variation in PAH abundance.  However, the offset between the models
and the SINGS sample in the IRAC diagrams is almost perpendicular to
the impact of these changes. This suggests that the PAH template itself
is incorrect, leading to an offset in the IRAC3/IRAC4 ($5.8 \um / 8.0
\um$) colour. This is interesting, as this offset exists for both the
\citet{draineli07} and \citet{grovesetal08sparam} PAH templates, both
of which have been matched to \emph{Spitzer}-IRS spectra of the PAH
features \citep{draineetal07, grovesetal08sparam}. Hence, it is
uncertain if either the templates or the observations are incorrect. It
is also possible emission from an AGN in some of the SINGS galaxies 
affects these fluxes, but the lack of clear differences between e.g.
Seyferts and star-forming galaxies argue against this possibility.

The MIPS colours seem reasonable in Figure~\ref{plot_fluxratios}. While
the $8.0 \um / 24 \um$ and $24 \um / 70 \um$ model colours are somewhat
offset, they still fall into reasonable ranges based on the
observations. Any offset that does exist is most likely due to the lack
of stochastic dust in \mcrx \ (as can be seen in
Figure~\ref{plot_dl07-template-sed}). This also explains the offset
spatially-resolved colours in Figure~\ref{plot_bendo}. Similarly, the
$70 \um / 160 \um$ colours of the models match those of the SINGS
sample.

At yet longer wavelengths, we have the greatest offset from the SINGS
galaxies with the $160 / 850 \um$ discrepancy, whose origin is
puzzling. The SINGS and SLUGS samples have very different far-IR
characteristics, with the SINGS galaxies being seemingly deficient in
cold dust. It is possible this is because of different sample selection
criteria, in particular that the low detection rate at $850 \um$ for
the SLUGS galaxies is biasing their sample to galaxies with large
amounts of cold dust \citep{willmeretal09}. However, it is important to
remember that the SINGS sample is also not a statistically unbiased
sample.

That the intensity of the radiation heating the dust grains is at the
origin of the different $160 / 850 \um$ values, is (at least partially)
supported by our simulations.  \citet{draineetal07} found that the dust
emission SEDs of the SINGS galaxies could be fit by having most dust
heated by an intensity slightly larger than the local ISRF (typically
2--$5 \> U$, their Figure 7) plus a mass of dust at higher intensities
scaling as $dM / dU \sim U^{ - 2 }$. In no case did their SED fits
include dust heated by intensities lower than $1 \> U$. In contrast,
the same fits performed on the SLUGS galaxies yield lower minimum
intensities and a significantly lower fraction of dust heated by higher
intensities.

In our Sbc galaxy, the diffuse ISM dust that sees intensities higher
than $\sim 5 \> U$ also declines roughly as $U^{ - 2 }$, but $60 \%$ of
the dust mass is heated by intensities $< 1 \> U$, very different from
the SINGS results.  If all dust heated by intensities $< 5 \> U$ is
instead assumed to emit as if the intensity was $5 \> U$, the $160 /
850 \um$ flux ratio is increased.  Removing the emission from dust in
low intensities lowers the $850 \um$ emission by $50 \%$, but the $160
/ 850 \um$ flux ratio is still a factor of 2 lower than the SINGS
galaxies.  Thus, the presence of dust at $< 1 \> U$ can only be one of
the factors contributing to the discrepancy with the SINGS galaxies.

Another factor is that the simulations presented here are missing dust
at much higher intensities. In the fits by \citet{draineetal07}, the
$U^{ - 2 }$ distribution continues up to values of $10^5$ -- $10^7 \>
U$, while the maximum intensity in the diffuse ISM is $10^2 \> U$. The
\mapiii \ models are supposed to model the regions of higher intensity
but, with the time-averaged models currently used, the PDRs have a
radiation intensity of $\sim 100 \> U$, the \hii\ regions themselves
intensities $\sim 10^3 \> U$. These intensities are not as high as
those found by \citet{draineetal07}. The time-averaging means that the
highest-intensity early times are diluted. Another factor is that the
\mapiii \ models used here currently do not include ultra-compact \hii\
regions, that represent the dust around the individual stars (rather
than the cluster as a whole). At early times, each individual star is
expected to be surrounded by a very compact \hii\ region where dust can
absorb a significant fraction of the ionizing photons
\citep{churchwell02, dopitaetal03}. As seen in
\citet{grovesetal08sparam}, these contribute significantly to the hot
dust emission (i.e. high $U$).  Including these ultra-compact \hii\
regions is a topic that will be addressed in future versions of \mcrx .

It is an open question why the low-end heating intensities in the
simulations should differ so much from the SINGS galaxies. As discussed
above, a lack of \emph{higher} heating intensities in the diffuse ISM
of the simulations is to be expected due to their limited resolution
and should be taken partly into account by the \mapiii \ models.
However, explaining a lack of dust at \emph{low} intensities is more
difficult. The only obvious missing source of cold dust in the
simulations is dense, self-shielded clumps without internal sources,
which are currently not included in the model. This means that the cold
dust in the simulations must be in regions with little starlight,
presumably at large radii in the disk. What drives the differences in
far-IR SEDs between the SINGS and SLUGS samples is not clear, but
investigating how the distribution of heating intensities depends on
different properties of the simulated galaxies is a promising avenue
for progress.

Finally, the full SEDs in Figure~\ref{plot_sings-bestfit} show not only
the offsets on an SED basis, giving a nice physical picture for the
simulation SEDs, but also the issues in matching the simulation SEDs to
the observations. For example, it is possible to match the IR or UV
separately using the models, but matching the full SED is difficult,
with many of the match optical-UV galaxies actually having different
peaks of their IR (different average dust temperature).

The thing to remember throughout all these comparisons, especially the
SED matching, is that these are simulations \emph{based} on
observations of real galaxies, not modelling individual real galaxies.
Hence, to find small mismatches between the simulations and
observations is quite reasonable. However, being based on real galaxies
(such as the Sbc galaxies representing late-type spirals), we should
expect these to fall within the same regions of parameter space as the
SINGS galaxies. Yet disentangling the issues arising from the
hydrodynamic simulations, the modelling of the source light
(Starburst99 and \mapiii ), the \mcrx \ radiative transfer, and even
issues with selection effects of various observed samples are complex,
making such comparisons useful, but ultimately difficult to interpret.

\section{Summary \& Future Expansions}
\label{section_summary}

We have presented in this work an updated version of the radiative
transfer code \mcrx \ that is able, given a hydrodynamic model of a
galaxy, to model the 2-dimensional UV-IR spectrum of a galaxy, that can
be used to create images in any bands or emission lines, or integrated
galaxy SEDs.  We have detailed how this code creates such models, using
the outputs from the stellar population synthesis code Starburst99, the
radiative transfer code \mapiii , and including a polychromatic ray
tracing algorithm and dust heating algorithm. Through various tests we
have validated the output of this code, demonstrating its convergence,
consistency and conservation of energy.

Using \mcrx \ and hydrodynamic galaxy simulations of
\citet{coxetal05methods}, we have created 2D SEDs for 7 different
galaxy models at different evolutionary times and inclinations. We have
used these galaxies as both demonstrations of the \mcrx \ capabilities
and as test cases, comparing the outputs with the multiwavelength SINGS
\citep{daleetal06sings} and SLUGS \citep{willmeretal09} datasets.

This comparison showed an overall good match creating similar colours
to the observed SINGS sample, with only slight offsets. The exception
is the $160 \um / 850 \um$ colour, which the simulations underestimate
by a factor $\sim 5$ compared to the SINGS sample, but which agrees
better with the SLUGS sample. These large differences between different
observed samples makes it difficult to evaluate whether this
discrepancy is a failure of our model or an expression of some
selection effect in the observed samples.

We have also shown the sensitivity of these results to the ``free
parameters'' of the \mcrx \ model, demonstrating the relative
insensitivity to uncertain parameters, such as the photodissociation
covering factor, and the deterministic sensitivity to other parameters
that are expected to vary, such as the PAH fraction.

In all, we have shown in this paper both the viability and veracity of
the code, which can be used in theory with any hydrodynamic galaxy
simulation to produce the spectra of galaxies.

However, there are clearly still several issues or areas that remain
outstanding with the code that we still plan to address.  Some of the
direct problems we have already discussed in the main text. One of the
current issues in the model is the lack of treatment for the cold phase
ISM. As discussed in Section \ref{section_interface}, these dense
clumps are currently ignored within the current version of \mcrx \ due
to their typically low filling-factor. Yet to extend the code to treat
all galaxy situations, a treatment is needed, with our current idea
being to treat these dense clumps using a ``megagrains'' formalism
\citep{hobsonpadman93}.

Another issue is the current time-averaging of the \mapiii \ particles.
While the resolution of the stellar particles used here mean that such
time-averaging is a reasonable approximation, as the treatment of the
stellar particles gets better, this time averaging approximation breaks
down, and we will need to include the full time-resolution available
from the \citet{grovesetal08sparam} models. This will also lead to a
finer sampling of the emission lines, allowing time-dependence of
emission line strengths to be checked.

Another clear lack is the small range of galaxy simulations dealt with
in this work. While representative of ``normal'' galaxies, these
simulations do not represent the extreme situations which test the
\mcrx \ code. \mcrx \ is also being used for simulations of gas-rich
galaxy mergers at high redshift \citep[e.g.][]{narayananetal08,
youngeretal08}.  In these, the complex geometries, high star-formation
rates and high dust columns all provide more stringent tests of the
models. The larger optical depths will also require a more stringent
evaluation with respect to the double counting issue and overlapping
\mapiii \ particles, and due to the much larger filling factor of dense
clumps in such simulations, the treatment of the multiphase ISM clumps
must be included.  Detailed comparisons of such simulations with
observed luminous infrared galaxies is planned in a future paper.

In addition, these mergers also lead into one of the features currently
lacking in \mcrx , active galactic nuclei (AGN). These strong nuclear
sources can clearly contribute to, if not dominate, a galaxies SED.
Many hydrodynamic simulations \citep[e.g.][]{springeletal04,
dimatteoetal05} include these, needing the AGN feedback to help blow
out the gas and quench the star formation to make the colours of the
merger remnant consistent with observed early-type galaxies.  However,
due to the difficulties in disentangling the galaxy emission from the
`AGN' emission, the interface problem between the AGN ``source
particle'' and the galaxy ISM will be even more severe than for the
\mapiii \ models used here and, unlike for stars, a simple UV-IR
spectral model for AGN does not exist.

One issue not mentioned throughout the text is that of the spectral
resolution of the models. Currently, the spectrum is divided into $\sim
1000$ bins, with higher resolution in the UV-optical, with fine
emission line resolution, and becoming much coarser in the IR. This
resolution is basically defined by the radiative transfer within the
\mapiii \ models. Yet much higher spectral resolution is possible,
thanks to the polychromatic rays of \mcrx , and runs with more than
13,000 wavelengths covering the UV/optical wavelengths, enough to
clearly resolve stellar absorption lines, have been tried successfully.
With this high resolution, including kinematic effects in the radiation
transfer calculation is a natural addition that would make it possible
to compare simulations to increasingly common IFU observations of
kinematics in severely dust-obscured merging galaxies. The
polychromatic algorithm used by \mcrx \ lends itself naturally to the
inclusion of kinematics, and such an improvement is planned.

Finally, the treatment of the dust grains themselves, always an
unknown, is an ongoing work. Our current idea is to allow a more
physical basis for the dust -- linking closely the details of the local
environment, such as metallicity, radiation field (including hard
spectrum AGN), clumpy versus diffuse ISM, etc., with the details of the
dust, such as the PAH fraction \citep[as done
within][]{grovesetal08sparam}, or dust size distribution (i.e. Milky
Way versus LMC).  In the future this could be expanded to track dust
types separately (ie silicaceous versus carbonaceous dust) following
the pollution events of the ISM (ie SN versus AGB star enrichment).

\mcrx \ is currently a state-of-the-art radiative transfer code,
suitable for the production of realistic UV-IR SEDs and images from
hydrodynamic simulations, and with these improvements its applicability
will only increase.

%NLX% exclude from vocabulary builder
\section*{Acknowledgements}
%NLX% end exclude from vocabulary builder

The authors are grateful for the hospitality of the Max-Planck Institut
f\"{u}r Astrophysik in Garching, where part of this paper was written.
We also thank Tom Abel, Fabio Governato, Chris Hayward, Desika
Narayanan, Jennifer Lotz, Yuexing Li, Greg Novak, and Joel Primack for
enduring long discussions about various aspects of our model, and Chris
Willmer for providing an electronic sample of the SLUGS galaxies.

PJ was supported by Spitzer Theory Grant 30183 from the Jet Propulsion
Laboratory and by programs HST-AR-10678/10958/11758, provided by NASA
through grants from the Space Telescope Science Institute, which is
operated by the Association of Universities for Research in Astronomy,
Incorporated, under NASA contract NAS5-26555.

This research used computational resources of the NASA Advanced
Supercomputing Division (NAS) and the National Energy Research
Scientific Supercomputing Center (NERSC), which is supported by the
Office of Science of the US Department of Energy.

%NLX% exclude from vocabulary builder
%\bibliographystyle{another-apj}
\bibliographystyle{mn2e} %\bibliographystyle{../../../bib/astronat/apj/apj}
%\bibliography{patriks,../../../thesis/tj}
\bibliography{patriks} \label{lastpage}

\end {document}